\documentclass[12pt]{report}
\usepackage{latexsym}
\usepackage[cp1251]{inputenc}
\usepackage[english]{babel}
\usepackage[unicode]{hyperref}

\def\pu#1#2{
\mbox{$\!\!\begin{picture}(0,0)
\put (-#1,-#2){\line(1,0){#1}}
\put (-#1,-#2){\line(0,1){#2}}
\put (#1,#2){\line(-1,0){#1}}
\put (#1,#2){\line(0,-1){#2}}
\put (-#1,#2){\line(1,0){#1}}
\put (-#1,#2){\line(0,-1){#2}}
\put (#1,-#2){\line(-1,0){#1}}
\put (#1,-#2){\line(0,1){#2}}
\end{picture}$}
}

\def\ak{
  \begin{picture}(18,20)
  \put (-5,0){\vector(1,1){30}}
  \put (-5,0){\vector(1,-1){30}}
  \end{picture} }

\def\be#1{\begin{equation}\label{#1}}
\def\ee{\end{equation}}
\def\re#1{(\ref{#1})}

\def\bn{\begin{enumerate}}
\def\en{\end{enumerate}}
\def\bi{\begin{itemize}}
\def\ei{\end{itemize}}
\def\i{\item}

\def\l#1{\langle #1 \rangle}
\def\okr#1{U(#1)}

\def\c#1{\left\{\begin{array}{lllll}#1\end{array}\right\}}
\def\d#1{\left[\begin{array}{lllll}#1\end{array}\right]}
\def\b#1{\left(\begin{array}{lllll}#1\end{array}\right)}
\def\ra#1{\;\mathop{\to}\limits^{#1}\;}

\def\by{\begin{array}{llllllllllllll}}
\def\ey{\end{array}}
\def\been{\begin{itemize}}
\def\enen{\end{itemize}}
\def\beenum{\begin{enumerate}}
\def\enenum{\end{enumerate}}
\def\eam{\mathbin{{\mathop{=}\limits^{\mbox{\scriptsize def}}}}}

\def\beenum{\begin{enumerate}}
\def\enenum{\end{enumerate}}
\def\ba{\left\{\begin{array}{llllllllll}}
\def\ea{\end{array} \right\}}
\def\bau{\left[\begin{array}{llllllllll}}
\def\eau{\end{array} \right]}
\def\beq{\begin{equation}}

{\bgroup%
\setcounter{equation}{#1}%
\begin{eqnarray}}{\end{eqnarray}\egroup}

\def\blackbox{\vrule height 7pt width 7pt depth 0pt}

\def\been{\begin{enumerate}}

\def\beit{\begin{itemize}}

\def\blackbox{\vrule height 7pt width 7pt depth 0pt}

\def\diagrw#1{{
  \def\normalbaselines{\baselineskip20pt \lineskip3pt \lineskiplimit3pt }
  \matrix{#1}}}

\def\eam{\buildrel \rm def \over =}

\def\enen{\end{enumerate}}

\def\enit{\end{itemize}}

\def\oc{\mathop{\approx}\limits^{+}}

\def\pleftdouble#1#2
{
\begin{picture}(20,8)
  \put (25,5){\vector(-1,0){30}}
  \put (11,9){\makebox(1,1){$\scriptstyle {#1}$}}
  \put (25,1){\vector(-1,0){30}}
  \put (11,-3){\makebox(1,1){$\scriptstyle {#2}$}}
  \end{picture}
}

\def\pmiddleright#1{
  \begin{picture}(35,8)
  \put (-5,3){\vector(1,0){45}}
  \put (17,8){\makebox(1,1){$\scriptstyle #1$}}
  \end{picture} }

\def\pright#1{
  \begin{picture}(20,8)
  \put (-5,3){\vector(1,0){30}}
  \put (9,8){\makebox(1,1){$\scriptstyle #1$}}
  \end{picture} }

\def\prightdouble#1#2
{
\begin{picture}(20,8)
  \put (-5,5){\vector(1,0){30}}
  \put (11,10){\makebox(1,1){$\scriptstyle {#1}$}}
  \put (-5,1){\vector(1,0){30}}
  \put (11,-5){\makebox(1,1){$\scriptstyle {#2}$}}
  \end{picture}
}

\def\pssearr#1
{
\begin{picture}(20,20)
  \put (-8,20){\vector(1,-2){37}}
  \put (10,-10){\makebox(1,1)[l]{$\scriptstyle {#1}$}}
\end{picture}
}

\def\ra#1{\mathop{\to}\limits^{#1}}

\def\pright#1{
  \begin{picture}(20,8)
  \put (-5,3){\vector(1,0){30}}
  \put (9,7){\makebox(1,1){$\scriptstyle #1$}}
  \end{picture} }

\def\eam{\buildrel \rm def \over =}

\def\eam{\buildrel \rm def \over =}

\def\pleftdouble#1#2
{
\begin{picture}(20,8)
  \put (25,5){\vector(-1,0){30}}
  \put (11,9){\makebox(1,1){$\scriptstyle {#1}$}}
  \put (25,1){\vector(-1,0){30}}
  \put (11,-3){\makebox(1,1){$\scriptstyle {#2}$}}
  \end{picture}
}

\def\pmiddleright#1{
  \begin{picture}(35,8)
  \put (-5,3){\vector(1,0){45}}
  \put (17,8){\makebox(1,1){$\scriptstyle #1$}}
  \end{picture} }

\def\pright#1{
  \begin{picture}(20,8)
  \put (-5,3){\vector(1,0){30}}
  \put (9,8){\makebox(1,1){$\scriptstyle #1$}}
  \end{picture} }

\def\prightdouble#1#2
{
\begin{picture}(20,8)
  \put (-5,5){\vector(1,0){30}}
  \put (11,10){\makebox(1,1){$\scriptstyle {#1}$}}
  \put (-5,1){\vector(1,0){30}}
  \put (11,-5){\makebox(1,1){$\scriptstyle {#2}$}}
  \end{picture}
}

\def\pssearr#1
{
\begin{picture}(20,20)
  \put (-8,20){\vector(1,-2){37}}
  \put (10,-10){\makebox(1,1)[l]{$\scriptstyle {#1}$}}
\end{picture}
}

\def\ra#1{\mathop{\to}\limits^{#1}}

\def\bc{\begin{center}\begin{tabular}{l}}
\def\ec{\end{tabular}\end{center}}

\def\modnop#1{\mathop{#1}\limits_{n}}

\def\redeq{\;\mathop{\approx}\limits^{r}\;}

\def\bi{\begin{itemize}}
\def\pa{\,|\,}
\def\oc{\;\mathop{\approx}\limits^{+}\;}

\def\mor#1#2#3{\by #1&\pright{#2}&#3\ey}
\def\ei{\end{itemize}}
\def\bn{\begin{enumerate}}
\def\en{\end{enumerate}}
\def\i{\item}

\def\bigset#1#2{\left\{\by #1 \left| \by #2 \ey\right\}\ey\right.}

\def\redeq{\;\mathop{\approx}\limits^{r}\;}

\def\oc{\mathop{\approx}\limits^{+}}

\def\bi{\begin{itemize}}
\def\pa{\,|\,}

\def\mor#1#2#3{\by #1&\pright{#2}&#3\ey}
\def\ei{\end{itemize}}
\def\bn{\begin{enumerate}}
\def\en{\end{enumerate}}
\def\i{\item}
\def\bigset#1#2{\left\{\by #1 \left| \by #2 \ey\right\}\ey\right.}

\newcounter{theorem}

\title{Theory of Processes}
\author{A.M.Mironov}
\date{ }
\begin{document}

\maketitle
\tableofcontents
\newpage

\section*{Foreword}

This book is based on author's lectures 
on the theory of processes
for students of Faculty of Mathematics and Mechanics 
and Faculty of Computational Mathematics and Cybernetics 
of Moscow State University.

The book gives a detailed exposition of basic concepts 
and results of a theory of processes.
The presentation of theoretical concepts and 
results is accompanied with illustrations of their application 
to solving various problems of verification of processes. 
Some of these examples are taken from the books \cite{b72} 
and \cite{b75}. 

Along with well-known results there are presented 
author's results related to verification of processes
with message passing, and there are given examples 
of an application of these results.




\chapter{Introduction}

\section{A subject of theory of processes}

{\bf Theory of processes} is a branch 
of mathematical theory of systems,
which studies mathematical models 
of behavior of dynamic systems, called {\bf processes}.

Informally, a {\bf process} is a model of a behavior, 
which performs {\bf actions}. Such actions may be, for example
\bi
\i reception or transmission of any objects, or
\i transformation of these objects.
\ei

The main advantages of theory of processes 
as a mathematical apparatus designed to
modeling 
and analysis of dynamic systems, are as follows.

\bn
\i An apparatus of 
   theory of processes is well suited 
   for formal description 
   and analysis of behavior of 
   {\bf distributed dynamic systems},
   i.e. such systems, which consist 
   of several interacting components, with the following properties:
   \bi
   \i all these components work in parallel, and
   \i interaction of these components occurs by 
      sending signals or messages from one component 
      to other component.
   \ei
   The most important example of a distributed dynamic
   systems is a computer system. 
   In this system 
   \bn
   \i one class of components is determined 
      by a set of computer programs, 
      that are executed in this system,
   \i other class of components is associated with 
      a hardware platform, on the base of which 
      the computer programs are executed, 
   \i the third class of components represents a set 
       information resources (databases, knowledge bases, 
      electronic libraries, etc.) which
      are used for the operation of this system
   \i also it can be taken into account a class of components 
      associated with the human factor.
   \en
\i Methods of theory of processes allow to analyse 
   with acceptable complexity models with 
   very large and even infinite sets of states.
   This is possible due to methodology of symbolic
   transformation of expressions 
   which are symbolic representation of processes.

   The most important examples of models 
   with an infinite set of states are models 
   of computer programs with variables, 
   domains of which have very large size.
   In many cases, models of such programs can be 
   analyzed more easily,
   if domains of some variables in these models
   are represented as infinite sets.
   For example, a domain of variables 
   of the type \verb"double" is a finite 
   set of real numbers, 
   but this set is very large, 
   and in many cases it is puprosely 
   to replace this finite domain by
   an infinite domain of all real numbers
   In some cases a representation of an analyzed program 
   as a model with an infinite set of states
   greatly simplifies a reasoning about this program.
   An analysis of a model of this program with a finite
   (but very large) set of states with use of methods
   based on explicit or symbolic representation of a set 
   of states can have very high computational complexity, 
   and in some cases a replacement 
   \bi
   \i the problem of an analysing of original finite model 
   \i on the problem of an analysing of the corresponding 
      infinite model by methods which are based 
      on symbolic transformations of expressions 
      describing this model
   \ei
   can provide a substantial gain 
   in computational complexity.
\i Methods of theory of processes
   are well suited for investigation of 
   {\bf hierarchical} systems, i.e. such systems
   that have a multilevel structure.
   Each component of such systems is considered as a subsystem, 
   which, in turn, 
   may consist of several subcomponents.
   Each of these subcomponents can interact 
   \bi
   \i with other subcomponents, and 
   \i with higher-level systems.
   \ei
\en

The main sources of problems and objects of results
of the theory of processes are distributed computer systems.

Also the theory of processes can be used 
for modeling  and analysis of behavior of systems of 
different nature, 
most important examples of which are 
{\bf organizational systems}. 
These systems include
\bi
\i enterprise performance management systems,
\i state organizations,
\i system of organization of commertial processes
   (for example,
   management system of commercial transactions, auctions, etc.)
\ei

The processes relating to behavior of 
such systems
are called {\bf business-processes}.

\section{Verification of processes}

The most important class of problems, 
whose solution intended theory of processes, 
is related to the problem of verification of processes. 

The problem of {\bf verification} of a process 
consists of a constructing a formal proof that 
analyzed process has the required properties. 

For many processes this problem is extremely important. 
For instance, the safe operation of such systems as 
\bi 
\i control systems of nuclear power stations, 
\i medical devices with computer control 
\i board control systems of aircrafts and spacecrafts
\i control system of secret databases 
\i systems of e-business 
\ei 
is impossible without a satisfactory solution 
of the problem of verification of correctness and security 
properties of such systems.
A violation of these properties 
in such systems may lead to significant 
damage to the economy and the human security. 

The exact formulation of the problem of verification 
consist of the following parts. 

\bn
\i Construction of a process $P$, which is a
   mathematical model of behavior of analyzed system. 
\i Representation inspected properties 
   in the form of 
   a mathematical object $S$ 
   (called a {\bf specification}). 
\i Construction of a {\bf mathematical proof}
   of a statement that the process $P$
   satisfies the specification $S$. 
\en 

\section{Specification of processes}

A {\bf specification} of a process represents 
a description of properties of this process 
in the form of some mathematical object. 

An example of a specification is 
the requirement of reliability of data transmission 
through the unreliable medium. 
It does not specify how exactly should be provided
this ensured reliability. 

For example, the following objects 
can be used as a specification. 
\bn 
\i A logical formula which expresses
   a requirement for an analysed process. 

   For example, such a requirement may be a condition 
   that if the process has received some request, then
   the process will give response to this request
   after a specified time. 
   
\i Representation of an analyzed process
   on a higher level of abstraction. 

   This type of specifications can be used 
   in multi-level designing of processes: 
   for every level of designing of a process
   an implementation of the process at this level 
   can be considered as a specification 
   for implementation of this process at the next level 
   of designing. 
   
\i A reference process, 
   on which it is assumed that 
   this process has a given property. 

   In this case, the problem of verification consists of
   a construction of a proof of equivalence of 
   a reference process and an analysed processes. 
\en 

In a construction of specifications 
it should be guided the following principles. 
\bn 
\i A property of a process can be expressed 
   in different specification languages (SL), and 
\bi 
\i on one SL it can be expressed in a simple form, and
\i on another SL it can be expressed in a complex form.
\ei 
For example, a specification that describes a
relationship between input and output values 
for a program that computes the decomposition of an integer 
into prime factors, has 
\bi 
\i a complex form in the language of predicate logic, but 
\i a simple form, if this specification is express 
   in the form of a standard program. 
\ei 
Therefore, for representation of properties of processes
in the form of specifications it is important 
to choose a most appropriate  SL, 
which allows to write this specification 
in a most clear and simple form.

\i If a property of a process initially was expressed 
in a natural language, then in translation of this prorerty
to a corresponding formal specification 
it is important to ensure consistency between 
\bi 
\i a natural-language description of this property, and 
\i its formal specification, 
\ei 
because in case of non-compliance of this condition 
results of verification will not have a sense. 
\en 

\chapter{The concept of a process} \label{ponjatieprocessa}

\section{Representation of behavior of dynamic systems 
in the form of processes}

One of possible methods of mathematical modeling 
of a behavior of dynamic systems 
is to present a behavior of these systems 
in the form of {\bf processes}. 

A process usually does not take into account 
all details of a behavior of an analyzed system. 

A behavior can be represented by different processes reflecting 
\bi 
\i different degrees of abstraction in the model of 
   this behavior, and 
\i different levels of detailization of 
   actions executable by a system. 
\ei 

If a purpose of constructing of a process 
for representation of behavior of a system 
is to check properties of this behavior, 
then a choice of level of detailization of 
the system's actions must be dependent on  
the analyzed properties. 
The construction of a process for representation
of a behavior of an analyzed system should take 
into account the following principles. 
\bn 
\i A description of the process should not 
   be excessively detailed, because
   as excessive complexity of this description 
   can cause significant computational problems 
   in formal analysis of this process. 

\i A description of the process should not 
   be overly simplistic, it should 
   \bi 
   \i to reflect those aspects of a 
      behavior of the simulated system, 
      that are relevant to analyzed properties, and 
   \i preserve all those properties of the behavior 
      of this system, 
      that are interesting for analysis 
   \ei 
   because if this condition does not hold, then 
   an analysis of such a process will not make sense. 
\en 

\section{Informal concept of a
process and examples of processes}

Before formulating a precise definition of a process, 
we give an informal explanation of a concept of a process, 
and consider simplest examples of processes. 

\subsection{Informal concept of a process}

As it was stated above, we understand a 
process as a model of a behavior of a dynamic system,
on some level of abstraction. 

A {\bf process} can be thought as a graph $P$, 
whose components have the following sense. 

\bi 
\i Nodes of the graph $P$ are called {\bf states}
   and represent situations (or classes of situations), 
   in which a simulated system
   can be at different times of its functioning. 

   One of the states is selected, it is called 
   an {\bf initial state} of the process $P$. 

\i Edges of the graph $P$ have labels.
   These labels represent
   {\bf  actions}, which may be executed by the
   simulated system. 

\i An {\bf execution} of the process $P$ 
   is described by a walking along 
   the edges of the graph $P$ 
   from one state to another. 
   The execution starts from the initial state. 

   A label of each edge represents an action 
   of the process, 
   executed during the transition from the 
   state at the beginning of the edge 
   to the state at its end. 
\ei 

\subsection{An example of a process} \label{sdfjsdjgklfjd}

As a first example of a process, 
consider a process representing the simplest model 
of behavior of a vending machine. 

We shall assume that this machine has 
\bi 
\i a coin acceptor, 
\i a button, and 
\i a tray for output of goods. 
\ei 

When a customer wants to buy a good, he 
\bi 
\i drops a coin into the coin acceptor, 
\i presses the button
\ei and then the good appears in the tray. 

Assume that our machine 
sells chocolates for 1 coin per each. 

We describe actions of this machine. 
\bi 
\i On the initiative of the customer, 
   in the machine may occur the following actions: 
   \bi 
   \i an input of the coin in the coin acceptor, and 
   \i a pressing of the button. 
   \ei 
\i In response, the machine can perform 
   reaction: an output of a chocolate on the tray. 
\ei 

Let us denote the actions by short names:
\bi
\i an input of a coin we denote by 
   $\mbox{{\it in\_coin}}$,
\i a pressing of the button by 
   $\mbox{{\it pr\_but}}$, and
\i an output of a chocolate by $\mbox{{\it out\_choc}}$.
\ei

Then the process of our vending machine 
has the following form:
$$\by
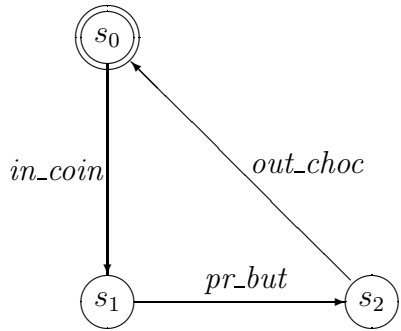
\begin{picture}(100,120)

\put(0,100){\oval(20,20)}
\put(0,100){\oval(24,24)}
\put(0,100){\makebox(0,0){$s_0$}}

\put(0,0){\oval(20,20)}
\put(0,0){\makebox(0,0){$s_1$}}

\put(100,0){\oval(20,20)}
\put(100,0){\makebox(0,0){$s_2$}}

\put(0,90){\vector(0,-1){80}}
\put(10,0){\vector(1,0){80}}
\put(92,8){\vector(-1,1){83}}

\put(-2,50){\makebox(0,0)[r]{
$\mbox{\it in\_coin}$}}

\put(50,3){\makebox(0,0)[b]{
$\mbox{\it pr\_but}$}}

\put(50,53){\makebox(0,0)[l]{
$\mbox{\it out\_choc}$}}

\end{picture}\\
\vspace{5mm}\ey
$$

This diagram explains how the vending machine does work: 
\bi 
\i at first, the machine is in the state $s_0$, 
   in this state the machine expects an input of a coin 
   in the coin acceptor \\
   (the fact that the state $s_0$ is initial, 
   shown in the diagram by a double circle
   around the identitifier of this state)
\i when a coin appears, the machine
   goes to the state $s_1$ 
   and waits for pressing the button 
\i after pressing the button the machine 
   \bi 
   \i goes to the state $s_2$, 
   \i outputs a chocolate, and 
   \i returns to the state $s_0$. 
   \ei 
\ei 

\subsection{Another example of a process}
\label{sdfjsdjgklfjd1}

Consider a more complex example of a vending machine, 
which differs from the previous one that sells two 
types of goods: tea and coffee, 
and the cost of tea is 1 ruble, 
and the cost of coffee is 2 rubles. 

The machine has two buttons: 
one for tea, and another for coffee. 

Buyers can pay with coins in 
denominations of 1 ruble and 2 ruble.
These coins will be denoted by the symbols
$\mbox{\it coin\_1}$ and $\mbox{\it coin\_2}$, respectively.

If a customer dropped in a coin acceptor a coin $\mbox{\it coin\_1}$,
then he can only buy a tea. 
If he dropped a coin $\mbox{\it coin\_2}$, then 
he can buy a coffee or two of tea. 
Also it is possible to buy a coffee, dropping in a coin acceptor 
a couple of coins $\mbox{\it coin\_1}$.

A process of such vending machine has the following form:
$$\by
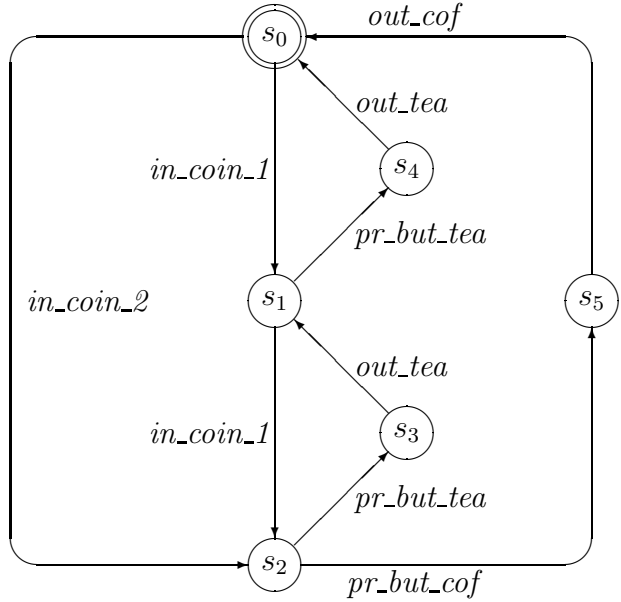
\begin{picture}(0,230)

\put(0,200){\oval(20,20)}
\put(0,200){\oval(24,24)}
\put(0,200){\makebox(0,0){$s_0$}}

\put(0,100){\oval(20,20)}
\put(0,100){\makebox(0,0){$s_1$}}

\put(0,0){\oval(20,20)}
\put(0,0){\makebox(0,0){$s_2$}}

\put(50,50){\oval(20,20)}
\put(50,50){\makebox(0,0){$s_3$}}

\put(50,150){\oval(20,20)}
\put(50,150){\makebox(0,0){$s_4$}}

\put(120,100){\oval(20,20)}
\put(120,100){\makebox(0,0){$s_5$}}

\put(0,190){\vector(0,-1){80}}
\put(0,90){\vector(0,-1){80}}


\put(-2,150){\makebox(0,0)[r]{
$\mbox{\it in\_coin\_1}$}}

\put(-98,100){\makebox(0,0)[l]{
$\mbox{\it in\_coin\_2}$}}

\put(-2,50){\makebox(0,0)[r]{
$\mbox{\it in\_coin\_1}$}}

\put(7.5,7.5){\vector(1,1){35.5}}

\put(26,25){\makebox(0,0)[l]{
$\mbox{\it pr\_but\_tea}$}}

\put(26,75){\makebox(0,0)[l]{
$\mbox{\it out\_tea}$}}

\put(26,125){\makebox(0,0)[l]{
$\mbox{\it pr\_but\_tea}$}}

\put(26,175){\makebox(0,0)[l]{
$\mbox{\it out\_tea}$}}

\put(50,-2){\makebox(0,0)[t]{
$\mbox{\it pr\_but\_cof}$}}

\put(50,202){\makebox(0,0)[b]{
$\mbox{\it out\_cof}$}}

\put(43,57.5){\vector(-1,1){35.5}}

\put(7.5,107.5){\vector(1,1){35.5}}

\put(43,157.5){\vector(-1,1){34}}

\put(120,10){\vector(0,1){80}}
\put(120,110){\line(0,1){80}}
\put(110,10){\oval(20,20)[br]}
\put(110,190){\oval(20,20)[tr]}
\put(110,200){\vector(-1,0){98}}
\put(110,0){\line(-1,0){100}}

\put(-100,10){\line(0,1){180}}
\put(-90,10){\oval(20,20)[bl]}
\put(-90,190){\oval(20,20)[tl]}
\put(-90,200){\line(1,0){78}}
\put(-90,0){\vector(1,0){80}}

\end{picture}\\
\vspace{1mm}
\ey
$$

For a formal definition of a process 
we must clarify a concept of an action. 
This clarification is presented 
in section \ref{sdfgsdfgw5etrwerw}.

\section{Actions} \label{sdfgsdfgw5etrwerw}

To define a process $P$, 
which is a behavior model 
of a dynamic system, 
it must be specified 
a set $Act(P)$ of {\bf actions}, 
which can be performed by the process $P$. 

We shall assume that actions 
of all processes 
are elements of a certain universal set 
$Act$ of all possible actions,
that can be performed
by any process, i.e., for every process $P$
$$Act(P) \subseteq Act$$ 

A choice of the set $Act(P)$ 
of actions of the process $P$ 
depends on a purpose of a modeling. 
In different situations, for a 
representation of a model of an 
analyzed system 
in the form of a process 
it may be choosen 
different sets of actions. 

We shall assume that the set $Act$ 
of actions is subdivided
on the following 3 classes. 
\bn 
\i {\bf Input actions}, 
   which are denoted 
   by symbols of the form
   $$\alpha?$$ 
   The action $\alpha?$ 
   is interpreted as an input 
   of an object with the name $\alpha$. 
\i {\bf Output actions}, 
   which are denoted
   by symbols of the form
   $$\alpha!$$ 
   The action $\alpha!$ 
   is interpreted as an output
   of an object with the name $\alpha$. 
   
\i {\bf Internal} (or {\bf invisible}) 
   actions, which are denoted by the symbol 
   $\tau$. 

   An action of the process $P$
   is said to be {\bf internal}, if this 
   action does not related with an interaction
   of this process with its {\bf environment}, 
   i.e. with processes which are external
   with respect to the process $P$, and 
   with which it can interact.

   For example, an internal action 
   can be due to the interaction 
   of components of $P$. 

   In fact, internal actions 
   may be different, but 
   we denote all of them by 
   the same symbol $\tau$.   
   This reflects our desire 
   not to distinguish between
   all internal actions, 
   because they are not observable
   outside the process $P$. 
\en 

Let $\mbox{\it Names}$ be a set of all 
names of all objects, which can be used 
in input or output actions.
The set $Names$ is assumed to be infinite. 

The set $Act$ of all actions, 
which can be executed by processes, 
is a disjoint union of the form
\be{actlsdfg4e} \by Act &=&&
\{\alpha?\mid \alpha\in Names\}&\cup&\\
&&\cup&
\{\alpha!\mid \alpha\in Names\}&\cup&\\&&\cup&
\{\tau\}\ey\ee

Objects, which can be used in input 
or output actions, may have 
different nature 
(both material and not material).
For example, they may be 
\bi 
\i material resources, 
\i people 
\i money 
\i information 
\i energy
\i etc. 
\ei 

In addition, the concept of an input 
and an output can have a virtual character, 
i.e. the words {\it input} and {\it output}
may only be used as a metaphor, 
but in reality no input or output of any 
real object may not occur. 
Informally, we consider a non-internal 
action of a process $P$ as
\bi 
\i an {\bf input action}, 
   if this action was caused by 
   a process from an environment of $P$, and 
\i an {\bf output action}, 
   if it was caused by $P$.
\ei 

For each name $\alpha\in Names$
the actions $\alpha?$ and $\alpha!$ 
are said to be {\bf complementary}. 

We shall use the following notation. 
\bn 
\i For every action $a\in Act\setminus
   \{\tau\}$ 
   the symbol $\bar a$ denotes an action,
   which is complementary to $a$, i.e.
   $$\overline{\alpha?}\eam \alpha!,\quad
   \overline{\alpha!}\eam \alpha?$$ 
\i For every action 
   $a\in Act\setminus \{\tau\}$
   the string 
   $name(a)$ denotes the
   name specified in the action $a$, i.e.
   $$name(\alpha?)\eam name(\alpha!)\eam 
   \alpha$$
\i For each subset 
   $L\subseteq Act\setminus \{\tau\}$
   \bi
   \i $\overline{L}\eam \{\overline{a}\mid 
   a\in L\}$
   \i $names(L)\eam \{name(a)\mid a\in L\}$
   \ei
\en

\section{Definition of a process} \label{asdfasdf3rt4wertweywey}

A {\bf process} is a triple $P$ of the form
\be{amfhgjklsdfh}
P = (S, s^0, R)\ee
whose components have the following meanings. 
\bi 
\i $S$ is a set whose elements 
   are called {\bf states} of the process $P$.
\i $s^0\in S$ is a selected state, called 
   an {\bf initial state} of the process $P$. 
\i $R$ is a subset of the form 
   $$R\subseteq S\times Act\times S$$ 
   Elements of $R$ are called {\bf   
   transitions}. 

   If a transition from $R$ has 
   the form $(s_1,a,s_2)$, then
   \bi 
   \i we say that this transition 
      is a transition from the state 
      $s_1$ to the state $s_2$ 
      with an execution 
      of the action $a$, 
   \i states $s_1$ and 
      $s_2$ are called a {\bf start} 
      and an {\bf end} of this transition, 
      respectively, and the action $a$ 
      is called a {\bf label}
      of this transition, and 
   \i sometimes, in order to improve 
      a visibility, we will 
      denote this transition by the diagram
      \be{wert5dfgssetg}
      \diagrw{s_1&\pright{a}&s_2}\ee 
   \ei 
\ei 

An {\bf execution} of a process 
$P=(S, s^0, R)$
is a generation of a sequence 
of transitions of the form 
$$\diagrw{s^0&\pright{a_0}&s_1&\pright{a_1}&s_2&\pright{a_2}&\ldots}$$
with an execution of 
actions $a_0,a_1,a_2\ldots$, 
which are labels of these transitions.

At every step $i\geq 0$ of this execution
\bi 
\i the process $P$ is in some state 
   $s_i$ ($s_0=s^0$), 
\i if there is at least one transition 
   from $R$ starting at $s_i$,    
   then the process $P$
   \bi 
   \i non-deterministically chooses 
      a transition from $R$
      starting at 
      $s_i$, labeled such 
      action $a_i$, 
      which can be executed
      at the current time \\ 
      (if there is no such transitions, 
      then
      the process suspends 
      until at least one
      such transition will occur) 
   \i performs the action $a_i$, and then 
   \i goes to the state $s_{i+1}$, 
      which is the end of the selected
      transition 
   \ei 
\i if $R$ does not contain 
   transitions starting at $s_i$, 
   then the process 
   completes its work. 
\ei 

The symbol $Act(P)$ denotes the set 
of all actions in
$Act\setminus \{\tau\}$, which 
can be executed by the process $P$, 
i.e.
$$Act(P)\eam \{a\in Act\setminus \{\tau\}\mid
\exists \;(\diagrw{s_1&\pright{a}&s_2})\in R\}$$ 

Process \re{amfhgjklsdfh} 
is said to be {\bf finite}, 
if its components $S$ and $R$ 
are finite sets.

A finite process can be represented 
graphically as a diagram, in which 
\bi 
\i each state is represented by 
   a circle in the diagram, 
   and an identifier 
   of this state can be written
   in this circle
\i each transition is represented by 
   an arrow connecting 
   beginning of this transition 
   and its end,       
   and a label of this transition 
   is written 
   on this arrow
\i an initial state is indicated in some way 
   \\
   (for example, instead of the usual circle, 
   a double circle is drawn)
\ei 
Examples of such diagrams contain in sections
\ref{sdfjsdjgklfjd} and \ref{sdfjsdjgklfjd1}.

\section{A concept of a trace}

Let $P=(S,s^0,R)$ be a process. 

A {\bf  trace} of the process $P$ 
is a finite or infinite sequence 
$$a_1,a_2,\ldots \qquad$$
of elements of $Act$, 
such that there is a sequence of states 
of the process $P$ $$s_0,s_1,s_2,\ldots$$ 
with the following properties: 
\bi 
\i $s_0$ coincides with the initial state 
   $s^0 $ of $P$ 
\i for each $i\geq 1$ the set $R$ 
   contains the transition 
   $$\diagrw{s_i&\pright{a_i}&s_{i+1}}$$ 
\ei 

A set of all traces of the process $P$ 
we shall denote by $Tr(P)$. 

\section{Reachable and unreachable states}

Let $P$ be a process of the form 
\re{amfhgjklsdfh}.

A state $s$ of the process $P$ is said to be
{\bf reachable}, if $s = s^0$, 
or there is a sequence of transitions 
of $P$, having the form 
$$
\diagrw{s_0&\pright{a_1}&s_1},\quad
\diagrw{s_1&\pright{a_2}&s_2},\quad\ldots\quad
\diagrw{s_{n-1}&\pright{a_n}&s_n}
$$
in which $n\geq 1, \;s_0=s^0$ and $s_n=s$. 

A state is said to be {\bf unreachable}, 
if it is not reachable. 

It is easy to see that after removing of all 
\bi
\i unreachable states from $S$, and 
\i transitions from $R$ 
   which does contain these unreachable states 
\ei  
the resulting process $P'$ 
(which is referred as
a {\bf  reachable part} of the process $P$) 
will represent 
exactly the same behavior, which is 
represented by the process $P$. 
For this reason, we consider 
such processes $P$ and $P'$ as equal. 

\section{Replacement of states}
\label{zamena}

Let 
\bi 
\i $P$ be a process of the form 
   \re{amfhgjklsdfh}, 
\i $s$ be a state from $S$ 
\i $s'$ be an arbitrary element, 
   which does not belong to the set $S$. 
\ei 
Denote by $P'$ a process, 
which is obtained from $P$ 
by replacement $s$ on $s'$ 
in the sets and $S$ $R$, 
i.e. every transition in $R$ of the form 
$$\diagrw{s&\pright{a}&s_1}\quad \mbox{or}\quad
\diagrw{s_1&\pright{a}&s}$$
is replaced by a transition 
$$\diagrw{s'&\pright{a}&s_1}\quad \mbox{or}
\quad
\diagrw{s_1&\pright{a}&s'}$$
respectively. 

As in the previous section, it is easy 
to see that $P$ and $P'$
represent the same behavior, 
and for this reason, we can consider 
such processes $P$ and $P'$ as equal. 

It is possible to replace not only 
one state, but arbitrary subset of states 
of the process $P$.
Such a replacement can be represented as
an assignment of a bijection of the form
\be{fere5}f:S\to S'\ee
and the result of such replacement is by definition a process $P'$ of the form
\be{iso2}P' = (S', (s')^0, R')\ee
where 
\bi 
\i $(s')^0 \eam f(s^0)$, and 
\i for each pair $s_1, s_2\in S$ 
   and each $a\in Act$ 
   $$
   (\diagrw{s_1&\pright{a}&s_2})\;\in R\quad
   \Leftrightarrow\quad
   (\diagrw{f(s_1)&\pright{a}&f(s_2)})\;\in R'.
   $$ 
\ei 
Since the processes $P$ and $P'$ 
represent the same behavior, 
we can treat them as equal. 

In the literature on the theory 
of processes such processes $P$ and $P'$ 
sometimes are said to be {\bf isomorphic}. 
Bijection \re{fere5} with the above 
properties is called an {\bf 
isomorphism} between $P$ and $P'$.
The process $P'$ is said to be
an {\bf isomorphic copy} of the process $P$.

\chapter{Operations on processes} \label{operatsii}

In this chapter we define several algebraic operations 
on the set of processes.

\section{Prefix action}

The first such operation is called a {\bf prefix action}, 
this is an unary operation denoted by ``$a.$'', 
where $a$ is an arbitrary element of $Act$. 

Let $P=(S, s^0, R)$ be a process and $a\in Act$.

An effect of the operation $a.$ 
on the process $P$ results to the process, 
which has the following components:
\bi
\i a set of states of $a.P$ is obtained from $S$
   by an adding a new state $s\not\in S$
\i an initial state of $a.P$ is the added state $s$
\i a set of transitions of $a.P$ is obtained from $R$
   by adding a new transition of the form
   $$\diagrw{s&\pright{a}&s^0}$$
\ei
The resulting process is denoted by $$a.P$$

We illustrate an effect of this operation 
on the  example of a vending machine 
presented at section \ref{sdfjsdjgklfjd}. 
Denote the process, which represents a 
behavior of this automaton, 
by $P_{\mbox{\scriptsize\rm vm}}$. 

Extend the set of actions of the vending machine 
by a new input action $$\mbox{{\it enable}}\,?$$
which means an enabling of this machine. 

The process $\mbox{\it enable}\,?.\;P_{\mbox{\scriptsize\rm vm}}$
represents a behavior of the new vending machine, 
which in the initial state can not 
\bi 
\i accept coins, 
\i perceive pressing the button, and
\i output chocolates. 
\ei 

The only thing that he can is to be enabled. 
After that, its behavior will be no 
different from that of the original machine. 

A graph representation of
$\mbox{\it enable}\,?.\;P_{\mbox{\scriptsize\rm vm}}$
has the following form: 
$$\by
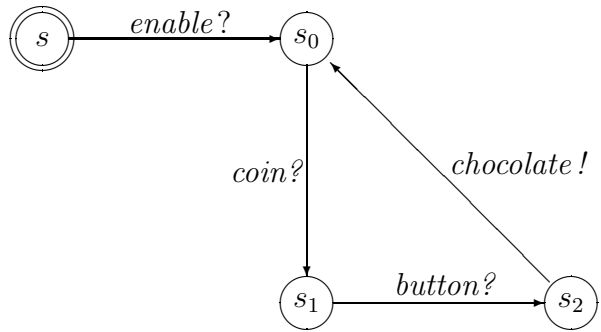
\begin{picture}(0,120)

\put(-100,100){\oval(20,20)}
\put(-100,100){\oval(24,24)}
\put(-100,100){\makebox(0,0){$s$}}

\put(0,100){\oval(20,20)}
\put(0,100){\makebox(0,0){$s_0$}}

\put(0,0){\oval(20,20)}
\put(0,0){\makebox(0,0){$s_1$}}

\put(100,0){\oval(20,20)}
\put(100,0){\makebox(0,0){$s_2$}}

\put(0,90){\vector(0,-1){80}}
\put(-90,100){\vector(1,0){80}}
\put(10,0){\vector(1,0){80}}
\put(92,8){\vector(-1,1){83}}

\put(-50,102){\makebox(0,0)[b]{
$\mbox{\it enable}\,?$}}

\put(-2,50){\makebox(0,0)[r]{
$\mbox{\it coin?}$}}

\put(50,3){\makebox(0,0)[b]{
$\mbox{\it button?}$}}

\put(50,53){\makebox(0,0)[l]{
$\mbox{\it chocolate\,!}$}}

\end{picture}\\
\vspace{5mm}\ey
$$

\section{Empty process}

Among all the processes, there is one the most simple. 
This process has only one state, and has no transitions. 
To indicate such a process 
we use a constant (i.e. a 0-ary operation) {\bf 0}. 

Returning to examples with vending machines,
it can be said that the process {\bf 0} represents 
a behavior of a broken vending machine, 
that is such a machine, which can not execute any action. 

By applying the operations of prefix action
to the process ${\bf 0}$ it is possible
to define a behavior of more complex machines. 
Consider, for example, the following process: 
$$P=\mbox{{\it coin}}\,?.\mbox{{\it button}}\,?.
\mbox{{\it chocolate}}\,!.\,\mbox{\bf 0}$$

A graph representation of this process is as follows: 
$$\by
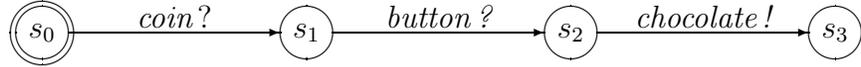
\begin{picture}(300,0)

\put(0,0){\oval(20,20)}
\put(0,0){\oval(24,24)}
\put(0,0){\makebox(0,0){$s_0$}}
\put(100,0){\makebox(0,0){$s_1$}}
\put(200,0){\makebox(0,0){$s_2$}}
\put(300,0){\makebox(0,0){$s_3$}}

\put(100,0){\oval(20,20)}
\put(200,0){\oval(20,20)}
\put(300,0){\oval(20,20)}

\put(10,0){\vector(1,0){80}}
\put(110,0){\vector(1,0){80}}
\put(210,0){\vector(1,0){80}}

\put(50,2){\makebox(0,0)[b]{$\mbox{{\it coin}}\,?$}}
\put(150,2){\makebox(0,0)[b]{$\mbox{\it button\,?}$}}
\put(250,2){\makebox(0,0)[b]{$\mbox{\it chocolate\,!}$}}

\end{picture}\\
\vspace{5mm}\ey
$$

This process defines a behavior of a vending machine, 
which serves exactly one customer, and after this breaks. 

\section{Alternative composition}\label{sgjkdsflgertyyyy}

Next operation on processes is a binary operation, 
which is called an {\bf alternative composition}. 

This operation is used in the case 
when, having a pair of processes $P_1$ and $P_2$,
we must construct a process $P$, 
which will operate 
\bi 
\i either as the process $P_1$, 
\i or     as the process $P_2$, 
\ei 
and the choice of a process, 
according to which $P$ will operate,
can be determined 
\bi 
\i either by $P$ itself, 
\i or by an environment in which $P$ does operate. 
\ei 

For example, if $P_1$ and $P_2$ have the form 
\be{fgbhdsfgsdf}\by P_1=\alpha\,?\;. \;P'_1\\
P_2=\beta\,?\;. \;P'_2\ey\ee
and at the initial time an environment of $P$
\bi 
\i can give $P$ the object $\alpha$, but 
\i can not give $P$ the object $\beta$
\ei 
then $P$ will choose a behavior which
is only possible in this situation, 
i.e. will operate according to the process $P_1$.

Note that in this case it is chosen such a process, 
first action in which can be executed in the current time. 
After choosing of $P_1$, and execution of the action $\alpha\,?$, 
the process $P$ is obliged to continue its work 
according to this choice, i.e. it will operate like $P'_1$. 
It is possible, that after execution of the action $\alpha?$
\bi 
\i $P$ will not be able to execute any action, 
   working in accordance with $P'_1$
\i though at this time
   $P$ will be able to execute an action, 
   working in accordance with $P'_2$.
\ei 

But at this time $P$ can not change his choice 
(i.e. can not choose $P'_2$ instead of $P'_1$). 
$P$ can only wait until it will be possible to work 
in accordance with $P'_1$. 

If in the initial time 
the environment can give $P$ both $\alpha$ and $\beta$, 
then $P$ chooses a process whereby it will work, 
\bi 
\i non-deterministically (i.e., arbitrarily), or 
\i subject to some additional factors. 
\ei 

The exact definition of the operation of alternative 
composition is as follows. 

Let $P_1$ and $P_2$ be processes of the form 
$$P_i=(S_i, s^0_i, R_i)\quad (i=1,2)$$
and the sets of states $S_1$ and $S_2$
have no common elements. 

An {\bf alternative composition} of processes
$P_1$ and $P_2$ is a process 
$$P_1+P_2=(S, s^0, R)$$ 
whose components are defined as follows. 
\bi 
\i $S$ is obtained by adding to the union $S_1\cup S_2$
   a new state $s^0$, which is an
   initial state of $P_1 + P_2$
\i $R$ contains all transitions from 
   $R_1$ and $R_2$, and
\i for each transition in $R_i$ ($i = 1,2$) 
   $$\diagrw{s_i^0&\pright{a}&s}$$
   $R$ contains the transition
   $$\diagrw{s^0&\pright{a}&s}$$
\ei 

If $S_1$ and $S_2$ have common elements, 
then to define a process $P_1 + P_2$
you first need to replace in $S_2$
those states that are also in $S_ 1$
on new states,
and also modify accordingly $R_2$ and $s^0_2$. 

Consider, for example, vending machine 
which sells two types of drinks: cola and fanta, and 
\bi 
\i if a customer puts in a coin 
   $\mbox{\it coin\_1}$, then
   the machine issues a glass of cola, and
\i if a customer puts in a coin 
   $\mbox{\it coin\_2}$, then
   a machine gives a glass of fanta
\ei 
with the machine is broken immediately after 
the sale of one glass of a drink. 

A behavior of this automaton is described by 
the following process: 
\be{gsdfgsdfgsdfgsdf4}
\by P_{\mbox{\scriptsize\rm drink}}&=&
\mbox{\it coin\_1}\,?\;.\;\mbox{{\it cola}}\,!\;.
\;{\bf 0}&+\\&+&
\mbox{\it coin\_2}\,?\;.\;\mbox{{\it fanta}}\,!\;.\;{\bf 0}\ey
\ee

Consider a graph representation of process 
\re{gsdfgsdfgsdfgsdf4}. 

Graph representation of terms in the sum 
\re{gsdfgsdfgsdfgsdf4} have the form 
$$\by
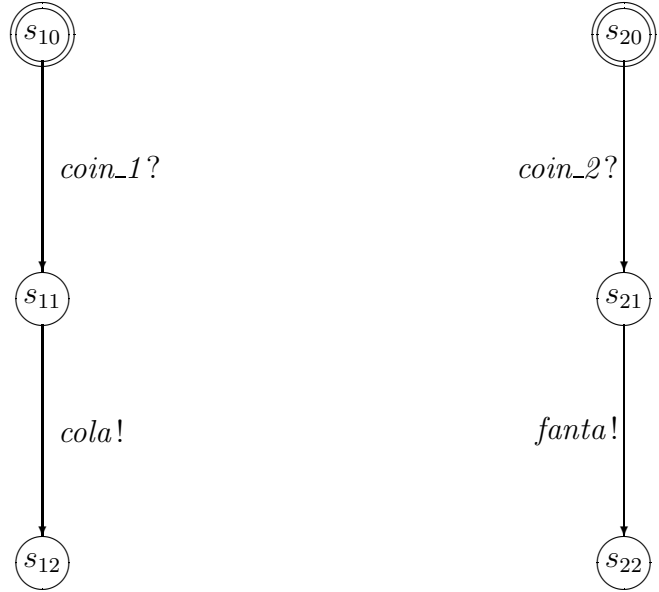
\begin{picture}(0,220)

\put(-110,200){\oval(20,20)}
\put(-110,200){\oval(24,24)}
\put(-110,200){\makebox(0,0){$s_{10}$}}

\put(110,200){\oval(20,20)}
\put(110,200){\oval(24,24)}
\put(110,200){\makebox(0,0){$s_{20}$}}

\put(-110,100){\oval(20,20)}
\put(-110,100){\makebox(0,0){$s_{11}$}}

\put(110,100){\oval(20,20)}
\put(110,100){\makebox(0,0){$s_{21}$}}

\put(-110,0){\oval(20,20)}
\put(-110,0){\makebox(0,0){$s_{12}$}}

\put(110,0){\oval(20,20)}
\put(110,0){\makebox(0,0){$s_{22}$}}

\put(-110,90){\vector(0,-1){80}}
\put(-110,190){\vector(0,-1){80}}

\put(110,90){\vector(0,-1){80}}
\put(110,190){\vector(0,-1){80}}

\put(-108,150){\makebox(0,0)[l]{
$\mbox{\it coin\_1}\,?$}}

\put(-108,50){\makebox(0,0)[l]{
$\mbox{{\it cola}}\,!$}}

\put(108,150){\makebox(0,0)[r]{
$\mbox{\it coin\_2}\,?$}}

\put(108,50){\makebox(0,0)[r]{
$\mbox{{\it fanta}}\,!$}}

\end{picture}\\
\;
\ey
$$

According to a definition of an alternative composition,
a graph representation of process \re{gsdfgsdfgsdfgsdf4}
is obtained by adding to the previous diagram 
a new state and the corresponding transitions, 
to result in the following diagram: 
$$\by
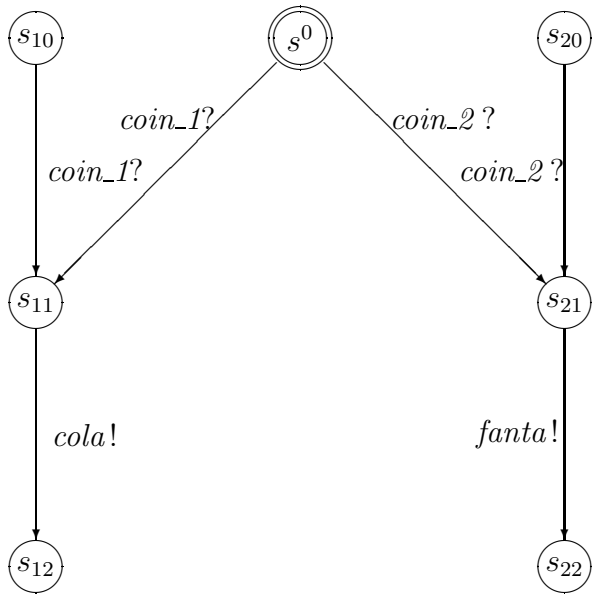
\begin{picture}(0,220)

\put(-100,200){\oval(20,20)}
\put(-100,200){\makebox(0,0){$s_{10}$}}

\put(0,200){\oval(20,20)}
\put(0,200){\oval(24,24)}
\put(0,200){\makebox(0,0){$s^0$}}

\put(100,200){\oval(20,20)}
\put(100,200){\makebox(0,0){$s_{20}$}}

\put(-100,100){\oval(20,20)}
\put(-100,100){\makebox(0,0){$s_{11}$}}

\put(100,100){\oval(20,20)}
\put(100,100){\makebox(0,0){$s_{21}$}}

\put(-100,0){\oval(20,20)}
\put(-100,0){\makebox(0,0){$s_{12}$}}

\put(100,0){\oval(20,20)}
\put(100,0){\makebox(0,0){$s_{22}$}}

\put(-100,90){\vector(0,-1){80}}
\put(-100,190){\vector(0,-1){80}}

\put(100,90){\vector(0,-1){80}}
\put(100,190){\vector(0,-1){80}}


\put(9,  191){\vector(1,-1){84}}
\put(-9, 191){\vector(-1,-1){84}}

\put(-100,150){\makebox(0,0)[l]{
$\mbox{\it coin\_1}?$}}
\put(100,150){\makebox(0,0)[r]{
$\mbox{\it coin\_2}\;?$}}

\put(-32,170){\makebox(0,0)[r]{
$\mbox{\it coin\_1}?$}}
\put(30,170){\makebox(0,0)[l]{
$\mbox{\it coin\_2}\;?$}}

\put(-98,50){\makebox(0,0)[l]{
$\mbox{{\it cola}}\,!$}}

\put(98,50){\makebox(0,0)[r]{
$\mbox{{\it fanta}}\,!$}}

\end{picture}\\
\;
\ey
$$

Since the states $s_{10}$ and $s_{20}$ are unreachable, 
it follows that it is possible to delete them 
and transitions associated with them, 
resulting in a diagram
$$\by
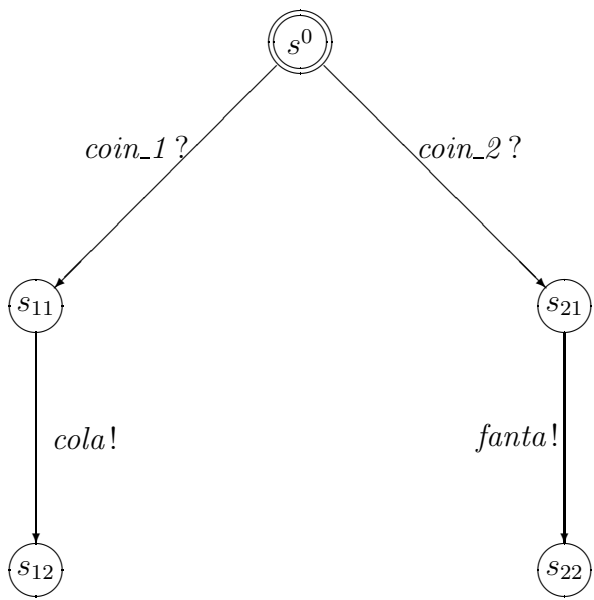
\begin{picture}(0,220)

\put(0,200){\oval(20,20)}
\put(0,200){\oval(24,24)}
\put(0,200){\makebox(0,0){$s^0$}}

\put(-100,100){\oval(20,20)}
\put(-100,100){\makebox(0,0){$s_{11}$}}

\put(100,100){\oval(20,20)}
\put(100,100){\makebox(0,0){$s_{21}$}}

\put(-100,0){\oval(20,20)}
\put(-100,0){\makebox(0,0){$s_{12}$}}

\put(100,0){\oval(20,20)}
\put(100,0){\makebox(0,0){$s_{22}$}}

\put(-100,90){\vector(0,-1){80}}

\put(100,90){\vector(0,-1){80}}


\put(9,  191){\vector(1,-1){84}}
\put(-9, 191){\vector(-1,-1){84}}

\put(-42,160){\makebox(0,0)[r]{
$\mbox{\it coin\_1}\;?$}}
\put(40,160){\makebox(0,0)[l]{
$\mbox{\it coin\_2}\;?$}}

\put(-98,50){\makebox(0,0)[l]{
$\mbox{{\it cola}}\,!$}}

\put(98,50){\makebox(0,0)[r]{
$\mbox{{\it fanta}}\,!$}}

\end{picture}\\
\;
\ey
$$
which is the desired graph representation 
of process \re{gsdfgsdfgsdfgsdf4}. 

Consider another example. 
We describe an exchange machine, 
which can enter 
banknotes in denominations of 100 dollars. 
The machine shall issue 
\bi 
\i either 2 banknotes on 50 dollars, 
\i or 10 banknotes on 10 dollars 
\ei 
and the choice of method of an exchange
is carried regardless of the wishes of the customer. 
Just after one session of an exchange 
the machine is broken. 

$$\by P_{\mbox{\scriptsize\rm exchange}}\;=\\=\;
\mbox{\it 1\_on\_1000}\;?\;.(\mbox{{\it 2\_on\_500}}
\;!\;.{\bf 0}\;+\;
\mbox{{\it 10\_on\_100}}\;!\;.{\bf 0})\ey$$

These two examples show that 
the operation of an alternative composition 
can be used to describe at least 
two fundamentally different situations. 
\bn 
\i First, it can express a dependence 
   of system behavior from the behavior 
   of its environment. 

   For instance, in the case of a vending machine 
   $P_{\mbox{\scriptsize\rm drink}}$
   a behavior of the machine is determined 
   by an action of a purchaser, namely
   by a denomination of a coin, 
   which a purchaser introduced into the machine. 

   In this case, a process representing 
   a behavior of the simulated vending machine
   is {\bf deterministic}, i.e.
   its behavior is uniquely determined by input actions. 

\i In the second, on an example of a machine
   $P_{\mbox{\scriptsize\rm exchange}}$
   we see that for the same input action is possible 
   different response of the machine. 

   This is an example of a {\bf nondeterminism},
   i.e. an uncertainty of a behavior of a system. 

   Uncertainty in a behavior of systems 
   can occur by at least two reasons. 
   \bn 
   \i First, a behavior of systems may depend on 
      {\bf random factors}. 

These factors can be, for example, 
\bi 
\i failures in hardware, 
\i conflicts in a computer network 
\i absence of banknotes of required 
   value at an ATM 
\i or anything else 
\ei 
   \i Second, a model is always some abstraction 
   or simplification of a real system, and
some of the factors 
influencing a behavior of this system may 
be eliminated from a consideration. 

   \en 
In particular, on the example of 
$P_{\mbox{\scriptsize\rm exchange}}$ 
we see that a real reason of choosing 
of a variant of behavior of the machine 
can be not taken into account 
in the process, 
which is a model of a behavior of this machine. 
\en 

One can schematically represent the above variants of
using alternative composition as follows: 
$$\diagrw{&&\by\mbox{Dependence}\\
\mbox{on the input}\\
\mbox{data}
\ey\cr
\by \mbox{Alternative }\\
\mbox{composition}
\ey&\ak&&&\by\mbox{Random}\\\mbox{factors}\ey\cr
&&\by\mbox{Nondeter-}\\\mbox{minism}
\ey&\ak\cr
&&&&\by\mbox{Unknown}
\\\mbox{factors}\ey}$$

\section{Parallel composition}

The operation of parallel composition is used for 
building models of behavior of dynamic systems, 
composed of several communicating components. 

Before giving a formal definition of this operation, 
we will discuss the concept of parallel working 
of a pair of systems $Sys_1$ and $Sys_2$, 
which we consider as components of a system $Sys$, i.e.
\be{gdgsdfgsdfgsd}Sys \eam \{Sys_1, Sys_2\}\ee

Let processes $P_1$ and $P_2$ represent
behaviors of the systems $Sys_1$ and $Sys_2$
respectively. 

When the system $Sys_i \; \; (i = 1,2)$ 
works as a part of the system $Sys$, 
its behavior is described by the same process $P_i$.

Denote by $\{P_1, P_2\}$ a process, 
describing a behavior of \re{gdgsdfgsdfgsd}. 
The purpose of this section is to find
an explicit description of $\{P_1, P_2\}$
(i.e. to define a sets of its states and transitions).

Here to simplify the exposition, 
we identify the concepts 
\begin{center} ``a process $P$'', and \end{center}
\begin{center} ``a system whose behavior 
is described by a process $P$'' 
\end{center}

As noted above, an execution of arbitrary process 
can be interpreted as 
a bypassing of a graph corresponding to this process, 
with an execution of actions that are labels of passable edges. 

We shall assume that in passage of each edge 
$\diagrw{s&\pright{a}&s'}$
\bi 
\i a transition from $s$ to $s'$ occurs instantaneously, and 
\i an execution of the action $a$
   occurs precisely at the time of this transition. 
\ei 

In fact, an execution of each action 
occurs within a certain period of time, 
but we shall assume that for each traversed edge 
$\diagrw{s&\pright{a}&s'}$
\bi 
\i before the completion of an execution of the action $a$
   the process $P$ is in the state $s$, and 
\i after the completion of an execution of $a$
   the process $P$ instantly transforms into the state $s'$. 
\ei 

Since an execution of various actions 
has different durations, then we will assume 
that the process $P$ is in each state 
an indefinite period of time during its execution.

Thus, an execution of the process $P$
consists of an alternation of the following 
two activities: 
\bi 
\i waiting for an indefinite period of time 
   in one of the states, and 
\i instantaneous transition from one state to another. 
\ei 

Waiting in one of the states can occur 
\bi 
\i not only because there is 
   an execution of some action at this time, 
\i but also because the process $P$
   can not 
   perform any action at this time.
\ei 
For example, if 
\bi 
\i $P=\alpha\,?.\,P'$, and 
\i in the initial time there is no a process who can 
   give $P$ an oblect with the name $\alpha$
\ei 
then $P$ would wait until some process will give him 
an oblect with the name $\alpha$. 

As we know, for each process 
\bi 
\i its actions are either input, 
   or output, or internal, and 
\i each input or output action 
   is a result of a communication
   of this process with other 
   process. 
\ei 
Each input or output action 
of the process $P_i$ $(i = 1,2)$ 
\bi 
\i either is a result of communication
   of $P_i$ with a
   process outside of  
   the set $\{P_1,P_2\}$, 
\i or is a result of communication
   of $P_i$ with
   the process $P_j$, where 
   $j\in \{1,2\}\setminus \{i\}$. 
\ei 

From the point of view of the process 
$\{P_1, P_2\}$, 
actions of the second type 
are internal actions 
of this process, because they 
\bi 
\i are not a result 
   of a communication of the 
   process $\{P_1, P_2\}$ 
   with its environment, and 
\i are the result of communication 
   between 
   the components of this process. 
\ei 
Thus, each step of the process 
$\{P_1, P_2\}$ 
\bi 
\i[(a)] either is a
   result of a comminication 
   of one 
   of the processes $P_i$
   $(i = 1,2)$ 
   with a 
   process 
   outside of $\{P_1, P_2\}$, 
\i[(b)] or is an internal action 
   of $P_1$ or $P_2$, 
\i[(c)] or is an 
   internal action, which is a result 
   of a communication of 
   $P_1$ and $P_2$, 
   and this communication 
   has the following form:
   \bi 
   \i one of these processes, say
      $P_i$, passes to 
      another process 
      $P_j\quad(j\in 
      \{1,2\}\setminus \{i\})$
      some object, and 
   \i the process $P_j$
      at the same time 
      takes this object 
      from the process $P_i$ 
   \ei 
   (This kind of a communication 
   is called a
   {\bf synchronous communication}, 
   or a {\bf handshaking}). 
\ei 

Each possible variant  of 
a behavior of the
process $P_i \; \; (i = 1,2)$
can be associated with a {\bf thread}
denoted by the symbol $\sigma_i$.
A {\bf thread} is a vertical line, 
on which there are drawn 
points with labels, where 
\bi 
\i labels of points represent 
   actions executed by the
   process $P_i $, and 
\i labelled points are arranged 
   in a chronological order, 
   i.e. 
   \bi 
   \i first point is labelled by
      a first action of 
      the process $P_i $, 
   \i second point (which is 
      located under the first point)
      is labelled by a 
      second action of the 
      process $P_i$, 
   \i etc. 
   \ei 
\ei 

For each labelled point $p$
on the thread, we denote 
by $act(p)$
a label of this point. 

Assume that there is drawn on a plane 
a couple of parallel threads
\be{gsdfdsfgdsf354}\sigma_1\;\; \sigma_2\ee
where $\sigma_i\;\;(i=1,2)$
represents a possible variant 
of a behavior of the process $P_i$
in the process $\{P_1,P_2\}$. 

Consider those labelled points 
on the threads from
\re{gsdfdsfgdsf354}, 
which correspond to actions of 
the type (c), 
i.e. to communications of
processes $P_1$ and $P_2$.  
Let $p$ be one of such points, 
and, for example, it is
on the thread $\sigma_1$. 

According to the definition 
of a communication, 
at the same time, 
in which there is executed the
action $act(p)$, 
the process $P_2$ executes 
a complementary action, 
i.e. there is a point $p'$
on the thread $\sigma_2$, 
such that 
\bi 
\i $act(p')= \overline{act(p)}$, and 
\i actions $act(p)$ and $act(p')$
   execute at the same time. 
\ei 

Note that 
\bi 
\i in the thread $\sigma_2$ may be 
   several points with the label 
   $\overline{act(p)}$, 
   but exactly one of these points 
   corresponds to the action, 
   which is executed jointly 
   with the action corresponding to the
   point $p$, and 
\i in the thread $\sigma_1$ may be 
   several points with the label 
   $act(p)$, 
   but exactly one of these points 
   corresponds to the action, 
   which is executed jointly 
   with the action corresponding to the
   point $p'$.
\ei 
Transform our diagram of threads 
\re{gsdfdsfgdsf354} as follows: 
for each pair of points 
$p,p'$ with the above properties 
\bi 
\i join the points $p$ and $p'$ by an
   arrow, \bi\i
   the start of which is 
   the one of these points, which 
   has a label of the form 
   $\alpha\,!$, 
   and \i
   the end of which 
   is another of these points\ei
\i draw a label 
   $\alpha$
   on this arrow, and 
\i replace labels of the points 
   $p$ and $p'$ on $\tau$. 
\ei 
The arrow from $p$
to $p'$ is called a
{\bf synchronization arrow}. 
Such arrows usually are
drawn horizontally.

After such changes 
for all pairs of points, 
which are labelled by 
actions of the type (c), 
we will obtain a diagram, which 
is called a {\bf Message 
Sequence Chart (MSC)}. 
This diagram represents 
one of possible variants 
of execution of the process
$\{P_1,P_2\}$. 

We shall denote 
a set of all MSCs, 
each of which corresponds to 
some variant of execution 
of the process $\{P_1,P_2\}$, as
$$Beh\{P_1,P_2\}$$

Consider the following example
of a process of the form $\{P_1,P_2\}$:
\bi 
\i $P_1$ is a model of 
   a vending machine,
   whose behavior is given by 
   \be{ghjsdfkwerltjkherwlgjkr1}
   P_1 = \mbox{\it coin}\,?.\; 
   \mbox{\it chocolate}\,!. 
   \mbox{\bf 0}\ee 
   (i.e., the machine
   gets a coin, 
   gives a chocolate, and then breaks) 
\i $P_2 $ is a model of a customer,
   whose behavior is given by 
   \be{ghjsdfkwerltjkherwlgjkr2}
   P_2 = \mbox{\it coin}\,!.\; 
   \mbox{\it chocolate}\,?. 
   \mbox{\bf 0}\ee
   (i.e., the customer drops a coin, 
   receives a chocolate, 
   and then ceases to function 
   as customer).
\ei 

Threads of these processes have the form 
$$\by
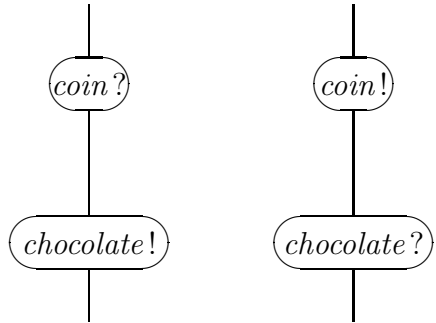
\begin{picture}(0,120)

\put(-50,120){\line(0,-1){20}}
\put(-50,90){\oval(30,20)}
\put(-50,90){\makebox(0,0){$\mbox{\it coin}\,?$}}
\put(-50,80){\line(0,-1){40}}
\put(-50,30){\oval(60,20)}
\put(-50,30){\makebox(0,0){$\mbox{\it chocolate}\,!$}}
\put(-50,20){\line(0,-1){20}}

\put(50,120){\line(0,-1){20}}
\put(50,90){\oval(30,20)}
\put(50,90){\makebox(0,0){$\mbox{\it coin}\,!$}}
\put(50,80){\line(0,-1){40}}
\put(50,30){\oval(60,20)}
\put(50,30){\makebox(0,0){$\mbox{\it chocolate}\,?$}}
\put(50,20){\line(0,-1){20}}

\end{picture}\\
\;
\ey
$$

If all actions on 
these threads are actions 
of the type (c), 
then this diagram 
can be transformed 
into the following MSC: 

$$\by
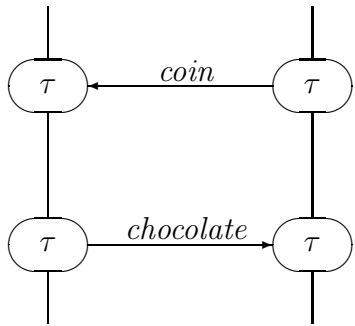
\begin{picture}(0,120)

\put(-50,120){\line(0,-1){20}}
\put(-50,90){\oval(30,20)}
\put(-50,90){\makebox(0,0){$\tau$}}
\put(35,90){\vector(-1,0){70}}
\put(0,92){\makebox(0,0)[b]{
$\mbox{\it coin}$}}
\put(-35,30){\vector(1,0){70}}
\put(0,32){\makebox(0,0)[b]{
$\mbox{\it chocolate}$}}
\put(-50,80){\line(0,-1){40}}
\put(-50,30){\oval(30,20)}
\put(-50,30){\makebox(0,0){$\tau$}}
\put(-50,20){\line(0,-1){20}}

\put(50,120){\line(0,-1){20}}
\put(50,90){\oval(30,20)}
\put(50,90){\makebox(0,0){$\tau$}}
\put(50,80){\line(0,-1){40}}
\put(50,30){\oval(30,20)}
\put(50,30){\makebox(0,0){$\tau$}}
\put(50,20){\line(0,-1){20}}

\end{picture}\\
\;
\ey
$$

However, it is possible 
the following variant of execution 
of the process $\{P_1,P_2\}$: 
\bi 
\i first actions of $P_1$ and $P_2$
   are of the type (c), i.e. 
   the customer drops
   a coin, 
   and the machine accepts the coin
\i second action of automaton $P_1$
   is a communication 
   with a process that is external 
   with respect to $\{P_1,P_2\}$ \\
   (that is, for example, 
   a thief walked up to the machine, 
   and took a chocolate, 
   before than the customer 
   $P_2$ was able to take it)
\ei 
In this situation, 
the customer can not execute
a second action as an internal 
action of $\{P_1,P_2\}$. 
According to a description of the process $P_2$, in this case 
two variants of behavior of the customer 
are possible.  
\bn 
\i The customer will be 
   in a state of endless waiting.

   The corresponding MSC 
   has the form 

$$\by
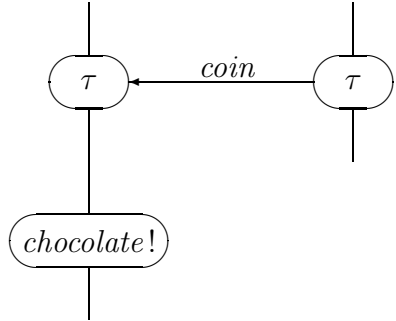
\begin{picture}(0,120)

\put(-50,120){\line(0,-1){20}}
\put(-50,90){\oval(30,20)}
\put(-50,90){\makebox(0,0){$\tau$}}
\put(35,90){\vector(-1,0){70}}
\put(0,92){\makebox(0,0)[b]{
$\mbox{\it coin}$}}
\put(-50,80){\line(0,-1){40}}
\put(-50,30){\oval(60,20)}
\put(-50,30){\makebox(0,0){$\mbox{\it chocolate}\,!$}}
\put(-50,20){\line(0,-1){20}}

\put(50,120){\line(0,-1){20}}
\put(50,90){\oval(30,20)}
\put(50,90){\makebox(0,0){$\tau$}}
\put(50,80){\line(0,-1){20}}

\end{picture}\\
\;
\ey
$$

\i The customer will be able 
   successfully complete its work. 

   This would be the case 
   if some process external to 
   to $\{P_1,P_2\}$ will 
   give a chocolate to the customer. 

   The corresponding MSC has the form 
   $$\by
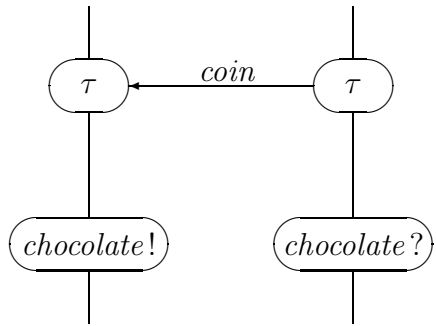
\begin{picture}(0,120)

\put(-50,120){\line(0,-1){20}}
\put(-50,90){\oval(30,20)}
\put(-50,90){\makebox(0,0){$\tau$}}
\put(35,90){\vector(-1,0){70}}
\put(0,92){\makebox(0,0)[b]{
$\mbox{\it coin}$}}
\put(-50,80){\line(0,-1){40}}
\put(-50,30){\oval(60,20)}
\put(-50,30){\makebox(0,0){$\mbox{\it chocolate}\,!$}}
\put(-50,20){\line(0,-1){20}}

\put(50,120){\line(0,-1){20}}
\put(50,90){\oval(30,20)}
\put(50,90){\makebox(0,0){$\tau$}}
\put(50,80){\line(0,-1){40}}
\put(50,30){\oval(60,20)}
\put(50,30){\makebox(0,0){$\mbox{\it chocolate}\,?$}}
\put(50,20){\line(0,-1){20}}

\end{picture}\\
\;
\ey
$$
\en 

Now consider the general question: 
how a process of the form 
$\{P_1,P_2\}$ can be defined explicitly, 
i.e. in terms of states and transitions.

At first glance, 
this question is incorrect, 
because 
$\{P_1,P_2\}$ must be a model of
a {\bf parallel} execution of 
the processes $P_1$ and $P_2$, 
in which 
\bi 
\i it can be 
   possible a simultaneous
   execution of actions
   by both processes 
   $P_1$, $P_2$, 
\i and, therefore, 
   the process $\{P_1,P_2\}$
   can execute such actions, which 
   are pairs of actions from 
   the set $Act$, 
   which 
   can not
   belong to the set $Act$
   (by assumption). 
\ei 

Note on this, that 
absolute simultaneity 
holds only for those pairs of actions
that generate an internal action 
of the process $\{P_1,P_2\}$
of the type (c). 

For all other pairs of actions 
of the processes $P_1$ and $P_2$, 
even if they occurred simultaneously
(in terms of external observer),
we can assume without loss of generality, that one of them
happened a little earlier or 
a little later than another. 

Thus, we can assume that 
the process $\{P_1,P_2\}$ 
executes consequentially,
i.e. under any variant
of an execution of the process
$\{P_1,P_2\}$ 
actions executed by them form 
some linearly ordered sequence
\be{gsdfgsdft356}tr = (act_1,act_2,\ldots)\ee
in which the actions are 
ordered by the time 
of their execution: 
at first it was executed 
$act_1$, then - $act_2$, etc. 

Because each possible 
variant of an execution
of the process $\{P_1,P_2\}$ 
can be represented by a MSC, 
then we can assume that sequence
\re{gsdfgsdft356}
can be obtained by 
some {\it linearization} 
of this MSC 
(i.e., by ``pulling'' it in a chain). 

For a definition of 
a linearization of a MSC 
we introduce some auxiliary 
concepts and notations. 

Let $C$ be a MCS. Then 
\bi 
\i $Points(C)$ denotes 
   a set of all points 
   belonging to the MSC $C$, 
\i for each point $p \in Points(C)\quad
   act(p)$ 
   denotes an action, 
   ascribed to the point $p$
\i for each pair of points 
   $p, p '\in Points(C)$
   the formula
   $$p\to p'$$ 
   means that 
   one of the following conditions
   does hold: 
   \bi 
   \i $p$ and $p'$ are in the same
      thread, and $p'$ is lower 
      than $p$, or 
   \i there is a synchronization 
      arrow from $p$
      to $p'$
   \ei 
\i for each pair of points 
   $p, p'\in Points(C)$
   the formula
   $$p\leq p'$$
   means that either $p=p'$, or
   there is a sequence of points 
   $p_1,\ldots, p_k$, such that 
   \bi 
   \i $p=p_1,\;\; p'=p_k$
   \i for each $i=1,\ldots, k-1 
      \quad p_i\to p_{i+1}$
   \ei 
\ei 

The relation $\leq$ on points of a MSC 
can be regarded as a relation of 
a chronological order, i.e. 
the formula $p \leq p'$
can be interpreted as stating that 
\bi 
\i the points $p$ and $p'$
   are the same or connected 
   by a synchronization arrow \\
   (i.e. actions in $p$ and $p'$
   coincide) 
\i or an action in the $p'$
   occurred later than 
   there was an action in the $p$. 
\ei 

The exact definition of a linearization of a MSC has the following form. 

Let 
\bi
\i $C$ be a MSC, 
\i $tr$ be a sequence of actions 
   of the form \re{gsdfgsdft356}, and
\i $Ind(tr)$ be 
   a set of indices of 
   elements of the sequence $tr$, i.e.
   $$Ind(tr) = \{1,2,\ldots\}$$
   (this set can be finite or infinite)
\ei

The sequence $tr$ is called 
a {\bf linearization} of the MSC $C$, 
if there is a surjective mapping 
$$lin: Points(C)\to Ind(tr)$$
satisfying the following conditions. 
\bn 
\i for each pair $p,p'\in Points(C)$
   $$p\leq p'\quad\Rightarrow\quad
	 lin(p)\leq lin(p')$$
\i for each pair $p,p'\in Points(C)$
   the equality
   $$lin(p)=lin(p')$$
   holds if and only if 
   \bi 
   \i $p = p'$, or 
   \i there is a 
      synchronization arrow 
      from $p$ to $p'$
   \ei 
\i $\forall \,p\in Points(C)\quad
   act(p)=act_{lin(p)}$.
\en 
i.e. the mapping $lin$
\bi 
\i preserves the chronological order 
\i identifies those points of the 
   MSC $C$, 
   which correspond to 
   one action of $\{P_1,P_2\}$, and 
\i does not identify any other points. 
\ei 

Denote by $Lin(C)$ the set of 
all linearizations of the MSC $C$. 

Now the problem 
of explicit description 
of the process $\{P_1,P_2\}$
can be formulated as follows: 
construct a process $P$, satisfying 
the condition
\be{gsdfgdf55rf}Tr(P) = 
\bigcup\limits_{C\in Beh\{P_1,P_2\}}Lin(C)\ee
i.e. in the process $P$
should be represented 
all linearizations 
of any possible joint 
behavior of processes 
$P_1$ and $P_2$. 

Condition \re{gsdfgdf55rf} 
is justified by the following consideration: because we do not know 
\bi 
\i how clocks
   in the processes $P_1$ and $P_2$
   are related,
   and 
\i what is a length of a stay 
   in each state in which 
   these processes fall 
\ei 
then we must take into account
every possible order 
of an execution of 
actions, 
which does not contradict to 
the relation of a chronological order. 

Begin the construction of 
a process $P$, 
satisfying condition 
\re{gsdfgdf55rf}. 

Let the processes $P_1$ and $P_2$
have the form 
$$P_i=(S_i, s^0_i, R_i)\quad
(i=1,2)$$

Consider any linearization $tr$ 
of an arbitrary MSC from 
$Beh\{P_1,P_2\}$
$$tr\;=\;(\;a_1,\;a_2,\;\ldots\;)$$

Draw a line, which will be interpreted as a scale of time. 
Select on this line
points $p_1$, $p_2$, $\ldots$
labelled by the actions 
$a_1, a_2, \ldots$ respectively, 
such that these actions are located 
on the line
in the same order in which they are 
listed in $tr$.

Let the symbols $I_0, I_1, I_2,\ldots$
denote the following sections 
of this line: 
\bi 
\i $I_0$ is the set of all points 
   of the line 
   before the point $p_1$, i.e.
   $$I_0\eam\;\;]-\infty, p_1[$$
\i for each $i\geq 1$  the 
   plot $I_i$
   consists of  points between $p_i$
   and $p_{i+1}$, i.e. 
   $$I_i\eam\;\;]p_i,p_{i+1}[$$
\ei 
Each of these sections $I_i$
can be interpreted 
as an interval of time during which 
the process $P$
does not perform any action,
i.e. at times between 
$p_i$ and $p_{i+1}$
the processes $P_1$ and $P_2$
are in fixed states $(s_1)_i $
and $(s_2)_i$, respectively. 

Denote by $s_i$ the pair 
$((s_1)_i, (s_2)_i)$. 
This pair can be interpreted 
as a state of the process $P$, 
in which he is 
at each time 
from the interval $I_i$. 

By the definition of the sequence $tr$, 
we have one of two situations. 
\bn 
\i The action $a_i$
   has a type (a) or (b), i.e. 
   $a_i$ was executed by 
   one of the processes 
   included in $P$. 

   There are two cases. 

   \bn 
   \i The action $a_i$ was executed 
      by the  process $P_1$. 

   In this case we have the following relation 
   between the states $s_i$ and $s_{i+1}$:
   \bi
   \i $\diagrw{(s_{1})_i&\pright{a_i}&(s_{1})_{i+1}}\;\;\in R_{1}$
   \i $(s_{2})_{i+1} = (s_{2})_{i}$
   \ei
   
   \i The action $a_i$ was executed 
   by the process $P_2$. 

   In this case we have the following relation 
   between the states $s_i$ and $s_{i+1}$:
   \bi
   \i $\diagrw{(s_{2})_i&
      \pright{a_i}&(s_{2})_{i+1}}\;\;\in R_{2}$
   \i $(s_{1})_{i+1} = (s_{1})_{i}$
   \ei
   \en 
\i The action $a_i$ is of the type (c). 

   In this case we have the following relation 
   between the states $s_i$ and $s_{i+1}$:
   \bi
   \i $\diagrw{(s_1)_i&\pright{a}&(s_1)_{i+1}}\;\;\in R_1$
   \i $\diagrw{(s_2)_i&\pright{\overline{a}}&(s_2)_{i+1}}
	   \;\;\in R_2$
   \ei
   for some $a\in Act\setminus \{\tau\}$. 
\en 

The above properties of the sequence 
$tr$
can be reformulated as follows: 
$tr$ is a trace of the process
\be{dsfgsdfgsdfger}(S, s^0, R)\ee
whose components are defined 
as follows: 
\bi 
\i $S\eam S_1\times S_2\eam \{(s_1,s_2)\mid s_1\in S_1, s_2\in S_2\}$
\i $s^0\eam (s_1^0, s_2^0)$
\i for \bi \i each transition $\diagrw{s_1&\pright{a}&s'_1}$
   from $R_1$, and 
   \i each state $s\in S_2$
   \ei 
   $R$ contains the transition 
 $$\diagrw{(s_1,s)&\pright{a}&(s'_1,s)}$$
\i for \bi \i each transition 
   $\diagrw{s_2&\pright{a}&s'_2}$
   from $R_2$, and 
   \i each state $s\in S_1$\ei 
   $R$ contains the transition 
 $$\diagrw{(s,s_2)&\pright{a}&(s,s'_2)}$$
\i for each pair of transitions with complementary labels 
   $$\by
   \diagrw{s_1&\pright{a}&s'_1}\quad\in R_1\\
\diagrw{s_2&\pright{\overline{a}}&s'_2}\quad\in R_2\ey$$
   $R$
   contains the transition 
$$\diagrw{(s_1,s_2)&\pright{\tau}&(s'_1,s'_2)}$$
\ei 

It is easy to show the converse: 
each trace of process
\re{dsfgsdfgsdfger}
is a linearization of some 
MSC $C$ from the set $Beh\{P_1,P_2\}$. 

Thus, an explicit representation 
of the process $P=\{P_1, P_2\}$
can be defined as process
\re{dsfgsdfgsdfger}. 
This process is called 
a {\bf parallel composition}
of the processes $P_1$ and $P_2$, 
an is denoted as
$$P_1 \pa P_2$$

We give an example of the process 
$P_1 \pa P_2$, in the case 
where the processes $P_1$ and $P_2$
represent behaviors of 
a vending machine and a customer 
(see \re{ghjsdfkwerltjkherwlgjkr1} and 
\re{ghjsdfkwerltjkherwlgjkr2}). 

A graph representation 
of these processes have the form 

$$\by
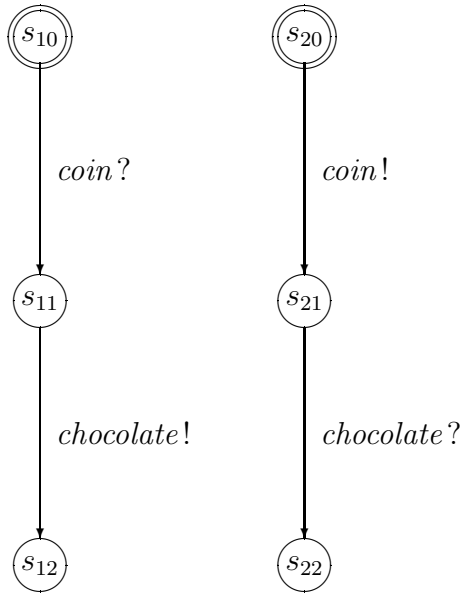
\begin{picture}(0,220)

\put(-50,200){\oval(20,20)}
\put(-50,200){\oval(24,24)}
\put(-50,200){\makebox(0,0){$s_{10}$}}

\put(50,200){\oval(20,20)}
\put(50,200){\oval(24,24)}
\put(50,200){\makebox(0,0){$s_{20}$}}

\put(-50,100){\oval(20,20)}
\put(-50,100){\makebox(0,0){$s_{11}$}}

\put(50,100){\oval(20,20)}
\put(50,100){\makebox(0,0){$s_{21}$}}

\put(-50,0){\oval(20,20)}
\put(-50,0){\makebox(0,0){$s_{12}$}}

\put(50,0){\oval(20,20)}
\put(50,0){\makebox(0,0){$s_{22}$}}

\put(-50,90){\vector(0,-1){80}}
\put(-50,190){\vector(0,-1){80}}

\put(50,90){\vector(0,-1){80}}
\put(50,190){\vector(0,-1){80}}

\put(-48,150){\makebox(0,0)[l]{
$\mbox{\it coin}\,?$}}

\put(-48,50){\makebox(0,0)[l]{
$\mbox{{\it chocolate}}\,!$}}

\put(52,150){\makebox(0,0)[l]{
$\mbox{\it coin}\,!$}}

\put(52,50){\makebox(0,0)[l]{
$\mbox{{\it chocolate}}\,?$}}

\end{picture}\\
\;
\ey
$$

A graph representation of 
the process $P_1 | P_2$
has the form
$$\by
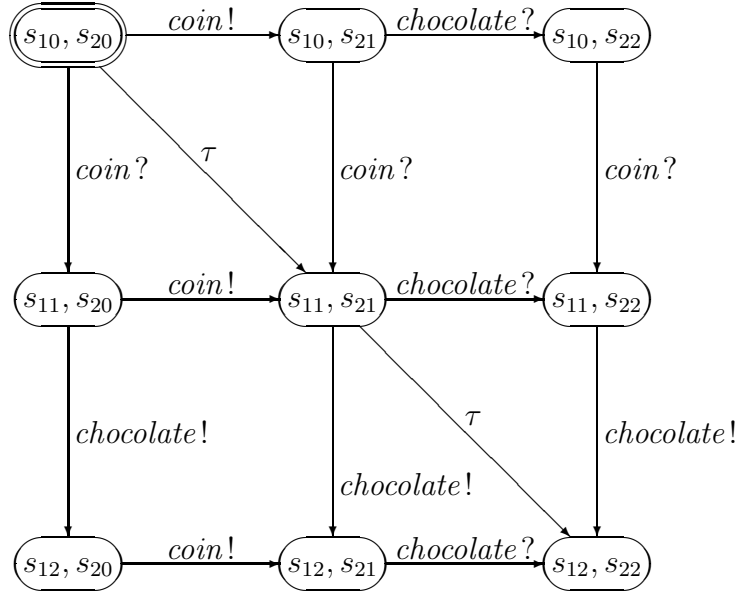
\begin{picture}(200,220)

\put(0,200){\oval(40,20)}
\put(0,200){\oval(44,24)}
\put(0,200){\makebox(0,0){$(s_{10},s_{20})$}}

\put(0,100){\oval(40,20)}
\put(0,100){\makebox(0,0){$(s_{11},s_{20})$}}

\put(0,0){\oval(40,20)}
\put(0,0){\makebox(0,0){$(s_{12},s_{20})$}}

\put(100,200){\oval(40,20)}
\put(100,200){\makebox(0,0){$(s_{10},s_{21})$}}

\put(100,100){\oval(40,20)}
\put(100,100){\makebox(0,0){$(s_{11},s_{21})$}}

\put(100,0){\oval(40,20)}
\put(100,0){\makebox(0,0){$(s_{12},s_{21})$}}

\put(200,200){\oval(40,20)}
\put(200,200){\makebox(0,0){$(s_{10},s_{22})$}}

\put(200,100){\oval(40,20)}
\put(200,100){\makebox(0,0){$(s_{11},s_{22})$}}

\put(200,0){\oval(40,20)}
\put(200,0){\makebox(0,0){$(s_{12},s_{22})$}}

\put(0,188){\vector(0,-1){78}}
\put(2,150){\makebox(0,0)[l]{$\mbox{\it coin}\,?$}}
\put(100,190){\vector(0,-1){80}}
\put(102,150){\makebox(0,0)[l]{$\mbox{\it coin}\,?$}}
\put(200,190){\vector(0,-1){80}}
\put(202,150){\makebox(0,0)[l]{$\mbox{\it coin}\,?$}}

\put(0,90){\vector(0,-1){80}}
\put(2,50){\makebox(0,0)[l]{$\mbox{\it chocolate}\,!$}}
\put(100,90){\vector(0,-1){80}}
\put(102,30){\makebox(0,0)[l]{$\mbox{\it chocolate}\,!$}}
\put(200,90){\vector(0,-1){80}}
\put(202,50){\makebox(0,0)[l]{$\mbox{\it chocolate}\,!$}}

\put(22,200){\vector(1,0){58}}
\put(50,202){\makebox(0,0)[b]{$\mbox{\it coin}\,!$}}
\put(120,200){\vector(1,0){60}}
\put(150,202){\makebox(0,0)[b]{$\mbox{\it chocolate}\,?$}}
\put(20,100){\vector(1,0){60}}
\put(50,102){\makebox(0,0)[b]{$\mbox{\it coin}\,!$}}
\put(120,100){\vector(1,0){60}}
\put(150,102){\makebox(0,0)[b]{$\mbox{\it chocolate}\,?$}}
\put(20,0){\vector(1,0){60}}
\put(50,2){\makebox(0,0)[b]{$\mbox{\it coin}\,!$}}
\put(120,0){\vector(1,0){60}}
\put(150,2){\makebox(0,0)[b]{$\mbox{\it chocolate}\,?$}}

\put(12,188){\vector(1,-1){78}}
\put(50,155){\makebox(0,0)[l]{$\tau$}}
\put(110,90){\vector(1,-1){80}}
\put(150,55){\makebox(0,0)[l]{$\tau$}}

\end{picture}\\
\;
\ey
$$

Note that a size of the set 
of states of 
$P_1 | P_2$
is equal to 
a product of sizes 
of sets of states 
of $P_1$ and $P_2$.
Thus, 
a size of a description 
of the process $P_1 \pa P_2$
may substantially exceed the total complexity 
of sizes of descriptions of its components, $P_1$ and $P_2$. 
This may make impossible 
to analyze this process, if it is represented in an explicit form, 
because of its high complexity.

Therefore, in practical problems 
of an analysis of processes 
of the form $P_1 \pa P_2$, 
instead of an explicit 
construction of $P_1\pa P_2$
there is constructed a process,
in which each MSC from 
$Beh\{P_1,P_2\}$
\bi\i is not represented by all possible linearizations, but
\i is represented 
by at least one linearization. \ei
A complexity of such process 
can be significantly less in 
comparison with a complexity 
of the process $P_1|P_2$. 

A construction of a process 
of this kind makes sense, 
for example, if an analyzed property 
$\varphi$ of the process $P_1\pa P_2$
has the following quality: 
for arbitrary $C\in Beh\{P_1,P_2\}$
\bi 
\i if $\varphi$ holds
   {\it for one of linearizations}
   of $C$,
\i then $\varphi$ holds
   {\it for all 
   linearizations} of $C$. 
\ei 

Typically, a process 
in which each MSC from 
$Beh\{P_1,P_2\}$ 
is represented by 
at least one linearization, 
is constructed as a certain 
{\bf subprocess} of the process
$P_1|P_2$, 
i.e. is obtained from $P_1|P_2$
by removing of some states 
and associated transitions. 
Therefore, such processes 
are said to be {\bf reduced}. 

The problem of constructing of 
reduced processes 
is called a 
{\bf partial order reduction}.
This problem has been intensively studied by many leading experts 
in the field of verification. 

Consider, for example, 
a reduced process 
for the above process 
$P_1 \pa P_2$, 
consisting of 
a vending machine 
and the customer. 

$$\by
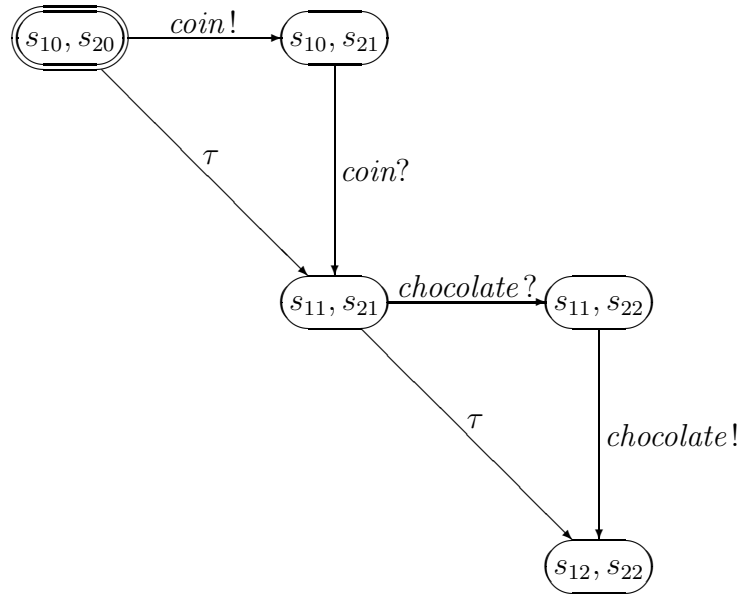
\begin{picture}(200,220)

\put(0,200){\oval(40,20)}
\put(0,200){\oval(44,24)}
\put(0,200){\makebox(0,0){$(s_{10},s_{20})$}}

\put(100,200){\oval(40,20)}
\put(100,200){\makebox(0,0){$(s_{10},s_{21})$}}

\put(100,100){\oval(40,20)}
\put(100,100){\makebox(0,0){$(s_{11},s_{21})$}}

\put(200,100){\oval(40,20)}
\put(200,100){\makebox(0,0){$(s_{11},s_{22})$}}

\put(200,0){\oval(40,20)}
\put(200,0){\makebox(0,0){$(s_{12},s_{22})$}}

\put(100,190){\vector(0,-1){80}}
\put(102,150){\makebox(0,0)[l]{$\mbox{\it coin}?$}}
\put(200,90){\vector(0,-1){80}}
\put(202,50){\makebox(0,0)[l]{$\mbox{\it chocolate}\,!$}}

\put(22,200){\vector(1,0){58}}
\put(50,202){\makebox(0,0)[b]{$\mbox{\it coin}\,!$}}
\put(120,100){\vector(1,0){60}}
\put(150,102){\makebox(0,0)[b]{$\mbox{\it chocolate}\,?$}}

\put(12,188){\vector(1,-1){78}}
\put(50,155){\makebox(0,0)[l]{$\tau$}}
\put(110,90){\vector(1,-1){80}}
\put(150,55){\makebox(0,0)[l]{$\tau$}}

\end{picture}\\
\;
\ey
$$
In conclusion, we note that 
the problem of analyzing of 
processes 
consisting of several communicating
components, 
most often arises in situations 
where such 
components 
are computer programs 
and hardware devices
of a computer system.
A communication 
between programs 
in such system
is implemented
by {\bf mediators}, i.e.
by certain processes which can 
communicate synchronously
with programs. 

Communications between 
programs are usually implemented 
by the following two ways. 

\bn 
\i {\bf Communication 
   through shared memory.} 

   In this case, mediators are 
   memory cells 
   accessed by both programs. 

   A communication in this case
   can be implemented 
   as follows: 
   one program writes an information 
   in these cells, 
   and other program 
   reads contents of cells. 

\i {\bf Communicaton 
   by sending messages.}

   In this case, 
   a mediator is a channel,
   which can be used by
   programs for 
   the following actions: 
   \bi 
   \i sending a message 
      to the channel, and 
   \i receiving of a message 
      from the channel. \ei 
   The channel may be 
   implemented as 
   a buffer 
   storing several messages. 
   Messages in the channel can be 
   organized on the principle of queue 
   (i.e., messages leave
   the channel in the same order 
   in which they had come). 
\en 

\section{Restriction}

Let \bi\i $P = (S, s^0, R)$ be a process,
and \i $L$ be a subset of the set $Names$.\ei

A {\bf restriction} of $P$ with respect to $L$
is the process
$$P\setminus L = (S, s^0, R')$$
which is obtained from $P$ by removing of those
transitions that have labels with the names from $L$, i.e.
$$R' \eam \bigset{(\diagrw{s&\pright{a}&s'})
\in R}{a=\tau,\;\;\mbox{or} \\ name(a)\not\in L}$$

As a rule, the operation of a restriction is used
together with the operation of parallel composition, 
for representation of processes that
\bi
\i consist of several communicating components, and
\i a communication between these components
   must satisfy certain restrictions.
\ei

For example, let processes $P_1$ and $P_2$
represent a behavior of a vending machine
and a customer respectively, which were
discussed in the previous section.

We would like to describe a process, 
which is a model of such parallel execution 
of processes $P_1 $and $P_2$,
at which these processes can execute actions
associated with buying and selling of a chocolate
only jointly.

The desired process can be obtained by an application 
to the process $P_1 | P_2$ the operation 
of a restriction with respect to 
the set of names of all actions
related to buying and selling of a chocolate.
This process is described by the expression
\be{getg577}P\eam (P_1|P_2)
\setminus
\{\mbox{{\it coin}}, \mbox{{\it chocolate}}\}\ee

A graph representation of process \re{getg577}
has the form
$$\by
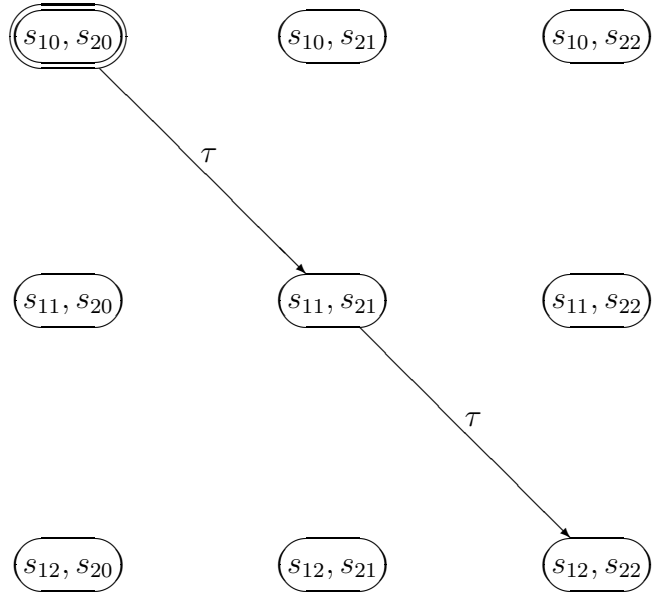
\begin{picture}(200,220)

\put(0,200){\oval(40,20)}
\put(0,200){\oval(44,24)}
\put(0,200){\makebox(0,0){$(s_{10},s_{20})$}}

\put(0,100){\oval(40,20)}
\put(0,100){\makebox(0,0){$(s_{11},s_{20})$}}

\put(0,0){\oval(40,20)}
\put(0,0){\makebox(0,0){$(s_{12},s_{20})$}}

\put(100,200){\oval(40,20)}
\put(100,200){\makebox(0,0){$(s_{10},s_{21})$}}

\put(100,100){\oval(40,20)}
\put(100,100){\makebox(0,0){$(s_{11},s_{21})$}}

\put(100,0){\oval(40,20)}
\put(100,0){\makebox(0,0){$(s_{12},s_{21})$}}

\put(200,200){\oval(40,20)}
\put(200,200){\makebox(0,0){$(s_{10},s_{22})$}}

\put(200,100){\oval(40,20)}
\put(200,100){\makebox(0,0){$(s_{11},s_{22})$}}

\put(200,0){\oval(40,20)}
\put(200,0){\makebox(0,0){$(s_{12},s_{22})$}}

\put(12,188){\vector(1,-1){78}}
\put(50,155){\makebox(0,0)[l]{$\tau$}}
\put(110,90){\vector(1,-1){80}}
\put(150,55){\makebox(0,0)[l]{$\tau$}}

\end{picture}\\
\;
\ey
$$

After removing unreachable states we get a 
process with the following graph representation:
$$\by
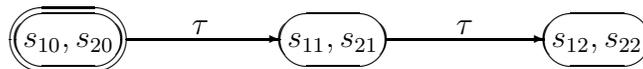
\begin{picture}(200,0)

\put(0,0){\oval(40,20)}
\put(0,0){\oval(44,24)}
\put(0,0){\makebox(0,0){$(s_{10},s_{20})$}}

\put(100,0){\oval(40,20)}
\put(100,0){\makebox(0,0){$(s_{11},s_{21})$}}

\put(200,0){\oval(40,20)}
\put(200,0){\makebox(0,0){$(s_{12},s_{22})$}}

\put(22,0){\vector(1,0){58}}
\put(50,2){\makebox(0,0)[b]{$\tau$}}
\put(120,0){\vector(1,0){60}}
\put(150,2){\makebox(0,0)[b]{$\tau$}}

\end{picture}\\
\;
\ey
$$

Consider another example.
Change a definition of a vending machine and a customer:
let them also to send a signal indicating successful 
completion of their work.
For example, these processes may have the following form:
$$\by
P_1\eam \mbox{{\it coin}}?.\mbox{{\it chocolate}}\,!.
\mbox{{\it clank}}\,!.{\bf 0}\\
P_2\eam \mbox{{\it coin}}!.\mbox{{\it chocolate}}\,?.\mbox{{\it hurrah}}\,!.{\bf 0}\ey$$

In this case, a graph representation of process \re{getg577}, 
after a removal of unreachable states, has the form 
$$\by
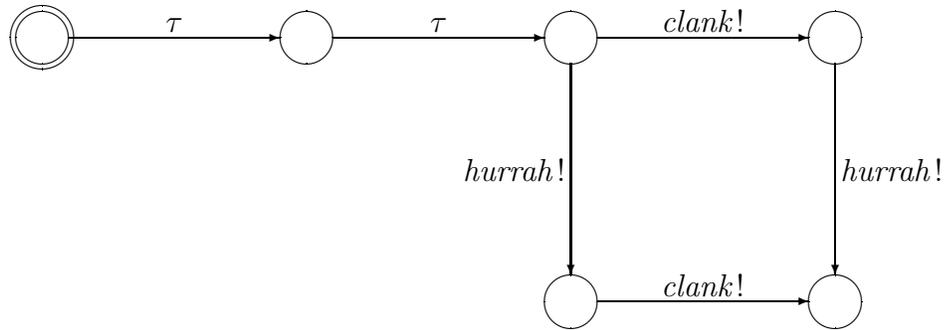
\begin{picture}(300,120)
\put(0,100){\oval(20,20)}
\put(0,100){\oval(24,24)}
\put(100,100){\oval(20,20)}
\put(200,100){\oval(20,20)}
\put(200,0){\oval(20,20)}
\put(300,100){\oval(20,20)}
\put(300,0){\oval(20,20)}
\put(10,100){\vector(1,0){80}}
\put(50,102){\makebox(0,0)[b]{$\tau$}}
\put(110,100){\vector(1,0){80}}
\put(150,102){\makebox(0,0)[b]{$\tau$}}
\put(200,90){\vector(0,-1){80}}
\put(198,50){\makebox(0,0)[r]{$\mbox{\it hurrah}\,!$}}
\put(300,90){\vector(0,-1){80}}
\put(302,50){\makebox(0,0)[l]{$\mbox{\it hurrah}\,!$}}
\put(210,100){\vector(1,0){80}}
\put(250,102){\makebox(0,0)[b]{$\mbox{\it clank}\,!$}}
\put(210,0){\vector(1,0){80}}
\put(250,2){\makebox(0,0)[b]{$\mbox{\it clank}\,!$}}
\end{picture}\\
\;
\ey
$$

This process allows execution only 
those non-internal actions that are not related to 
buying and selling a chocolate.

Note that in this case \bi\i in process \re{getg577}
a nondeterminism is present, although 
\i in the components of $P_1$ and $P_2$ 
a nondeterminism is absent.\ei
The cause of a nondeterminism in \re{getg577}
is our incomplete knowledge about the simulated system:
because we do not have a precise knowledge about a duration
of actions $\mbox{\it clank}\,!$ and 
$\mbox{\it hurrah}\,!$, then the model of the
system should allow any order of execution of these actions.

\section{Renaming}

The last operation that we consider is 
an unary operation, which is called a {\bf renaming}.

To define this operation, it is 
necessary to define a mapping of the form
\be{xcfgvsf}f: Names \to Names\ee

An effect of the operation of renaming
on process $P$ is 
changing labels of transitions of $P$:
\bi
\i any label of the form $\alpha\,?$
   is replaced on 
   $f(\alpha)\,?$, and
\i any label of the form $\alpha\,!$
   is replaced on 
   $f(\alpha)\,!$
\ei
The resulting process is denoted by $P[f]$.

We shall refer any mapping of the form \re{xcfgvsf}
also as a {\bf renaming}.

If a renaming $f$ acts non-identically only on the names
$$\alpha_1,\ldots, \alpha_n$$
and maps them to the names
$$\beta_1,\ldots, \beta_n$$
respectively, then the process $P[f]$ 
can be denoted also as
$$P[\beta_1/\alpha_1,\ldots, \beta_n/\alpha_n]$$

The operation of renaming can be used, for example, 
in the following situation: this operation
allows to use several copies of a process $P$ 
as different components in constructing 
of a more complex process $P'$.
Renaming serves for prevention of 
collisions between names of actions
used in different occurrences of $P$ in $P'$.

\section{Properties of operations on processes} \label{archnatella}

In this section we give some elementary properties
of defined above operations on processes. 
All these properties have a form of equalities. 
For the first two properties, we give their proof,
other properties are listed without comments 
in view of their evidence.

Recall (see section \ref{zamena}),
that we consider two processes as equal, if 
\bi 
\i they are isomorphic, or
\i one of these processes can be obtained 
   from another by removing some of 
   unreachable states
   and transitions which contain unreachable states. 
\ei

\bn
\i Operation $+$ is associative, 
   i.e. for any processes $P_1$, $P_2$ and $P_3$
   the following equality holds:
   \be{sdgfsdg44676}(P_1+P_2)+P_3 = P_1+(P_2+P_3)\ee

   Indeed, let the processes $P_i \; (i = 1,2,3)$
   have the form
   \be{vid1456}P_i=(S_i, s^0_i, R_i)\quad
   (i=1,2,3)\ee
   and
   their sets of states $S_1, S_2$ and $S_3$
   are pairwise disjoint. 
   Then both sides 
   of equality \re{sdgfsdg44676}
   are equal to the process
   $P=(S, s^0, R)$,
   whose components are defined as follows:
\bi
\i $S\eam S_1\cup S_2\cup S_3 \cup \{s^0\}$, 
   where $s^0 $ is a new state\\
   (which does not belong to $S_1, S_2$ and $S_3$)
\i $R$ contains all transitions from
   $R_1, R_2$ and $R_3$
\i for each transition from $R_i$ ($i = 1,2,3$)
   of the form
   $$\diagrw{s_i^0&\pright{a}&s}$$
   $R$ contains the transition
   $\diagrw{s^0&\pright{a}&s}$
\ei

The property of associativity of the operation $+$
allows to use expressions of the form
\be{asfdfasd}P_1+\ldots+ P_n\ee
because for any parenthesization 
of the expression \re{asfdfasd}
we shall get one and the same process.

A process, which is a value of expression \re{asfdfasd}
can be described explicitly as follows.

Let the processes $P_i\;(i=1,\ldots, n)$ have the form 
   \be{nvid1456}P_i=(S_i, s^0_i, R_i)\quad
   (i=1,\ldots, n)\ee
   with the sets of states $S_1, \ldots, S_n$
   are pairwise disjoint. Then
   a process, which is a value of the expression 
   \re{asfdfasd},
   has the form
   $$P=(S, s^0, R)$$
   where the components $S, s^0, R$
   are defined as follows:
\bi
\i $S\eam S_1\cup \ldots \cup S_n \cup \{s^0\}$, 
   where $s^0$ is a new state\\
   (which does not belong to $S_1\ldots, S_n$)
\i $R$ contains all transitions from
   $R_1, \ldots, R_n$
\i for each transition from $R_i$ ($i=1,\ldots, n$)
   of the form
   $$\diagrw{s_i^0&\pright{a}&s}$$
   $R$ contains the transition
   $\diagrw{s^0&\pright{a}&s}$
\ei

\i The operation $\pa$ is associative, 
   i.e. for any processes
   $P_1$, $P_2$ and $P_3$
   the following equality holds:
   \be{dfgsdgre6r46}(P_1\pa P_2)\pa P_3 = P_1\pa (P_2\pa P_3)\ee

   Indeed, let the processes $P_i \; (i = 1,2,3)$
   have the form \re{vid1456}.
   Then both sides of \re{dfgsdgre6r46}
   are equal to the process
   $P = (S, s^0, R)$
   whose components
   are defined as follows:

\bi
\i $S\eam S_1\times S_2\times S_3
   \eam \\\eam \{(s_1,s_2,s_3)\mid s_1\in S_1, s_2\in S_2, s_3
   \in S_3\}$
\i $s^0\eam (s_1^0, s_2^0, s_3^0)$
\i for 
   \bi 
   \i each transition 
      $\diagrw{s_1&\pright{a}&s'_1}$ from $R_1$, and 
   \i each pair of states $s_2 \in S_2, s_3 \in S_3$
   \ei
   $R$ contains the transition
   $$\diagrw{(s_1,s_2, s_3)&\pright{a}&(s'_1,s_2,s_3)}$$
\i for 
   \bi 
   \i each transition $\diagrw{s_2&\pright{a}&s'_2}$
      from $R_2$, and 
   \i each pair of states $s_1\in S_1, s_3\in S_3$
   \ei
   $R$ contains the transition
   $$\diagrw{(s_1,s_2, s_3)&\pright{a}&(s_1,s'_2,s_3)}$$
\i for \bi \i each transition 
   $\diagrw{s_3&\pright{a}&s'_3}$
   from $R_3$, and \i each pair of states 
   $s_1\in S_1, s_2\in S_2$
   \ei
   $R$ contains the transition
   $$\diagrw{(s_1,s_2, s_3)&\pright{a}&(s_1,s_2,s'_3)}$$

\i for 
   \bi 
   \i each pair of transitions with complementary labels
   $$\by
   \diagrw{s_1&\pright{a}&s'_1}\quad\in R_1\\
   \diagrw{s_2&\pright{\overline{a}}&s'_2}\quad\in R_2\ey$$
   and
   \i each state $s_3 \in S_3$
   \ei
   $R$ contains the transition
   $$\diagrw{(s_1,s_2,s_3)&\pright{\tau}&(s'_1,s'_2,s_3)}$$
\i for 
   \bi 
   \i each pair of transitions with complementary labels
   $$\by
   \diagrw{s_1&\pright{a}&s'_1}\quad\in R_1\\
   \diagrw{s_3&\pright{\overline{a}}&s'_3}\quad\in R_3\ey$$
   and
   \i each state $s_2 \in S_2$
   \ei
   $R$ contains the transition
   $$\diagrw{(s_1,s_2,s_3)&\pright{\tau}&(s'_1,s_2,s'_3)}$$
\i for \bi \i each pair of transitions with complementary labels
   $$\by
   \diagrw{s_2&\pright{a}&s'_2}\quad\in R_2\\
   \diagrw{s_3&\pright{\overline{a}}&s'_3}\quad\in R_3\ey$$
   and
   \i each state $s_1 \in S_1$
   \ei
   $R$ contains the transition
   $$\diagrw{(s_1,s_2,s_3)&\pright{\tau}&(s_1,s'_2,s'_3)}$$
\ei

The property of associativity of the operation $\pa$
allows to use expressions of the form
\be{paasfdfasd}P_1 \pa \ldots \pa P_n\ee
because for any parenthesization 
of the expression \re{paasfdfasd}
we shall get one and the same process.

A process, which is a value of expression \re{paasfdfasd}
can be described explicitly as follows.

Let the processes $P_i \; (i = 1, \ldots, n)$
have the form \re{nvid1456}.
Then
   a process, which is a value of the expression 
   \re{paasfdfasd},
   has the form
   $$P=(S, s^0, R)$$
   where the components $S, s^0, R$
   are defined as follows:
\bi
\i $S\eam S_1\times \ldots \times S_n
   \eam \\\eam \{(s_1,\ldots,s_n)\mid s_1\in S_1, \ldots, s_n
   \in S_n\}$
\i $s^0\eam (s_1^0, \ldots, s_n^0)$
\i for \bi
   \i each $i\in \{1,\ldots, n\}$
   \i each transition $\diagrw{s_i&\pright{a}&s'_i}$
   from $R_i$, and \i each list of states
   $$s_1,\ldots, s_{i-1}, s_{i+1},\ldots, s_n$$
   where $\forall \; j\in \{1,\ldots, n\}\quad
   s_j\in S_j$
   \ei
   $R$ contains the transition
   $$\diagrw{(s_1,\ldots, s_n)&
   \pright{a}&(s_1,\ldots, s_{i-1}, s'_i,
   s_{i+1},\ldots, s_n)}$$
\i for \bi
   \i each pair of indices $i,j\in \{1,\ldots, n\}$, 
   where $i<j$
   \i each pair of transitions with complementary labels\
   of the form
   $$\by
   \diagrw{s_i&\pright{a}&s'_i}\quad\in R_i\\
   \diagrw{s_j&\pright{\overline{a}}&s'_j}\quad\in R_j\ey$$
   and
   \i each list of states
   $$s_1,\ldots, s_{i-1}, s_{i+1},\ldots,
   s_{j-1}, s_{j+1},\ldots,
   s_n$$
   where $\forall \; k\in \{1,\ldots, n\}\quad
   s_k\in S_k$
   \ei
   $R$ contains the transition
   $$\diagrw{(s_1,\ldots,s_n)&\pright{\tau}&
   \b{s_1,\ldots, s_{i-1}, s'_i, s_{i+1},\ldots,
   s_{j-1}, s'_j, \\s_{j+1},\ldots,
   s_n}}$$
\ei

\i The operation $+$ is commutative, i.e. for any processes
   $P_1$ and $P_2$
   the following equality holds:
   $$P_1+P_2 = P_2+P_1$$
\i The operation $\pa$ is commutative, i.e. for 
   any processes $P_1$ and $P_2$   the following equality holds:
   $$P_1\pa P_2 = P_2\pa P_1$$
\i ${\bf 0}$ is a neutral element with respect
   to the operation $\pa$: $$P \pa {\bf 0} = P$$
   The operation $+$ has a similar property, 
   in this property there is used a concept of 
   strong equivalence of processes (defined below)
   instead of equality of processes .
   This property, as well as the property 
   of idempotency of the operation $+$
   are proved in section \ref{dopsvojstva} 
   (theorem \ref{thidempo}).
\i ${\bf 0}  \setminus L = {\bf 0}$
\i ${\bf 0}[f] = {\bf 0}$
\i $P \setminus L = P$, if
   $L\cap names(Act(P))  = \emptyset$.

   (recall that $Act (P)$
   denotes a set of actions
   $a\in Act\setminus \{\tau\}$, such that
   $P$ contains a transition with the label $a$)
\i \label{metkka1}
   $(a.P) \setminus L =
   \left\{
   \by {\bf 0}, \;\;\mbox{if $a\neq \tau$ and } name(a) \in L \\
   a.(P \setminus L),\;\;\mbox{otherwise}\ey\right.$
\i $(P_1 + P_2) \setminus L = (P_1 \setminus L) + (P_2 \setminus L)$
\i $(P_1 \pa P_2) \setminus L = (P_1 \setminus L) \pa 
   (P_2 \setminus L)$,
   if $$L \cap names(Act(P_1) \cap \overline{Act(P_2)}) = 
   \emptyset$$
\i $(P \setminus L_1) \setminus L_2 = P\setminus (L_1 \cup L_2)$
\i $P[f] \setminus L = (P \setminus f^{-1}(L))[f]$
\i $P[id] = P$, where $id$ is an identity function
\i $P[f] = P[g]$, if restrictions of functions
   $f$ and $g$ on the set $names(Act(P))$ are equal.
\i $(a.P)[f]  = f(a). (P[f])$
\i $(P_1 + P_2)[f] = P_1[f] + P_2[f]$
\i $(P_1\pa P_2)[f] = P_1[f] \pa P_2[f]$,
   if a restriction of $f$ on the set
   $$names(Act(P_1) \cup Act(P_2))$$
   is an injective mapping.
\i $(P \setminus L)[f] = P[f] \setminus f(L)$, if the mapping
   $f$ is an injective mapping.
\i $P[f][g] = P[g \circ f]$
\en

\chapter{Equivalences of processes}
\label{weqiro23rui34iop}

\section{A concept of an equivalence 
of processes}
\label{mett6}

The same behavior can be represented 
by different processes. 
For example, consider two processes: 
$$
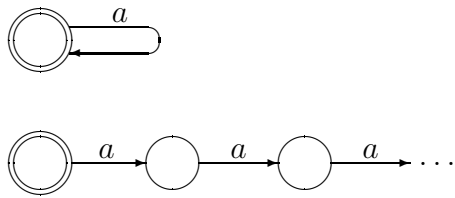
\begin{picture}(150,20)
\put(0,0){\oval(20,20)}
\put(0,0){\oval(24,24)}
\put(11,5){\line(1,0){29}}
\put(40,-5){\vector(-1,0){29}}
\put(40,0){\oval(10,10)[r]}
\put(30,7){\makebox(0,0)[b]{$a$}}
\end{picture}
$$

$$\by
\begin{picture}(150,20)
\put(0,0){\oval(20,20)}
\put(0,0){\oval(24,24)}
\put(12,0){\vector(1,0){28}}
\put(25,2){\makebox(0,0)[b]{$a$}}
\put(50,0){\oval(20,20)}
\put(60,0){\vector(1,0){30}}
\put(75,2){\makebox(0,0)[b]{$a$}}
\put(100,0){\oval(20,20)}
\put(110,0){\vector(1,0){30}}
\put(125,2){\makebox(0,0)[b]{$a$}}
\put(150,0){\makebox(0,0)[c]{$\ldots$}}
\end{picture}\\\;\\\ey
$$

The first process has only one state, 
and the second has infinite set of states,
but these processes represent 
the similar behavior, 
which consists of a perpetual 
execution of the actions $a$. 

One of important problems in the theory
of processes consists of a finding 
of an appropriate definition 
of equivalence of processes, 
such that processes are equivalent 
according to this definition
if and only if they represent a similar
behavior.

In this chapter 
we present several 
definitions 
of equivalence of processes. 
In every particular situation 
a choice of an appropriate variant 
of the concept of 
equivalence of processes
should be determined by 
a particular understanding 
(i.e. related to this situation) 
of a similarity 
of a behavior of processes.

In sections \ref{trassovajaa} 
and \ref{gjsdfkglty56yfgdf} 
we introduce
concepts of trace equivalence 
and strong equivalence of processes. 
These concepts are used 
in situations where all actions 
executing in the processes 
that have equal status. 

In sections \ref{defequisdfgval43} 
and \ref{defequisdfgvalfghgj543}
we consider other variants of the concept 
of equivalence of processes: 
namely, observational equivalence and 
observational congruence. 
These concepts are used in situations 
when we consider the invisible action 
$\tau$ as negligible, i.e. when we 
assume that two traces are equivalent, 
if one of them can be obtained 
from another by insertions 
and/or deletions of $\tau$. 

With each possible definition 
of equivalence of processes 
there are related two natural problems. 
\bn 
\i Recognition for two given processes, 
   whether they are equivalent. 
\i Construction for a given process  
   $P$ such a process $P '$, which 
   is the least complicated 
   (for example, has a minimum 
   number of states) among all processes 
   that are equivalent to $P$. 
\en 

\section{Trace equivalence of processes}
\label{trassovajaa}

As mentioned above, we would like 
to consider two processes as equivalent, 
if they describe a same behavior. 
So, if we consider a behavior of a process
as a generation of a trace, then
one of necessary conditions of equivalence 
of processes $P_1$ and $P_2$ 
is coincidence of sets of their traces: 
\be{uslovietrass}Tr({P_1})= Tr({P_2})\ee

In some situations, condition 
\re{uslovietrass}
can be used as a definition of equivalence 
of $P_1$ and $P_2$. 

However, the following example shows that 
this condition does not reflect one important 
aspect of an execution of processes. 

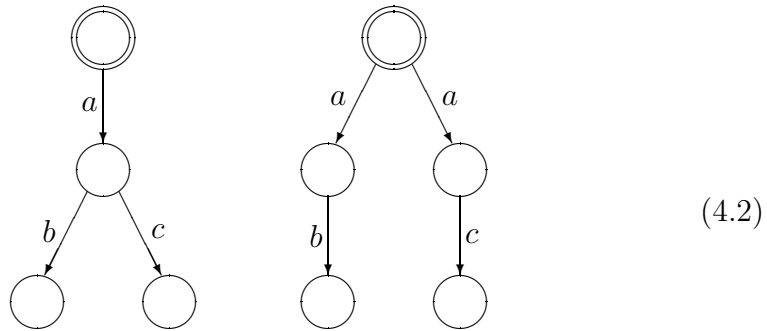
\be{ris3}
\by
\begin{picture}(0,100)

\put(0,100){\oval(20,20)}
\put(0,100){\oval(24,24)}

\put(0,50){\oval(20,20)}

\put(-25,0){\oval(20,20)}
\put(25,0){\oval(20,20)}

\put(0,88){\vector(0,-1){28}}

\put(-6,42){\vector(-1,-2){16}}
\put(6,42){\vector(1,-2){16}}

\put(-2,75){\makebox(0,0)[r]{$a$}}
\put(-18,27){\makebox(0,0)[r]{$b$}}
\put(18,27){\makebox(0,0)[l]{$c$}}

\end{picture}
\\
\vspace{5mm}\ey
\begin{picture}(50,80)
\end{picture}
\begin{picture}(50,80)
\end{picture}
\by
\begin{picture}(0,100)

\put(0,100){\oval(20,20)}
\put(0,100){\oval(24,24)}

\put(-25,50){\oval(20,20)}
\put(25,50){\oval(20,20)}

\put(-25,0){\oval(20,20)}
\put(25,0){\oval(20,20)}

\put(-7,90){\vector(-1,-2){15}}
\put(7,90){\vector(1,-2){15}}

\put(-25,40){\vector(0,-1){30}}
\put(25,40){\vector(0,-1){30}}

\put(-18,77){\makebox(0,0)[r]{$a$}}
\put(18,77){\makebox(0,0)[l]{$a$}}

\put(-27,25){\makebox(0,0)[r]{$b$}}
\put(27,25){\makebox(0,0)[l]{$c$}}
\end{picture}\\
\vspace{5mm}\ey
\ee

Sets of traces of these processes are equal: 
$$Tr(P_1)=Tr(P_2)=\{\varepsilon, a,ab,ac\}$$
(where $\varepsilon$ is an empty sequence). 

However, these processes have the 
following essential difference:
\bi 
\i in the left process, after execution of 
   a first action ($a$) 
   there is a possibility to 
   choose next action ($b$ or $c$), while 
\i in the right process, 
   after execution of 
   a first action there is no  such possibility:
   \bi \i 
   if a first transition occurred on the left edge, 
   then a second action 
   can only be the action $b$, and 
\i if a first transition occurred on the right edge, 
   then a second action 
   can only be the action $c$
\ei 
   i.e. a second action was predetermined 
   before execution of a first action. 
\ei 

If we do not wish to consider these 
processes as equivalent, then 
condition  \re{uslovietrass}
must be enhanced in some a way. 
One version of such enhancement
is described below. 
In order to formulate it,
define the notion of a trace from 
a state of a process. 

Each variant of an execution of a process
$P = (S, s^0, R)$ 
we interpret as a generation 
of a sequence of transitions 
\be{gsdf454}\diagrw{s_0&\pright{a_1}&s_1
&\pright{a_2}&s_2&\pright{a_3}&\ldots}\ee
starting from the initial state $s^0$
(i.e. $s_0 = s^0$). 

We can consider a generation of
sequence \re{gsdf454} 
not only from the initial state $s^0$, 
but from arbitrary state $s\in S$, 
i.e. consider a sequence of the form
\re{gsdf454}, in which $s_0 = s$. 
The sequence $(a_1, a_2, \ldots)$ 
of labels of these transitions we shall call 
a {\bf trace starting at $s$}.
A set of all such traces we denote 
by $Tr_s(P)$. 

Let $P_1$ and $P_2$ be processes 
of the form
$$P_i = (S_i, s_i^0, R_i)\qquad(i=1,2)$$

Consider a finite sequence of transitions
of $P_1$ of the form
\be{gsd1f454}\diagrw{s_1^0=s_0&\pright{a_1}&s_1
&\pright{a_2}&\ldots&\pright{a_n}&s_n}
\quad(n\geq 0)\ee
(the case $n=0$ corresponds to the empty 
sequence of transitions \re{gsd1f454}, 
in which $s_n = s_1^0$).

The sequence \re{gsd1f454}
can be considered as an initial 
phase of execution of the process $P_1$, 
and every trace from $Tr_{s_n}(P_1)$
can be considered
as a continuation of this phase. 

The processes $P_1$ and $P_2$ 
are said to be {\bf trace equivalent}, if 
\bi 
\i for each initial phase \re{gsd1f454}
of an execution of the process $P_1$ 
there is an initial phase of an 
execution of the process $P_2$ 
\be{gsd1fsddf454}\diagrw
{s_2^0=s'_0&\pright{a_1}&s'_1
&\pright{a_2}&\ldots&\pright{a_n}&s'_n}\ee
with the following properties:
\bi
\i \re{gsd1fsddf454} has the same trace
$a_1\ldots a_n$, as \re{gsd1f454}, and 
\i at the end of \re{gsd1fsddf454}
there is the same choice of further 
execution that at the end 
of \re{gsd1f454}, i.e. 
\be{gs5srdfertg4we4}
Tr_{s_n}(P_1) =
Tr_{s'_n}(P_2)\ee\ei
\i and a symmetrical condition holds: 
  for each sequence of transitions 
  of $P_2$ of the form 
\re{gsd1fsddf454} 
there is a sequence of transitions of
$P_1$ of the form 
\re{gsd1f454}, 
such that \re{gs5srdfertg4we4} holds. 
\ei 
These conditions have the following
disadvantage: they contain
\bi\i unlimited sets of sequences 
of transitions of the form \re{gsd1f454}
and \re{gsd1fsddf454}, and 
\i unlimited sets of traces from 
\re{gs5srdfertg4we4}. \ei
Therefore, checking of these conditions 
seems to be difficult even 
when the processes $P_1$ and $P_2$ are finite. 

There is a problem of finding of 
necessary and sufficient conditions 
of trace equivalency,
that can be algorithmically checked 
for given processes $P_1$ and $P_2$
in the case when these processes are finite. 

Sometimes there is considered
an equivalence between processes
which is obtained 
from the trace equivalence 
by a replacement of condition 
\re{gs5srdfertg4we4} 
on the weaker condition: 
$$Act(s_n)=Act(s'_n)$$
where for each state $s$ 
$Act(s)$ denotes a set 
all actions $a \in Act$, 
such that there is a transition starting at $s$ 
with the label $a$.

\section{Strong equivalence}
\label{gjsdfkglty56yfgdf}

Another variant of the concept of 
equivalence of processes 
is {\bf strong equivalence}. 
To define the concept of strong equivalence, 
we introduce auxiliary notations. 

After the process 
\be{gsdf54er3}P=(S,s^0,R)\ee
has executed its first action, and turn
to a new state $s^1$, its behavior will be
indistinguishable from a behavior of the 
process
\be{gsdf54er31}P'\;\eam \;(S,s^1,R)\ee
having the same components as $P$, 
except of an initial state. 

We shall consider the diagram
\be{ghfgdrtryery}\diagrw{P&\pright{a}&P'}\ee
as an abridged notation of the statement that 
\bi 
\i $P$ and $P '$ are processes of the form 
   \re{gsdf54er3}, and 
   \re{gsdf54er31} respectively, and 
\i $R$ contains the transition 
   $\diagrw{s^0&\pright{a}&s^1}$. 
\ei 
\re{ghfgdrtryery} 
can be interpreted as a statement 
that the process $P$ can 
\bi 
\i execute the action $a$, and then 
\i behave like the process $P'$. 
\ei 

A concept of strong equivalence 
is based on the following understanding 
of equivalence of processes: 
if we consider processes $P_1$ and 
$P_2$ as equivalent, then 
it must be satisfied 
the following condition: 
\bi \i if one of these 
processes $P_i$ can 
\bi \i execute some action $a \in Act$, 
  \i and then behave like some process 
   $P'_i$ \ei 
\i then the other 
process 
$P_j\quad (j\in \{1,2\}\setminus \{i\})$
also must be able 
\bi \i execute the same action $a$, 
\i and then behave like some process $P'_j$, 
which is equivalent to $P'_i $. \ei \ei 

Thus, the desired equivalence must be a 
a binary relation $\mu$ on the set of all 
processes, 
the following properties. 
\bi 
\i[(1)] 
   If $(P_1, P_2) \in \mu$, and 
   \be{wetrtwe1}
   \diagrw{P_1&\pright{a}&P'_1}\ee
   for some process $P'_1$, 
   then there is a
   process $P'_2$, such that 
   \be{wetrtwe2}
   \diagrw{P_2&\pright{a}&P'_2}\ee
   and 
   \be{gsdfgsdfgdfg}(P'_1, P'_2) \in \mu\ee
\i[(2)] symmetric property: 
   if $(P_1, P_2) \in \mu$, and 
   for some process $P'_2$ 
   \re{wetrtwe2} holds, 
   then there is a
   process $P'_1$, such that 
   \re{wetrtwe1}
   and \re{gsdfgsdfgdfg} hold. 
\ei 

Denote by the symbol
${\cal M}$ a set of all binary 
relations, which possess the above properties. 

The set ${\cal M}$ is nonempty: 
it contains, for example, 
a diagonal relation, which 
consists of all pairs of the form $(P, P)$, 
where $P$ is an arbitrary process. 

The question naturally arises:
which of the relations from ${\cal M}$
can be used for a definition 
of strong equivalence?

We suggest the most simple answer 
to that question: 
we will consider $P_1$ and $P_2$ 
as strongly equivalent if and only if
there exists at least one relation
$\mu\in{\cal M}$, which 
contains the pair $(P_1, P_2)$. 

Thus, we define the desired relation
of strong equivalence on the set 
of all processes as the union 
of all relations from ${\cal M}$. 
This relation is denoted by $\sim$. 

It is not so difficult to prove that 
\bi 
\i $\sim\; \in\; {\cal M}$, and 
\i $ \sim$ is an equivalence relation, 
   because 
   \bi 
   \i reflexivity of $\sim$ follows 
    from the fact that the diagonal relation 
   belongs to ${\cal M}$, 
   \i symmetry of $\sim$ follows
     from the fact that if $\mu\in{\cal M}$, 
then $\mu^{-1}\in {\cal M}$
   \i transitivity of $\sim$ follows 
   from the fact that if 
$\mu_1 \in{\cal M}$ and 
$\mu_2 \in{\cal M}$,  then
$\mu_1 \circ \mu_2 \in{\cal M}$. 
   \ei 
\ei 

If processes $P_1$ and $P_2$ 
are strongly equivalent, 
then this fact is denoted by
$$P_1 \sim P_2$$ 

It is easy to prove that if processes 
$P_1$ and $P_2$ 
are strongly equivalent 
they they are trace equivalent. 

To illustrate the concept 
of strong equivalence
we give a couple of examples.
\bn 
\i The processes 
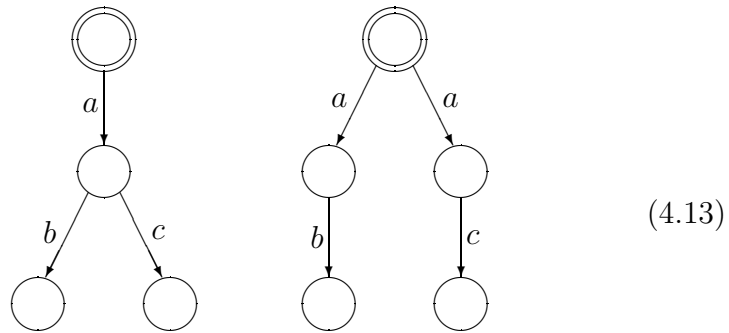
\be{ris5467}
\by
\begin{picture}(0,100)

\put(0,100){\oval(20,20)}
\put(0,100){\oval(24,24)}

\put(0,50){\oval(20,20)}

\put(-25,0){\oval(20,20)}
\put(25,0){\oval(20,20)}

\put(0,88){\vector(0,-1){28}}

\put(-6,42){\vector(-1,-2){16}}
\put(6,42){\vector(1,-2){16}}

\put(-2,75){\makebox(0,0)[r]{$a$}}
\put(-18,27){\makebox(0,0)[r]{$b$}}
\put(18,27){\makebox(0,0)[l]{$c$}}

\end{picture}
\\
\vspace{5mm}\ey
\begin{picture}(50,80)
\end{picture}
\begin{picture}(50,80)
\end{picture}
\by
\begin{picture}(0,100)

\put(0,100){\oval(20,20)}
\put(0,100){\oval(24,24)}

\put(-25,50){\oval(20,20)}
\put(25,50){\oval(20,20)}

\put(-25,0){\oval(20,20)}
\put(25,0){\oval(20,20)}

\put(-7,90){\vector(-1,-2){15}}
\put(7,90){\vector(1,-2){15}}

\put(-25,40){\vector(0,-1){30}}
\put(25,40){\vector(0,-1){30}}

\put(-18,77){\makebox(0,0)[r]{$a$}}
\put(18,77){\makebox(0,0)[l]{$a$}}

\put(-27,25){\makebox(0,0)[r]{$b$}}
\put(27,25){\makebox(0,0)[l]{$c$}}
\end{picture}\\
\vspace{5mm}\ey
\ee
   are not strongly equivalent, because 
   they are not trace equivalent.
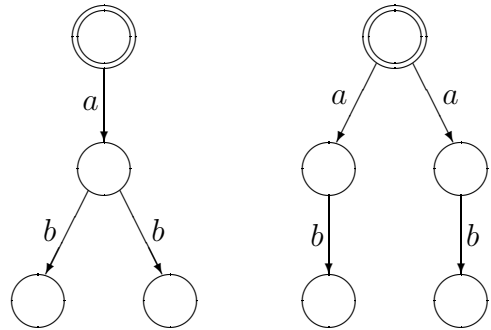
\i Processes 
$$
\by
\begin{picture}(0,100)

\put(0,100){\oval(20,20)}
\put(0,100){\oval(24,24)}

\put(0,50){\oval(20,20)}

\put(-25,0){\oval(20,20)}
\put(25,0){\oval(20,20)}

\put(0,88){\vector(0,-1){28}}

\put(-6,42){\vector(-1,-2){16}}
\put(6,42){\vector(1,-2){16}}

\put(-2,75){\makebox(0,0)[r]{$a$}}
\put(-18,27){\makebox(0,0)[r]{$b$}}
\put(18,27){\makebox(0,0)[l]{$b$}}

\end{picture}\\
\vspace{5mm}\ey
\begin{picture}(50,80)
\end{picture}
\begin{picture}(50,80)
\end{picture}
\by
\begin{picture}(0,100)

\put(0,100){\oval(20,20)}
\put(0,100){\oval(24,24)}

\put(-25,50){\oval(20,20)}
\put(25,50){\oval(20,20)}

\put(-25,0){\oval(20,20)}
\put(25,0){\oval(20,20)}

\put(-7,90){\vector(-1,-2){15}}
\put(7,90){\vector(1,-2){15}}

\put(-25,40){\vector(0,-1){30}}
\put(25,40){\vector(0,-1){30}}

\put(-18,77){\makebox(0,0)[r]{$a$}}
\put(18,77){\makebox(0,0)[l]{$a$}}

\put(-27,25){\makebox(0,0)[r]{$b$}}
\put(27,25){\makebox(0,0)[l]{$b$}}
\end{picture}
\\
\vspace{5mm}
\ey
$$
are strongly equivalent. 
\en 

\section{Criteria of strong equivalence}

\subsection{A logical criterion 
of strong equivalence} \label{logcrit}

Let $Fm$ be a set 
of {\bf formulas} defined as follows. 
\bi 
\i The symbols $\top$ and $\bot$ 
   are formulas  from $Fm$. 
\i If $\varphi\in Fm$, 
   then $\neg \varphi\in Fm$. 
\i If $\varphi\in Fm$ and $\psi\in Fm$, 
  then $\varphi \wedge \psi\in Fm$. 
\i If $\varphi\in Fm$, 
   and $a \in Act$, then 
   $\langle a \rangle \varphi\in Fm$. 
\ei 

Let $P$ be a process, and $\varphi \in Fm$. 
A {\bf value} 
of the formula $\varphi$ on the process $P$ 
is an element $P(\varphi)$ of the set 
$\{0,1\}$ defined as follows. 
\bi
\i $P(\top) \eam 1,\; \;P(\bot) \eam 0$
\i $P(\neg \varphi) \eam 1 - P(\varphi)$
\i $P(\varphi\wedge \psi) \eam P(\varphi)\cdot 
P(\psi)$
\i $P(\langle a \rangle \varphi)  \eam
    \left\{\by 1,&
   \mbox{if there is a process } P':\\
   &\diagrw{P&\pright{a}&P'},\; P'(\varphi)=1
   \\
   0,&\mbox{otherwise}
	 \ey\right.$
\ei

A {\bf theory} of the process $P$ is a subset
$Th (P)\subset Fm$, 
defined as follows: 
$$Th(P) = 
\{\varphi\in Fm\mid P(\varphi)=1\}$$

\refstepcounter{theorem}
{\bf Theorem \arabic{theorem}\label{th5}}.

Let $P_1$ and $P_2$ be finite processes. 
Then 
$$P_1 \sim P_2\quad\Leftrightarrow\quad
Th(P_1) = Th(P_2)$$

{\bf Proof.}

Let $P_1 \sim P_2$. The statement that
for each $\varphi\in Fm$ the equality
$P_1(\varphi)=P_2(\varphi)$
holds, can be proven by induction 
on the structure of $\varphi$. 

Prove the implication ``$\Leftarrow$''. 
Suppose that 
\be{gjklsdfgjsdfgkljwe5}
Th(P_1) = Th(P_2)\ee
Let $\mu$ be a binary relation on the set 
of all processes, defined as follows:
$$\mu\eam \{(P_1, P_2)\mid Th(P_1) 
= Th(P_2)\}$$

We prove that $\mu$ satisfies 
the definition of strong equivalence. 
Let this does not hold, 
that is, for example, for some $a\in Act$
\bi 
\i[(a)] there is a process $P'_1$, such that 
   $$\diagrw{P_1&\pright{a}&P'_1}$$
\i[(b)] but there is no a process $P'_2$, 
   such that 
   \be{gdjksflgerge}
   \diagrw{P_2&\pright{a}&P'_2}\ee
   and $Th(P'_1) = Th(P'_2)$. 
\ei 
Condition (b) can be satisfied
in two situations: 
\bi 
\i[1.] There is no a process $P'_2$, 
   such that \re{gdjksflgerge} holds.
\i[2.] There exists a process $P'_2$, 
   such that \re{gdjksflgerge} holds, 
   but for each such process $P'_2$
   $$Th(P'_1) \neq Th(P'_2)$$
\ei 
We show that in both these situations 
there is a formula $\varphi\in Fm$, 
such that
$$P_1(\varphi)=1,\quad P_2(\varphi)=0$$
that would be contrary to assumption 
\re{gjklsdfgjsdfgkljwe5}.

\bn 
\i If the first situation holds, 
   then we can take as $\varphi$ 
   the formula $\langle a \rangle \top$. 
\i Assume that the second situation holds. 
   Let 
   $$P'_{2,1},\ldots, P'_{2,n}$$
   be a list of all processes $P'_2$
   satisfying \re{gdjksflgerge}.

   By assumption, for each 
   $i=1,\ldots, n$, 
   the inequality 
   $$Th(P'_1) \neq Th(P'_{2,i})$$
   holds, i.e. for each $i = 1, \ldots, n$ 
   there is a formula $\varphi_i$, 
   such that 
    $$P'_1(\varphi_i)=1,\qquad P'_{2,i}(\varphi_i)=0$$
In this situation, we can take as $\varphi$ 
the formula 
$\langle a \rangle (\varphi_1
\wedge\ldots\wedge \varphi_n)$. 
$\quad \blackbox$ \en 

For example, let $P_1$ and $P_2$ be 
processes \re{ris5467}. 
As stated above, these processes are not 
strongly equivalent. 
The following formula can be taken as 
a justification of the statement that 
$P_1\not\sim P_2$:
$$\varphi\eam \langle a \rangle
(\langle b \rangle \top \wedge \langle c \rangle  
\top)$$
It is easy to prove that 
$P_1(\varphi)=1$ and 
$P_2(\varphi)=0$. 

There is a problem of finding 
for two given processes $P_1$ and $P_2$ 
a list of formulas of a smallest size 
$$ \varphi_1, \ldots, \varphi_n $$ 
such that 
$P_1 \sim P_2$ if and only if 
$$\forall\; i=1,\ldots, n 
\qquad P_1(\varphi_i) = P_2(\varphi_i)$$

\subsection{A criterion of 
strong equivalence, based 
on the notion of a bisimulation} 
\label{bimoodeer}

\refstepcounter{theorem}
{\bf Theorem \arabic{theorem}\label{teorrrr}}.

Let $P_1$ and $P_2$ 
be a couple of processes of the form
$$P_i=(S_i, s^0_i, R_i)\qquad(i=1,2)$$
Then $P_1 \sim P_2$ if and only if 
there is a relation 
$$\mu\subseteq S_1\times S_2$$
satisfying the following conditions. 
\bi 
\i[0.] $(s_1^0,s_2^0)\in \mu$.
\i[1.] For each pair $(s_1,s_2)\in \mu$
and each transition from $R_1$ of the form 
$$\diagrw{s_1&\pright{a}&s'_1}$$
there is a transition from $R_2$ of the form
$$\diagrw{s_2&\pright{a}&s'_2}$$
such that $(s'_1,s'_2)\in \mu$.
\i[2.] For each pair $(s_1,s_2)\in \mu$
and each transition from $R_2$ of the form 
$$\diagrw{s_2&\pright{a}&s'_2}$$
there is a transition from $ R_1$ of the form 
$$\diagrw{s_1&\pright{a}&s'_1}$$
such that $(s'_1,s'_2)\in \mu$.
\ei 
A relation $\mu$, satisfying these conditions, 
is called a {\bf bisimulation (BS)} 
between $P_1$ and $P_2$. 

\section{Algebraic properties of strong 
equivalence} \label{dopsvojstva}

\refstepcounter{theorem}
{\bf Theorem \arabic{theorem}\label{th6}}.

Strong equivalence is a congruence, 
i.e.,  if $P_1 \sim P_2$, then
\bi
\i for each $a\in Act \quad
   a.P_1\sim a.P_2$
\i for each process $P\quad
   P_1+P\sim P_2+P$
\i for each process $P\quad
   P_1|P\sim P_2|P$
\i for each $L\subseteq Names\quad
   P_1\setminus L\sim
   P_2\setminus L$
\i for each renaming $f\quad
   P_1[f] \sim
   P_2[f]$
\ei

{\bf Proof.}

As it was stated in section \ref{bimoodeer}, 
the statement 
$$P_1 \sim P_2$$ is equivalent to 
the statement that there is a BS 
$\mu$ between $P_1$ and $P_2$. 
Using this $\mu$, we construct a BS 
for justification of each of the foregoing 
relationships. 
\bi 
\i Let $s^0_{(1)}$ 
                     and $s^0_{(2)}$ 
   be initial 
   states of the processes
   $a.P_1$ and $a.P_2$ respectively. 

   Then the relation
   $$\{(s^0_{(1)}, s^0_{(2)})\} 
   \;\cup \;\mu$$ 
   is a 
   BS between $a.P_1$ and $a.P_2$. 
\i Let 
   \bi \i $s^0_{(1)}$ and $s^0_{(2)}$ be
   initial states of $P_1+P$ and $P_2+P$
   respectively, and 
   \i $S$ be a set of states of the process $P$. 
   \ei 
   Then \bi \i the relation 
   $$\{(s^0_{(1)}, s^0_{(2)})\} \;\cup \;
   \mu\;\cup \;
   Id_{S}
   $$ 
   is a BS 
   between $P_1 + P$ and $P_2 + P$, 
   and \i the relation 
   $$\{((s_1,s),(s_2,s))\mid (s_1,s_2)\in 
\mu,\;q\in S\}$$
   is a BS between 
   $P_1 | P$ and $P_2 | P$. \ei 
\i The relation $\mu$ is a BS 
   \bi 
   \i between $P_1 \setminus L$ 
      and $P_2 \setminus L$, and 
   \i between $P_1 [f]$ and 
     $P_2 [f]$. $ \quad \blackbox$ 
   \ei 
\ei 

\refstepcounter{theorem}
{\bf Theorem \arabic{theorem}\label
{thidempo}}.

Each process 
$P = (S, s^0, R)$ 
has the following properties. 

\bn
\i $P + {\bf 0} \sim P$
\i $P + P \sim P$
\en

{\bf Proof.}

\bn 
\i Let
   $s^0_0$ be an initial state 
   of the process $P + {\bf 0}$. 

Then the relation  
$$\{(s^0_0, s^0)\} \;\cup\; Id_S$$
is a BS between $P + {\bf 0}$ and $P$.

\i By definition of the operation
   ``$+$'', processes in the left side of the
   statement $P + P \sim P$
   should be considered as two disjoint 
   isomorphic copies of $P$ 
   of the form
      $$P_{(i)}=(S_{(i)}, s^0_{(i)}, R_{(i)})
      \quad(i=1,2)$$
   where $S_{(i)} = \{s_{(i)}\mid s\in S\}$.

Let  $s^0_0$ be an initial state 
of the process $P + P$. 

Then the relation
$$\{(s^0_0, s^0)\}\;\cup\;
\{(s_{(i)}, s)\mid s\in S,\; i=1,2\}$$
is a BS between $P + P$ and $P$.
$\blackbox$

\en 

Below for 
\bi 
\i each process $P = (S, s^0, R)$, 
and \i each state $s \in S$ 
\ei 
we denote by $P(s)$ 
the process $(S, s, R)$, 
which is obtained from $P$ 
by a replacement of an initial state. 
\\ 

\refstepcounter{theorem}
{\bf Theorem \arabic{theorem}\label
{thidempo1}}.

Let $P = (S, s^0, R)$ be a process, 
and a set of all its transitions, starting 
from $s^0$, has the form 
   $$\{\diagrw{s^0&\pright{a_i}&s^i}
\mid i=1,\ldots, n\}$$
   Then 
   \be{ptreigjerk}P\; \sim\;
   a_1. P_1 + \ldots + a_n. P_n\ee
   where for each $i=1,\ldots, n$ 
$$P_i\;\eam\;
   P(s^i)
   \;\eam\;(S, s^i,R)$$

{\bf Proof.}

\re{ptreigjerk} holds because there is 
a BS between left and right sides 
of \re{ptreigjerk}.

For a construction of this BS 
we replace all the processes 
$P_i$ in the right side of \re{ptreigjerk}
on their disjoint copies, 
i.e. we can consider that 
for each $i = 1, \ldots, n$ 
\bi
\i the process $P_i$ has the form
   $$P_i=(S_{(i)}, s^i_{(i)},R_{(i)})$$
   where all the sets $S_{(1)}, \ldots, S_{(n)}$
   are disjoint, and
\i a corresponding bijection 
   between $S$  and $S_{(i)}$ 
   maps each state $s \in S$ 
   to a state, denoted by the symbol 
   $s_{(i)}$. 
\ei

   Thus, we can assume that 
   each summand $a_i.P_i$ in
   the right side of \re{ptreigjerk}
   has the form
   $$\by
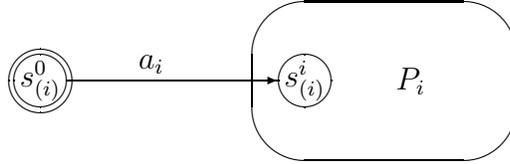
\begin{picture}(150,60)

\put(0,30){\oval(20,20)}
\put(0,30){\oval(24,24)}
\put(0,30){\makebox(0,0){$s^0_{(i)}$}}

\put(10,30){\vector(1,0){80}}

\put(40,33){\makebox(0,0)[b]{
$a_i$}}

\put(130,30){\oval(100,60)}
\put(140,30){\makebox(0,0){$P_i$}}

\put(100,30){\oval(20,20)}
\put(100,30){\makebox(0,0){$s^i_{(i)}$}}

\end{picture}
\ey
$$
and sets of states of these summands 
are pairwise disjoint. 

According to the definition of the operation $+$, 
the right side of \re{ptreigjerk} has the form 

$$\by
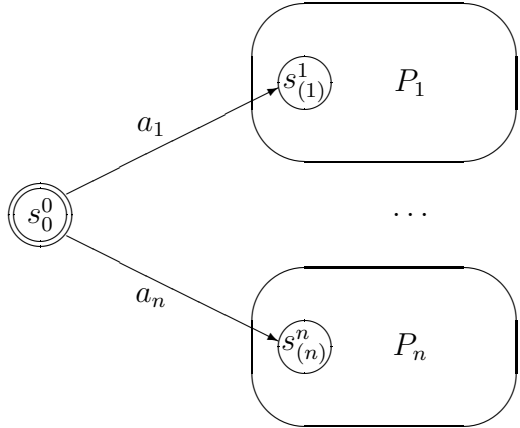
\begin{picture}(150,160)

\put(0,80){\oval(20,20)}
\put(0,80){\oval(24,24)}
\put(0,80){\makebox(0,0){$s^0_0$}}

\put(10,88){\vector(2,1){80}}
\put(10,72){\vector(2,-1){80}}

\put(40,110){\makebox(0,0)[b]{
$a_1$}}

\put(40,52){\makebox(0,0)[t]{
$a_n$}}

\put(130,130){\oval(100,60)}
\put(140,130){\makebox(0,0){$P_1$}}

\put(140,80){\makebox(0,0){$\ldots$}}

\put(100,130){\oval(20,20)}
\put(100,130){\makebox(0,0){$s^1_{(1)}$}}

\put(130,30){\oval(100,60)}
\put(140,30){\makebox(0,0){$P_n$}}

\put(100,30){\oval(20,20)}
\put(100,30){\makebox(0,0){$s^n_{(n)}$}}

\end{picture}
\ey
$$

BS between left and right sides 
of \re{ptreigjerk} has be defined, for example,
as the relation
$$\{(s^0, s^0_0) \} \; \cup \; 
\{(s, s_{(i)}) \mid s \in S, \; i = 1, \ldots, n \}
\quad \blackbox $$ 

\refstepcounter{theorem}
{\bf Theorem \arabic{theorem}\label{thidempo2}
(expansion theorem)}.

Let $P$ be a process of the form
\be{archttter}P\;=\;P_1\pa\ldots\pa P_n\ee
where for each $i \in \{1,\ldots, n\}$
the process $P_i$ has the form 
\be{sfgvsdgklsdfjkgsd}P_i = \sum\limits_{j=1}^{n_i}a_{ij}.\;P_{ij}\ee

Then $P$ is strongly equivalent to a sum of 
\been 
\i all processes of the form 
 \be{ssllaagg1}a_{ij}.\;\Big(
 P_1 \pa \ldots \pa P_{i-1}\pa
 P_{ij} \pa
 P_{i+1}\pa\ldots\pa P_{n}\Big)\ee
\i and all processes of the form 
 \be{ssllaagg2}\tau.\left(\by
 P_{1} \pa \ldots \pa P_{i-1} \pa
 P_{ik}  \pa
 P_{i+1} \pa \ldots \\ \ldots \pa P_{j-1}\pa
 P_{jl}  \pa
 P_{j+1} \pa \ldots \pa P_{n}
 \ey \right)
 \ee
 where $1\leq i < j \leq n,\quad
 a_{ik}, a_{jl} \neq \tau$, and 
 $a_{ik} = \overline{a_{jl}}$. 
\enen 

{\bf Proof.}

By theorem \ref{thidempo1}, 
$P$ is strongly equivalent to a sum, 
each summand of which 
corresponds to a transition
starting from the initial state $s^0$ of 
the process $P$.
For each transition of $P$ of the form
$$\diagrw{s^0&\pright{a}&s}$$
this sum contains the summand
$a.P(s)$.

According to \re{sfgvsdgklsdfjkgsd}, 
for each $i=1,\ldots, n$
the process $P_i$ has the form 
$$\by
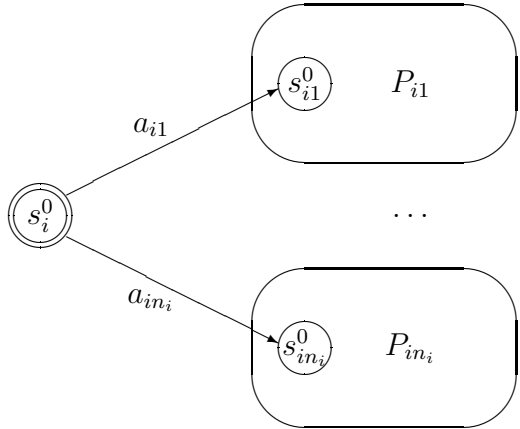
\begin{picture}(150,160)

\put(0,80){\oval(20,20)}
\put(0,80){\oval(24,24)}
\put(0,80){\makebox(0,0){$s^0_i$}}

\put(10,88){\vector(2,1){80}}
\put(10,72){\vector(2,-1){80}}

\put(40,110){\makebox(0,0)[b]{
$a_{i1}$}}

\put(40,52){\makebox(0,0)[t]{
$a_{in_i}$}}

\put(130,130){\oval(100,60)}
\put(140,130){\makebox(0,0){$P_{i1}$}}

\put(140,80){\makebox(0,0){$\ldots$}}

\put(100,130){\oval(20,20)}
\put(100,130){\makebox(0,0){$s^0_{i1}$}}

\put(130,30){\oval(100,60)}
\put(140,30){\makebox(0,0){$P_{in_i}$}}

\put(100,30){\oval(20,20)}
\put(100,30){\makebox(0,0){$s^0_{in_i}$}}

\end{picture}
\ey
$$
where $s_i^0, s^0_{i1},\ldots, s^0_{in_i}$ 
are initial states of the processes 
$$P_i,\;\; P_{i1},\;\;\ldots, \;\;P_{in_i}$$ 
respectively. 

Let
\bi 
\i $S_i$ be a set
   of states of the process $P_i$, and 
\i $S_{ij}$ (where $j=1,\ldots, n_i$) 
   be a set of states of the process $P_{ij}$. 
\ei 
We can assume that $S_i$ 
is a disjoint union of the form
\be{fsdgwegere34545}S_i = \{s_i^0\}\;\cup\; 
S_{i1}\;\cup \;\ldots \;\cup\;S_{in_i}\ee

According to the description of a process
of the form \re{archttter}, 
which is presented in item 2 of section 
\ref{archnatella}, 
we can assume that components of $P$ 
have the following form. 
\bi 
\i A set of states of the process $P$ 
   has the form
   \be{fdsdfgfsdfsdgsdgwegere34545}
   S_1\times \ldots \times S_n\ee
\i An initial state $s^0$ of $P$ is a list 
   $$(s^0_1,\ldots, s^0_n)$$
\i Transitions of $P$, starting from
    its initial state, are as follows. 
\bi 
\i Transitions of the form 
   \be{perehh1}\diagrw{s^0&\pright{a_{ij}}&
(s^0_1,\ldots, s^0_{i-1}, s^0_{ij},
   s^0_{i+1},\ldots, s^0_n)}\ee
\i Transitions of the form 
   \be{perehh2}\!\!\!\!\!\diagrw{s^0&\pright{\tau}&
   \left(\by s^0_1,\ldots, s^0_{i-1}, 
   s^0_{ik}, 
   s^0_{i+1},\ldots \\ \ldots
   s^0_{j-1}, s^0_{jl}, s^0_{j+1},\ldots,
   s^0_n\ey\right)}\ee
   where $1\leq i < j \leq n,\quad
   a_{ik}, a_{jl} \neq \tau$, and 
   $a_{ik} = \overline{a_{jl}}$. 
\ei 
\ei 
Thus, there is an one-to-one correspondence 
between
\bi
\i the set of transitions of the process $P$, 
   starting from $s^0$, and
\i the set of summands of the form 
   \re{ssllaagg1} and \re{ssllaagg2}. 
\ei

For the proof of theorem \ref{thidempo2}
it is enough to prove that 
\bi 
\i For each $i=1,\ldots, n$, and each 
    $j=1,\ldots, n_i$ 
   the following   equivalence holds:
   \be{fgksfldjgdfklwer}\by
   P(s^0_1,\ldots, s^0_{i-1}, s^0_{ij},
   s^0_{i+1},\ldots, s^0_n)
   \sim\\\sim
   \Big(
   P_1 \pa \ldots \pa P_{i-1}\pa
	P_{ij} \pa
   P_{i+1}\pa\ldots\pa P_{n}\Big)
   \ey\ee
\i for \bi \i any $i, j$, such that 
   $1\leq i<j\leq n$, 
   and \i any $k=1,\ldots, n_i,\;\;
   l=1,\ldots, n_j$ \ei 
   the following equivalence holds:
   \be{fgksfldjgdfklwer1}
   \by
   P
   \left(\by s^0_1,\ldots, s^0_{i-1}, s^0_{ik}, 
s^0_{i+1},\ldots \\ \ldots
   s^0_{j-1}, s^0_{jl}, s^0_{j+1},\ldots,
   s^0_n\ey\right)	  \sim \\ \sim
   \left(\by
   P_{1} \pa \ldots \pa P_{i-1} \pa
 P_{ik}  \pa
 P_{i+1} \pa \ldots \\ \ldots \pa P_{j-1}\pa
 P_{jl}  \pa
 P_{j+1} \pa \ldots \pa P_{n}
 \ey \right)
\ey\ee
\ei 

We shall prove only \re{fgksfldjgdfklwer}
(\re{fgksfldjgdfklwer1} can be proven similarly). 

A set of states of the process 
\be{syyrrttwewwfffff}\Big(
   P_1 \pa \ldots \pa P_{i-1}\pa
	P_{ij} \pa
   P_{i+1}\pa\ldots\pa P_{n}\Big)\ee
has the form
\be{fdfsdgsdgwegere34545}S_1\times \ldots 
\times
S_{i-1}\times S_{ij}\times S_{i+1}\times 
\ldots \times S_n\ee

\re{fsdgwegere34545} implies that 
$S_{ij}\subseteq S_i$, 
i.e. set \re{fdfsdgsdgwegere34545}
is a subset of set
\re{fdsdfgfsdfsdgsdgwegere34545}
of states of the process 
\be{syyrr1ttwewwfffff}
P(s^0_1,\ldots, s^0_{i-1}, s^0_{ij},
   s^0_{i+1},\ldots, s^0_n)\ee

We define the desired BS $\mu$ 
between processes 
\re{syyrrttwewwfffff} and 
\re{syyrr1ttwewwfffff}
as the diagonal relation 
$$\mu\eam\{(s,s)\mid s\in \re
{fdfsdgsdgwegere34545}\}$$
Obviously, 
\bi 
\i a pair of initial states of processes 
   \re{syyrrttwewwfffff} and 
   \re{syyrr1ttwewwfffff}
   belongs to $\mu$, 
\i each transition 
   of the process
   \re{syyrrttwewwfffff} is also a transition 
   of the process \re{syyrr1ttwewwfffff}, and 
\i if a start 
   of some transition of the process 
   \re{syyrr1ttwewwfffff} belongs to 
   the subset \re{fdfsdgsdgwegere34545}, 
   then the end of this transition also belongs 
   to the subset \re{fdfsdgsdgwegere34545}\\
   (to substantiate this claim we note that 
  for each transition of $P_i$, if its start
   belongs to $S_{ij}$, then its end
   also belongs to $S_{ij}$). 
\ei 
Thus, $\mu$ is a BS, and this proves 
the claim \re{fgksfldjgdfklwer}. 
$\blackbox $ \\ 

The following theorem 
is a strengthening 
of theorem \ref{thidempo2}. 
To formulate it, we will use the following 

notation. 
If $f: Names\to Names$ is a renaming, 
then the symbol $f$ denotes also a mapping 
of the form $$f: Act \to Act$$ 
defined as follows. 
\bi
\i $\forall\; \alpha\in Names\quad f(\alpha!) 
\eam f(\alpha)!$, 
   $f(\alpha?) \eam f(\alpha)?$
\i $f(\tau)\eam \tau$
\ei

\refstepcounter{theorem}
{\bf Theorem \arabic{theorem}\label{thidempo34}}.

Let $P$ be a process of the form 
$$P\;=\;\Big(P_1[f_1]\pa\ldots\pa P_n[f_n]\Big)
\setminus L$$
where for each $i \in \{1,\ldots, n\}$
$$P_i \sim \sum\limits_{j=1}^{n_i}a_{ij}.\;P_{ij}$$

Then $P$ is strongly equivalent to a sum of
\been 
\i all processes of the form 
 $$f_i(a_{ij}).\;\left(\by\left(\by
 P_1[f_1] \pa \ldots \\ \ldots \pa
 P_{i-1}[f_{i-1}]\pa
 P_{ij}[f_i] \pa
 P_{i+1}[f_{i+1}]\pa\ldots \\ \ldots
 \pa P_{n}[f_n]\ey\right)\setminus L\ey\right)
$$
 where $a_{ij}=\tau$ or 
 $name(f_i(a_{ij}))\not\in L$, and 
\i all processes of the form 
  $$\tau.\left(\by \left(\by
 P_{1}[f_1] \pa \ldots \\ \ldots
  \pa P_{i-1}[f_{i-1}] \pa
 P_{ik}[f_{i}]	\pa
 P_{i+1}[f_{i+1}] \pa \ldots \\ 
\ldots \pa P_{j-1}[f_{j-1}]
 \pa
 P_{jl}[f_{j}]	\pa
 P_{j+1}[f_{j+1}] \pa \ldots\\ \ldots \pa P_{n}[f_{n}]
 \ey \right) \setminus L\ey \right)
 $$
 where 
  $1\leq i < j \leq n,\quad
 a_{ik},a_{jl}\neq \tau$, and 
 $f_i(a_{ik}) = \overline{f_j(a_{jl})}$. 
\enen 

{\bf Proof.}

This theorem follows directly from 
\bi 
\i the previous theorem, 
\i theorem \ref{th6}, 
\i properties  6, 9, 10, 16 and 17 
   from section \ref{archnatella}, and 
\i the first assertion from theorem 
   \ref{thidempo}. 
\ei 

\section{Recognition of strong equivalence}

\label{raspozn11}

\subsection{Relation $\mu(P_1, P_2)$}

Let $P_1, P_2$ be a couple of processes 
of the form
$$P_i=(S_i, s^0_i, R_i)\qquad(i=1,2)$$

Define an operator $'$
on the set of all relations from 
$S_1$ to $S_2$, 
that maps each relation 
$\mu\subseteq S_1\times S_2$
to the relation 
$\mu'\subseteq S_1\times S_2$, 
defined as follows: 
$$\!\!\!\by \mu'
\eam \bigset{\!\!\!\by (s_1,s_2)\in\\\in  S_1
\times S_2\ey}
{\forall a\in Act\\
\forall s'_1\in S_1: (s_1\ra{a}s'_1)\in R_1\\
\hspace{3mm}\exists s'_2\in S_2: \left\{\by 
(s_2\ra{a}s'_2)\in R_2\\
(s'_1,s'_2)\in\mu\ey\right.\\
\forall s'_2\in S_2: (s_2\ra{a}s'_2)\in R_2\\
\hspace{3mm}\exists s'_1\in S_1: \left\{\by 
(s_1\ra{a}s'_1)\in R_1\\
(s'_1,s'_2)\in\mu\ey\right.}
\ey$$

It is easy to prove that 
for each $\mu\subseteq S_1\times S_2$
$$\by \mbox{$\mu$ 
satisfies conditions 1 and 2}\\
\mbox{from the definition of a BS}\ey
\quad\Leftrightarrow\quad
\mu\subseteq \mu'
$$
Consequently, 
$$\mbox{$\mu$ is a BS
between
$P_1$ and $P_2$
}\quad\Leftrightarrow\quad\left\{\by
(s^0_1,s^0_2)\in \mu\\
\mu\subseteq \mu'
\ey\right.
$$

It is easy to prove that the operator $'$ is 
monotone, i.e.
\begin{center}
if $\mu_1\subseteq \mu_2$,
then $\mu'_1\subseteq \mu'_2$.\end{center}

Let $\mu_{max}$ be a union 
of all relations from the set
\be{lfgsdougsdfghweui}
\{\mu\subseteq S_1\times S_2\mid 
\mu\subseteq \mu'\}\ee

Note that the relation $\mu_{max}$ 
belongs to the set
\re{lfgsdougsdfghweui}, 
since for every $\mu\in 
\re{lfgsdougsdfghweui}$ 
from
\bi 
\i the inclusion $\mu\subseteq 
(\bigcup\limits_{\mu\in
\re{lfgsdougsdfghweui}}\mu) = 
\mu_{max}$, and 
\i monotonicity of $'$ 
\ei 
it follows that for each $\mu\in
\re{lfgsdougsdfghweui}$
$$\mu\subseteq \mu'\subseteq
\mu'_{max}$$
So 
$\mu_{max}=\bigcup\limits_{\mu\in
\re{lfgsdougsdfghweui}}\mu\subseteq 
\mu'_{max}$, 
i.e. $\mu_{max} \in \re{lfgsdougsdfghweui}$. 

Note that the following equality holds
$$\mu_{max} = \mu'_{max}$$
because 
\bi\i the inclusion
$\mu_{max} \subseteq \mu'_{max}$, and 
\i  monotonicity of $'$ 
\ei imply 
the inclusion 
   $$\mu'_{max} \subseteq \mu''_{max}$$
   i.e. 
$\mu'_{max}\in \re{lfgsdougsdfghweui}$, 
whence, by virtue of maximality 
of $\mu_{max}$, 
we get the inclusion 
$$\mu'_{max} \subseteq \mu_{max}$$

Thus, the relation $\mu_{max}$ is 
\bi 
\i a greatest element 
   of the partially ordered set
   \re{lfgsdougsdfghweui}
   (where a partial order is the relation 
   of inclusion), and 
\i a greatest fixed point of the operator
   $'$. \ei 
We shall denote this relation by
\be{345dtoto}\mu(P_1,P_2)\ee

From theorem \ref{teorrrr} it follows 
that
$$P_1\sim P_2\quad\Leftrightarrow\quad
(s^0_1,s^0_2)\in \mu(P_1,P_2)$$

From the definition of the relation
$\mu(P_1,P_2)$ 
it follows that this relation consists of all pairs 
$(s_1,s_2)\in S_1\times S_2$, 
such that 
$$P_1(s_1)\sim P_2(s_2)$$

The relation 
$\mu(P_1,P_2)$ can be considered as 
a {\bf similarity measure} 
between $P_1$ and $P_2$. 

\subsection{A polynomial algorithm 
for recognizing of strong equivalence}
\label{raspozznnaavv}

Let $P_1$ and $P_2$ be processes of the form 
$$P_i=(S_i, s^0_i, R_i)\qquad(i=1,2)$$

If the sets $S_1$ and $S_2$ are finite, 
then the problem of checking of statement
\be{dfghsdfgsdfgwertwet}P_1\sim P_2\ee
obviously is algorithmically solvable: 
for example, you can iterate over all relations 
$\mu\subseteq S_1\times S_2$
and for each of them verify 
conditions 0, 1 and 2 
from the definition of BS. 
The algorithm finishes its work when 
\bi 
\i it is found a relation
   $\mu\subseteq S_1\times S_2$
   which satisfies 
   conditions of 0, 1 and 2 
   from the definition of BS, 
   in this case the algorithm
   gives the answer 
   $$P_1 \sim P_2$$ 
   or 
\i all relations 
   $\mu\subseteq S_1 \times S_2$ 
   are checked, 
   and none of them satisfy
   conditions of 0, 1 and 2 
   from the definition of BS.
   In this case, the algorithm 
   gives the answer 
   $$P_1 \not \sim P_2$$ 
\ei 

If $P_1 \not \sim P_2$, 
then the above algorithm 
will give the answer 
after checking of all relations 
from $S_1$ to $S_2$, 
the number of which is 
$$2^{|S_1|\cdot |S_2|}$$
(where for every finite set $S$
we denote by $|S|$
a number of elements of $S$),
i.e. this algorithm 
has exponential complexity. 

The problem of checking $P_1\sim P_2$
can be solved by more efficient 

algorithm, which has polynomial complexity. 
To construct such an algorithm, we consider 
the following sequence of relations from
$S_1$ to $S_2$:
\be{4twregsdfg46}
\{\mu_i\mid i\geq 1\}\ee
where $\mu_1\eam S_1\times S_2$, 
and $\forall \;i\geq 1\quad \mu_{i+1}\eam 
\mu_i'$. 

From \bi\i the inclusion 
$\mu_1\supseteq\mu_2$, and 

\i the monotonicity of the operator
$'$\ei 
it follows that 
$$\by
\mu_2=\mu'_1\supseteq\mu'_2=\mu_3\\
\mu_3=\mu'_2\supseteq\mu'_3=\mu_4\\
\mbox{etc.}\ey$$
Thus, the sequence \re{4twregsdfg46}  is 
monotone: 
$$\mu_1\supseteq\mu_2\supseteq\ldots$$
Since all members of sequence 
\re{4twregsdfg46}
are subsets of the finite set $S_1 \times S_2$, 
then this sequence can not decrease infinitely, 
it will be stabilized at some member, 
i.e. there is an index $i \geq 1$, such tha
$$
\mu_i=\mu_{i+1}=\mu_{i+2}=\ldots$$
We prove that the relation $\mu_i$
(where $i$ is the above index)
coincides with the relation $\mu(P_1,P_2)$. 
\bi 
\i Since $\mu_i=\mu_{i+1}=\mu_i'$, 
  i.e. $\mu_i$ is a fixed 
   point of the operator $'$, then 
   \be{fgsdfwert3434}\mu_i\subseteq \mu
(P_1,P_2)\ee
   since $\mu(P_1,P_2)$ is the largest 
   fixed point of the operator $'$. 
\i For each $j\geq 1$ 
   the inclusion 
   \be{dfgsdfg456545464}\mu(P_1,P_2)
   \subseteq \mu_j\ee 
   holds, 
   because 
   \bi 
   \i inclusion \re{dfgsdfg456545464}
      holds for $j = 1$, and 
   \i if inclusion \re{dfgsdfg456545464} holds
      for some $j$, then on the reason of 
      monotonicity of the operator $'$, 
      the following equalities hold: 
      $$\mu(P_1,P_2) =
      \mu(P_1,P_2)'\subseteq \mu'_j = 
      \mu_{j+1}$$
      i.e. inclusion
      \re{dfgsdfg456545464}
      holds for $j+1$.
   \ei 
   In particular, \re{dfgsdfg456545464} 
   holds for $j = i$. 
\ei 
The equality 
\be{fgsdfgsdfwert3434}\mu_i = \mu(P_1,P_2)
\ee
follows from \re{fgsdfwert3434} and 
\re{dfgsdfg456545464} for $j = i$.

Thus, the problem of checking of the 
statement $P_1\sim P_2$
can be solved by 
\bi 
\i finding a first member 
   $\mu_i$ of sequence 
   \re{4twregsdfg46}, 
   which satisfies the    condition 
   $\mu_i = \mu_{i+1}$, and 
\i checking the condition
   \be{fgsdgwet5454436}(s_1^0,s_2^0)\in 
   \mu_i\ee
\ei 
The algorithm gives the answer 
$$P_1\sim P_2$$
if and only if \re{fgsdgwet5454436} holds. 

For a calculation of terms of the sequence 
\re{4twregsdfg46} the following algorithm 
can be used.
This algorithm computes a relation $\mu'$
for a given relation 
$\mu\subseteq S_1\times S_2$.
$$
\by
\mu':=\emptyset\\
\mbox{{\bf loop for each } } (s_1,s_2)\in 
\mu\\
\left|\by
\mbox{include } := \top\\
\!\!\!	 \by
   \mbox{{\bf loop for each } } s'_1, 
a:\;\diagrw{s_1&\pright{a}&s'_1}\\
   \left|\by
   \mbox{found } := \bot\\
   \!\!\!	\by
	  \mbox{{\bf loop for each } } 
s'_2:\;\diagrw{s_2&\pright{a}&s'_2}\\
	  \left|\by
	  \mbox{found } :=  \mbox{found } 
\vee (s'_1,s'_2)\in \mu\\
	  \ey\right.\\
	  \mbox{{\bf end of loop}}
	  \ey\\
   \mbox{include } := \mbox{include } \wedge 
\mbox{found }
   \ey\right.\\
   \mbox{{\bf end of loop}}
   \ey	\\
\!\!\!	 \by
   \mbox{{\bf loop for each } } s'_2, 
a:\;\diagrw{s_2&\pright{a}&s'_2}\\
   \left|\by
   \mbox{found } := \bot\\
  \!\!\!	\by
	  \mbox{{\bf loop for each } } 
s'_1:\;\diagrw{s_1&\pright{a}&s'_1}\\
	  \left|\by
	  \mbox{found } :=  \mbox{found } 
\vee (s'_1,s'_2)\in \mu\\
	  \ey\right.\\
	  \mbox{{\bf end of loop}}
	  \ey\\
   \mbox{include } := \mbox{include } \wedge 
\mbox{found }
   \ey\right.\\
   \mbox{{\bf end of loop}}
   \ey\\
   \mbox{{\bf if } include {\bf then }} \mu':= 
\mu'\cup\{(s_1,s_2)\}
\ey\right.\\
\mbox{{\bf end of loop}}
\ey
$$

Note that this 
algorithm is correct only when 
$\mu'\subseteq \mu$
(which occurs 
in the case when this algorithm is used 
to calculate terms of the sequence 
\re{4twregsdfg46}). 
In a general situation the outer loop must 
have the form 
$$\mbox{{\bf loop for each } } 
(s_1,s_2)\in S_1\times S_2$$

Estimate a complexity of the algorithm. 

Let $A$ be the number
$$\max(|Act(P_1)|, |Act(P_2)|)+1$$
\bi 
\i The outer loop does no more than $|S_1| 
\cdot |S_2|$ iterations. 
\i Both loops contained in the external loop
   make max 
   $|S_1| \cdot |S_2| \cdot A$
   iterations. 
\ei 
Therefore, a complexity of this algorithm can 
be evaluated as
$$O(|S_1|^2 \cdot |S_2|^2 \cdot A)$$

Since for a calculation 
of a member $\mu_i$
of sequence \re{4twregsdfg46}, 
on which \re{4twregsdfg46} is stabilized, 
we must calculate not more than 
$|S_1|\cdot|S_2|$ members of 
this sequence, 
then, consequently, the desired relation 
$\mu_i=\mu(P_1,P_2)$
can be calculated during 
$$O(|S_1|^3 \cdot |S_2|^3 \cdot A)$$

\section{Minimization of processes} 
\label{minimi11}

\subsection{Properties of  relations 
of the form $\mu(P,P)$}

\refstepcounter{theorem}
{\bf Theorem \arabic{theorem}\label{th1}}.

For each process $P \eam (S, s^0, R)$
the relation $\mu(P, P)$ is an equivalence. \\ 

{\bf Proof}. 
\bn 
\i {\bf Reflexivity} of the relation 
   $\mu(P, P) $ 
   follows from the fact 
   that the diagonal relation 
   $$Id_S=\{(s,s)\mid s\in S\}$$
   satisfy conditions 1 and 2 
   from the definition of BS, i.e. 
   \begin{center}
   $Id_S\in \re{lfgsdougsdfghweui}$.
   \end{center}
\i {\bf  Symmetry} of the relation 
   $\mu(P, P)$ 
   follows from the fact that 
   if  a relation $\mu$ 
   satisfies conditions 1 and 2 
   from the definition of BS, then 
   the inverse relation 
   $\mu^{-1}$ 
   also satisfies these conditions, 
   that is, 
  \begin{center}
   if $\mu\in \re{lfgsdougsdfghweui}$, then
   $\mu^{-1} \in \re{lfgsdougsdfghweui}$.
   \end{center}
\i {\bf Transitivity} of the relation 
   $\mu(P, P)$ 
   follows from the fact that 
   the product $$\mu(P, P)\circ \mu(P, P)$$
   satisfies conditions 1 and 2 from
   the definition of BS, i.e. 
   $$\mu(P, P)\circ \mu(P, P)\subseteq
   \mu(P, P)\qquad\blackbox$$
\en 

Let $P_{\sim}$ be a process, whose 
components have the following form. 
\bi 
\i Its states are
   equivalence classes of the set $S$ 
   of states of $P$, 
   corresponding to the equivalence 
   $\mu(P, P)$. 
\i Its initial state is the class $[s^0]$, which 
   contains the initial state $s^0$ of $P$. 
\i A set of its transitions 
   consists of all transitions of the form 
   $$\diagrw{[s_1]&\pright{a}&[s_2]}$$
   where $\diagrw{s_1&\pright{a}&s_2}$ 
   is an arbitrary transition from $R$. 
\ei 
The process $P_{\sim}$ is said to be
a {\bf factor-process}
of the process $P$ with respect to 
the equivalence $\mu(P, P) $. \\ 

\refstepcounter{theorem}
{\bf Theorem \arabic{theorem}\label{th2}}.

For each process $P$ the relation
$$\mu\eam\{\;(s, [s])\;\mid s\in S\}$$
is BS between $P$ and $P_\sim $. \\ 

{\bf Proof.}

Check  
the properties 0, 1, 2
from the definition of BS
for the relation $\mu$. 

Property 0 holds by definition 
of an initial state of the process $P_\sim$. 

Property 1 holds by definition 
of a  set of transitions of $P_\sim $. 

Let us prove property 2. 
Let $P_\sim $ contains a transition 
$$\diagrw{[s]&\pright{a}&[s']}$$

Prove that there is a transition in $R$ of 
the form 
$$\diagrw{s&\pright{a}&s''}$$
such that $(s'',[s'])\in \mu$, i.e. 
$[s'']=[s']$, i.e. 
$$(s'',s')\in \mu(P, P)$$

From the definition of a set of transitions
of the process $P_\sim$ 
it follows that $R$ 
contains a transition of the form 
\be{sdfysdklg34rtwrer}\diagrw{s_1&\pright{a}&s_1'}\ee
where $[s_1]=[s]$ and $[s_1']=[s']$, i.e. 
$$\by (s_1, s)\in \mu(P, P)\quad\mbox{and}\\
	(s_1', s')\in \mu(P, P)\ey$$

Since $\mu(P, P)$ is a BS, then 
from
\bi 
\i $\re{sdfysdklg34rtwrer} \in R$, and 
\i $(s_1,s)\in \mu(P, P)$ 
\ei 
it follows that $R$ 
contains a transition of the form 
\be{sdfy4sdklg34rtwrrer}\diagrw{s&\pright{a}&s_1''}\ee
where $(s_1'', s_1')\in \mu(P, P)$. 

Since $\mu(P, P)$ is transitive, then 
from 
$$\by (s_1'',s_1')\in \mu(P, P)\quad\mbox{and}\\(s_1', s')\in \mu(P, P)\ey$$
it follows that
$$(s_1'', s')\in \mu(P, P)$$

Thus, as the desired state $s''$ it can taken 
the state $s_1''$. 
$ \blackbox $ \\ 

From theorem \ref{th2} it follows that 
for each process $P$ $$P \sim P_\sim$$ 

\subsection{Minimal processes 
with respect to $\sim$}

A process $P$ is said to be
{\bf minimal with respect to $\sim$}, if 
\bi 
\i each its state is reachable, and 
\i $\mu(P, P) = Id_S$ \\ 
  (where $S$ is a set of states of $P$). 
\ei 
Below minimal processes with respect to  
$\sim$ are called simply 
{\bf  minimal processes}. \\ 

\refstepcounter{theorem}
{\bf Theorem \arabic{theorem}\label{th3}}.

Let the processes $P_1$ and $P_2$ 
minimal, and 
$P_1 \sim P_2$. 

Then $P_1$ and $P_2$ are isomorphic. \\ 

{\bf Proof.} 

Suppose that $P_i\;(i=1,2)$ 
has the form 
$(S_i, s_i^0, R_i)$, 
and let 
$\mu\subseteq S_1\times S_2$ be 
BS between $P_1$ and $P_2$. 

Since $\mu^(-1)$ is also BS, 
and composition of BSs is BS, 
then
\bi 
\i $\mu\circ \mu^{-1}$ is BS between $P_1$ 

   and $P_1$ , and
\i $\mu^{-1}\circ \mu$ is BS between $P_2$ 
and $P_2$ 
\ei 
whence, using definition of the relations 
$\mu(P_i, P_i)$, 
and the definition of a minimal process, 
we get the inclusions 
\be{gsdfghsety544673434}
\by \mu\circ \mu^{-1} \subseteq \mu(P_1, 
P_1)= Id_{S_1}\\
  \mu^{-1}\circ \mu \subseteq \mu(P_2, P_2)= 
Id_{S_2}\ey\ee

Prove that the relation $\mu$ is functional, 
i.e. for each $s\in S_1$ there is a unique 
element $s '\in S_2$, such that 
$(s, s') \in\mu$. 
\bi 
\i If $s = s_1^0$, then we define
   $s'\eam s_2^0$. 
\i If $s\neq s_1^0$ then, 
   since every state in $P_1$ is
   reachable, then there is a path in $P_1$
   of the form
$$\diagrw{s_1^0&\pright{a_1}&\ldots&\pright{a_n}&s}$$

Since $\mu$ is BS, 
then there is a path in $P_2$ of the form
$$\diagrw{s_2^0&\pright{a_1}&\ldots&\pright{a_n}&s'}$$
and $(s, s') \in\mu$. 
\ei 
Thus, in both cases there is an element 
$s'\in S_2$, such that $(s, s') \in\mu$. 

Let us prove the uniqueness of the element 
$s'$ with the property $(s, s') \in\mu$. 

If there is an element $s''\in S_2$, such that
$(s,s'')\in \mu$, 
then $(s'',s)\in \mu^{-1}$, 
which implies 
$$(s'',s')\in \mu^{-1}\circ \mu = Id_{S_2}$$
so $s''=s'$. 

For similar reasons, 
the relation $\mu^{-1}$ is also functional. 

From conditions \re{gsdfghsety544673434}
it is easy to deduce bijectivity 
of the mapping, which corresponds to 
the relation $\mu$. 
By the definition of BS, this implies that
$P_1$ and $P_2$ are isomorphic. 
$ \blackbox $ \\ 

\refstepcounter{theorem}
{\bf Theorem \arabic{theorem}\label{th4}}.

Let \bi \i a process $P_2$ is obtained from 
a process $P_1$ by removing of unreachable 
states, and \i $P_3\eam(P_2)_\sim$.\ei
Then the process $P_3$ is minimal, and 
$$P_1\sim P_2\sim P_3$$

{\bf Proof}. 

Since each state of $P_2$ is reachable, 
then from the definition of transitions 
of a factor-process, 
it follows that each state of $P_3$ 
is also achievable. 

Now, we prove that 
\be{grfdgsdfg4w3t54w5}
\mu(P_3, P_3)=Id_{S_3}\ee
i.e. suppose that
$(s',s'')\in \mu(P_3, P_3)$, 
and prove that $s'=s''$. 

From the definition of a factor-process 
it follows that there are states 
$s_1,s_2\in S_2$, 
such that 
$$\by s'&=&[s_1] \\ s''&=&[s_2]\ey$$
where $[\cdot]$ 
denotes an equivalence class 
with respect to $\mu(P_2, P_2)$. 

From theorem \ref{th2} it follows that
$$\by(s_1, s')&\in& \mu(P_2, P_3)\\
(s'',s_2)&\in& \mu(P_3, P_2)\ey$$

Since a composition of BSs is also BS, 
then the composition 
\be{fgsdfkjlewrkteyt}
\mu(P_2, P_3)\circ
\mu(P_3, P_3) \circ
\mu(P_3, P_2)
\ee
is BS between $P_2$ and $P_2$, 
so 
\be{rtgsdfgewg}\re{fgsdfkjlewrkteyt}
\subseteq
\mu(P_2, P_2)
\ee
Since 
$(s_1,s_2)\in \re{fgsdfkjlewrkteyt}$, 
then, in view of \re{rtgsdfgewg}, we get: 
$$s'=[s_1]=[s_2]=s''$$

In conclusion, we note that 
\bi \i the statement 
    $P_1 \sim P_2$ is obvious, and 
\i the statement $P_2 \sim P_3$ 
follows from theorem \ref{th2}. 
$\blackbox$ \ei 

\subsection{An algorithm for minimizing 
of finite processes}

The algorithm 
described in section \ref{raspozznnaavv}
can be used to solve the problem of 
{\bf minimizing of finite processes}, 
which has the following form:
for a given finite process $P$ build 
a process $Q$ with the smallest number 
of states, which is strongly equivalent to $P$. 

To build the process $Q$, 
first there is constructed a process $P'$, 
obtained from $P$ by 
removing of unreachable states. 
The process $Q$ has the form $P'_\sim$. 

A set of states of the process $P'$ 
can be constructed as follows. 
Let $P$ has the form 
$$P=(S,s^0, R)$$
Consider the 
sequence of subsets of the set $S$
\be{gfsdgsdfg34rt5}S_0\subseteq S_1
\subseteq S_2\subseteq \ldots\ee
defined as follows. 
\bi 
\i $S_0\eam \{s^0\}$
\i for each $i\geq 0$ the set $S_{i+1}$ 
   is obtained from $S_i$ by adding 
   all states $s'\in S$, such that 
   $$\exists s\in S, \;\exists a\in Act:\;
   (\diagrw{s&\pright{a}&s'})\;\in R$$
\ei 
Since $S$ is finite, 
then the sequence \re{gfsdgsdfg34rt5}
can not increase infinitely. 
Let $S_i$ be a member of the sequence 
\re{gfsdgsdfg34rt5}, 
where this sequence is stabilized. 
It is obvious that \bi \i all states from $S_i$ 
are reachable, and 
\i all states from $S\setminus S_i$ are
unreachable. \ei 
Therefore, a set of states of the process $P'$ 
is the set $S_i$. 

Let $S'$ be a set of states of the process $P'$. 

Note that for a computation of the relation
$\mu(P', P')$ it is necessary to 
calculate no more than $|S'|$ 
members of sequence \re{4twregsdfg46}, 
because
\bi\i each relation in
   the sequence \re{4twregsdfg46}
is an equivalence
(since if a binary relation $\mu$ 
on the set of states of a 
process is an equivalence, then
the relation $\mu'$ is also an equivalence),
and\i 
\bi
\i each member of the sequence 
\re{4twregsdfg46} 
defines a partitioning of the set $S'$, 
and \i for each $i\geq 1$, if 
$\mu_{i+1}\neq \mu_i$, then 
a partitioning corresponding to 
$\mu_{i+1}$
is a refinement of a partitioning 
corresponding to $\mu_i$, \ei
and it is easy to show that 
a number of such refinements 
is no more than $|S'|$.\ei

\refstepcounter{theorem}
{\bf Theorem \arabic{theorem}\label{th9}}.

The process $P'_\sim $ 
has the smallest number of states among all 
finite processes 
that are strongly equivalent to $P$. \\ 

{\bf Proof}. 

Let \bi\i $P_1$ be a finite process, 
such that $P_1 \sim P$, and
\i $P'_1$ be a reachable part of $P_1$.\ei
As it was established above, 
$$P_1\sim P'_1 \sim (P'_1)_\sim$$
Since $P\sim P' \sim P'_\sim$
and $P\sim P_1$, then, consequently, 
\be{sdfasdfwrf}P'_\sim\sim (P'_1)_\sim\ee

As it was proved in theorem \ref{th4}, 
the processes 
$P'_\sim$ and $(P'_1)_\sim$
are minimal. 
From this and from 
\re{sdfasdfwrf}, by virtue of 
theorem \ref{th3} we get that the processes 
$P'_\sim$ and $(P'_1)_\sim$ are isomorphic. 
In particular, they have same number of states. 

Since 
\bi 
\i a number of states of the process 
   $(P'_1)_\sim$
   does not exceed a number 
  of states of the process 
   $P'_1$ (since states of the process 
   $(P'_1)_\sim$ are classes of 
   a partitioning of the 
   set of states of the process $P'_1$), and 
\i a number of states the process $P'_1$ 
   does not exceed a number of states of the process 
   $P_1$ (since a set of states 
   of the process $P'_1$ is a subset of a
   set of states of the process $P_1$)
\ei 
then, consequently, a number of states 
of the process 
$P'_\sim $ 
does not exceed a number 
of states of the process 
$P_1$. $\blackbox$ 

\section{Observational equivalence}
\label{defequisdfgval43}

\subsection{Definition of observational 
equivalence} \label{defobsequiv}

Another variant of the concept 
of equivalence of processes
is {\bf observational equivalence}. 
This concept is used in those 
situations where we 
consider the internal action 
$\tau$ as negligible, 
and consider two traces as the same, 
if one of them 
can be obtained from another by insertions 
and/or deletions of  internal actions 
$\tau$. 

For a definition of the concept of 
observable equivalence 
we introduce auxiliary notations. 

Let $P$ and $P '$ be processes. 

\bn 
\i The notation
   \be{naeghfgdrtryery}\diagrw{P&\pright{\tau^*}&P'}\ee
   means that
   \bi 
   \i either $P = P'$ 
   \i or there is a sequence of processes 
      $$P_1,\ldots, P_n\quad(n\geq 2)$$
      such that 
      \bi 
      \i $P_1 = P,\quad P_n = P'$
      \i for each $i=1,\ldots, n-1$
         $$\diagrw{P_i&\pright{\tau}&P_{i+1}}$$
      \ei 
   \ei 
   \re{naeghfgdrtryery} 
   can be interpreted as the statement that
   the process $P$ may 
   imperceptibly turn into a process $P'$.

\i For every action $a\in Act\setminus\{\tau\}$
   the notation
   \be{nadeghdffgdrtryery}
   \diagrw{P&\pright {a_\tau}&P'}\ee
   means that there are processes 
   $P_1$ and $P_2$ 
   with the following properties: 
   $$
   \diagrw{P&\pright{\tau^*}&P_1},\quad
   \diagrw{P_1&\pright{a}&P_2},\quad
   \diagrw{P_2&\pright{\tau^*}&P'}
   $$

   \re{nadeghdffgdrtryery} 
   can be interpreted as the statementthat the process $P$ may 
   \bi
   \i execute a sequence 
      of actions, such that 
      \bi
      \i the action $a$ 
         belongs to this sequence, and
      \i all other actions in this sequence are internal
      \ei
      and then 
   \i turn into a process $P'$.
   \ei
\en
If \re{nadeghdffgdrtryery} holds, then 
we say that the process $P$ may
\bi
\i {\bf observably execute} the action
   $a$, and then 
\i turn into a process $P'$. 
\ei

The concept of observational 
equivalence is based on the following 
understanding of equivalence of processes: 
if we consider processes $P_1$ and 
$P_2$ as equivalent, then they
must satisfy the following conditions. 
\bn 
\i 
\bi 
   \i If one of these processes $P_i$ 
      may imperceptibly turn into 
      some process $P'_i$, 
   \i then another
     process 
     $P_j\quad (j\in \{1,2\} \setminus \{i\})$
     also must be able 
     imperceptibly turn into some 
     process $P'_j$, 
     which is equivalent to $P'_i$. 
   \ei 
\i \bi 
   \i If one of these processes $P_i$ may
\bi 
\i observable execute some 
   action 
  $a\in Act\setminus \{\tau\}$, 
  and then 
\i turn into a process $P'_i$
\ei 
   \i then the other 
process $P_j\quad (j\in \{1,2\}\setminus \{i\})$ 
must be able 
\bi 
\i observably execute the same action $a$, 
   and then 
\i turn into a process $P'_j$, 
   which is equivalent to $P'_i$. 
\ei 
   \ei 
\en 

Using notations \re{naeghfgdrtryery} 
and \re{nadeghdffgdrtryery}, 
the above informally described 
concept of observational equivalence 
can be expressed formally
as a binary relation $\mu$ 
on the set of all processes, 
which has the following properties. 
\bi 
\i[(1)] 
   If $(P_1, P_2) \in \mu$, and 
   for some process $P'_1$ 
   \be{naewetrtwe1}\diagrw{P_1&\pright{\tau}&P'_1}\ee
   then there is a 
   process $P'_2$, such that 
   \be{naewetrtwe2}\diagrw{P_2&\pright{\tau^*}&P'_2}\ee
   and 
\be{naegsdfgsdfgdfg}(P'_1, P'_2) \in \mu\ee
\i[(2)] symmetric property: 
   If $(P_1, P_2) \in \mu$, and 
   for some process $P'_2$ 
   \be{nafghsdftrtwe1}\diagrw{P_2&\pright{\tau}&P'_2}\ee
   then there is a
   process $P'_1$, such that 
   \be{naewetrtwasdfasdfe2}\diagrw{P_1&\pright{\tau^*}&P'_1}\ee
   and \re{naegsdfgsdfgdfg}. 
\i[(3)] 
   If $(P_1, P_2) \in \mu$, and 
   for some process $P'_1$ 
   \be{naewetreeetwe1}\diagrw{P_1&\pright{a}&P'_1}\ee
   then there is a
   process $P'_2$, such that 
      \be{naewetrtweeeeee2}
    \diagrw{P_2&\pright{a_\tau}&P'_2}\ee
   and \re{naegsdfgsdfgdfg}. 
\i[(4)] symmetric property: 
   If $(P_1, P_2) \in \mu$, and 
   for some process $P'_2$ 
   \be{nafghsdftrtwqqwwe3edde1}
   \diagrw{P_2&\pright{a}&P'_2}\ee
   then there is a
   process $P'_1$, such that 
   \be{naewetrtwasdfasdfsdfsdferww44e2}
   \diagrw{P_1&\pright{a_\tau}&P'_1}\ee
   and \re{naegsdfgsdfgdfg}. 
\ei 

Let ${\cal M}_\tau$ be 
a set of all binary relations on the set of 
processes, 
which have the above properties. 

The set ${\cal M}_\tau$ is not empty: 
it contains, for example, 
the diagonal relation, which 
consists of all pairs $(P, P)$, 
where $P$ is an arbitrary process. 

As in the case of strong equivalence, 
the natural question arises about 
what kind of a relationship, 
within the set ${\cal M}_\tau$, 
can be used for a definition 
of the concept of observational equivalence. 

Just as in the case of strong equivalence, 
we offer the following answer to this question: 
we will consider $P_1$ and $P_2$ as
observationally equivalent if and only if 
there is a relation
$\mu\in{\cal M}_\tau$, that 
contains the pair $(P_1, P_2)$, 
i.e. we define 
a relation of observational
equivalence on the set of all processes 
as the union of all relations from${\cal M}_\tau$. 
This relation is denoted by 
the symbol $\approx$. 

It is easy to prove that 
\bi 
\i $\approx\; \in\; {\cal M}_\tau$,
\i $ \approx$ is an equivalence relation, because 
   \bi 
   \i reflexivity of $\approx$ 
   follows from the  
  fact that the diagonal relation 
  belongs to ${\cal M}_\tau$, 
   \i symmetry of $\approx$ follows from the fact that if $\mu\in {\cal M}_\tau$, 
then $\mu^{-1}\in {\cal M}_\tau$ 
   \i transitivity of $\approx$ follows from the fact that if $\mu_1\in {\cal M}_\tau$ and 
$\mu_2\in {\cal M}_\tau$, 
then 
$\mu_1\circ \mu_2\in {\cal M}_\tau$. 
   \ei 
\ei 

If processes $P_1$ and $P_2$ are 
observationally equivalent, then
this fact is indicated 
by $$P_1 \approx P_2$$ 

It is easy to prove that if processes 
$P_1$ and $P_2$ are
strongly equivalent, then they are 
observationally equivalent. 

\subsection{Logical criterion 
of observational equivalence}
\label{bimoodeerobs}

A {\bf logical criterion of observational equivalence} is similar to the analogous
criterion from section \ref{logcrit}. 
In this criteria it is used the same set $Fm$
of formulas. 
The notion of a value of a formula
on a process differs from the 
analogous notion 
in section \ref{logcrit} 
only for formulas of the form $\langle a \rangle \varphi$: 
\bi 
\i a value of the formula 
   $\langle \tau \rangle \varphi$ on 
   the process $P$ is equal to 
      $$\left\{\by 1,&
   \mbox{if there is a process } P':\\
   &\diagrw{P&\pright{\tau^*}&P'},\; P'(\varphi)=1
   \\
   0,&\mbox{otherwise}
	 \ey\right.$$
\i a value of the formula 
  $\langle a \rangle \varphi$ 
(where $a \neq \tau$) 
   on $P$ is    equal to
   $$\left\{\by 1,&
   \mbox{if there is a process } P':\\
   &\diagrw{P&\pright{a_\tau}&P'},\; P'(\varphi)=1
   \\
   0,&\mbox{otherwise}
	 \ey\right.$$
\ei

For each process $P$ the notation
$ Th_\tau (P)$ 
denotes a set of all formulas which 
have a value 1 on the process $P$
(with respect to the modified definition of
the notion of a value of a formula 
on a process). \\

\refstepcounter{theorem}
{\bf Theorem \arabic{theorem}
\label {th5obs}}.

Let $P_1$ and $P_2$ be finite processes.
Then 
$$P_1 \approx P_2\quad\Leftrightarrow\quad
Th_\tau(P_1) = Th_\tau(P_2)\quad\blackbox$$

As in the case of $\sim$, 
there is a problem of finding 
for two given processes $P_1$ and $P_2$ 
a list of formulas of a smallest size 
$$ \varphi_1, \ldots, \varphi_n $$ 
such that 
$P_1 \approx P_2$ if and only if 
$$\forall\; i=1,\ldots, n 
\qquad P_1(\varphi_i) = P_2(\varphi_i)$$

Using theorem \ref{th5obs}, 
we can easily prove that 
\be{mtaum}\mbox{for each process $P$}
\qquad P\;\approx\;\tau.P\ee

Note that, \bi\i according to \re{mtaum}, 
the following statement holds:
$${\bf 0}\;\approx\; \tau.\;{\bf 0}$$
\i however, the statement
\be{as4dfasdf}{\bf 0}+a.{\bf 0}\;\approx\; \tau.\;{\bf 0}+a.{\bf 0}\quad
\mbox{(where $a\neq \tau$)}\ee
does not hold, 
what is easy to see 
by considering the graph representation 
of left and right sides of \re{as4dfasdf}: 
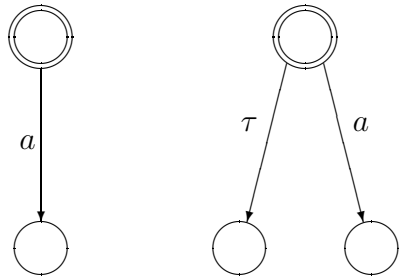
$$
\begin{picture}(120,120)

\put(0,100){\oval(20,20)}
\put(0,100){\oval(24,24)}

\put(0,20){\oval(20,20)}

\put(0,88){\vector(0,-1){58}}

\put(-2,60){\makebox(0,0)[r]{$a$}}

\put(100,100){\oval(20,20)}
\put(100,100){\oval(24,24)}

\put(75,20){\oval(20,20)}
\put(125,20){\oval(20,20)}

\put(93,90){\vector(-1,-4){15}}
\put(107,90){\vector(1,-4){15}}

\put(82,67){\makebox(0,0)[r]{$\tau$}}
\put(118,67){\makebox(0,0)[l]{$a$}}

\end{picture}
$$
A formula, which takes different 
values on these processes, 
may have, for example, the following form: 
$$ \neg \langle \tau \rangle \neg \langle a \rangle \top $$ 
\ei

Thus, the relation $\approx$ 
is not a congruence, 
as it does not preserve 
the operation $+$. 

Another example: if 
$a,b\in Act\setminus\{\tau\}$ 
and $a\neq b$, then 
$$a.{\bf 0}+b.{\bf 0}\;\;\not\approx\;\;
\tau.a.{\bf 0}+\tau.b.{\bf 0}
$$
although $a.{\bf 0}\approx \tau.a.{\bf 0}$
and $b.{\bf 0}\approx \tau.b.{\bf 0}$.

A graph representation of these processes 
has the form 
$$
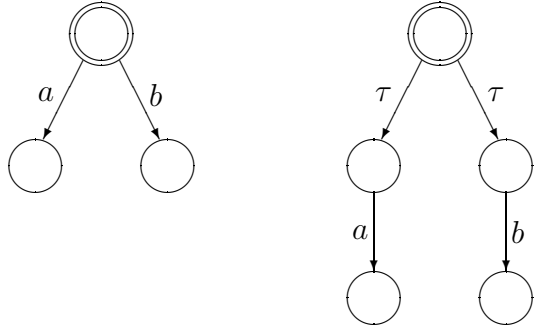
\begin{picture}(0,125)

\put(0,100){\oval(20,20)}
\put(0,100){\oval(24,24)}

\put(-25,50){\oval(20,20)}
\put(25,50){\oval(20,20)}

\put(-7,90){\vector(-1,-2){15}}
\put(7,90){\vector(1,-2){15}}

\put(-18,77){\makebox(0,0)[r]{$a$}}
\put(18,77){\makebox(0,0)[l]{$b$}}

\end{picture}
\hspace{45mm}
\begin{picture}(0,115)

\put(0,100){\oval(20,20)}
\put(0,100){\oval(24,24)}

\put(-25,50){\oval(20,20)}
\put(25,50){\oval(20,20)}

\put(-25,0){\oval(20,20)}
\put(25,0){\oval(20,20)}

\put(-7,90){\vector(-1,-2){15}}
\put(7,90){\vector(1,-2){15}}

\put(-25,40){\vector(0,-1){30}}
\put(25,40){\vector(0,-1){30}}

\put(-18,77){\makebox(0,0)[r]{$\tau$}}
\put(18,77){\makebox(0,0)[l]{$\tau$}}

\put(-27,25){\makebox(0,0)[r]{$a$}}
\put(27,25){\makebox(0,0)[l]{$b$}}
\end{picture}
$$
$\;$\\

The fact that these processes are not
observationally equivalent 
is substantiated by the formula 
$$\langle \tau \rangle \neg \langle a \rangle\top$$

\subsection{A criterion of observational
equivalence based on the concept 
of an observational BS}
\label{obbssbimoodeer}

For the relation $\approx$ 
there is an analog of the criterion 
based on the concept of BS 
(theorem \ref{teorrrr} in section 
\ref{bimoodeer}). 
For its formulation we shall  
introduce auxiliary notations. 

Let $P = (S, s^0, R)$ be a process, 
and $s_1, s_2$ be a pair of its states. 
Then 
\bi 
\i the notation
   $$\diagrw{s&\pright{\tau^*}&s'}$$
   means that
   \bi 
   \i either $s = s'$, 
   \i or there is a sequence of states 
$$s_1,\ldots, s_n\quad (n\geq 2)$$ 
 such that 
$s_1=s,\;\; s_n=s'$, and 
$\forall\,i=1,\ldots, n-1$
$$(\diagrw{s_i&
\pright{\tau}&s_{i+1}})\;\;\in R$$
   \ei 
\i the notation
   $$\diagrw{s&\pright{a_\tau}&s'}\quad\mbox{(where $a\neq \tau$)}$$
   means that there are
   states $s_1$ and $s_2$, such that 
   $$
   \diagrw{s&\pright{\tau^*}&s_1},\quad
   \diagrw{s_1&\pright{a}&s_2},\quad
   \diagrw{s_2&\pright{\tau^*}&s'}.
   $$
\ei 

\refstepcounter{theorem}
{\bf Theorem \arabic{theorem}
\label{teorrrrobs}}.

Let $P_1$ and $P_2$ be processes of the form
$$P_i=(S_i, s^0_i, R_i)\qquad(i=1,2)$$
Then
 $P_1 \approx P_2$ if and only if 
there is a relation 
$$\mu\subseteq S_1 \times S_2$$ 
satisfying the following conditions. 
\bi 
\i[0.] $(s_1^0,s_2^0)\in \mu$.
\i[1.] For each pair $(s_1,s_2)\in \mu$
and each transition from $R_1$ 
of the form 
$$\diagrw{s_1&\pright{\tau}&s'_1}$$
there is a state $s'_2\in S_2$,
such that 
$$\diagrw{s_2&\pright{\tau^*}&s'_2}$$
and 
\be{euriggepro}(s'_1,s'_2)\in \mu\ee
\i[2.] For each pair $(s_1,s_2)\in \mu$
and each transition from $R_2$ of the form 
$$\diagrw{s_2&\pright{\tau}&s'_2}$$
there is a state $s'_1\in S_1$,
such that 
$$\diagrw{s_1&\pright{\tau^*}&s'_1}$$
and \re{euriggepro}. 
\i[3.] For each pair $(s_1,s_2)\in \mu$
and each transition from $R_1$ of the form 
$$\diagrw{s_1&\pright{a}&s'_1}\quad(a\neq \tau)$$
there is a state $s'_2\in S_2$,
such that 
$$\diagrw{s_2&\pright{a_\tau}&s'_2}$$
and \re{euriggepro}. 
\i[4.] For each pair $(s_1,s_2)\in \mu$
and each transition from $R_2$ of the form 
$$\diagrw{s_2&\pright{a}&s'_2}\quad(a\neq \tau)$$
there is a state $s'_1\in S_1$,
such that 
$$
\diagrw{s_1&\pright{a_\tau}&s'_1}
$$
and \re{euriggepro}. 
\ei 
A relation $\mu$, satisfying these conditions, 
is called an {\bf observational BS (OBS)}between $P_1$ and $P_2$. 

\subsection{Algebraic properties 
of observational equivalence}
\label{aldffrrghghggdf}

\refstepcounter{theorem}
{\bf Theorem \arabic{theorem}\label{obsssth6}}.

The relation of observational equivalence 
preserves all operations on processes 
except for the operation $+$, 
i.e. if $P_1\approx P_2$, then
\bi
\i for each $a\in Act \quad
   a.P_1\approx a.P_2$
\i for each  process $P\quad
   P_1|P\approx P_2|P$
\i for each $L\subseteq Names\quad
   P_1\setminus L\approx
   P_2\setminus L$
\i for each renaming $f\quad
   P_1[f] \approx
   P_2[f]$
\ei

{\bf Proof.}

As it was established in section 
\ref{obbssbimoodeer}, 
the statement 
$P_1 \approx P_2$ is equivalent to 
the following statement:
there is an OBS 
$\mu$ between $P_1$ and $P_2$. 
Using this $\mu$, we construct OBSs
for justification of each of the foregoing statements. 
\bi 
\i Let $s^0_{(1)}$ and $s^0_{(2)}$
   be initial states of the processes 
   $a.P_1$ and $a.P_2$ respectively. 

   Then the relation 
   $$\{((s_1,s),(s_2,s))\mid (s_1,s_2)\in \mu,\;q\in S\}$$
  is an OBS between $P_1|P$ and $P_2|P$. 
\i Let $S$ be a set of states of the process $P$. 
   Then the relation 
   $$\{((s_1,s),(s_2,s))\mid (s_1,s_2)\in \mu,\;q\in S\}$$
   is an OBS between $P_1 | P$ and $P_2 | P$. 
\i the relation $\mu$ is an OBS 
   \bi
   \i between $P_1\setminus L$ and
	  $P_2\setminus L$, and
   \i between $P_1[f]$ and
	  $P_2[f]$.$\quad\blackbox$
   \ei
\ei 

\subsection{Recognition of observational equivalence and minimization of 
processes with respect to $\approx$}

The problems of
\bn 
\i recognition for two given finite processes, 
   whether they are 
   observationally equivalent, and 
\i construction for a given finite
   process $P$ such a process $Q$, 
   that
   has the smallest number of states 
   among all processes, which are
   observationally equivalent to $P$ 
\en 
can be solved on the base of a theory 
that is analogous to the theory contained 
in sections \ref{raspozn11} and \ref{minimi11}. 

We will not explain in detail this theory, 
because it is analogous to the theory 
for the case $\sim$. 
In this theory, for any pair of processes 
$$P_i=(S_i, s^0_i, R_i)\qquad(i=1,2)$$
also it is determined an operator $'$
on relations from $S_1$ to $S_2$, 
that maps each relation
$\mu\subseteq S_1\times S_2$ 
to the relation $\mu'_\tau$, 
such that 
$$\by \mbox{$\mu$ satisfies conditions
1, 2, 3, 4}\\
\mbox{from the definition of OBS}\ey
\quad\Leftrightarrow\quad
\mu\subseteq \mu'_\tau
$$
In particular, 
$$\mbox{$\mu$ is OBS
between
$P_1$ and $P_2$
}\quad\Leftrightarrow\quad\left\{\by
(s^0_1,s^0_2)\in \mu\\
\mu\subseteq \mu'_\tau
\ey\right.
$$

Let $\mu_\tau(P_1, P_2)$ be 
a union of all relations from the set 
\be{obslfgsdougsdfghweui}
\{\mu\subseteq S_1\times S_2\mid \mu\subseteq \mu'_\tau\}\ee

The relation $\mu_\tau(P_1, P_2)$
is the greatest element (with respect to 
an inclusion) of the set
\re{obslfgsdougsdfghweui}, 
and has the property 
$$P_1\approx P_2\quad\Leftrightarrow\quad
(s^0_1,s^0_2)\in \mu_\tau(P_1,P_2)$$

From the definition of the relation
$\mu_\tau(P_1,P_2)$
follows that it consists of all pairs 
$(s_1,s_2)\in S_1\times S_2$, 
such that 
$$P_1(s_1)\approx P_2(s_2)$$

The relation $\mu_\tau(P_1,P_2)$ 
can be considered 
as another similarity measure 
between $P_1$ and $P_2$. 

These is a polynomial algorithm 
of a computation of the relation 
$\mu_\tau(P_1,P_2)$.
This algorithm is similar 
to the corresponding algorithm 
from section \ref{raspozznnaavv}.
For constructing of this algorithm 
it should be considered 
the following consideration. 
For checking the condition 
$$\diagrw{s&\pright{\tau^*}&s'}$$
(where $s, s'$ are states of a process $P$)
it is enough to analyze sequences
of transitions of the form 
$$\diagrw{s&\pright{\tau}&s_1&\pright{\tau}&
s_2&\pright{\tau}&\ldots}$$
length of which does not exceed 
a number of states of the process $P$. 

\subsection{Other criteria of
equivalence of processes} 

For proving that processes 
$P_1$ and $P_2$ are strongly 
equivalent or observationally equivalent, 
the following criteria can be used. 
In some cases, use of these criteria 
for proving of an appropriate equivalence 
between $P_1$ and $P_2$
is much easier than all other methods. 

A binary relation $\mu$ on the set 
of processes is said to be
\bi
\i BS (mod $\sim$), 
   if $\mu\subseteq (\sim \mu\sim)'$
\i OBS (mod $\sim$),
   if $\mu\subseteq (\sim \mu\sim)'_\tau$
\i OBS (mod $\approx$),
   if $\mu\subseteq 
   (\approx \mu\approx)'_\tau$
\ei

It is easy to prove that 
\bi
\i if $\mu$ is BS (mod $\sim$), 
   then $\mu\;\subseteq\; \sim$, and
\i if $\mu$ is OBS (mod $\sim$ or mod $\approx$), then
   $\mu\;\subseteq\; \approx$.
\ei

Thus, to prove 
$P_1\sim P_2$ or $P_1\approx P_2$
it is enough to find a suitable 
\bi 
\i BS (mod $\sim$), or 
\i OBS (mod $\sim$ or mod $\approx$) 
\ei 
respectively, such that $$(P_1,P_2)\in \mu$$

\section{Observational congruence}
\label{defequisdfgvalfghgj543}

\subsection{A motivation of the concept 
of observational congruence}

As stated above, 
a concept of equivalence of processes 
can be defined not uniquely. 
In the previous sections 
have already been considered 
different types of equivalence of processes.
Each of these equivalences
reflects a certain point of view 
on what types of a behavior 
should be considered as equal. 

In addition to these concepts 
of equivalence of processes, 
it can be determined, for example, 
such concepts of equivalence, that
\bi 
\i take into account 
   a duration of an execution of actions, 
   i.e., in particular, 
   one of conditions of equivalence 
   of processes $P_1$ and $P_2$ 
   can be as follows: 
   \bi 
   \i if one of these processes $P_i $ 
      may, within a some period of time 
      imperceptibly turn into a process $P'_i$, 
   \i then the other        process 
      $P_j\quad (j\in \{1,2\}\setminus \{i\})$
      must be able 
      for approximately the same amount of time 
      imperceptibly turn into a process $P'_j$, 
      which is equivalent to $P'_i$ \\
      (where the concept of ``approximately 
      the same amount of time'' 
      can be clarified in different ways) 
   \ei 
\i or take into account 
   the property of {\bf fairness},
   i.e. processes can not be considered
   as equivalent, if 
   \bi
   \i   one of them is    fair, and 
   \i another is not fair
   \ei
   where one of possible
   definitions of fairness of processes
   is as follows: 
   a process is said to be {\bf fair}
   if there is no
   an infinite sequence of transitions of the form
   $$\diagrw{s_0&\pright{\tau}&s_1&
   \pright{\tau}&s_2&\pright{\tau}&\ldots}$$
   such that the state $s_0$ is reachable, 
   and for each $i\geq 0$
   $$Act(s_i)\setminus \{\tau\}\neq \emptyset$$

   Note that observational equivalence 
   does not take into account 
   the property of fairness:
   there are two processes $P_1$ and $P_2$, 
   such that 
   \bi
   \i $P_1 \approx P_2$, but 
   \i $P_1$ is fair, and
      $P_2$ is not fair.
   \ei
   For example 
   \bi 
   \i $P_1 =  a. {\bf 0}$, where $ a \neq \tau$, 
   \i $P_2 = a. {\bf 0}  \pa \tau^* $, where the      
      process $ \tau^*$ has one state and one 
      transition with a label $\tau$ 
   \ei
\i etc. 
\ei 

In every particular situation, 
a decision about which 
a concept of equivalence 
of processes is best used, 
essentially depends on the purposes 
for which this concept is intended. 

In this section we define another kind 
of equivalence of processes 
called an {\bf observational congruence}.
This equivalence is denoted by $\oc$. 
We define this equivalence, 
based on the following conditions 
that it must satisfy. 
\bn 
\i Processes that 
   are equivalent with respect to 
   $\oc$,  must be observationally equivalent. 
\i Let 
   \bi 
   \i a process $P$ is constructed 
      as a composition of processes 
      $$P_1, \ldots, P_n$$ 
      that uses operations 
      \be{sdfsw45sdd445}
       a.,\quad +, \quad\pa,\quad 
      \setminus L,\quad [f]\ee
   \i and we replace one of components
      of this composition 
      (for example, the process $P_i$), 
      on other process $P'_i$, 
      which is equivalent to $P_i$. 
   \ei
   A process which is obtained from $P$ 
   by this replacement, must be equivalent 
   to the original process $P$. 
\en 

It is easy to prove that an equivalence 
$\mu$ on the set of processes 
satisfies the above conditions if and only if 
\be{sdfgsdewewter}\left\{ \by
\mu\;\subseteq \;\approx\\
\mu\;\;\mbox{is a congruence}
 \\\mbox{$\;$}\qquad
\mbox{ with respect to operations
 \re{sdfsw45sdd445}}\ey\right.\ee

There are several equivalences
which satisfy 
conditions \re{sdfgsdewewter}.
For example, 
\bi 
\i ithe diagonal relation 
   (consisting of pairs of the form $ (P, P) $),    and 
\i strong equivalence $(\sim)$
\ei 
satisfy these conditions.

Below we prove that among 
all equivalences 
satisfying conditions \re{sdfgsdewewter}, 
there is greatest equivalence
(with respect to inclusion). 
It is natural to consider this 
equivalence as the desired equivalence 
($\oc$). 

\subsection{Definition of a concept of observational congruence}

To define a concept of observational 
congruence, we introduce 
an auxiliary notation. 

Let $P$ and $P '$ be a couple of processes. 
The notation
$$\diagrw{P&\pright{\tau^+}&P'}$$
means that there is 
a sequence of processes 
$$P_1,\ldots, P_n\quad(n\geq 2)$$
such that 
\bi
\i $P_1 = P,\quad P_n = P'$, and 
\i for each $i=1,\ldots, n-1$
   $$\diagrw{P_i&\pright{\tau}&P_{i+1}}$$
\ei

We shall say that 
processes $P_1$ and $P_2$ 
are in a relation of 
{\bf observational congruence}
and denote this fact by $$P_1 \oc P_2$$ 
if the following conditions hold. 
\bi 
\i[(0)] $P_1 \approx P_2$. 
\i[(1)] 
   If, a process $P'_1$ is such that
   \be{connaewetrtwe1}
   \diagrw{P_1&\pright{\tau}&P'_1}\ee
   then there is a process $P'_2$, such that 
   \be{connaewetrtwe2}
   \diagrw{P_2&\pright{\tau^+}&P'_2}\ee
   and 
\be{connaegsdfgsdfgdfg}
P'_1 \approx P'_2\ee
\i[(2)] Symmetrical condition: 
   if a process $P'_2$ is such that
   \be{connafghsdftrtwe1}
   \diagrw{P_2&\pright{\tau}&P'_2}\ee
   then there is a process $P'_1$, such that
   \be{connaewetrtwasdfasdfe2}
   \diagrw{P_1&\pright{\tau^+}&P'_1}\ee
   and \re{connaegsdfgsdfgdfg}.
\ei 

It is easy to prove that 
observational congruence is an equivalence relation. 

\subsection{Logical criterion 
of observational congruence}
\label{conbimoodeerobs}

A {\bf logical criterion of observational congruence} of two processes is produced 
by a slight modification of the logical criterion 
of observational equivalence 
from section \ref{bimoodeerobs}. 

A set of formulas $Fm^+$, 
which is used in this criterion, 
is an extension of the set of formulas $Fm$ 
from section \ref{bimoodeer}.
$Fm^+$ is obtained from $Fm$ 
by adding a modal connective
$\langle \tau^+ \rangle$.

The set $Fm^+$ is defined as follows.
\bi 
\i Every formula from $Fm$ 
   belongs to $Fm^+$.
\i For every formula $\varphi \in Fm$    
   the string
   $$ \langle \tau^+ \rangle \varphi $$ 
   is a formula from $Fm^+$. 
\ei 
For every formula $\varphi \in Fm^+$ 
and every process $P$ 
a {\bf value} of $\varphi$ on $P$ 
is denoted by $P(\varphi)$ and
is defined as follows. 
\bi 
\i If $\varphi \in Fm$, then $P(\varphi)$ 
   is defined as in section \ref{bimoodeerobs}. 
\i If $\varphi=\langle  
   \tau^+ \rangle \psi$,    
   where $ \psi \in Fm$, then 
   $$P(\varphi)\;\eam\;
   \left\{\by 1,&
   \mbox{if there is a process } P':\\
   &\diagrw{P&\pright{\tau^+}&P'},
   \; P'(\psi)=1
   \\
   0,&\mbox{otherwise}
   \ey\right.$$
\ei 
For each process $P$ we denote by
$Th^+_\tau (P)$ 
a set of all formulas $\varphi\in Fm^+$, 
such that $P(\varphi)=1$.\\

\refstepcounter{theorem}
{\bf Theorem \arabic{theorem}\label{conth5obs}}.

Let $P_1$ and $P_2$  be finite
processes. Then 
$$P_1 \oc P_2\quad\Leftrightarrow\quad
Th^+_\tau(P_1) = Th^+_\tau(P_2)\quad\blackbox$$

As in the case of $\sim$ and $\approx$, 
there is a problem of finding 
for two given processes $P_1$ and $P_2$ 
a list of formulas of a smallest size 
$$ \varphi_1, \ldots, \varphi_n \; 
\in Fm^+$$ 
such that 
$P_1 \oc P_2$ if and only if 
$$\forall\; i=1,\ldots, n 
\qquad P_1(\varphi_i) = P_2(\varphi_i)$$

\subsection{Criterion of
observational congruence 
based on the concept of observational BS}
\label{conobbssbimoodeer}

We shall use the following notation.
Let 
\bi 
\i $P$ be a process 
   of the form $(S, s^0,R)$,
   and 
\i $s_1,s_2$ be a pair of states from $S$. 
\ei 
Then the notation
$$\diagrw{s&\pright{\tau^+}&s'}$$
means that there is a sequence of states 
$$s_1,\ldots, s_n\quad (n\geq 2)$$  
such that 
$s_1=s,\;\; s_n=s'$, and 
for each $i=1,\ldots, n-1$
$$(\diagrw{s_i&\pright{\tau}&
s_{i+1}})\;\;\in R$$

\refstepcounter{theorem}
{\bf Theorem \arabic{theorem}
\label{conteorrrrobs}}.

Let $P_1, P_2$ 
be a pair of processes of the form
$$P_i=(S_i, s^0_i, R_i)\qquad(i=1,2)$$
The statement $P_1\oc P_2$ holds
if and only if 
there is a relation
$$\mu\subseteq S_1\times S_2$$
satisfying the following conditions. 
\bi 
\i[0.] $\mu$ is an OBS between 
   $P_1$ and $P_2$ \\
(the concept of an OBS is described 
in section \ref{obbssbimoodeer}). 
\i[1.] For each transition from $R_1$ 
  of the form 
$$\diagrw{s^0_1&\pright{\tau}&s'_1}$$
there is a state $s'_2\in S_2$, 
such that 
$$\diagrw{s^0_2&\pright{\tau^+}&s'_2}$$
and \be{coneuriggepro}(s'_1,s'_2)\in \mu\ee
\i[2.] For each transition from $R_2$ 
of the form 
$$\diagrw{s^0_2&\pright{\tau}&s'_2}$$
there exists a state $s'_1\in S_1$, 
such that 
$$\diagrw{s^0_1&\pright{\tau^+}&s'_1}
$$
and \re{coneuriggepro}. $\blackbox$
\ei 

Below the string $\mbox{\rm OBS}^+$
is an abbreviated notation of the phrase
\begin{center}
``an OBS satisfying conditions 1 and 2 
of theorem \ref{conteorrrrobs}''. 
\end{center}

\subsection{Algebraic properties 
of observational congruence} 
\label{sdfwerr4543534ergteg}

\refstepcounter{theorem}
{\bf Theorem \arabic{theorem}\label{conobsssth6}}.

The observational congruence 
is a congruence 
with respect to all operaions
on processes, i.e. if $P_1 \oc P_2$, then 
\bi
\i for each  $a\in Act \quad
   a.P_1\oc a.P_2$
\i for each process $P\quad
   P_1+P\oc P_2+P$
\i for each process $P\quad
   P_1|P\oc P_2|P$
\i for each $L\subseteq Names\quad
   P_1\setminus L\oc
   P_2\setminus L$
\i for each renaming $f\quad
   P_1[f] \oc
   P_2[f]$
\ei

{\bf Proof.}

As it was stated
in section \ref{conobbssbimoodeer}, 
the statement 
$P_1 \oc P_2$ holds if and only if
there is $\mbox{\rm OBS}^+$ $\mu$ 
between $P_1$ and $P_2$. 
Using this $\mu$, for each
of the above statements 
we shall justify this statement
by construction of corresponding 
$\mbox{\rm OBS}^+$. 

\bi 
\i Let $s^0_{(1)}$ and $s^0_{(2)}$  
  be initial    states of the processes
$a.P_1$ and $a.P_2$ 
respectively. 

   Then the relation 
   $$\{(s^0_{(1)}, s^0_{(2)})\} \;\cup \;
   \mu$$
   is $\mbox{\rm OBS}^+$
 between $a.P_1$ and $a.P_2$
\i Let \bi \i 
   $s^0_{(1)}$ and $s^0_{(2)}$ 
be    initial states of $P_1+P$ and $P_2+P$
   respectively, and 
   \i $S$ be denote a set of states 
   of the process $P$. 
   \ei 
   Then the relation 
   $$\{(s^0_{(1)}, s^0_{(2)})\} \;\cup \;
   \mu\;\cup \;
   Id_{S}$$
   is $\mbox{\rm OBS}^+$
   between $P_1+P$ and $P_2+P$.
\i Let 
   $S$ be a set of states 
   of the process $P$. 
   Then the relation 
   $$\{((s_1,s),(s_2,s))\mid (s_1,s_2)\in \mu,\;q\in S\}$$
   is $\mbox{\rm OBS}^+$
   between $P_1|P$ and $P_2|P$.
\i The relation $\mu$ is 
   $\mbox{\rm OBS}^+$
   \bi
   \i between $P_1\setminus L$ and
	  $P_2\setminus L$, and
   \i between $P_1[f]$ and
	  $P_2[f]$.$\quad\blackbox$
   \ei
\ei 

\refstepcounter{theorem}
{\bf Theorem \arabic{theorem}\label{conobsssth61er}}.

For any processes $P_1$ and $P_2$ 
$$P_1\approx P_2\quad \Leftrightarrow \quad
\left\{\by
P_1 \oc P_2&\mbox{or} \\
P_1 \oc \tau.P_2 &\mbox{or} \\
\tau.P_1 \oc P_2\ey\right.$$

{\bf Proof.}

The implication ``$\leftarrow$'' 
follows from
\bi
\i the inclusion 
   $\oc\;\subseteq\;\approx$, 
   and 
\i the fact that 
   \be{sdfssdfwwerr}
   \mbox{for any process $P$}\quad
   P\approx \tau.P\ee
\ei

Prove the implication `` $ \rightarrow $''. Suppose 
\be{sdffsad34}\mbox{$P_1\approx P_2$}\ee
and
\be{sdffsad341}\mbox{
it is not true that $P_1\oc P_2$}\ee

\re{sdffsad341} can occur, 
for example, in the following case: 
\be{preddkgsl}\by
\mbox{ there is a process $P'_1$, such that}\\
\mbox{ $\diagrw{P_1&\pright{\tau}&P'_1}$
	 }\ey
\ee
and
\be{preddkgs345l}\by
\mbox{ there is no a process $P'_2\approx P'_1,$}\\
\mbox{ such that 
$\diagrw{P_2&\pright{\tau^+}&P'_2}$}\ey
\ee

We shall prove that in this case 
$$P_1 \oc \tau.P_2$$

According to the definition of  
observational congruence,
we must prove that 
conditions (0), (1) and (2) from this 
definition are satisfied. 
\bi 
\i[(0)]: $P_1 \approx \tau.P_2$. 

  This condition follows from \re{sdffsad34} 
   and \re{sdfssdfwwerr}. 

\i[(1)]:
   if there is a process $P'_1$ such that
   \be{343678}\diagrw{P_1&\pright{\tau}
   &P'_1}\ee
   then there is a process  
   $P'_2 \approx P'_1$ such that
   \be{356678888}
   \diagrw{\tau.P_2&\pright{\tau^+}&P'_2}\ee
   From \re{sdffsad34}, \re{343678}, 
   and from the definition of observational    
   equivalence 
   it follows that these is a process 
   $P'_2 \approx P'_1$ such that
\be{tyyrthhy}\diagrw{P_2&\pright{\tau^*}&P'_2}\ee
   \re{356678888} 
   follows from 
   $\diagrw{\tau.P_2&\pright{\tau}&P_2}$
   and \re{tyyrthhy}. 
\i[(2)]:
   if there is a process $P'_2$ such that
   \be{c4ftrtwe1}
   \diagrw{\tau.P_2&\pright{\tau}&P'_2}\ee
   then there is a process $P'_1 \approx P'_2$ 
   such that
   $$\diagrw{P_1&\pright{\tau^+}&P'_1}$$
   From the definition 
   of the operation of prefix 
   actions and from \re{c4ftrtwe1}
   we get the equality $$P'_2 = P_2$$ 
   Thus, we must prove that 
\be{predfdgddkgs345l}\by
\mbox{for some process $P'_1 \approx P_2$}\\
\mbox{the formula
   $\diagrw{P_1&\pright{\tau^+}&P'_1}$
holds}\ey\ee

Let $P'_1$ be a process 
that is referred in the assumption 
\re{preddkgsl}. 
From the assumption \re{sdffsad34} 
we get
\be{predfdgddkgs34dsfgsdfg5l}\by
\mbox{there is a process
    $P'_2\approx P'_1$,}\\
   \mbox{such that
   $\diagrw{P_2&\pright{\tau^*}&P'_2}$}\ey\ee

   Comparing 
\re{predfdgddkgs34dsfgsdfg5l}
and \re{preddkgs345l}, 
we get the equality $P'_2 = P_2$, i.e., 
we have proved \re{predfdgddkgs345l}. 
\ei 

\re{sdffsad341} 
may be true also on the reason that
\bi 
\i there is a process $P'_2$, such that 
   $\diagrw{P_2&\pright{\tau}&P'_2}$, 
   and 
\i there is no a process 
   $P'_1 \approx P'_2$, such that 
   $$\diagrw{P_1&\pright{\tau^+}&P'_1}$$
\ei 

In this case, by similar reasoning 
it can be proven that 
$$\tau. P_1 \oc P_2\quad \blackbox$$

\refstepcounter{theorem}
{\bf Theorem \arabic{theorem}\label{conobsssthewrtgert376df1}}.

The relation $\oc$ coincides with the relation 
\be{fdjgkl543sdfgjsdfkl}
\{(P_1,P_2) \mid \forall\;
P\quad P_1 + P \approx P_2 + P\}\ee

{\bf Proof.}

The inclusion 
$\oc\;\subseteq \; \re{fdjgkl543sdfgjsdfkl}$
follows from the fact that 
\bi 
\i $\oc$ is a congruence 
   (i.e., in particular, $\oc$ 
   preserves the operation ``$+$''),  and
\i $\oc \;\subseteq \;\approx$.
\ei 

Prove the inclusion 
$$\re{fdjgkl543sdfgjsdfkl}\;\subseteq \; \oc$$

Let  $(P_1, P_2)\in \re{fdjgkl543sdfgjsdfkl}$. 

Since for each process $P$ 
the following statement holds
\be{dsdfg345435}
P_1 + P \approx P_2 + P
\ee
then, setting in \re{dsdfg345435} 
$P \eam {\bf 0}$, we get
\be{fgsdfg353678}P_1 + {\bf 0} 
\approx P_2 + {\bf 0}\ee

Since \bi \i for each process $P$ 
the following statement holds:
$$P + {\bf 0} \sim P$$ 
\i and , furthermore, 
$\sim\;\subseteq \;\approx$\ei 
then from \re{fgsdfg353678}
we get 
\be{dfk3l65666}P_1 \approx P_2\ee

If it is not true that $P_1 \oc P_2$, then 
from \re{dfk3l65666} on the reason of 
theorem \ref{conobsssth61er}
we get that
\bi 
\i either $P_1 \oc \tau.P_2$,
\i or $\tau.P_1 \oc P_2$
\ei 
Consider, for example, the case 
\be{vndiwe}P_1 \oc \tau.P_2\ee
(the other case is considered analogously). 

Since $\oc$ is a congruence, 
then from \re{vndiwe} it follows that 
for any process $P$ 
\be{vndiwe1}P_1+P \oc \tau.P_2+P\ee

From 
\bi
\i \re{dsdfg345435}, \re{vndiwe1}, and 
\i the inclusion $\oc\;\subseteq \;\approx$
\ei
it follows that for any process $P$ 
\be{vndiwe2}P_2+P \approx \tau.P_2+P\ee
Prove that 
\be{vndiwe3}P_2 \oc \tau.P_2\ee
\re{vndiwe3} equivalent to the following
statement: there is a process
$P'_2\approx P_2$, such that 
\be{msd557}
\diagrw{P_2 &\pright{\tau^+}& P'_2}\ee

Since the set $Names$ is infinite
(by an assumption from
section \ref{sdfgsdfgw5etrwerw}), 
then there is an action 
$b\in Act\setminus \{\tau\}$,
which does not occur in $P_2$.

Statement \re{vndiwe2}  must be true
in the case when $P$ has the form 
$b.{\bf 0}$, i.e. the following 
statement must be true:
\be{vn3dsdf343}P_2+b.{\bf 0} \approx \tau.P_2+b.{\bf 0}\ee

Since 
$$\diagrw{\tau.P_2+b.{\bf 0}&\pright{\tau}&P_2}$$
then 
\bi\i from \re{vn3dsdf343}, and
\i from the definition of the relation 
$\approx$ \ei
it follows that there is a process 
$P'_2\approx P_2$ such that
\be{madfjh3}\diagrw{P_2+b.{\bf 0}&
\pright{\tau^*}&P'_2}\ee

The case $P_2+b.{\bf 0} = P'_2$
is impossible, because 
\bi\i the left side of this equality
does contain the action $b$, 
and \i the right side of this equality
does not contain the action $b$.
\ei
Consequently, on the reason of \re{madfjh3}, 
we get the statement
\be{madfefd3jh3}\diagrw{P_2+b.{\bf 0}&\pright{\tau^+}&P'_2}\ee

From the definition of the operation $+$, it follows that \re{madfefd3jh3} 
is possible if and only if \re{msd557} holds. 

Thus, we have proved that 
there is a process $P'_2\approx P_2$ 
such that \re{msd557} holds, 
i.e. we have proved \re{vndiwe3}. 

\re{vndiwe} and \re{vndiwe3} imply 
that  $P_1 \oc P_2$. $ \blackbox $ \\

\refstepcounter{theorem}
{\bf Theorem \arabic{theorem}
\label{conobsssth6df1}}.

$\oc$ is the greatest congruence 
contained in $\approx$, 
i.e. for each congruence 
$\nu$ on the set of all processes 
the following implication holds: 
$$\nu \;\subseteq\; \approx 
\quad \Rightarrow \quad  
\nu\; \subseteq \;\oc$$

{\bf Proof.}

Prove that if $(P_1,P_2)\in \nu$,
then $P_1\oc P_2$.

Let $(P_1,P_2)\in \nu$.
Since $\nu$ is a congruence, then
\be{bfsdbsdfbsddb}\mbox{for each 
process $P$}
\quad(P_1+P,P_2+P)\in \nu\ee

If $\nu\;\subseteq \;\approx$, 
then from \re{bfsdbsdfbsddb}
it follows that 
\be{bfsdbtgysdfbsddb}
\mbox{for each process $P$}
\quad P_1+P \approx P_2+P\ee
According to theorem 
\ref{conobsssthewrtgert376df1},
\re{bfsdbtgysdfbsddb} implies that
$P_1 \oc P_2$. $ \blackbox$ \\

\refstepcounter{theorem}
{\bf Theorem \arabic{theorem}
\label{conobsssth6df13}}.

The relations $\sim$, $\approx$ and $\oc$ 
have the following property:
\be{sdjfkl35467}\sim\;\;\subseteq\;\;\oc\;\;\subseteq \;\;\approx\ee

{\bf Proof.}

The inclusion $\oc \;\subseteq\; \approx$
holds by definition of $\oc$.

The inclusion $\sim\;\subseteq \;\oc$
follows from
\bi \i the inclusion 
$\sim\;\subseteq \;\approx$, and 
\i from the fact that if processes 
$P_1, P_2$ are such that
$$P_1\sim P_2$$
then this pair of processes
satisfies conditions from
the definition of the relation $\oc$.
$\blackbox$\ei

Note that both inclusions in 
\re{sdjfkl35467} are proper:
\bi
\i $a.\tau.{\bf 0}\not\sim a.{\bf 0}$, but
   $a.\tau.{\bf 0}\oc a.{\bf 0}$
\i $\tau.{\bf 0}\oc\!\!\!\!\!/\; {\bf 0}$, but
   $\tau.{\bf 0}\approx {\bf 0}$
\ei

\refstepcounter{theorem}
{\bf Theorem \arabic{theorem}
\label{conobsssweth6df13}}.

\bn
 \i If $P_1\approx P_2$, then for each 
    $a\in Act$
    $$a.P_1 \oc a.P_2$$
    In particular, for each process $P$
    \be{atausdfklgj}
    a.\tau.P\oc a.P\ee
\i  For any process $P$
    \be{dfhjkerwer34343434}
    P + \tau.P \oc \tau.P\ee
\i  For any processes 
    $P_1$ and $P_2$, and any $a\in Act$
    \be{dfhjkerwer343434312344}
     a.(P_1 + \tau.P_2) + a.P_2 \oc a.(P_1 + 
          \tau.P_2)\ee
\i For any processes $P_1$ and $P_2$
   \be{dfhjkerwer34343431234sdfg4}
   P_1 + \tau.(P_1 + P_2) \oc 
    \tau.(P_1 + P_2)\ee
\en

{\bf Proof.}

For each of the above statements
we shall construct 
an $\mbox{\rm OBS}^+$ 
between its left and right sides.

\bn
\i As it was stated in 
   theorem \ref{teorrrrobs}
   (section \ref{obbssbimoodeer}),
   the statement $P_1\approx P_2$ 
   is equivalent to the statement that
   there is an $\mbox{\rm OBS}$
   $\mu$ between $P_1$ and $P_2$.

   Let $s^0_{(1)}$ and $s^0_{(2)}$  
   be initial
   states of the processes 
   $a.P_1$ and $a.P_2$  respectively.

   Then the relation
   $$\{(s^0_{(1)}, s^0_{(2)})\} 
   \;\cup \;
   \mu$$
   is an $\mbox{\rm OBS}^+$ 
   between $a.P_1$ and $a.P_2$.

   \re{atausdfklgj} follows from \bi
   \i the above statement, and
   \i the statement 
     $\tau.P\approx P$, which holds
     according to \re{mtaum}. \ei

\i Let $P$ has the form
   $$P = (S, s^0,R)$$
   and let 
   $S_{(1)}$ ? $S_{(2)}$  
   be duplicates of the set $S$
   in the processes $P$ and $\tau.P$    
   respectively, which contain in 
   the left side of the statement
   \re{dfhjkerwer34343434}. 
   Elements of these duplicates
   will be denoted by 
   $s_{(1)}$ and $s_{(2)}$
   respectively, where $s$ is 
   an arbitrary element of the set $S$.

   Let $s^0_{l}$ and $s^0_{r}$ be initial
   states of the processes in 
   the left and right sides of 
   \re{dfhjkerwer34343434}  respectively.
   Then the relation
   $$\{(s^0_{l}, s^0_{r})\} \;\cup \;
   \{(s_{(i)}, s)\mid s\in S, \; i=1,2\}$$
   is $\mbox{\rm OBS}^+$
 between
   left and right sides of the statement
   \re{dfhjkerwer34343434}.

\i Let $P_i=(S_i, s^0_i, R_i)\quad(i=1,2)$.
   We can assume that 
$S_1\cap S_2=\emptyset$.
   Let
   \bi
   \i $s^0_\tau$ be an initial state 
     of the process
    \be{dfjke5555}P_1 + \tau.P_2\ee
   \i $s^0$ be an initial state of the process
\be{dfjkeffr5555}a.(P_1 + \tau.P_2)\ee
Note that \re{dfjkeffr5555} coincides with 
the right side of \re{dfhjkerwer343434312344}.
   \ei
   The left side of 
   \re{dfhjkerwer343434312344}
   is strongly equivalent to the process 
   $P'$, which is obtained
   from \re{dfjkeffr5555} by adding the 
   transition
   $$\diagrw{s^0&\pright{a}&s^0_2}$$
   it is easily to make sure in this
   by considering the graph    
   representation of the process $P'$, 
   which has the form
   $$\by
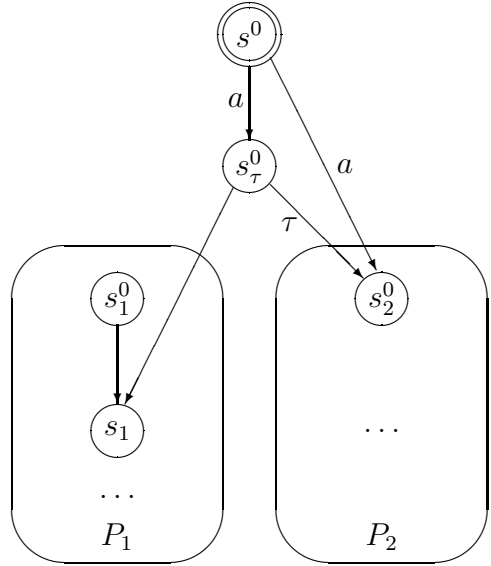
\begin{picture}(0,220)

\put(0,200){\oval(20,20)}
\put(0,200){\oval(24,24)}
\put(0,200){\makebox(0,0){$s^0$}}

\put(0,150){\oval(20,20)}
\put(0,150){\makebox(0,0){$s^0_\tau$}}

\put(50,100){\oval(20,20)}
\put(50,100){\makebox(0,0){$s^0_2$}}

\put(-50,100){\oval(20,20)}
\put(-50,100){\makebox(0,0){$s^0_1$}}

\put(-50,50){\oval(20,20)}
\put(-50,50){\makebox(0,0){$s_1$}}

\put(-50,25){\makebox(0,0){$\ldots$}}
\put(50,50){\makebox(0,0){$\ldots$}}

\put(50,60){\oval(80,120)}
\put(-50,60){\oval(80,120)}

\put(-50,5){\makebox(0,0)[b]{$P_1$}}
\put(50,5){\makebox(0,0)[b]{$P_2$}}

\put(8,191){\vector(1,-2){40}}
\put(8,143){\vector(1,-1){35}}
\put(-6,142){\vector(-1,-2){41}}
\put(-50,90){\vector(0,-1){30}}
\put(0,188){\vector(0,-1){28}}

\put(-2,175){\makebox(0,0)[r]{$a$}}
\put(33,150){\makebox(0,0)[l]{$a$}}
\put(12,128){\makebox(0,0)[l]{$\tau$}}
\end{picture}
\ey
$$

It is easy to prove that the process $P'$ is 
observationally congruent to the 
process \re{dfjkeffr5555}.
The sets of states of these processes can be considered as
duplicates $S_{(1)}$ and $S_{(2)}$
of one and the same set  $S$,
and $\mbox{\rm OBS}^+$ 
between $P'$ and \re{dfjkeffr5555}
has the form
\be{dfsdklgsdfkge4554}\{(s_{(1)}, s_{(2)})\mid s\in S\}\ee

Since
\bi
\i according to theorem 
  \ref{conobsssth6df13}, 
   we have the inclusion
   $\sim \;\subseteq \;\oc$, and
\i \re{dfjkeffr5555} coincides with the right part
    of    \re{dfhjkerwer343434312344},
\ei
then we have proved that 
the left and right sides of the statement
\re{dfhjkerwer343434312344}
are observationally congruent. $ \blackbox $

\i Reasonings in this case are similar to the
   reasonings in the previous case.
   We will not explain them in detail,
   only note that \bi \i left part of the statement
   \re{dfhjkerwer34343431234sdfg4}
   is strongly equivalent to 
   the process $P'$, which has the following 
   graph representation:
   $$\by
\begin{picture}(0,220)

\put(0,200){\oval(20,20)}
\put(0,200){\oval(24,24)}
\put(0,200){\makebox(0,0){$s^0$}}

\put(0,150){\oval(20,20)}
\put(0,150){\makebox(0,0){$s^0_{12}$}}

\put(50,100){\oval(20,20)}
\put(50,100){\makebox(0,0){$s^0_2$}}

\put(-50,100){\oval(20,20)}
\put(-50,100){\makebox(0,0){$s^0_1$}}

\put(-50,50){\oval(20,20)}
\put(-50,50){\makebox(0,0){$s_1$}}

\put(50,50){\oval(20,20)}
\put(50,50){\makebox(0,0){$s_2$}}

\put(-50,25){\makebox(0,0){$\ldots$}}
\put(50,25){\makebox(0,0){$\ldots$}}

\put(50,60){\oval(80,120)}
\put(-50,60){\oval(80,120)}

\put(-50,5){\makebox(0,0)[b]{$P_1$}}
\put(50,5){\makebox(0,0)[b]{$P_2$}}

\put(-6,189){\vector(-1,-3){43}}
\put(-6,142){\vector(-1,-2){41}}
\put(6,142){\vector(1,-2){41}}
\put(-50,90){\vector(0,-1){30}}
\put(50,90){\vector(0,-1){30}}
\put(0,188){\vector(0,-1){28}}

\put(2,175){\makebox(0,0)[l]{$\tau$}}

\end{picture}
\ey
$$

where 
\bi 
\i $s^0_1$ and $s^0_2$ are 
   initial states of the processes
   $P_1$ and $P_2$, and
\i $s^0_{12}$ is an initial state 
   of the process $P_1+P_2$
\ei

\i the right part of the statement
   \re{dfhjkerwer34343431234sdfg4}
   (which we denote by $P''$) 
   is obtained from $P'$ by removing
   of transitions of the form
   $$\diagrw{s^0&\pright{}&s_1}$$
\ei

It is easy to prove that $P'\oc P''$.
Sets of states of these processes 
can be considered as
duplicates $S_{(1)}$ and $S_{(2)}$
of one and the same set $S$,
and $\mbox{\rm OBS}^+$ 
between $P'$ and $P''$
has the form \re{dfsdklgsdfkge4554}. 
$\blackbox$
\en

\subsection{Recognition of
observational congruence}

To solve the problem of recognition
for two given finite processes, 
whether they are observationally congruent,
it can be used the following theorem. \\

\refstepcounter{theorem}
{\bf Theorem \arabic{theorem}\label{conth5fdobse4}}.

Let $P_1$ and $P_2$ be finite processes.
The statement
$$P_1 \oc P_2$$
holds if and only if
$$\left\{\by (s^0_1,s^0_2)\in \mu_\tau(P_1, P_2)\\
\mu_\tau(P_1, P_2) \; \mbox{is an 
\rm OBS}^+
\ey\right.\quad\blackbox$$

\subsection{Minimization of processes 
with respect to observational congruence}

To solve the problem of minimizing
of finite processes 
with respect to observational congruence
the following theorems can be used. \\

\refstepcounter{theorem}
{\bf Theorem \arabic{theorem}\label{conthfe2225obse4}}.

Let $P=(S, s^0, R)$ be a process.

Define a {\bf factor-process} $P_\approx$  
of the process $P$ with respect to 
the equivalence $\mu_\tau(P, P)$,
as a process with the following components.
\bi
\i States of $P_{\approx}$ are
   equivalence classes of the set $S$
   with respect to the 
   equivalence $\mu_\tau(P, P)$.
\i An initial state of $P_\approx$  
   is the class $[s^0]$.
\i Transitions of the process
   $P_\approx$ have the form
   $$\diagrw{[s_1]&\pright{a}&[s_2]}$$
   where $\diagrw{s_1&\pright{a}&s_2}$ is
   an arbitrary transition from $R$.
\ei
Then $P\oc (P_\approx). \quad \blackbox$\\

\refstepcounter{theorem}
{\bf Theorem \arabic{theorem}\label{conthfe2225o4bse4}}.

Let $P'$ be a process which 
is obtained from a process $P$ 
by removing of unreachable states. 
Then $P'_\approx$ has the smallest number 
of states among all processes that are 
observationally congruent to 
$P.\quad \blackbox$

\chapter{Recursive definitions of processes}
\label{recur}

In some cases, it is more convenient 
to describe a process by a recursive definition, 
intsead of explicit description of  sets 
of its states and transitions. 
In the present chapter we introduce a method
of description of  processes by 
recursive definitions.

\section{Process expressions}
\label{ppprocvyr}

In order to formulate a notion 
of recursive description of a process
we introduce a notion 
of a {\bf process expression}.

A set $PE$ of
{\bf process expressions (PE)}
is defined inductively, i.e. we define
\bi
\i elementary PEs, and
\i rules for constructing new PEs 
   from existing ones.
\ei

Elementary PEs have the following form.
\begin{description}
\item[process constants:]$\;$\\ 
We assume that there is given 
a countable set of process constants, 
and each of them is associated with 
a certain process, which is called
a {\bf value} of this constant.

Each process constant is a PE.

There is a process constant,
whose value is the empty process {\bf 0}.
This constant is denoted by the same symbol 
${\bf 0}$.

\item[process names:] $\;$\\
   We assume that there is given 
   a countable set
   {\bf of process names}, and
each process name is a PE.
\end{description}

Rules for constructing new PEs 
from existing ones have the 
following form.

\begin{description}
\item[prefix action:] $\;$\\
   For each $a\in Act$ and each PE $P$ 
   the string $a.P$ is a PE.
\item[choice:] $\;$\\
   For any pair of PEs $P_1, P_2$
   the string $P_1 + P_2$
   is a PE.
\item[parallel composition:] $\;$\\
   For any pair of PEs $P_1, P_2$
   the string $P_1 \pa P_2$
   is a PE.
\item[restriction:] $\;$\\
   For each subset $L \subseteq Names$
   and each PE $P$
   the string $P\setminus L$ is a PE.
\item[renaming:] $\;$\\
   For each renaming $f$
   and each PE $P$
   the string $P[f]$ is a PE.
\end{description}

\section{A notion of a recursive definition 
of  processes}
\label{gjklefgmsdfgsdgrrr}

A {\bf recursive definition (RD) 
of processes}
is a list of formal equations 
of the form
\be{spu}\left\{\by
A_1 = P_1 \\\ldots \\A_n = P_n
\ey \right.\ee
where 
\bi\i $A
_1,\ldots, A_n$ are different 
process names, and
\i $P_1,\ldots, P_n$ are PEs,
satisfying the following condition: 
for every $i=1,\ldots, n$ 
each process name, which 
has an occurrence in $P_i$, 
coincides with one
of the names of $A_1,\ldots, A_n$.
\ei

We shall assume that for each process
name $A$ there is a unique RD such that
$A$ has an occurrence in this RD.

In section \ref{smysslerlwe} 
we define a correspondence, 
which associates with each PE $P$ 
some process $[\![P]\!]$.
To define this correspondence, 
we shall give first
\bi
\i a notion of an {\bf embedding} of processes, 
   and
\i a notion of a {\bf  limit} of a sequence 
   of embedded processes.
\ei

\section{Embedding of processes}

Let $P_1$ and $P_2$ be processes of the form
\be{sadgsdfhgfd}
P_i=(S_i,s^0_i, R_i)\quad(i=1,2)\ee

The process $P_1$ is said to be 
{\bf embedded} to the process $P_2$, 
if there is an injective mapping 
$f: S_1 \to S_2$, such that
\bi
\i $f(s_1^0)=s_2^0$, and
\i for any $s',s''\in S_1$ and 
   any $a\in Act$
    $$(s'\ra{a}s'')\in R_1   \quad\Leftrightarrow\quad
   (f(s')\ra{a}f(s''))\in R_2
   $$
\ei

For each pair of processes $P_1, P_2$
the notation
$$P_1 \hookrightarrow P_2$$
is an abridged notation of the statement that
$P_1$  is embedded to $P_2$.

If the processes $P_1$ and  $P_2$ 
have the form \re{sadgsdfhgfd}, 
and $P_1 \hookrightarrow P_2$, 
then we can identify $P_1$ 
with its image in $P_2$, i.e. 
we can assume that 
\bi
\i $S_1\subseteq S_2$
\i $s^0_1 = s^0_2$
\i $R_1\subseteq R_2$.
\ei

\refstepcounter{theorem}
{\bf Theorem \arabic{theorem}\label{thwe4}}.
Let $P_1 \hookrightarrow P_2$.
Then
\bi
\i $a.P_1 \hookrightarrow  a.P_2$
\i $P_1+P \hookrightarrow  P_2+P$
\i $P_1\pa P \hookrightarrow  P_2\pa P$
\i $P_1\setminus L \hookrightarrow	P_2\setminus L$
\i $P_1[f]\hookrightarrow  
   P_2[f].\quad\blackbox$
\ei

Below we consider expressions
which are built from \bi\i processes, 
and \i symbols of operations on processes 
($a.$, $+$, $\pa$, $\setminus L$, $[f]$).\ei
We call such expressions 
as {\bf expressions over processes}.
For each expression over processes 
it is defined a process which is 
a value of this expression.
In the following reasonings we shall denote
an expression over the process and its value 
by the same symbol. \\

\refstepcounter{theorem}
{\bf Theorem \arabic{theorem}
\label{thwe4wed34}}.

Let
\bi
\i $P$ be an expression over processes, 
\i $P_1, \ldots, P_n$ be 
   a list of all processes occurred in $P$
\i $P'_1, \ldots, P'_n$ be a list of processes
   such that 
   $$\forall \, i=1,\ldots, n \quad P_i \hookrightarrow P'_i$$
\i $P'$ be an expression which is obtained
   from $P$ by a replacement 
   for each $i=1,\ldots, n$
   each occurrence of the process $P_i$ to
   the corresponding process $P'_i$.
\ei
Then $P\hookrightarrow  P'$.\\

{\bf Proof}.

This theorem is proved by induction 
on a structure of the expression $P$.
We prove that
for each subexpression $Q$ of the 
expression $P$ 
\be{we23423556566}Q \hookrightarrow Q'\ee 
where
$Q'$ is a subexpression 
of the expression $P'$,
which corresponds to the subexpression $Q$.

\begin{description}
\i[base of induction:] $\;$\\
If $Q = P_i$, then
$Q'= P'_i$, and \re{we23423556566}
holds by assumption.
\i[inductive step:] $\;$\\
From theorem \ref{thwe4}
it follows that for each subexpression $Q$ 
of the expression $P$
the following implication holds:
if for each proper subexpression $Q_1$ 
of $Q$ the following statement holds
$$Q_1 \hookrightarrow Q'_1$$
then \re{we23423556566} holds.
\end{description}
Thus, \re{we23423556566} holds
for each subexpression $Q$ of $P$.
In particular, \re{we23423556566} 
holds for $P$. $\blackbox$

\section{A limit of a sequence 
of embedded processes}

Let $\{P_k \mid k \geq 0\}$ be 
a sequence of processes, such that
\be{asdwe333}\forall k\geq 0\quad
P_k\;\hookrightarrow \;P_{k+1}\ee

A sequence $\{P_k \mid k \geq 0 \}$
satisfying condition \re{asdwe333}
is called 
a {\bf a sequence of embedded processes}.

Define a process 
$\lim\limits_{k\to\infty}P_k$, 
which is called a {\bf limit} 
of the sequence of embedded processes
$\{P_k \mid k \geq 0\}$.

Let the processes $P_k\;(k\geq 0)$
have the form
$$P_k=(S_k, s^0_k, R_k)$$
On the reason of \re{asdwe333}, 
we can assume that $\forall k\geq 0$
\bi
\i $S_k\;\subseteq \;S_{k+1}$
\i $s^0_k=s^0_{k+1}$
\i $R_k\;\subseteq \;R_{k+1}$
\ei
i.e. the components of the processes
$P_k\;(k\geq 0)$ have the 
following properties:
\bi
\i $S_0\;\subseteq\;S_1\;\subseteq\;S_2
   \;\subseteq\;\ldots$
\i $s^0_0=s^0_1=s^0_2=\ldots$
\i $R_0\;\subseteq\;R_1\;\subseteq\;R_2
\;\subseteq\;\ldots$
\ei

The process $\lim\limits_{k\to\infty}P_k$
has the form
$$(\bigcup\limits_{k \geq 0}S_k, s^0_0, 
\bigcup\limits_{k \geq 0}R_k)$$

It is easy to prove that for each 
$k\geq 0$
$$P_k\hookrightarrow \lim\limits_{k\to\infty}P_k$$

\refstepcounter{theorem}
{\bf Theorem \arabic{theorem}\label{thwe411}}.

Let $\{P_k\mid k\geq 0\}$ and 
$\{Q_k\mid k\geq 0\}$ be 
sequences of embedded processes.

Then
\bi
\i $\lim\limits_{k\to\infty}(a.P_k)=
   a.(\lim\limits_{k\to\infty}P_k)$
\i $\lim\limits_{k\to\infty}(P_k+Q_k) =
(\lim\limits_{k\to\infty}P_k)+
(\lim\limits_{k\to\infty}Q_k)$
\i $\lim\limits_{k\to\infty}(P_k\pa Q_k) =
(\lim\limits_{k\to\infty}P_k)\pa
(\lim\limits_{k\to\infty}Q_k)$
\i $\lim\limits_{k\to\infty}(P_k\setminus L) =
(\lim\limits_{k\to\infty}P_k)\setminus L$
\i $\lim\limits_{k\to\infty}(P_k[f]) =
(\lim\limits_{k\to\infty}P_k)[f]\quad\blackbox$
\ei

Let 
\bi
\i $P$ be a PE, 
\i $A_1$, $\ldots$, $A_n$
   be a list of all process names
   occurred in $P$.
\ei
Then for every $n$--tuple 
of processes $P_1$, $\ldots$, $P_n$
the notation
$$P(P_1/A_1,\ldots, P_n/A_n)$$
denotes an expression over processes
(as well as its value)
obtained from $P$ by replacement 
for each $i=1,\ldots, n$
each occurrence of the 

process name $A_i$
on the corresponding process $P_i$. \\

\refstepcounter{theorem}
{\bf Theorem \arabic{theorem}\label{thwe4we1111d34}}.

Let \bi\i $P$ be a PE, and 
\i $A_1$, $\ldots$, $A_n$
be a list of all process names
occurred in $P$.\ei
Then for every list of sequences of 
embedded processes of the form
$$\{P_1^{(k)}\mid k\geq 0\}, \quad
\ldots\quad
    \{P_n^{(k)}\mid k\geq 0\}$$
the following equality holds:
$$\by P((\lim\limits_{k\to\infty}P_1^{(k)})/A_1,\ldots,
	(\lim\limits_{k\to\infty}P_n^{(k)})/A_n)=\\=
	\lim\limits_{k\to\infty} P(P_1^{(k)}/A_1,\ldots, P_n^{(k)}/A_n)
	\ey$$

{\bf Proof}.

This theorem is proved 
by induction on the structure 
of the PE $P$, 
using theorem \ref{thwe411}.
$ \blackbox$

\section{Processes defined by process
expressions}
\label{smysslerlwe}

In this section we describe a rule
which associates with each PE $P$ 
a process $[\![P]\!]$, which is 
defined by this PE.

If $P$ is a process constant, then 
$[\![P]\!]$ is a value of this constant.

If $P$ has one of the following forms
$$a.P_1,\quad P_1+P_2,\quad P_1\pa P_2, 
\quad P_1\setminus L,\quad P_1[f]$$
then $[\![P]\!]$ is a result of 
applying of the corresponding
operation to the process $P_1$
or to the pair of processes $(P_1, P_2)$,
i.e.
$$\by
\mbox{$\;$}[\![a.P]\!]\eam a.[\![P]\!]\\
\mbox{$\;$}[\![P_1+P_2]\!]\eam [\![P_1]\!]+[\![P_2]\!]\\
\mbox{$\;$}[\![P_1\pa P_2]\!]\eam [\![P_1]\!]\pa [\![P_2]\!]\\
\mbox{$\;$}[\![P\setminus L]\!]\eam [\![P]\!]\setminus L\\
\mbox{$\;$}[\![P\;[f]\;]\!]\eam [\![P]\!]\;[f]
\ey$$

We now describe a rule that associates 
processes with process names.

Let $\{A_i = P_i\mid i=1,\ldots,n\}$ be a RD.

Define a sequence of lists of processes
\be{wew23435435}\{(P_1^{(k)},
\ldots, P_n^{(k)})\mid k\geq 0\}\ee
as follows:
\bi
\i $P_1^{(0)} \eam{\bf 0},\;\; \ldots,\;\; P_n^{(0)} \eam {\bf 0}$
\i if the processes 
   $P_1^{(k)}$, $\ldots$, $P_n^{(k)}$
   are already defined, 
   then for each $i=1,\ldots, n$
   $$P_{i}^{(k+1)}\eam P_i(P_1^{(k)}/A_1, \ldots, P_n^{(k)}/A_n)$$
\ei

We prove that for each $k\geq 0$
and each $i=1,\ldots, n$
\be{sdwsewe4333}P_i^{(k)} \hookrightarrow P_i^{(k+1)}\ee
The proof will proceed by induction on $k$.

\begin{description}
\i[base of induction:] $\;$\\
If $k = 0$, then by definition
$P_i^{(0)}$ coincides 
with the process ${\bf 0}$,
which can be embedded in any process.

\i[inductive step:] $\;$\\
Suppose that for each $i=1,\ldots, n\quad
P_i^{(k-1)} \hookrightarrow P_i^{(k)}$.

By definition of the processes 
from the set \re{wew23435435},
the following equalities hold:
$$\by
P_i^{(k)}=P_i(P_1^{(k-1)}/A_1, \ldots, P_n^{(k-1)}/A_n)\\
P_i^{(k+1)}=P_i(P_1^{(k)}/A_1, \ldots, P_n^{(k)}/A_n)\ey$$

The statement
$P_i^{(k)} \hookrightarrow P_i^{(k+1)}$
follows from theorem \ref{thwe4wed34}.
$\blackbox$

\end{description}

Define for each $i=1,\ldots, n$
the process $[\![A_i]\!]$ as the limit
$$[\![A_i]\!]\eam 
\lim\limits_{k\to\infty}P_i^{(k)}$$

From theorem \ref{thwe4we1111d34}
it follows that
for each  $i=1,\ldots, n$ 
the following chain of equalities holds:
$$\by
P_i([\![A_1]\!]/A_1,\ldots, [\![A_n]\!]/A_n) = \\=
P_i((\lim\limits_{k\to\infty}P_1^{(k)})/A_1,\ldots,
(\lim\limits_{k\to\infty}P_n^{(k)})/A_n)=\\=
\lim\limits_{k\to\infty}P_i(
P_1^{(k)}/A_1,\ldots,P_n^{(k)}/A_n
)=\\=
\lim\limits_{k\to\infty}(P_i^{(k+1)})=[\![A_i]\!]
\ey$$
i.e. the list of processes 
$$[\![A_1]\!],\ldots, [\![A_n]\!]$$
is a solution of the system of equations, 
which corresponds to the RD 
$$\left\{\by
A_1 = P_1 \\\ldots \\A_n = P_n
\ey \right.$$
(variables of this system of 
equations are the process names
$A_1$, $\ldots$, $A_n$).

\section{Equivalence of RDs}

Suppose that there is given a couple of RDs
of the form
\be{sdsw}\left\{\by
A^{(1)}_1 = P^{(1)}_1 \\\ldots 
\\A^{(1)}_n = P^{(1)}_n
\ey \right.\quad\mbox{and}\quad
\left\{\by
A^{(2)}_1 = P^{(2)}_1 \\\ldots
\\A^{(2)}_n = P^{(2)}_n
\ey \right.\ee

For each $n$-tuple of processes 
$Q_1$, $\ldots$, $Q_n$ 
the string
$$P^{(j)}_i(Q_1,\ldots,Q_n)$$
denotes the following
expression on processes (and its value):
$$P^{(j)}_i(Q_1/A^{(j)}_1,\ldots,
Q_n/A^{(j)}_n)\qquad
(i=1,\ldots, n;\;  j=1,2)$$

Let $\mu$ be an equivalence 
on the set of all processes. 

RDs \re{sdsw} are said to be 
{\bf equivalent} 
with respect to $\mu$, if for
\bi 
\i each $n$--tuple of processes 
   $Q_1$, $\ldots$, $Q_n$, and 
\i each $ i = 1, \ldots, n $ 
\ei 
the following statement holds:
$$\Big(\; P^{(1)}_i(Q_1,\ldots,Q_n)
\;,\;
P^{(2)}_i(Q_1,\ldots,Q_n)
\;\Big)\;\in \mu$$

\refstepcounter{theorem}
{\bf Theorem \arabic{theorem}\label{th0ssr5e4}}.

Let $\mu$ be a congruence on the set 
   of all processes.

For every couple of RDs 
of the form \re{sdsw}, which are
equivalent with respect to $\mu$, 
the processes defined by these RDs, i.e. 
$$\{[\![A^{(1)}_i]\!]\mid i=1,\ldots, n\}\quad\mbox{and}\quad
  \{[\![A^{(2)}_i]\!]\mid i=1,\ldots, n\}$$
are also equivalent with respect to $\mu$, 
i.e. 
$$\forall \;i=1,\ldots,n\quad
\Big(\;[\![A^{(1)}_i]\!]
\;,\;
[\![A^{(2)}_i]\!]
\;\Big)\;\in \mu\quad\blackbox$$

\section{Transitions on $PE$} 

There is another way of defining 
of a correspondence between PEs 
and processes. 
This method is related to 
the concept of transitions on the set  $PE$. 
Every such transition 
is a triple of the form $(P,a,P')$, 
where $P,P'\in PE$, and $a\in Act$.
We shall represent a transition $(P,a,P')$ 
by the diagram
\be{wewrtyu}\diagrw{P&\pright{a}&P'}\ee

We shall define the set of transitions on $PE$
inductively, i.e. 
\bi
\i some transitions 
   will be described explicitly, and
\i other transitions will be described 
   in terms of inference rules.
\ei

In this section we assume that 
each process is a value of some 
process constants. 

Explicit transitions are defined as follows.
\bn 
\i if $P$ is a process constant, then
   $$\diagrw{P&\pright{a}&P'}$$
   where $P'$ is a process constant, 
   such that 
   \bi 
   \i values of $P$ and $P'$ 
      have the form 
      $$(S,s^0,R)\quad\mbox{and}
      \quad(S,s^1,R)
      $$
      respectively, and 
   \i $R$ contains the transition 
      $\diagrw{s^0&\pright{a}&s^1}$
\ei
\i $\mor{a.P}{a}{P}$, for any $a.P\in PE$
\en

Inference rules for constructing of new 
transitions on $PE$ from existing ones 
are defined as follows.

\bn
\i if 
   $\diagrw{P&\pright{a}&P'}$, then 
   \bi
   \i $\mor{P+Q}{a}{P'}$, and
   \i $\mor{Q+P}{a}{P'}$
   \i $\mor{P\pa Q}{a}{P'\pa Q}$, and
   \i $\mor{Q\pa P}{a}{Q\pa P'}$
   \i if $L\subseteq Names$,
	  $a\neq \tau$, and 
    $name(a)\not\in L$, then
	  $$\mor{P\setminus L}{a}    {P'\setminus L}$$
   \i for each renaming $f$
     $$\mor{P[f]}{f(a)}{P'[f]}$$
   \ei
\i if $a\neq\tau$, then from
   $$\mor{P_1}{a}{P'_1}
  \quad\mbox{and}\quad
   \mor{P_2}{\bar a}{P'_2}$$
   it follows that
  $$\mor{P_1\pa P_2}{\tau}{P'_1\pa P'_2}$$
\i For each RD \re{spu}
	  and
   each $i\in \{1,\ldots,n\}$
   \be{sys}\by
   \mbox{{if}}& \mor{P_i}{a}{P'}
   \\
   \mbox{{then}}& \mor{A_i}{a}{P'}
   \ey\ee
\en

For each PE $P \in PE$ 
a process $[\![P]\!]$, 
which corresponds to this PE, 
has the form 
$$(PE, P, {\cal R})$$
where ${\cal R}$ is a set of all transitions
on $PE$.\\

\refstepcounter{theorem}
{\bf Theorem \arabic{theorem}\label{thwe4we111we1d3swewee4}}.

For each RD \re{spu}
and each $i = 1, \ldots, n$ 
the following statement holds
$$[\![A_i]\!]\sim P_i([\![A_1]\!]/A_1,\ldots, [\![A_n]\!]/A_n)$$
(i.e. the list of processes 
$[\![A_1]\!],\ldots, [\![A_n]\!]$
is a solution (with respect to $\sim$)
of the system of equations 
which corresponds to RD \re{spu}.
$\blackbox$ 

\section{A method of a proof 
of equivalence of processes
with use of RDs} 
\label{asdfw3434r3r34r34}

One of possible methods for proof
of an equivalence ($\sim$ or $\oc$) 
between two processes consists of 
a construction of an appropriate RD 
such that both of these processes are 
components with the same numbers 
of some solutions of a system 
of equations related to this RD. 

The corresponding equivalences
are substantiated by theorem 
\ref{thwe4we111we1d3se4}. 

To formulate this theorem, 
we introduce the following 
auxiliary notion. 

Let $\mu$ be a binary relation of the 
set of all processes, and let 
there is given an RD of the form \re{spu}. 

A list of processes, defined by the RD, 
is said to be unique up to $\mu$, 
if for each pair of lists of processes 
$$(Q^{(1)}_1, \ldots, Q^{(1)}_n)\quad\mbox{and}\quad
(Q^{(2)}_1, \ldots, Q^{(2)}_n)$$
which satisfies to the condition
$$
\by \forall i=1,\ldots, n\\
\;\vspace{5mm}\by
\big(\;[\![Q^{(1)}_i]\!]\;,\;
P_i(Q^{(1)}_1/A_1, \ldots, 
Q^{(1)}_n/A_n)\;\big)\;\in \mu\\
\big(\;[\![Q^{(2)}_i]\!]\;,\;
P_i(Q^{(2)}_1/A_1, \ldots, 
Q^{(2)}_n/A_n)\;\big)\;\in \mu\ey\ey$$
the following statement holds:
   $$\forall i=1,\ldots, n\quad
   \Big(\;[\![Q^{(1)}_i]\!]\;,\; 
   [\![Q^{(2)}_i]\!]\;\Big)\;\in \mu
   $$

\refstepcounter{theorem}
{\bf Theorem \arabic{theorem}\label{thwe4we111we1d3se4}}.

Let there is given a RD of the form \re{spu}. 

\bn 
\i If each occurrence of each 
   process name 
   $A_i$ in each PE $P_j$ is contained in 
   a subexpression of the form $a.Q$, then
   a list of processes, which is 
   defined by this RD, is 
   unique up to $\sim$. 
\i If 
   \bi 
   \i each occurrence of each 
         process name $A_i$ 
      in each PE $P_j$ is contained in 
      a subexpression of the form 
      $a.Q$, where $a\neq\tau$, and 
   \i each occurrence of each 
      process name $A_i$ 
     in each PE $P_j$ is contained 
      only in subexpressions of 
      the forms $a.Q$ and $Q_1 + Q_2$ 
   \ei 
   then   a list of processes, 
   defined by this RD, is 
   unique up to $\oc$. 
   $\blackbox$ 
\en 

\section{Problems related to RDs} 
\label{sdjkflgw3459333}

\bn 
\i Recognition of existence 
   of finite processes that are equivalent 
   (with respect to 
   $\sim$, $\approx$, $\oc$) 
   to processes of the form $[\![A]\!]$. 
\i Construction of algorithms 
   for finding minimal 
   processes which are equivalent to 
   processes of the form $[\![A]\!]$ 
   in the case 
   when these processes are finite. 
\i Recognition of equivalence of
   processes of the form 
   $[\![A]\!]$ \\
   (these processes can be infinite, 
   and methods from chapter 
   \ref{weqiro23rui34iop}
   are not appropriate for them). 
\i Recognition of equivalence of RDs. 
\i Finding necessary and sufficient conditions 
   of uniqueness of a list of processes 
   which is defined by a RD
   (up to $\sim$, $\oc$). 
\en

\chapter{Examples of a proof 
of properties of processes}

\section{Flow graphs}

In this section we describe a notion of a flow graph, 
which is intended to enhance a visibility and 
to facilitate an understanding 
of a relationship between components of complex processes.
Each example of a complex process, which is considered in this 
book, will be accompanied by a flow graph,
which corresponds to this process.

Let $P_1, \ldots, P_n$ be a list of processes.

A {\bf structural composition} 
of the processes $P_1$, $\ldots$, $P_n$
is an expression $SC$ over processes, 
such that 
\bi 
\i $SC$ contains 
   only processes from the list
   $P_1$, $\ldots$, $P_n$, and
\i each symbol of an operation,
   which consists in $SC$,
   is a symbol of one of the 
   following operations:
   \bi
   \i parallel composition,
   \i restriction,
   \i renaming.
   \ei
\ei

Each structural composition $SC$
can be associated with a diagram, 
which is called a {\bf flow graph (FG)} 
of $SC$.

A FG of a structural composition $SC$ 
is defined by induction 
on a structure of $SC$ as follows.
\bn
\i If $SC$ consists of only a process $P_i$, 
   then 
   FG of $SC$ is an oval, inside of which 
   it is written an identifier of this process.
   
   On the border of this oval 
   it is drawn circles, 
   which are called {\bf ports}. 

   Each port corresponds to some 
   input or output action $a\in 
   Act(P_i)$, and
   \bi 
   \i an identifier of this action 
      is written near of the port, 
      as a label of the port, 
   \i if $a$ is an input action,
      then the port is white,
   \i if $a$ is an input action
      then the port is black.
   \ei 

    For every $a\in Act(P_i)
    \setminus \{\tau\}$ 
    there is 
    a unique port on the oval,
    such that its label is $a$.
\i If $SC = SC_1\pa SC_2$, then
   a FG of $SC$ is obtained 
   by a disjoint union of 
   FGs of $SC_1$ and $SC_2$, 
   with drawing of labelled arrows 
   on the disjoint union: for 
   \bi
   \i every black port $p_1$ 
      on one of these FGs, 
      and 
   \i every white port $p_2$ 
      on another of these FGs, 
      such that
      labels of these ports 
      are complementary actions
   \ei
   it is drawn an arrow
   from $p_1$ to $p_2$ with a label
   $name(a)$, where $a$ is a label of $p_1$.
\i If $SC = SC_1\setminus L$, then
   a FG of $SC$ is obtained from 
   a FG of $SC_1$ by a removal of 
   labels of ports, whose names 
   belong to $L$.
\i If $SC = SC_1\,[f]$, then
   a FG of $SC$ is obtained from 
   a FG of $SC_1$ by a corresponfing
   renaming of labels of ports.
\en

If $P$ is a process which is 
equal to a value of 
a structural composition $SC$, 
then the notation $FG(P)$ denotes 
a FG of $SC$.

\section{Jobshop}
\label{masterskaja}

Consider a model of a jobshop, 
which employs two workers, 
who use for working one mallet. 

A behavior of each worker in the jobshop
is described by the following process $Jobber$

$$\by
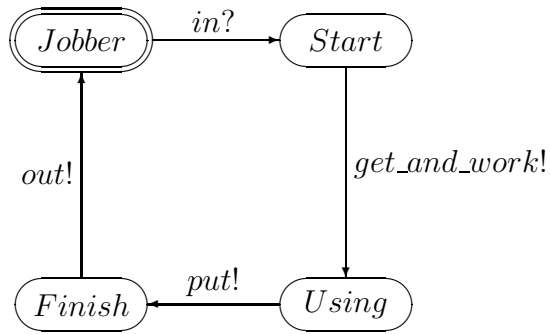
\begin{picture}(100,120)

\put(0,100){\makebox(0,0)[c]{$Jobber$}}
\put(100,100){\makebox(0,0)[c]{$Start$}}
\put(100,0){\makebox(0,0)[c]{$Using$}}
\put(0,0){\makebox(0,0)[c]{$Finish$}}

\put(0,100){\oval(50,20)}
\put(0,100){\oval(54,24)}
\put(0,0){\oval(50,20)}
\put(100,100){\oval(50,20)}
\put(100,0){\oval(50,20)}

\put(27,100){\vector(1,0){48}}
\put(75,0){\vector(-1,0){50}}
\put(100,90){\vector(0,-1){80}}
\put(0,10){\vector(0,1){78}}

\put(50,103){\makebox(0,0)[b]{$in?$}}
\put(50,3){\makebox(0,0)[b]{$put!$}}
\put(103,50){\makebox(0,0)[l]

{$get\_and\_work!$}}
\put(-3,50){\makebox(0,0)[r]{$out!$}}

\end{picture}
\\
\vspace{5mm}
\ey
$$
where
\bi 
\i the actions $in\,?$ and $out\,!$ 
   are used for interaction of a worker
   with a client, and denote 
   \bi 
   \i receiving of a material, and 
   \i issuance of a finished product
   \ei 
   respectively,
\i actions $get\_and\_work\,!$ 
   and $put\,!$ 
   are used for interaction of a worker
   with a mallet and denote 
   \bi 
   \i taking a mallet and 
      working with it, and 
   \i returning the mallet
   \ei 
   respectively.
\ei 

The action
$get\_and\_work\,!$ 
consists of several elementary actions.
We do not detail them
and combine them in one action. 

According to the definition 
of the process $Jobber$, 
a worker works as follows:
\bi 
\i at first he accepts a material 
\i then he takes the mallet and works 
\i then he puts the mallet 
\i then he gives the finished product
\i and all these actions are repeated.
\ei 

A behavior of the mallet 
we present using the following process 
$Mallet$: 

$$\by
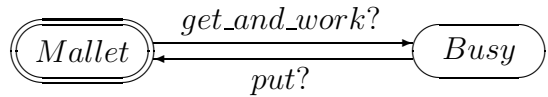
\begin{picture}(0,20)

\put(-75,0){\makebox(0,0)[c]{$Mallet$}}
\put(75,0){\makebox(0,0)[c]{$Busy$}}

\put(-75,0){\oval(50,20)}
\put(-75,0){\oval(54,24)}
\put(75,0){\oval(50,20)}

\put(-48,3){\vector(1,0){98}}
\put(50,-3){\vector(-1,0){98}}

\put(0,6){\makebox(0,0)[b]{$get\_and\_work?
$}}
\put(0,-6){\makebox(0,0)[t]{$put?$}}

\end{picture}
\\
\vspace{5mm}
\ey
$$
(note that the object ``mallet'' 
and the process ``Mallet'' are 
different concepts). 

A behavior of the jobshop 
is described by the process 
$Jobshop$:
$$Jobshop=
(Jobber\; \pa \;Jobber\; \pa \;Mallet)\setminus 
L$$
where $L=\{get\_and\_work,\; put\}$.

A flow graph of the process $Jobshop$
has the following form. 
$$\by
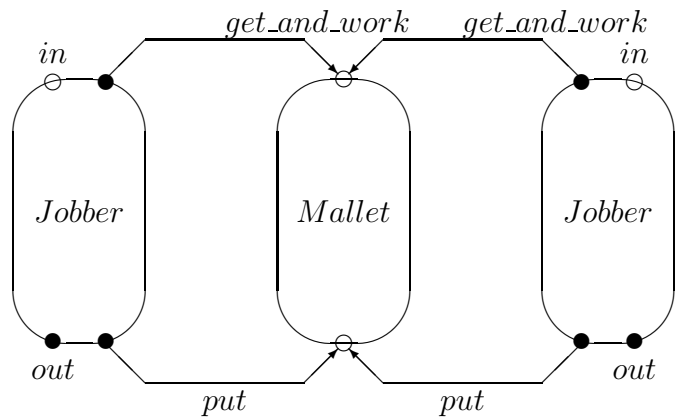
\begin{picture}(0,140)

\put(0,50){\oval(50,100)}
\put(100,50){\oval(50,100)}
\put(-100,50){\oval(50,100)}
\put(-100,50){\makebox(0,0)[c]{$Jobber$}}
\put(0,50){\makebox(0,0)[c]{$Mallet$}}
\put(100,50){\makebox(0,0)[c]{$Jobber$}}

\put(-90,99){\circle*{6}}
\put(-90,1){\circle*{6}}

\put(90,99){\circle*{6}}
\put(90,1){\circle*{6}}

\put(0,100){\circle{6}}
\put(0,0){\circle{6}}

\put(-110,99){\circle{6}}
\put(110,99){\circle{6}}
\put(-110,1){\circle*{6}}
\put(110,1){\circle*{6}}

\put(-90,100){\line(1,1){15}}
\put(-90,0){\line(1,-1){15}}
\put(90,100){\line(-1,1){15}}
\put(90,0){\line(-1,-1){15}}

\put(-75,115){\line(1,0){60}}
\put(-75,-15){\line(1,0){60}}
\put(75,115){\line(-1,0){60}}
\put(75,-15){\line(-1,0){60}}

\put(15,115){\vector(-1,-1){13}}
\put(15,-15){\vector(-1,1){13}}
\put(-15,115){\vector(1,-1){13}}
\put(-15,-15){\vector(1,1){13}}

\put(-45,118){\makebox(0,0)[b]

{$get\_and\_work$}}
\put(45,118){\makebox(0,0)[b]

{$get\_and\_work$}}
\put(-45,-18){\makebox(0,0)[t]{$put$}}
\put(45,-18){\makebox(0,0)[t]{$put$}}

\put(-110,106){\makebox(0,0)[b]{$in$}}
\put(110,106){\makebox(0,0)[b]{$in$}}
\put(-110,-6){\makebox(0,0)[t]{$out$}}
\put(110,-6){\makebox(0,0)[t]{$out$}}

\end{picture}
\\
\vspace{5mm}
\ey
$$

We now introduce the notion 
of an {\bf abstract worker}, 
about whom we know that 
he cyclically
\bi 
\i accepts a material and 
\i gives finished products 
\ei 
but nothing is known 
about details of his work. 

A behavior of the {\bf abstract worker}
we describe by the following process 
$Abs\_Jobber$:

$$\by
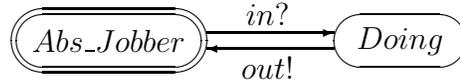
\begin{picture}(100,20)

\put(-10,0){\makebox(0,0)[c]{$Abs\_Jobber$}}
\put(100,0){\makebox(0,0)[c]{$Doing$}}

\put(-10,0){\oval(70,20)}
\put(-10,0){\oval(74,24)}
\put(100,0){\oval(50,20)}

\put(27,3){\vector(1,0){48}}
\put(75,-3){\vector(-1,0){48}}

\put(50,6){\makebox(0,0)[b]{$in?$}}
\put(50,-6){\makebox(0,0)[t]{$out!$}}

\end{picture}
\\
\vspace{5mm}
\ey
$$

A behavior of an {\bf abstract jobshop}
we describe by the following process 
$Abs\_Jobshop$:
$$Abs\_Jobshop = Abs\_Jobber \; \pa \; 
Abs\_Jobber$$

The process $Abs\_Jobshop$ is used 
as a {\bf specification} of the jobshop.
This process describes a behavior of 
the jobshop without details of its 
implementation. 

Prove that the process $Jobshop$
meets its specification, i.e. 
\be{dertkgdlsetgeugji}
Jobshop \oc Abs\_Jobshop
\ee

The process $Abs\_Jobshop$ is 
a parallel composition 
of two processes $Abs\_Jobber$. 
In order to avoid conflicts with the notations, 
we choose different identifiers to refer the
states of these processes. 

Suppose, for example, that
these processes have the form 
$$\by
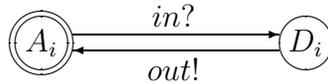
\begin{picture}(100,20)

\put(0,0){\makebox(0,0)[c]{$A_i$}}
\put(100,0){\makebox(0,0)[c]{$D_i$}}

\put(0,0){\oval(20,20)}
\put(0,0){\oval(24,24)}
\put(100,0){\oval(20,20)}

\put(12,3){\vector(1,0){78}}
\put(90,-3){\vector(-1,0){78}}

\put(50,6){\makebox(0,0)[b]{$in?$}}
\put(50,-6){\makebox(0,0)[t]{$out!$}}

\end{picture}
\\
\vspace{5mm}
\ey
$$
where $i=1,2$.

Parallel composition 
of these processes has the form 

$$\by
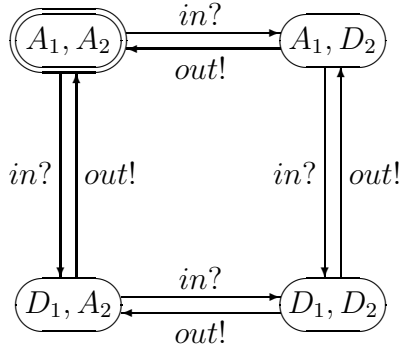
\begin{picture}(100,120)

\put(0,100){\makebox(0,0)[c]{$A_1,A_2$}}

\put(0,0){\makebox(0,0)[c]{$D_1,A_2$}}

\put(100,0){\makebox(0,0)[c]{$D_1,D_2$}}

\put(100,100){\makebox(0,0)[c]{$A_1,D_2$}}

\put(0,100){\oval(40,20)}
\put(0,100){\oval(44,24)}

\put(100,0){\oval(40,20)}
\put(0,0){\oval(40,20)}
\put(100,100){\oval(40,20)}

\put(22,103){\vector(1,0){58}}
\put(80,97){\vector(-1,0){58}}

\put(50,106){\makebox(0,0)[b]{$in?$}}
\put(50,94){\makebox(0,0)[t]{$out!$}}

\put(20,3){\vector(1,0){60}}
\put(80,-3){\vector(-1,0){60}}

\put(50,6){\makebox(0,0)[b]{$in?$}}
\put(50,-6){\makebox(0,0)[t]{$out!$}}

\put(-3,88){\vector(0,-1){78}}
\put(3,10){\vector(0,1){78}}

\put(-6,50){\makebox(0,0)[r]{$in?$}}
\put(6,50){\makebox(0,0)[l]{$out!$}}

\put(97,90){\vector(0,-1){80}}
\put(103,10){\vector(0,1){80}}

\put(94,50){\makebox(0,0)[r]{$in?$}}
\put(106,50){\makebox(0,0)[l]{$out!$}}

\end{picture}
\\
\vspace{5mm}
\ey
$$

Applying to this process the procedure 
of minimization with respect to 
observational equivalence, 
we get the process 

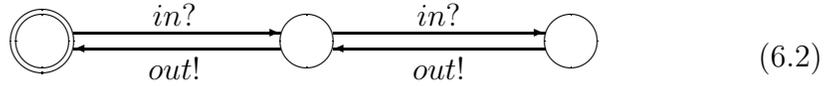
\be{sdafasdfdasfww}\by
\begin{picture}(200,20)

\put(0,0){\oval(20,20)}
\put(0,0){\oval(24,24)}
\put(100,0){\oval(20,20)}
\put(200,0){\oval(20,20)}

\put(12,3){\vector(1,0){78}}
\put(90,-3){\vector(-1,0){78}}

\put(110,3){\vector(1,0){80}}
\put(190,-3){\vector(-1,0){80}}

\put(50,6){\makebox(0,0)[b]{$in?$}}
\put(50,-6){\makebox(0,0)[t]{$out!$}}

\put(150,6){\makebox(0,0)[b]{$in?$}}
\put(150,-6){\makebox(0,0)[t]{$out!$}}

\end{picture}
\\
\vspace{5mm}
\ey
\ee

The process $Jobshop$ has 
$4\cdot4\cdot 2=32$ states, 
and we do not present it here
because of its bulkiness.
After a minimization of this 
process with respect to 
observational equivalence, 
we get a process, which is isomorphic 
to process \re{sdafasdfdasfww}.
This means that the following statement holds:
\be{sdfsdw4ewew}Jobshop \;\approx\; 
Abs\_Jobshop\ee
Because there is no transitions 
with a label $\tau$, starting from
initial states of processes 
\begin{center}$Jobshop$ and
$Abs\_Jobshop$\end{center}
then on the reason of \re{sdfsdw4ewew}
we conclude that \re{dertkgdlsetgeugji} holds. 

\section{Dispatcher}

Suppose that 
\bi
\i there is some company 
  which consists of several groups:
  $G_1$, $\ldots$, $G_n$, and
\i there is a special room in the building, 
   where the company does work, such that
   any group $G_i\quad(i\in \{1, \ldots, n\})$ 
   can use this room to conduct their workshops. 
\ei

There is a problem of non-conflictual use 
of the room by the groups
$G_1$, $\ldots$, $G_n$. 
This means that when one of the groups 
conducts a workshop in the room, 
other groups should be banned 
to hold their workshops in this room. 

This problem can be solved by use 
of a special process, which is called 
a {\bf dispatcher}. 

If any group $G_i$ wants to hold a workshop 
in this room, then $G_i$ should 
send the dispatcher a request 
to provide a right to use the room
for the workshop.

If the dispatcher knows that at this time
the room is busy, then he don't allows
$G_i$ to use this room.

When the room becomes free, 
the dispatcher sends $G_i$ a notice
that he allows to the group $G_i$
use this room.

After completion the workshop, 
the group $G_i$ must send the dispatcher 
a notice that the 
room is free. 

Consider a description of this system 
in terms of the theory of processes. 

A behavior of the dispatcher is described
by the process $D$, 
a graph representation of which 
consists of the following subgraphs:
for each $i = 1, \ldots, n$ 
it contains the subgraph
$$\by
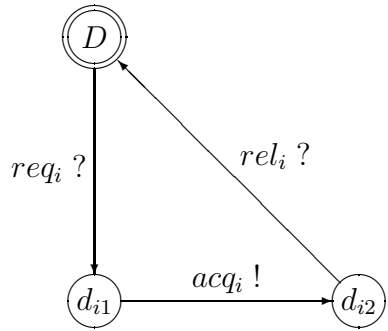
\begin{picture}(100,120)

\put(0,100){\makebox(0,0)[c]{$D$}}
\put(0,0){\makebox(0,0)[c]{$d_{i1}$}}
\put(100,0){\makebox(0,0)[c]{$d_{i2}$}}

\put(0,100){\oval(20,20)}
\put(0,100){\oval(24,24)}
\put(0,0){\oval(20,20)}
\put(100,0){\oval(20,20)}

\put(0,88){\vector(0,-1){78}}
\put(10,0){\vector(1,0){80}}
\put(93,7){\vector(-1,1){84}}

\put(-3,50){\makebox(0,0)[r]{$req_i\;?$}}
\put(50,3){\makebox(0,0)[b]{$acq_i\;!$}}
\put(55,55){\makebox(0,0)[l]{$rel_i\;?$}}

\end{picture}
\\
\vspace{0mm}
\ey
$$
i.e. 
$$D\sim
\sum\limits_{i=1}^n
req_i?.\;acq_i!.\;rel_i?.\;D
$$

Actions from $ Act(D)$
have the following meanings: 
\bi 
\i $req_i\,?$ is a receiving of a request
   from the group $G_i$
\i $acq_i\,!$ is a sending $G_i$
   of a notice that $G_i$ may use the room
\i $rel_i\,?$ is a receiving 
   a message that $G_i$ released
   the room. 
\ei 

In the following description 
of a behavior of each group $G_i$
\bi
\i we shall describe only an interaction 
   of $G_i$ \bi\i with the dispatcher, 
   and \i with the room\ei and 
\i will not deal with other functions
   of $G_i$. 
\ei

We shall denote
\bi 
\i a beginning of a workshop   in the room by the 
   action $start\,!$, and 
\i a completion of the meeting by
   the action of $finish\,!$. 
\ei 

A behavior of the group $G_i$
we describe by a process $G_i$, 
which has the following graph representation: 
$$\by
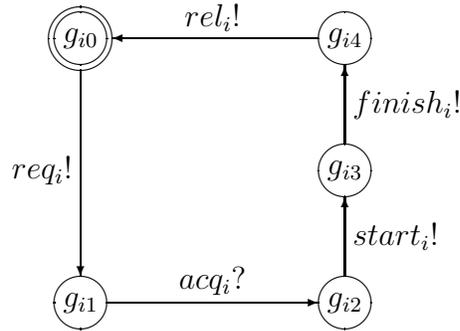
\begin{picture}(100,120)

\put(0,100){\makebox(0,0)[c]{$g_{i0}$}}
\put(0,0){\makebox(0,0)[c]{$g_{i1}$}}
\put(100,0){\makebox(0,0)[c]{$g_{i2}$}}
\put(100,50){\makebox(0,0)[c]{$g_{i3}$}}
\put(100,100){\makebox(0,0)[c]{$g_{i4}$}}

\put(0,100){\oval(20,20)}
\put(0,100){\oval(24,24)}
\put(0,0){\oval(20,20)}
\put(100,0){\oval(20,20)}
\put(100,100){\oval(20,20)}
\put(100,50){\oval(20,20)}

\put(0,88){\vector(0,-1){78}}
\put(10,0){\vector(1,0){80}}
\put(100,10){\vector(0,1){30}}
\put(100,60){\vector(0,1){30}}
\put(90,100){\vector(-1,0){78}}

\put(-3,50){\makebox(0,0)[r]{$req_i!$}}
\put(50,3){\makebox(0,0)[b]{$acq_i?$}}
\put(103,25){\makebox(0,0)[l]{$start_i!$}}
\put(103,75){\makebox(0,0)[l]{$finish_i!$}}
\put(50,103){\makebox(0,0)[b]{$rel_i!$}}

\end{picture}
\\
\vspace{0mm}
\ey
$$
i.e.
$G_i\sim
req_i!.\;acq_i?.\;start!.\;finish!. \;rel_i!.\;G_i$.

A joint behavior of the dispatcher 
and the groups can be described as the 
following process $Sys$: 
$$Sys = (D \pa G_1 \pa \ldots \pa G_n)
\setminus L$$
where 
$L=\{req_i, acq_i, rel_i\mid i=1,\ldots, n\}$.

A flow graph of the process $Sys$
for $n=2$ has the following form

$$\by
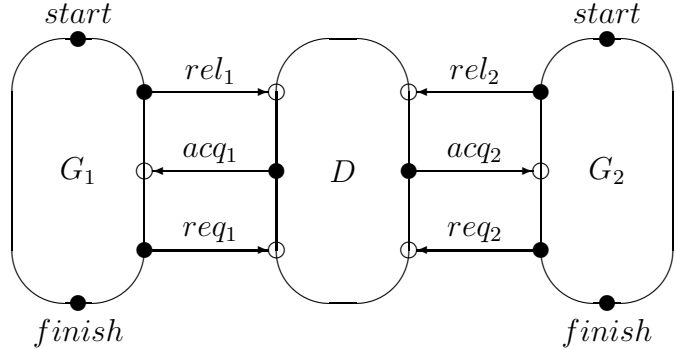
\begin{picture}(0,120)

\put(0,50){\oval(50,100)}
\put(100,50){\oval(50,100)}
\put(-100,50){\oval(50,100)}
\put(-100,50){\makebox(0,0)[c]{$G_1$}}
\put(0,50){\makebox(0,0)[c]{$D$}}
\put(100,50){\makebox(0,0)[c]{$G_2$}}

\put(-75,80){\circle*{6}}
\put(-75,50){\circle{6}}
\put(-75,20){\circle*{6}}

\put(75,80){\circle*{6}}
\put(75,50){\circle{6}}
\put(75,20){\circle*{6}}

\put(-25,80){\circle{6}}
\put(-25,50){\circle*{6}}
\put(-25,20){\circle{6}}

\put(25,80){\circle{6}}
\put(25,50){\circle*{6}}
\put(25,20){\circle{6}}

\put(-100,100){\circle*{6}}
\put(100,100){\circle*{6}}
\put(-100,0){\circle*{6}}
\put(100,0){\circle*{6}}

\put(-72,80){\vector(1,0){44}}
\put(-72,20){\vector(1,0){44}}
\put(-28,50){\vector(-1,0){44}}

\put(72,80){\vector(-1,0){44}}
\put(72,20){\vector(-1,0){44}}
\put(28,50){\vector(1,0){44}}

\put(-50,83){\makebox(0,0)[b]{$rel_1$}}
\put(-50,53){\makebox(0,0)[b]{$acq_1$}}
\put(-50,23){\makebox(0,0)[b]{$req_1$}}

\put(50,83){\makebox(0,0)[b]{$rel_2$}}
\put(50,53){\makebox(0,0)[b]{$acq_2$}}
\put(50,23){\makebox(0,0)[b]{$req_2$}}

\put(-100,106){\makebox(0,0)[b]{$start$}}
\put(100,106){\makebox(0,0)[b]{$start$}}
\put(-100,-6){\makebox(0,0)[t]{$finish$}}
\put(100,-6){\makebox(0,0)[t]{$finish$}}

\end{picture}
\\
\vspace{5mm}
\ey
$$

We now show that the processes 
which represent a behavior 
of the dispatcher and the groups 
indeed provide a conflict-free regime 
of use of the room.

The conflict-free property is that
\bi\i after a start of a workshop in the room
of any group
(i.e. after an execution the action $start!$
by this group), 
and \i before a completion of this workshop 
\ei
there is no another group which also 
may hold a workshop in this room
(i.e. which also can execute the action 
$start!$) until the first group has completed 
its workshop 
(i.e. until it has executed the action $finish!$).

Define a process $Spec$ as follows: 
$$\by
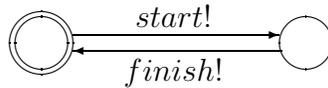
\begin{picture}(100,20)

\put(0,0){\oval(20,20)}
\put(0,0){\oval(24,24)}
\put(100,0){\oval(20,20)}

\put(12,3){\vector(1,0){78}}
\put(91,-3){\vector(-1,0){79}}

\put(50,6){\makebox(0,0)[b]{$start!$}}
\put(50,-6){\makebox(0,0)[t]{$finish!$}}

\end{picture}
\\
\vspace{0mm}
\ey
$$
i.e. $Spec\sim start!.\;finish!.\;Spec$.

The conflict-free property of 
the regime of use of the room 
is equivalent to the following statement:
\be{8}
Sys\approx Spec
\ee

To prove this statement, 
we transform the process $Sys$, 
applying several times the expansion theorem: 
$$\by
Sys\sim \\ \sim\sum\limits_{i=1}^n\tau.
\left(\by acq_i!.\; rel_i?.\; D \pa G_1 \pa \ldots 
\\ \ldots \pa acq_i?.\; start!.\; finish!.\; rel_i!.\; 
G_i \pa \ldots \\ \ldots \pa G_n\ey\right)
\setminus L\sim\\
\sim\sum\limits_{i=1}^n\tau.\tau.
\left(\by rel_i?.\;D \pa G_1 \pa \ldots \\ \ldots 
\pa start!.\; finish!.\; rel_i!.\; G_i \pa \ldots \\ 
\ldots \pa G_n\ey\right)\setminus L\sim\\
\sim\sum\limits_{i=1}^n\tau.\tau.start!.\;
\left(\by rel_i?.\;D \pa G_1 \pa \ldots \\ \ldots 
\pa finish!.\; rel_i!.\; G_i \pa \ldots \\ \ldots \pa 
G_n\ey\right)\setminus L\sim\\
\sim\sum\limits_{i=1}
^n\tau.\tau.start!.\;finish!.\;
\left(\by rel_i?.\;D \pa G_1 \pa \ldots \\ \ldots 
\pa rel_i!.\;G_i \pa \ldots \\ \ldots \pa 
G_n\ey\right)\setminus L\sim\\
\sim\sum\limits_{i=1}^n\tau.\tau.start!.\;finish!.\;\tau.
\underbrace{
\left(\by D \pa G_1 \pa \ldots \\ \ldots \pa G_i 
\pa \ldots \\ \ldots \pa G_n\ey\right)\setminus 
L}
\limits_{Sys}=\\
=\sum\limits_{i=1}
^n\tau.\tau.start!.\;finish!.\;\tau.Sys
\ey$$

Using the rules
$$P+P\sim P\quad
\mbox{and}\quad
\alpha.\tau.P\oc \alpha.P$$ 
we get the statement
$$Sys\oc \tau.start!.\;finish!.\;Sys$$

We now consider the equation 
\be{dsfwewewe22}X = 
\tau.start!.\;finish!.\;X\ee

According to theorem 
\ref{thwe4we111we1d3se4} 
from section \ref{asdfw3434r3r34r34}, 
there is a unique (up to $\oc$) 
solution of equation \re{dsfwewewe22} . 

As shown above, the process $Sys$ is 
a solution of \re{dsfwewewe22} 
up to $\oc$. 

The process $\tau.Spec$ is also a solution 
of \re{dsfwewewe22} up to $\oc$, 
because
$$\by\tau.Spec \sim \tau. start!.\;finish!.\; Spec 
\oc \\
\oc \tau. start!.\;finish!.\; (\tau.Spec)
\ey$$

Consequently, the following statement hold:
$$Sys \oc \tau.Spec$$ 
This statement implies \re{8}.

\section{Scheduler}

Suppose that there are $n$ processes 
\be{sdfsde3333}P_1,\ldots, P_n\ee
and for each $i=1,\ldots, n$ 
the set $Act(P_i)$ contains
two special actions: 
\bi
\i the action $\alpha_i?$, which 
   can be interpreted as a signal
   \be{nnrr}\mbox{{\it $P_i$ starts its regular 
    session}}\ee
\i the action $\beta_i?$, which 
   can be interpreted as a signal
   \be{zzrr}\mbox{{\it $P_i$ completes its 
regular session}}\ee
\ei

We assume that \bi\i all the names 
\be{sed3}\alpha_1,\ldots, 
\alpha_n,\beta_1,\ldots, \beta_n\ee
are different, and \i 
$\forall\;i=1,\ldots, n$ each 
name from 
$$names(Act(P_i))\setminus \{\alpha_i, 
\beta_i\}$$
does not belong to the set \re{sed3}. 
\ei
Let $L$ be the set \re{sed3}.

For each $i=1,\ldots, n$
the actions from the set
$$Act(P_i)\setminus \{\alpha_i?,\beta_i?\}$$
are said to be {\bf proper actions} of the
process $P_i$.

An arbitrary trace of each process $P_i$ 
may contain any quantity of 
the actions $\alpha_i?$ and $\beta_i?$ in any 
order. 

We would like to create a new process $P$, 
in which all the processes 
$P_1$, $\ldots$, $P_n$
would work together, 
and this joint work should 
obey certain regime. 

The process $P$ must have the form 
$$P=(P_1\pa \ldots \pa P_n\pa Sch)
\setminus L$$
where the process $Sch$ 
\bi
\i is called a {\bf scheduler}, and 
\i is designed for an establishing of a 
   required regime of an execution 
   of the processes $P_1$, $\ldots$, $P_n$.
\ei

Non-internal actions, which may be executed
by the process $Sch$, must belong to the set
\be{sed31}\{\alpha_1!,\ldots, 
\alpha_n!,\beta_1!,\ldots, \beta_n!\}\ee
By the definition of the process $P$, 
for each $i=1,\ldots, n$
\bi 
\i the actions $\alpha_i?$ and $\beta_i?$
   can be executed 
   by the process $P_i\in \re{sdfsde3333}$ 
   within the process $P$
   only simultaneously with an   execution of 
complementary actions 
   by the process $Sch$, and 
\i an execution of these actions 
   will be invisible outside the process $P$. 
\ei 

Informally speaking, 
each process $P_i$, 
which is executed within 
the process $P$, 
may start or complete 
its regular session if and only if 
the scheduler $Sch$ allows him to do it. 

A regime, which must be respected
by the processes $P_1$, $\ldots$, $P_n$,
during their execution 
within the process $P$, 
consists of the following two conditions. 
\bn 
\i For each $i=1,\ldots, n$ 
   an arbitrary trace of the process $P_i$, 
   which is executed within the process $P$, 
   should have the     form 
    $$
    \alpha_i?\; \ldots \;\beta_i?\; \ldots
    \alpha_i?\; \ldots \;\beta_i?\;\ldots$$
(where the dots represent proper actions of the 
process $P_i$), 
i.e. an execution of the process $P_i$ 
should be a sequence of sessions of the form
$$\alpha_i?\;\ldots\;\beta_i?\;\ldots$$
where each session 
\bi 
\i starts with an execution of 
   the action $\alpha_i?$ 
\i then several proper actions of $P_i$ 
   are executed, 
\i after a completion of the session
   the action $\beta_i?$ is executed, and
\i then $P_i$ can execute
   some proper actions\\
   (for example, these actions can be 
   related to a preparation 
   to the next session). 
\ei 
\i The processes $P_1$, $\ldots$, $P_n$
    are obliged to start their new sessions 
    in rotation, i.e.    
\bi 
\i at first, only $P_1$ may start its first session
\i then, $P_2$ may start its first session
\i $\ldots$ 
\i then, $P_n$ may start its first session
\i then, $P_1$ may start 
   its second session
\i then, $P_2$ may start 
   its second session
\i etc. 
\ei 
\en 
Note that we do not require that 
each process $P_i$ may receive a 
permission to start its $k$-th session 
only after the previous process $P_{i-1}$ 
completes its $k$-th session. 
However, we require that each process 
$P_i$ may receive a permission 
to start a new session, 
only if $P_i$ executed the action $\beta_i?$ 
(which 
signalizes a completion 
of a previous session of $P_i$). 

Proper actions of the processes $P_1$,
$\ldots$, $P_n$ can be executed in 
arbitrary order, and it is allowably
an interaction of these processes 
during their execution within the process $P$. 

The described regime can be formally
expressed as the following two conditions 
on an arbitrary trace $$tr\in Tr(Sch)$$
In these conditions we shall use the following 
notation: if 
$$tr\in Tr(Sch)\quad\mbox{and}\quad 
M\subseteq Act$$
then $tr\pa_M$ denotes 
a sequence of actions, which is 
derived from $tr$ by a removal of 
all actions which do not belong to $M$. 

Conditions which describe the above 
regime have the following form:
\be{isserpr1}\by \forall\; tr\in Tr(Sch),\;
\forall \;i=1,\ldots, n\\
tr \pa _{\{\alpha_i,\beta_i\}}\;=\;
   (\alpha_i!\;\beta_i!\;\alpha_i!\;\beta_i!
\;\alpha_i!\;\beta_i!\;\ldots)\ey\ee
and
\be{isserpr2}\by \forall\;tr\in Tr(Sch)\\
tr \pa _{\{\alpha_1,\ldots, \alpha_n\}}\;=\;
   (\alpha_1!\;\ldots \;\alpha_n!\;
   \alpha_1!\;\ldots \;\alpha_n!\;\ldots
   )\ey\ee

These conditions can be expressed 
as observational equivalence 
of certain processes. 

To define these processes, 
we introduce auxiliary notations. 

\bn 
\i Let $a_1\ldots a_n$ be a sequence 
of actions from $Act$. 
Then the string
$$(a_1\ldots a_n)^*$$
denotes a process which has the following 
graph representation
$$\by
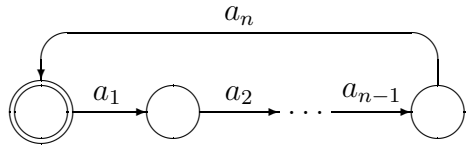
\begin{picture}(150,40)

\put(0,0){\oval(20,20)}
\put(0,0){\oval(24,24)}

\put(50,0){\oval(20,20)}
\put(150,0){\oval(20,20)}

\put(12,0){\vector(1,0){28}}
\put(60,0){\vector(1,0){30}}
\put(110,0){\vector(1,0){30}}

\put(150,10){\line(0,1){10}}
\put(140,20){\oval(20,20)[tr]}
\put(10,20){\oval(20,20)[tl]}
\put(10,30){\line(1,0){130}}
\put(0,20){\vector(0,-1){8}}

\put(75,33){\makebox(0,0)[b]{$a_n$}}

\put(25,3){\makebox(0,0)[b]{$a_1$}}
\put(75,3){\makebox(0,0)[b]{$a_2$}}
\put(125,3){\makebox(0,0)[b]{$a_{n-1}$}}
\put(100,0){\makebox(0,0)[c]{$\ldots$}}

\end{picture}
\\
\vspace{0mm}
\ey
$$
\i Let $P$ be a process, and
   \be{wewer3333}
   \{a_1,\ldots, a_k\}\subseteq
   Act \setminus \{\tau\}\ee
   be a set of actions.
   
   The string
   \be{sdfgsdwe44}
    hide\;(P,a_1,\ldots, a_k)\ee
   denotes the process 
     $$(\;P\pa(\,\overline{a_1}\,)^*\pa
   \ldots\pa (\,\overline{a_k}\,)^*\;)
   \setminus names(\{a_1,\ldots, a_k\})$$

\en 

Process \re{sdfgsdwe44} 
can be considered as a process, 
which is obtained from $P$ by 
a replacement on $\tau$
of all labels of transitions of $P$, 
which belong to the set \re{wewer3333}.

Using these notations, 
\bi
\i condition \re{isserpr1} 
   can be expressed as 
   follows: for each $i=1,\ldots n$
   \be{fdgddfd343}
   \!\!\!\!\by hide\left(\by Sch,& \alpha_1!,\ldots, 
   \alpha_{i-1}!,
   \alpha_{i+1}!,\ldots, \alpha_n! \\
   &\beta_1!,\ldots, \beta_{i-1}!,
   \beta_{i+1}!,\ldots, \beta_n!\ey\right)
   \approx\\	\approx (\alpha_i!.\;\beta_i!)^*\ey\ee and
\i condition \re{isserpr2} can be expressed 
   as follows: 
   \be{fdgddd3fsd43}hide\;
   (Sch, \beta_1!, \ldots, \beta_n!)
   \approx 
   (\alpha_1!.\;\ldots \alpha_n!)^*\ee
\ei   

It is easy to see that 
there are several schedulers 
that satisfy these conditions. 
For example, the following schedulers satisfy 
these conditions:
\bi
\i $Sch \;=\;(\alpha_1!\;\beta_1!
\;\ldots\;\alpha_n!\;\beta_n!)^*$
\i $Sch \;=\;(\alpha_1!\;\ldots \; \alpha_n!
\;\beta_1!\;\ldots \; \beta_n!)^*$
\ei

However, these schedulers impose 
too large restrictions
on an execution of the 
processes $P_1,\ldots, P_n$.

We would like to construct such 
a scheduler that allows
a maximal freedom 
of a joint execution 
of the processes $P_1,\ldots, P_n$
within the process $P$. 

This means that if at any time 
\bi 
\i the process  $P_i$ has an intention 
   to execute an action 
   $a\in \{\alpha_i?,\beta_i?\}$, and 
\i this intention of the process $P_i$ 
   does not contradict to the regime 
   which is described above 
\ei 
then the scheduler 
should not prohibit $P_i$ 
to execute this action 
at the current time, 
i.e. the action $\overline a$
must be among actions, 
which the scheduler can execute
at the current time.

The above informal description 
of a maximal freedom of an execution
of a scheduler can be formally clarified 
as follows: 
\bi 
\i each state $s$ of the scheduler 
   be associated with a pair $(i, X)$, where
   \bi 
   \i $i\in \{1,\ldots, n\}$, 
      $i$ is a number of a process, 
      which has the right to start 
      its regular session at the current 
      time 
   \i $X\subseteq \{1,\ldots, n\}
      \setminus\{i\}$,      
     $X$ is a set of active processes
      at the current time\\
      (a process is said to be active, if
      it started its regular session, but
      does not completed it yet)
    \ei
\i an initial state of the scheduler 
    is associated with a pair $(1,\emptyset)$
\i a set of transitions of the scheduler 
    consists of 
\bi
\i transitions of the form
   $$\diagrw{s&\pright{\alpha_i!}&s'}$$
   where
   \bi
   \i $s$ is associated with  $(i,X)$
   \i $s'$ is associated with  
      $(next(i), X\cup \{i\})$,
   where\\ 
   $next(i) \eam \left\{\by  i+1,& \mbox{if 
$i<n$, and} \\
   1,&\mbox{if $i=n$}\ey\right.$
   \ei
\i and transitions of the form
   $$\diagrw{s&\pright{\beta_j!}&s'}$$
   where \bi\i $s$ is associated with  
   $(i,X)$,\i $s'$ is associated with  
    $(i, X\setminus \{j\})$, 
     where $j\;\in X$\ei
\ei
\ei
The above description of properties 
of a required scheduler 
can be considered as its definition, i.e.
we can define a required scheduler 
as a process $Sch_0$ with the 
following components:
\bi 
\i a set of its states is the set of pairs 
   of the form 
   $$\{(i,X)\in \{1,\ldots, n\}
   \times \;{\cal P} (\{1,\ldots, n\})\mid
    i\not\in X\}$$
\i an initial state and transitions of  $Sch_0$
   are defined as it was described above.
\ei 

The definition of the scheduler $Sch_0$ 
has a significant deficiency: 
a size of the set of states of $Sch_0$ 
exponentially depends
on the number of processes 
\re{sdfsde3333}, 
that does not allow quickly
modify such scheduler 
in the case when the set 
of processes \re{sdfsde3333} 
is changed.

We can use $Sch_0$ only as an reference, 
with which we will compare other schedulers. 

To solve the original problem 
we define another scheduler $Sch$. 
We will describe it 
\bi 
\i not by explicit description of its states 
   and transitions, but
\i by setting of a certain expression, 
   which describes $Sch$ in terms 
   of a composition of several simple 
   processes. 
\ei 

In the description of the scheduler $Sch$ 
we shall use new names 
$\gamma_1$, $\ldots,$ $\gamma_n$.
Denote the set of these names by the 
symbol $\Gamma$. 

Process $Sch$ is defined as follows: 
\be{srfg345343}Sch \eam (Start \pa C_1 \pa 
\ldots \pa C_n)\setminus \Gamma\ee
where 
\bi
\i $Start \eam \gamma_1!.\;{\bf 0}$
\i for each $i=1,\ldots,n$ 
   the process $C_i$ is called a {\bf cycler}
   and has the form
   $$\by
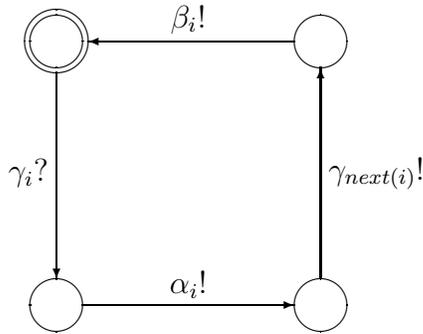
\begin{picture}(100,120)

\put(0,100){\oval(20,20)}
\put(0,100){\oval(24,24)}
\put(0,0){\oval(20,20)}
\put(100,0){\oval(20,20)}
\put(100,100){\oval(20,20)}

\put(0,88){\vector(0,-1){78}}
\put(10,0){\vector(1,0){80}}
\put(100,10){\vector(0,1){80}}
\put(90,100){\vector(-1,0){78}}

\put(-3,50){\makebox(0,0)[r]{$\gamma_i?$}}
\put(103,50){\makebox(0,0)[l]{$\gamma_{next(i)}!$}}
\put(50,103){\makebox(0,0)[b]{$\beta_i!$}}
\put(50,3){\makebox(0,0)[b]{$\alpha_i!$}}

\end{picture}
\\
\vspace{0mm}
\ey
$$
\ei

A flow graph of $Sch$ in the case $n=4$ 
has the following form: 
$$\by
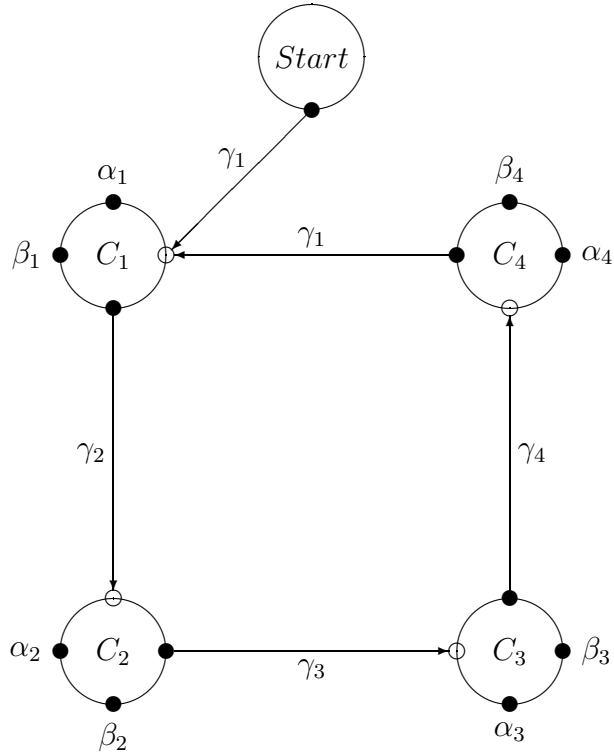
\begin{picture}(150,255)

\put(0,150){\oval(40,40)}
\put(0,0){\oval(40,40)}
\put(150,150){\oval(40,40)}
\put(150,0){\oval(40,40)}

\put(75,225){\oval(40,40)}
\put(75,205){\circle*{6}}

\put(75,205){\vector(-1,-1){53}}

\put(0,127){\vector(0,-1){104}}
\put(23,0){\vector(1,0){104}}
\put(150,23){\vector(0,1){104}}
\put(127,150){\vector(-1,0){104}}

\put(0,130){\circle*{6}}
\put(0,20){\circle{6}}
\put(0,-20){\circle*{6}}
\put(-20,0){\circle*{6}}
\put(20,0){\circle*{6}}

\put(130,0){\circle{6}}
\put(170,0){\circle*{6}}
\put(150,20){\circle*{6}}
\put(150,-20){\circle*{6}}

\put(150,130){\circle{6}}
\put(150,170){\circle*{6}}
\put(170,150){\circle*{6}}
\put(130,150){\circle*{6}}

\put(20,150){\circle{6}}
\put(0,170){\circle*{6}}
\put(-20,150){\circle*{6}}

\put(75,225){\makebox(0,0)[c]{$Start$}}

\put(0,150){\makebox(0,0){$C_1$}}
\put(0,0){\makebox(0,0){$C_2$}}
\put(150,0){\makebox(0,0){$C_3$}}
\put(150,150){\makebox(0,0){$C_4$}}

\put(0,177){\makebox(0,0)[b]{$\alpha_1$}}
\put(150,177){\makebox(0,0)[b]{$\beta_4$}}
\put(0,-27){\makebox(0,0)[t]{$\beta_2$}}
\put(150,-27){\makebox(0,0)[t]{$\alpha_3$}}
\put(-27,0){\makebox(0,0)[r]{$\alpha_2$}}
\put(-27,150){\makebox(0,0)[r]{$\beta_1$}}
\put(177,0){\makebox(0,0)[l]{$\beta_3$}}
\put(177,150){\makebox(0,0)[l]{$\alpha_4$}}

\put(-3,75){\makebox(0,0)[r]{$\gamma_2$}}
\put(153,75){\makebox(0,0)[l]{$\gamma_4$}}
\put(75,-3){\makebox(0,0)[t]{$\gamma_3$}}
\put(75,153){\makebox(0,0)[b]{$\gamma_1$}}

\put(45,183){\makebox(0,0)[b]{$\gamma_1$}}

\end{picture}
\\
\vspace{10mm}
\ey
$$

We give an informal explanation
of an execution of the process $Sch$.

The cycler $C_i$ is said to be
\bi 
\i {\bf disabled} 
   if it is in its initial state, and 
\i {\bf enabled}, 
   if it is not in its initial state.
\ei 

The process $Start$ enables
the first cycler $C_1$ and then ``dies''. 

Each cycler $C_i$ is responsible 
for an execution of the process $P_i$. 
The cycler $C_i$ 
\bi 
\i enables the next cycler 
   $C_{next(i)}$   
   after he gave a permission to the process 
   $P_i$ to start a regular session, and 
\i becomes disabled 
    after he gave a permission to the process 
   $P_i$ to complete a regular session. 
\ei 

Prove that process \re{srfg345343} satisfies 
condition \re{fdgddd3fsd43} 
(we omit checking of condition 
\re{fdgddfd343}). 

According to the definition of 
process \re{sdfgsdwe44}, 
condition \re{fdgddd3fsd43} has the form 
\be{fdgddffgdddd343}\by
(Sch \pa (\beta_1?)^* \pa \ldots \pa 
(\beta_n?)^*)
\setminus B\approx (\alpha_1!.\;\ldots \alpha_n!)^*\ey\ee
where $B=\{\beta_1,\ldots, \beta_n\}$.

Let $Sch'$ be the left side 
of \re{fdgddffgdddd343}.

Prove that 
\be{sdfskl4fgsd}Sch'\oc \tau. 
\alpha_1!.\;\ldots \alpha_n!.\;Sch'\ee
Hence by the uniqueness property 
(with respect to $\oc$) 
of a solution of the equation
$$X\;=\;\tau. \alpha_1!.\;\ldots 
\alpha_n!.\;X$$
we get the statement
$$Sch' \;\oc\;
(\tau\; \alpha_1!\;\ldots \alpha_n!\;)^*
$$
which implies \re{fdgddffgdddd343}. 

We will convert the left side 
of the statement
\re{sdfskl4fgsd} 
so as to obtain the right side 
of this statement. 
To do this, we will use 
properties 8, 11 and 12 
of operations on processes, 
which are contained in 
section \ref{archnatella}. 
We recall these properties: 
\bi
\i $P \setminus L = P$, if
   $L\cap names(Act(P))  = 
   \emptyset$
\i $(P_1 \pa P_2) \setminus L = 
   (P_1 \setminus L) \pa 
   (P_2 \setminus L)$,
   if $$L \cap names(Act(P_1) 
   \cap \overline{Act(P_2)}) = 
   \emptyset$$
\i $(P \setminus L_1) \setminus L_2 
   = P\setminus (L_1 \cup L_2)=
	(P \setminus L_2) \setminus L_1$
\ei
Using these properties, it is possible
to convert the left side of
\re{sdfskl4fgsd} as follows. 
\be{sdfjklwerwqopei}\by Sch'=\\
=(Sch \pa (\beta_1?)^* \pa \ldots \pa 
(\beta_n?)^*)\setminus B=\\
=\b{((Start \pa C_1 \pa \ldots \pa C_n)
\setminus \Gamma) \pa\\
\pa (\beta_1?)^* \pa \ldots \pa (\beta_n?)^*}
\setminus B=\\
=(Start \pa C'_1 \pa \ldots \pa C'_n)\setminus 
\Gamma
\ey\ee
where 
$$C'_i=(C_i \pa (\beta_i?)^*)\setminus 
\{\beta_i\}$$

Note that for each $i=1,\ldots, n$ 
the following statement holds:
\be{fglrtl554}
C_i'\;\oc\;\gamma_i?.\;\alpha_i!.\;\gamma_{next(i)}!.\;C'_i\ee

Indeed, by the expansion theorem,  
$$\by C'_i = (\big(\gamma_i?.\; \alpha_i!.\;
\gamma_{next(i)}!.\; \beta_i!.\; C_i\big) \pa 
(\beta_i?)^*)\setminus \{\beta_i\}
\sim\\
\sim \gamma_{i}?.\; \alpha_i!.\;  \gamma_{next(i)}!.\;\tau.C'_i\;\oc \;
\mbox{right side of \re{fglrtl554}}
\ey$$

Using this remark and 
the expansion theorem, 
we can continue the chain of equalities
\re{sdfjklwerwqopei}
as follows: 
\be{sdfjklwerwqopseei}\by
(Start \pa C'_1 \pa C'_2 \pa \ldots \pa C'_n)
\setminus \Gamma \oc \\
\oc (\underbrace{\gamma_1!.\;{\bf 0}}
\limits_{=Start}
 \pa \underbrace{\gamma_1?.\;\alpha_1!.\;\gamma_2!.\;C'_1}
\limits_{\oc C'_1} \pa C'_2
 \pa \ldots \pa C'_n)\setminus \Gamma \sim \\
\sim \tau.\;({\bf 0}
 \pa \alpha_1!.\;\gamma_2!.\;C'_1  \pa C'_2
 \pa \ldots \pa C'_n)\setminus \Gamma=\\
=\tau.\;(\alpha_1!.\;\gamma_2!.\;C'_1 \pa  
C'_2
 \pa \ldots \pa C'_n)\setminus \Gamma\sim\\
\sim \tau.\;\alpha_1!.\;(\gamma_2!.\;C'_1 \pa  
C'_2
 \pa \ldots \pa C'_n)\setminus \Gamma\oc\\
\oc \tau.\;\alpha_1!.\;(\gamma_2!.\;C'_1 \pa
 \underbrace{\gamma_2?.\;\alpha_2!.\;\gamma_3!.\;C'_2}
\limits_{\oc C'_2}
 \pa \ldots \pa C'_n)\setminus \Gamma\sim\\
\sim \tau.\;\alpha_1!.\;\tau.\;(C'_1 \pa
 \alpha_2!.\;\gamma_3!.\;C'_2
 \pa \ldots \pa C'_n)\setminus \Gamma\sim 
\ldots \sim\\
\sim \tau.\;\alpha_1!.\;\tau.\;\alpha_2!.\;\ldots 
\tau.\;\alpha_n!.\;
(C'_1 \pa \ldots \pa \gamma_1!.\;C'_n)
\setminus \Gamma\oc\\
\oc \tau.\;\alpha_1!.\;\ldots \alpha_n!.\;
(C'_1 \pa \ldots \pa \gamma_1!.\;C'_n)
\setminus \Gamma\oc\\
\oc \tau.\;\alpha_1!.\;\ldots \alpha_n!.\;
(\;
\underbrace{\gamma_1?.\;\alpha_1!.\;\gamma_2!.\;C'_1}
\limits_{\oc C'_1}
 \pa \ldots \pa \gamma_1!.\;C'_n\;)\setminus 
\Gamma\sim\\
\sim \tau.\;\alpha_1!.\;\ldots 
\alpha_n!.\;\underbrace{\tau.\;
(\alpha_1!.\;\gamma_2!.\; C'_1 \pa \ldots \pa 
C'_n)\setminus \Gamma}
\ey\ee

The underlined expression on the last 
line of the chain coincides 
with an expression on the 
fourth line of the chain, which is 
observationally congruent to $Sch'$. 

We have found that the last expression 
of the chain \re{sdfjklwerwqopseei}
is observationally congruent 
to the left side and
to the right side of \re{sdfskl4fgsd}.

Thus, the statement \re{sdfskl4fgsd}
is proven. $\blackbox$ \\

A reader is provided as an exercise 
the following problems.
\bn
\i To prove 
   \bi 
   \i condition \re{fdgddfd343}, and 
   \i the statement $Sch\approx Sch_0$,
   \ei 
\i To define and verify
   a scheduler that manages 
   a set $P_1$, $\ldots$, $P_n$
   of {\bf processes with priorities}, 
   in which each process $P_i$ 
   is associated with a certain {\bf priority}, 
   representing a number $p_i\in [0,1]$, 
   where $$\sum_{i=1}^n p_i=1$$

   The scheduler must implement 
   a regime of a joint execution 
   of the processes $P_1$, $\ldots$, $P_n$ 
   with the following properties:
   \bi 
   \i for each $i=1,\ldots, n$
      a proportion of a number 
      of sessions which are completed 
      by the process $P_i$, 
      relative to the total number 
      of sessions which are completed 
      by all processes 
      $P_1$, $\ldots$, $P_n$,  
      must asymptotically approximate to $p_i$
      with an infinite increasing
      of a time of an execution of 
      of the processes $P_1$, $\ldots$, $P_n$
   \i this scheduler should provide 
      a maximal freedom of an 
      execution of the 
      processes  $P_1,\ldots, P_n$.
   \ei 
\en

\section{Semaphore}

Let $P_1,\ldots, P_n$ be a list of processes, 
and for each $i=1,\ldots, n$ 
the process $P_i$ has the following form:
$$
P_i\;=\;(\alpha_i?\;a_{i1}\;\ldots\;a_{ik_i}
\;\beta_i?)^*$$
where 
\bi 
\i $\alpha_i?$ and $\beta_i?$ are special 
   actions representing 
   signals that \bi\i $P_i$ 
   started an execution of 
   a regular session, and 
   \i $P_i$ 
   completed an execution 
   of 
   a regular session
   \ei 
  respectively, and 
\i $a_{i1},\ldots,a_{ik_i}$ are proper
   actions of the process $P_i$. 
\ei 

We would like to create such a process $P$, 
in which all the processes  $P_1$, $\ldots$, 
$P_n$ would work together, 
and this joint work should 
obey the following regime: 
\bi 
\i if at some time of an execution of
   the process $P$
   any process $P_i$ started 
   its regular session (by an execution
   of the action $\alpha_i\,?$)
\i then this session must be 
   {\bf uninterrupted}
   i.e. all subsequent action 
   of the process $P$ 
   shall be actions of the process $P_i$, 
   until $P_i$ complete this session
   (by an execution
   of the action $\beta_i\,?$). 
\ei 

This requirement can be expressed 
in terms of traces: 
each trace of the process $P$ 
must have the form 
$$
\alpha_{i}?\;a_{i1}\;\ldots\;a_{ik_i}\;\beta_i?
\;
\alpha_{j}?\;a_{j1}\;\ldots\;a_{jk_j}
\;\beta_i?\;
\ldots
$$
i.e. each trace $tr$ of the process $P$ 
must be a concatenation of traces
$$tr_1\;\cdot\;tr_2\;\cdot\;tr_3\;\ldots$$
where each trace $tr_i$ in this concatenation 
represents a session of any process 
from the list $P_1$, $\ldots$, $P_n$.

A required process 
$P$ we define as follows: 
$$P\eam
\;(\;P_1[f_1]\;\pa \ldots \pa \;P_n[f_n] \;\pa\; 
Sem\;)
\setminus\{\pi,\varphi\}$$
where 
\bi 
\i $Sem$ is a special process
   designed to establish
   the required regime 
   of an execution of the 
    processes $P_1$, $\ldots$, $P_n$,
    this process 
   \bi\i is called a {\bf semaphore}, 
   and \i has the form 
   $$Sem\;=\;(\;\pi! \;\varphi!\;)^*$$\ei
   \i $f_i:  \alpha_i \mapsto \pi, \;\beta_i
   \mapsto \varphi$
\ei 

A {\bf specification} of the process 
$P$ is represented by the following 
statement:\be{sdfsadfsdaf}\by
P&\oc&
\tau.
{a_{11}.\ldots a_{1k_1}}
.\;P\;+\ldots+\\
&+&\tau.
{a_{n1}.\ldots a_{nk_n}}
.\;P\ey
\ee
A proof that the process $P$ 
meets this specification, 
is performed by means of the
expansion theorem: 
$$\by
P\;=\;(\;P_1[f_1]\;\pa \ldots \pa \;P_n[f_n] 
\;\pa\; Sem\;)
\setminus\{\pi,\varphi\} \sim\\
\sim \;\b{\pi?.a_{11}.\ldots.a_{1k_1}.
\varphi?.P_1[f_1]\;\pa \ldots \pa\\
\pa \;
\pi?.a_{n1}.\ldots.a_{nk_n}\varphi?.P_n[f_n] 
\;\pa\\\pa\; \pi!.\;\varphi!.\;Sem}
\setminus\{\pi,\varphi\} \sim\\
\sim \;\tau.\b{a_{11}.\ldots.a_{1k_1}.
\varphi?.P_1[f_1]\;\pa \ldots \pa\\
\pa \;
\pi?.a_{n1}.\ldots.a_{nk_n}\varphi?.P_n[f_n] 
\;\pa\\\pa\; \varphi!.\;Sem}
\setminus\{\pi,\varphi\} +\\+\ldots+\\
+ \tau.\b{\pi?.a_{11}.\ldots.a_
{1k_1}.\varphi?.P_1[f_1]\;\pa \ldots \pa\\
\pa \;
a_{n1}.\ldots.a_{nk_n}\varphi?.P_n[f_n] 
\;\pa\\\pa\; \varphi!.\;Sem}
\setminus\{\pi,\varphi\} \sim\\
\sim\ldots\sim\\\sim
\tau.
{a_{11}.\ldots a_{1k_1}}.\tau.\;P\;+\ldots
+\tau.
{a_{n1}.\ldots a_{nk_n}}.\tau.\;P \oc\\\oc
\tau.
{a_{11}.\ldots a_{1k_1}}.\;P\;+\ldots+
\tau.
{a_{n1}.\ldots a_{nk_n}}.\;P\;\;\blackbox
\ey
$$

Finally, pay attention to the following aspect. 
The prefix ``$\tau.$'' in each summand 
of the right side of \re{sdfsadfsdaf} 
means that a choice of a variant 
of an execution of the process $P$ 
at the initial time is determined 
\bi 
\i not by an environment 
   of the process $P$, but
\i by the process $P$ itself. 
\ei 
If this prefix was absent, 
then it would mean that a choice 
of a variant of an execution
of the process $P$ 
at the initial time
is determined by an environment 
of the process $P$.

\chapter{Processes with a message passing}

\section{Actions with a message passing}

The concept of a process 
which was introduced and studied
in previous chapters, 
can be generalized in different ways.

One of such generalizations consists of 
an addition to actions from $Act$ some
{\bf parameters} (or {\bf modalities}),
i.e. there are considered processes 
with actions of the form
$$(a,p)$$
where $a\in Act$, and $p$ is a parameter
which may have the following meanings: 
\bi 
\i a complexity (or a cost) of an execution 
   of the action $a$ 
\i a priority (or a desirability, or a plausibility)
   of the action $a$ 
  with respect to other actions 
\i a time (or an interval of time) 
   at which the action $a$ was executed
\i a probability of an execution of the 
   action $a$ 
\i or anything else. 
\ei 

In this chapter we consider  
a variant of such generalization, 
which is related to an addition
of {\bf messages}
to actions from $Act$.
These messages are transmitted 
together with an execution 
of the actions.

Recall our informal interpretation 
of the concept of an execution 
of an action:
\bi 
\i the action $\alpha\,!$ is executed 
   by sending of an object
   whose name is $\alpha$, and
\i the action $\alpha\,?$ is executed 
   by receiving of an object
   whose name is $\alpha$.
\ei 
We generalize this interpretation as follows. 
We shall assume that processes can
send or receive not only objects, 
but also pairs of the form
\begin{center}
(object, message)
\end{center}
i.e. an action may have the form 
\be{spssdfgsd343434546}\alpha\,!
\,v\quad\mbox{and}\quad \alpha\,?\,v\ee
where $\alpha \in Names$, 
and $v$ is a {\bf message},
that can be
\bi 
\i a string of symbols,
\i a material resource,
\i a bill, 
\i etc. 
\ei 

An execution of the actions 
$\alpha\,!\,v$ and $\alpha\,?\,v$, 
consists of sending or receiving 
the object $\alpha$ with the message $v$.

Recall that such entities as
\bi
\i a transferred object, and 
\i receiving and sending of objects
\ei
can have a virtual character 
(more details see in section 
\ref{sdfgsdfgw5etrwerw}). 

For a formal description 
of processes that can execute 
actions of the form 
\re{spssdfgsd343434546}, 
we generalize the concept of a process. 

\section{Auxiliary concepts} 
\label{vspomogatpon}

\subsection{Types, variables, values and 
constants}

We assume that there is given 
a set $Types$  of {\bf types}, 
and each type $t \in Types$ is associated with a 
set $D_t$ of 
{\bf values} of the type $t$. 

Types can be denoted by identifiers. 
We shall use the following identifiers:
\bi 
\i the type of integers 
   is denoted by \verb"int"
\i the type of boolean values 
    (0 and 1) is denoted by \verb"bool" 
\i the type of messages is denoted by 
   \verb"mes"
\i the type of lists of messages
   is denoted by \verb"list".
\ei 

Also, we assume that there are 
given the following sets. 
\bn
\i The set $Var$, whose elements 
   are called {\bf variables}.

   Every variable $x\in Var$ 
   \bi
   \i is associated with a type $t(x)\in Types$, 
      and
   \i can be associated with {\bf values} from the set 
      $D_{t(x)}$, i.e. at different times 
      the variable $x$ 
      can be associated with 
      various elements of the set $D_{t(x)}$.
   \ei
\i The set $Con$, whose elements 
   are called {\bf constant}. 

   Every constant $c\in Con$
   is associated with 
   \bi
   \i a type $t(c)\in Types$, and
   \i a value  $[\![c]\!]\in D_{t(c)}$,
      which is said to be an {\bf interpretation}
      of the constant $c$.
   \ei 
\en

\subsection{Functional symbols}
\label{vspomogatponfs}

We assume that there is given a set 
of {\bf functional symbols (FSs)}, 
and each FS $f$ is associated with
\bi 
\i a {\bf functional type} $t(f)$, which has the form
   \be{sdfgwe354w3t3}(t_1,\ldots, t_n)\to t\ee
   where $t_1,\ldots, t_n, t\in Types$, and 
\i a function 
   $$[\![f]\!]:D_{t_1}\times\ldots\times 
    D_{t_n}\to D_t$$ 
   which is called 
   an {\bf interpretation} of the FS $f$. 
\ei 
Examples of FSs: 
$$+,\quad -, \quad \cdot, \quad 
{\bf head},\quad {\bf tail}, \quad [\;]$$
where 
\bi 
\i the FSs $+$ and $-$ have the functional type 
   $$(\verb"int", \verb"int") \to \verb"int"$$ 
   the functions $[\![+]\!]$ and $[\![-]\!]$
   are the corresponding arithmetic operations
\i the FS $\cdot$ has the functional type 
  $$(\verb"list", \verb"list") 
   \to  \verb"list"$$ 
   the function $[\![\cdot]\!]$ maps
   each pair of lists $(u,v)$ to 
   their {\bf concatenation} (which
   is obtained by writing $v$ on the 
   right from $u$)
\i the FS {\bf head} has the functional type 
   $$\verb"list" \to \verb"mes"$$ 
   the function $[\![{\bf head}]\!]$ 
   maps each nonempty list
   to its first element \\
   (a value of $[\![{\bf head}]\!]$ 
   on an empty list can be any) 
\i the FS {\bf tail} has the functional type 
   $$\verb"list" \to \verb"list"$$
   the function  $[\![{\bf tail}]\!]$
   maps each nonempty list
   $u$ to the list which 
   is derived from $u$ by a removing 
   of its first element \\
   (a value of $[\![{\bf tail}]\!]$ 
   on an empty list can be any) 
\i the FS $[\;]$ has the functional type
   $$\verb"mes"\to \verb"list"$$
   the function $[\![\;[\;]\;]\!]$ 
   maps each message to the
   list which consists only of this message 
\i the FS ${\bf length}$  has the functional type 
   $$\verb"list"\to \verb"int"$$
   the function $[\![{\bf length}]\!]$  
   maps each list to its length \\
   (a length of a list is a number 
   of messages in this list)
\ei 

\subsection{Expressions} 
\label{vyrazhenija}

{\bf Expressions} consist of variables,
constants, and FSs, and are constructed 
by a standard way.
Each expression $e$ has a type 
$t(e)\in Types$, which is defined 
by a structure of this expression. 

Rules of constructing of expressions 
have the following form. 
\bi 
\i Each variable or constant 
   is an expression of the type 
   that is associated with
   this variable or constant. 
\i If 
   \bi 
   \i $f$ is a FS of the functional type 
      \re{sdfgwe354w3t3}, and 
   \i $e_1,\ldots, e_n$ are expressions 
      of the types $t_1,\ldots, t_n$ 
      respectively 
   \ei 
   then the list
   $f(e_1,\ldots, e_n)$
   is an expression of the type $t$. 
\ei 

Let $e$ be an expression.
If each variable $x$ occurred in $e$
is associated with a value $\sigma(e)$, 
then the expression $e$ can be associated
with a value $\sigma(e)$ which is defined 
by a standard way:
\bi
\i if $e=x\in Var$, then $\sigma(e)\eam    \sigma(x)$\\
   (the value $\sigma(x)$ is assumed to be given)
\i if $e=c\in Con$, then $\sigma(e)\eam  [\![c]\!]$\i 

if $e = f(e_1,\ldots, e_n)$, then 
   $$\sigma(e)\eam [\![f]\!](\sigma(e_1),\ldots, 
   \sigma(e_n))$$
\ei

Below we shall use the following notations. 
\bi 
\i The symbol ${\cal E}$ 
   denotes the set of all expressions.
\i The symbol ${\cal B}$ 
   denotes the set of       
   expressions of the type \verb"bool". 

   Expressions from ${\cal B}$ are called 
   {\bf formulas}. 

   In constructing of formulas 
   may be used boolean connectives 
   ($\neg, \wedge,\vee$, etc.) 
   interpreted by a standard way. 

   The symbol $\top$ 
   denotes a true formula, 
   and the symbol $\bot$ 
   denotes a false formula.

   Formulas of the form 
   $\wedge(b_1,b_2)$, $\vee (b_1,b_2)$, etc. 
   we shall write in a more familiar form 
   $b_1\wedge b_2$, $b_1\vee b_2$, etc. 

   In some cases, formulas of the form 
   $$b_1\wedge\ldots\wedge b_n\quad\mbox
   {and}\quad b_1\vee\ldots\vee b_n$$
   will be written in the form 
   {\def\arraystretch{1}
   $$\c{b_1\\\ldots\\b_n}\quad\mbox{and}   
   \quad\d{b_1\\\ldots\\b_n}$$
   }
   respectively. 
\i Expressions of the form $+(e_1,e_2)$, 
   $-(e_1,e_2)$ and $\cdot(e_1,e_2)$
   will be written
   in a more familiar form 
   $e_1+e_2$, $e_1-e_2$ and $e_1\cdot e_2$.
\i Expressions of the form ${\bf head}(e)$, 
   ${\bf  tail}(e)$, $[\;](e)$, and
   ${\bf length}(e)$ 
   will be written in the form $\hat e$, $e'$, 
   $[e]$ and $|e|$, respectively. 
\i A constant of the type \verb'list', 
    such that $[\![c]\!]$ is an empty list,
    will be denoted by the symbol 
    $\varepsilon$. 
\ei 

\section{A concept of a process with 
a message passing}

In this section we present 
a concept of a process with 
a message passing. 
This concept is derived from 
the original concept of a process 
presented in section 
\ref{asdfasdf3rt4wertweywey}
by the following modification. 
\bi 
\i Among components 
   of a process $P$ 
   there are the following
   additional components:
   \bi 
   \i the component $X_P$, which is called 
      a {\bf set of variables} 
      of the process $P$, and 
   \i the component $I_P$, which is called 
      an {\bf initial condition}
      of the process $P$. 
   \ei 
\i Transitions are labelled not by actions, 
   but by {\bf operators}.
\ei 

Before giving a formal definition 
of a process with a message passing,
we shall explain a meaning 
of the above concepts. 

For brevity, in this chapter 
we shall call processes 
with a message passing 
simply as {\bf processes}. 

\subsection{A set of variables of a process}

\label{sdfretgrf}

We assume that each process $P$ 
is associated with a set of variables 
$$X_P \subseteq Var$$ 

At any time 
$i$ of an execution of a process $P$ 
($i = 0,1,2, \ldots$) 
each variable $x \in X_P$
is associated with a 
{\bf value} 
$\sigma_i(x)\in D_{t(x)}$.
Values of the variables may be modified during 
an execution of the process.

An {\bf evaluation} of variables 
from $X_P$ is a family 
$\sigma$ of values 
associated with these variables, i.e. 
$$\sigma=\{\sigma(x)\in D_{t(x)}\mid x\in X_P\}$$

The notation $Eval (X_P)$ 
denotes a set of all evaluations 
of variables from $X_P$. 

For each time $i\geq 0$
of an execution of a process $P$ 
the notation $\sigma_i$ denotes an evaluation 
of variables from $X_P$ at this time.

Below we shall assume that for each 
process $P$ all expressions referring 
to the process $P$, contain 
variables only from the set $X_P$. 

\subsection{An initial condition}

Another new component of a 
process $P$ is a formula $I_P \in {\cal B}$, 
which is called an {\bf initial condition}. 
This formula expresses a condition on 
evaluation $\sigma_0$ of variables from $X_P$
at initial time of an execution of $P$: 
$\sigma_0$ must satisfy the condition 
$$\sigma_0(I_P)=1$$

\subsection{Operators} 
\label{spssdfgsdfgw5etrwerw}

The main difference between the new 
definition of a process 
and the old one is that 
\bi 
\i in the old definition 
   a label of each transition is an {\bf action}
   which is executed by a process, 
   when this transition is 
   performed, 
   and 
\i in the new definition 
   a label of each transition is an {\bf operator}
   i.e.  a {\bf scheme of an action},
   which takes a specific form only 
   when this transition is 
   performed.
\ei 

In a definition of an operator 
we shall use the set $Names$, 
which was introduced in section 
\ref{sdfgsdfgw5etrwerw}. 

A set of all operators
is divided into the following four classes. 
\bn 
\i {\bf Input operators}, 
   which have the form
   \be{op1wer23r4}\alpha\,?\,x\ee
   where $\alpha\in Names$ and $x\in Var$.

   An action corresponding 
   to the operator \re{op1wer23r4}
   is executed by 
   \bi
   \i an input to a process
     an object of the form $(\alpha, v)$, 
     where
     \bi
     \i $\alpha$ is a name referred 
        in \re{op1wer23r4}, and
     \i  $v$ is a message
     \ei 
    and 
   \i a record of the message $v$
      in the variable $x$
   \ei
   i.e. after an execution of this action 
   a value of the variable $x$ becomes 
   equal to $v$. 

\i {\bf Output operators},  
   which have the form
   \be{asdfgeewr354335537}
   \alpha\,!\,e\ee
   where $\alpha\in Names$ and
   $e\in {\cal E}$.

   An action corresponding 
   to the operator \re{asdfgeewr354335537}
   is executed by 
   an output 
   an object of the form $(\alpha, v)$
   from a process, 
     where
     \bi
     \i $\alpha$ is a name referred 
        in \re{asdfgeewr354335537}, and
     \i  $v$ is a value of the expression $e$
         on a current evaluation of variables
         of the process.
     \ei 
\i {\bf Assignments} 
    (first type of internal operators), 
   which have the form
   \be{sdfdwf34rf34f34f}x\;:=\;e\ee
   where \bi\i $x\in Var$,
   and \i $e\in {\cal E}$, where $t(e)=t(x)$\ei

   An action corresponding 
   to the operator \re{sdfdwf34rf34f34f}
   is executed by an updating of a value
   associated with the variable $x$: 
   after an execution of this operator 
   this value becomes equal 
   to a value of the expression $e$    
   on a current evaluation of 
   variables of the process.

\i {\bf Conditional operators} 
   (second type of internal operators), 
   which have the form
   $$\langle b\rangle$$
   where $b\in {\cal B}$.

   An action corresponding 
   to the operator $\langle b\rangle$
   is executed by a calculation of a 
   value of the formula $b$ on a current
   evaluation of variables of the process, and
   \bi
   \i if this value is 0, then an execution 
      of the whole action is impossible, and
   \i if this value is 1, then the execution 
      is completed.
   \ei
\en 

The set of all operators
is denoted by the symbol ${\cal O}$.

\subsection{Definition of a process} 
\label{asdffqe23rq23rtgtg}

A {\bf process} is a 5-tuple $P$ of the 
form 
\be{spsamfhgjklsdfh}
P = (X_P, I_P, S_P, s^0_P, R_P)\ee
whose components have the following
meanings: 
\bn
\i $X_P\subseteq Var$ 
   is a set of {\bf variables}
   of the process $P$
\i $I_P\in {\cal B}$ is a formula, called 
   an {\bf initial condition} of the process $P$ 
\i $S_P$ is a set of {\bf states} 
   of the process $P$
\i $s^0_P \in S_P$ is an {\bf initial state}
\i $R_P$ is a subset of the form 
   $$R_P\subseteq S_P\times {\cal O}\times 
   S_P$$
   Elements of $R_P$ are called 
   {\bf transitions}.
\en

If a transition from $R_P$ has the form
$(s_1,op,s_2)$, then we denote it as 
$$\diagrw{s_1&\pright{op}&s_2}$$
and say that 
\bi 
\i the state $s_1$ is a 
   {\bf start} of this transition, 
\i the state $s_2$ is an {\bf end} 
   of this transition, 
\i the operator $op$ is 
   a {\bf label} of this transition. 
\ei 

Also, we assume 
that for each process $P$ 
the set $X_P$ contains 
a special variable $at_P$, 
which takes values in $S_P$. 

\subsection{An execution of a process}
\label{sadfwasf43355}

Let $P$ be a process of the form
\re{spsamfhgjklsdfh}.

An {\bf execution} of the process $P$
is a bypass of the set $S_P$ of its states
\bi
\i starting from the initial state $s^0_P$, 
\i through transitions from $R_P$, and
\i with an execution of operators which are
   labels of visited transitions.
\ei

More detail: 
at each step $i\geq 0$ 
of an execution
\bi 
\i the process $P$ is 
    in located at some state $s_i$ \\($s_0=s^0_P$)
\i there is defined 
   an evaluation $\sigma_i \in Eval(X_P)$ \\
   ($\sigma_0(I_P)$ must be equal to $1$) 
\i if there is a transition from $R_P$    
   starting at $s_i$, then the process 
   \bi 
   \i selects a transition starting at $s_i$, 
      which is labelled by such an operator $op_i$
      that can be executed 
      at current step ($i$), \\
      (if there is no such transitions, 
      then the process $P$ suspends 
      until such transition
      will appear) 
   \i executes the operator $op_i$, and then 
   \i moves to a state  $s_{i+1}$ which is 
      an end of the selected transition 
   \ei 
\i if there is no a transition in $R_P$ starting
   in $s_i$, then the process 
   completes its work. 
\ei 

For each $i\geq 0$ an evaluation
$\sigma_{i+1}$ is determined 
\bi 
\i by the evaluation $\sigma_i$, and 
\i by the operator $op_i$, 
   which is executed at $i$-th step
   of an execution of the process $P$. 
\ei 
A relationship between $\sigma_i$, $\sigma_{i+1}$, 
and $op_i$ has the following form: 
\bn 
\i if $op_i = \alpha\,?\,x$, and 
   at an execution
   of this operator it was inputted
   a message $v$, then 
   $$\by \sigma_{i+1}(x)=v\\
   \forall y\in X_P\setminus \{x, at_P\}\qquad
   \sigma_{i+1}(y)=\sigma_i(y) \ey$$
\i if $op_i = \alpha\,!\,e$, then
   at an execution
   of this operator it is outputted
   the message 
   $$\sigma_i(e)$$
   and values of variables from    
   $X_P\setminus \{at_P\}$
   are not changed: 
   $$\forall x\in X_P\setminus \{at_P\}\qquad
   \sigma_{i+1}(x)=\sigma_i(x)$$
\i if $op_i=\;(x:=e)$, then 
   $$\by \sigma_{i+1}(x)=\sigma_i(e)\\
   \forall x\in X_P\setminus \{x,at_P\}\qquad
   \sigma_{i+1}(x)=\sigma_i(x) \ey
   $$
\i if $op_i=\;\langle b\rangle$ and 
   $\sigma_i(b)=1$, then
   $$\forall x\in X_P\setminus \{at_P\}\qquad
   \sigma_{i+1}(x)=\sigma_i(x)$$
\en 

We assume that for each $i\geq 0$
a value of the variable $at_P$ on an 
evaluation $\sigma_i$ 
is equal to a state $s\in S_P$, 
at which the process $P$ 
is located on step $i$, i.e.
\bi 
\i $\sigma_0(at_P)=s^0_P$
\i $\sigma_1(at_P)=s_1$, 
   where $s_1$ is an end of first transition 
\i $\sigma_2(at_P)=s_2$, 
   where $s_2$ is an end of 
   second transition 
\i etc. 
\ei 

\section{Representation of processes
by flowcharts}

In order to increase a visibility, 
a process can be represented 
by a flowchart. 

A language of flowcharts 
is originated in programming, 
where use of this language 
can greatly facilitate 
a description and understanding 
of algorithms and programs. 

\subsection{The notion of a flowchart}

A {\bf flowchart} is a directed graph, 
each node $n$ of which 
\bi
\i is associated with an {\bf operator} 
   $op(n)$, and 
\i is depicted as one of the following
    geometric figures: 
   a rectangle, an oval,  or a circle, 
   inside of which a label 
   indicating  $op(n)$
   can be contained .
\ei

An operator $op(n)$ can have one of the
following forms.

\begin{description}
\i[initial operator:] \be{nachalo}
\by
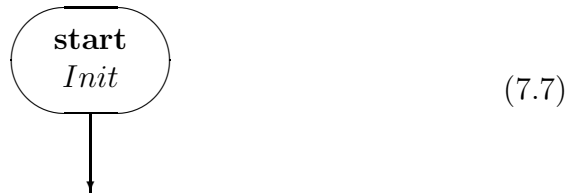
\begin{picture}(0,50)
\put(0,20){\oval(60,40)}
\put(0,20){\makebox(0,0){$\begin{array}{cc} 

{\bf start}
\\Init
\ey$}}
\put(0,0){\vector(0,-1){30}}
\end{picture}\\
\vspace{5mm}\ey
\ee
where $Init\in {\cal B}$ is a formula, 
called an {\bf initial condition}.
\i[assignment operator:]

  \be{priss}
  \by
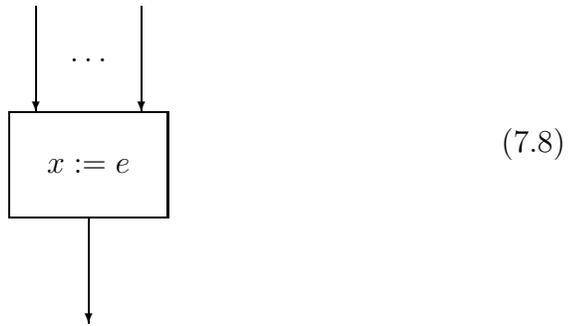
   \begin{picture}(0,120)
   \put(-20,110){\vector(0,-1){40}}
   \put(0,90){\makebox(0,0){$\ldots$}}
   \put(20,110){\vector(0,-1){40}}
   \put(-30,70){\line(1,0){60}}
   \put(-30,30){\line(0,1){40}}
   \put(30,30){\line(0,1){40}}
   \put(-30,30){\line(1,0){60}}
   \put(0,50){\makebox(0,0){$\begin{array}{cc}
   x:=e
   \ey$}}
   \put(0,30){\vector(0,-1){40}}
   \end{picture}
   \ey
   \ee
where 
\bi
\i $x\in Var$,
\i $e\in {\cal E}$, where $t(e)=t(x)$
\ei

\i[conditional operator:]
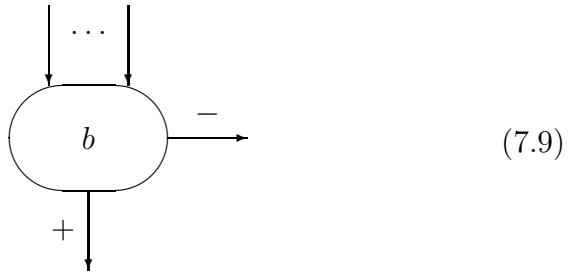
\be{pproo}
\by
\begin{picture}(0,100)
   \put(-15,100){\vector(0,-1){30}}
   \put(0,90){\makebox(0,0){$\ldots$}}
   \put(15,100){\vector(0,-1){30}}
\put(0,50){\oval(60,40)}
\put(0,50){\makebox(0,0){$b$}}
\put(0,30){\vector(0,-1){30}}
\put(30,50){\vector(1,0){30}}
\put(-5,15){\makebox(0,0)[r]{$+$}}
\put(45,55){\makebox(0,0)[b]{$-$}}
\end{picture}
\ey
\ee
where $b\in {\cal B}$. 
\i[sending operator:]
\be{possllwe}
  \by
   
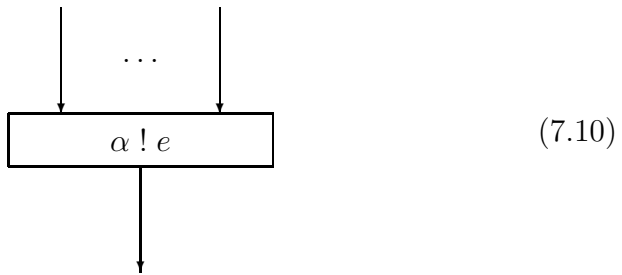
\begin{picture}(0,110)
   \put(-30,100){\vector(0,-1){40}}
   \put(0,80){\makebox(0,0){$\ldots$}}
   \put(30,100){\vector(0,-1){40}}
   \put(-50,60){\line(1,0){100}}
   \put(-50,40){\line(1,0){100}}
   \put(-50,40){\line(0,1){20}}
   \put(50,40){\line(0,1){20}}
   \put(0,50){\makebox(0,0){$\alpha\;!\;e
   $}}
   \put(0,40){\vector(0,-1){40}}
   \end{picture}
   \ey
\ee
where 
\bi 
\i $\alpha \in Names$ is a name \\
   (for example, it can be
   a destination of a message
   which will be sent), and 
\i $e\in {\cal E}$ is an expression 
   whose value is a message
   which will be sent. 
\ei 
\i[receiving operator:]
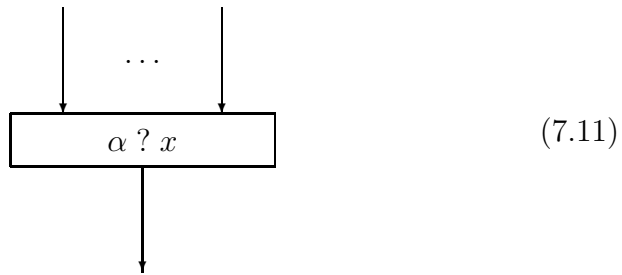
\be{polusp}
\by
   \begin{picture}(0,110)
   \put(-30,100){\vector(0,-1){40}}
   \put(0,80){\makebox(0,0){$\ldots$}}
   \put(30,100){\vector(0,-1){40}}
   \put(-50,60){\line(1,0){100}}
   \put(-50,40){\line(1,0){100}}
   \put(-50,40){\line(0,1){20}}
   \put(50,40){\line(0,1){20}}
   \put(0,50){\makebox(0,0){$\alpha\;?\;x
   $}}
   \put(0,40){\vector(0,-1){40}}
   \end{picture}
   \ey
\ee
where 
\bi 
\i $\alpha \in Names$ is a name \\
   (for example, it can be
   an expected source of a message
   which will be received), and 
\i $x\in Var$ is a variable
   in which a received message
   will be recorded.
\ei 
\i[choice:]
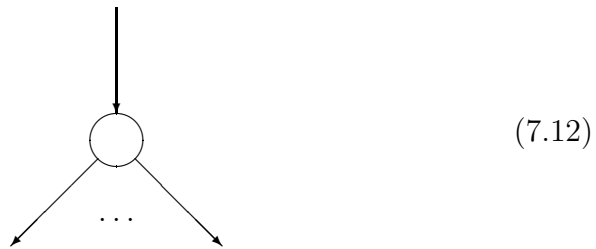
\be{pussiii}
\by
\begin{picture}(0,110)
\put(0,50){\oval(20,20)}
\put(0,100){\vector(0,-1){40}}
\put(0,20){\makebox(0,0){$\ldots$}}
\put(-7,43){\vector(-1,-1){33}}
\put(7,43){\vector(1,-1){33}}
\end{picture}
\ey
\ee

\i[join:]
\be{sorepussiii}
\by
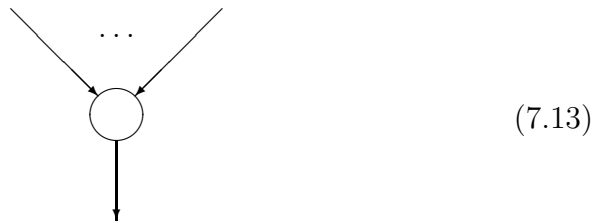
\begin{picture}(0,100)
\put(0,50){\circle{20}}
\put(0,40){\vector(0,-1){30}}
\put(0,80){\makebox(0,0){$\ldots$}}
\put(-40,90){\vector(1,-1){33}}
\put(40,90){\vector(-1,-1){33}}
\end{picture}
\ey
\ee

Sometimes 
\bi 
\i a circle representing this operator, and 
\i ends of some edges leading to this circle  
\ei 
are not pictured. That is, for example, 
a fragment of a flowchart of the form 
$$\by
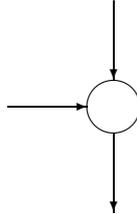
\begin{picture}(0,90)
\put(0,50){\circle{20}}
\put(0,40){\vector(0,-1){30}}
\put(0,90){\vector(0,-1){30}}
\put(-40,50){\vector(1,0){30}}
\end{picture}
\ey
$$
can be pictured as follows: 
$$\by
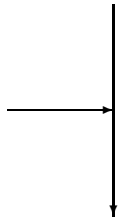
\begin{picture}(0,90)
\put(0,90){\vector(0,-1){80}}
\put(-40,50){\vector(1,0){40}}
\end{picture}
\ey
$$

\i[halt:]
\be{ghjkgkjhgkjhgjk}
\by
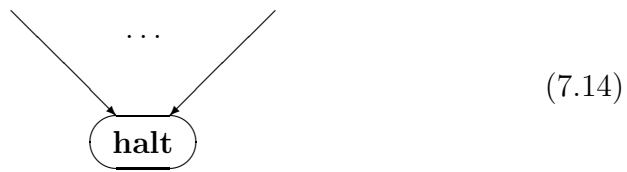
\begin{picture}(0,60)
\put(-50,50){\vector(1,-1){40}}
\put(0,40){\makebox(0,0){$\ldots$}}
\put(50,50){\vector(-1,-1){40}}
\put(0,0){\oval(40,20)}
\put(0,0){\makebox(0,0){${\bf halt}$}}
\end{picture}\\\mbox{$\;$}
\ey
\ee
\end{description}

Flowcharts must meet the following conditions: 
\bi 
\i a node of the type \re{nachalo}
   can be only one\\ (this node 
   is called a {\bf start node})
\i there is only one edge outgoing from 
   nodes of the types \re{nachalo}, 
   \re{priss}, \re{possllwe}, 
   \re{polusp}, \re{sorepussiii}
\i there are one or two edges outgoing 
   from nodes of the type \re{pproo}, and 
   \bi 
   \i if there is only one edge outgoing 
   from a node of the type \re{pproo}, then  
   this edge has the label ``$+$'', and
   \i if there are two edges outgoing 
   from a node of the type \re{pproo}, then  
   \bi\i one of them has the label 
      ``$+$'', and   
   \i another has the label ``$-$''.\ei
   \ei 
\i there is only one edge leading to a 
   node of the type \re{pussiii} 
\i there is no edges outgoing from a node
   of the type \re{ghjkgkjhgkjhgjk}
\ei 

\subsection{An execution of a flowchart}

An {\bf execution} of a flowchart
is a sequence of transitions 
\bi
\i from one node to another along edges, 
\i starting from a start node $n_0$, and
\i with an execution of operators 
   which correspond to visited nodes.
\ei

More detail: each step $i\geq 0$
of an execution of a flowchart
is associated with some node 
$n_i$ which is called {\bf current node}, 
and 
\bi 
\i if $n_i$ is not of the type
   \re{ghjkgkjhgkjhgjk}, 
   then after an execution of an operator
   corresponded to the node $n_i$
   it is performed a transition
   along an edge 
   outgoing from $n_i$ to a node 
   which will be current node at 
   next step of an execution
\i if $n_i$ is of the type 
   \re{ghjkgkjhgkjhgjk}, 
   then an execution of the flowchart is 
   completed. 
\ei 

Let $X$ be a set of all variables occurred in 
the flowchart.

At each step $i$ of an execution
($i=0,1,\ldots$) each variable $x\in X$
is associated with a value $\sigma_i(x)$.

The family $\{\sigma_i(x)\mid x\in X\}$
\bi
\i is denoted by $\sigma_i$, and 
\i is called an {\bf evaluation} 
   of variables of the flowchart
   at $i$--th step of its execution. 
\ei
The evaluation $\sigma_0$ must
meet the initial condition $Init$, 
i.e. the following statement 
must be true:
$$\sigma_0(Init)=1$$

An operator $op(n_i)$ associated with 
current node $n_i$ is executed as follows. 
\bi 
\i If $op(n_i)$ has the type \re{priss}, then
   the value $\sigma_i(e)$ is recorded in $x$ 
   i.e. 
   $$\by\sigma_{i+1}(x)\eam \sigma_i(e)\\
   \forall \;y\in X\setminus\{x\}
    \quad\sigma_{i+1}(y)\eam \sigma_i(y)
    \\\ey$$
\i If $op(n_i)$ has the type \re{pproo}
   then 
   \bi
   \i if $\sigma_i(b)=1$, then 
      a transition along 
      an edge outgoing from 
      $n_i$ with a label ``$+$'' is performed 
   \i if $\sigma_i(b)=0$, and there is 
      an edge outgoing from 
      $n_i$ with a label ``$-$'', then
      a transition along 
      this edge is performed 
   \i if $\sigma_i(b)=0$, and there is no
      an edge outgoing from 
      $n_i$ with a label ``$-$'',
      then an execution of $op(n_i)$
      is impossbile.
   \ei
\i If $op(n_i)$ has the type \re{possllwe} 
   then an execution of this operator
   consists of a sending the object
   \be{afdgre4y6rw}(\alpha\,,\,\sigma_i(e))\ee
   if it is possible. 

   If a sending the object \re{afdgre4y6rw}
   is impossible, then an execution 
   of $op(n_i)$ is impossbile.

\i If $op(n_i)$ has the type \re{polusp}
   then an execution of this operator
   consists of 
   \bi
   \i a receiving the object
   \be{afdgre4y6rw2}(\alpha\,,\,v)\ee
   (if it is possible), and \i a recording
   of $v$ in the variable $x$, 
   i.e. $$\by \sigma_{i+1}(x)\eam v\\
   \forall y\in X\setminus \{y\}\quad
   \sigma_{i+1}(y)\eam \sigma_i(y)\ey$$
   \ei 

   If a receiving the object \re{afdgre4y6rw2}
   is impossible, then an execution 
   of $op(n_i)$ is impossbile.

\i If current node $n_i$ is associated 
   with an operator of the type \re{pussiii},  
   then 
   \bi 
   \i among nodes which are ends of 
      edges outgoing from $n_i$
      it is selected a node $n$
      labelled by such an operator, 
      which can be executed
      at current time, and 
   \i it is performed a transition 
      to the node $n$.
   \ei 
   If there are several operators which 
   can be executed at current time,
   then a selection of the node $n$
   is performed non-deterministically.

\i an operator of the type
  \re{ghjkgkjhgkjhgjk} completes
   an execution of the flowchart. 
\ei 

\subsection{Construction of a process 
defined by a flowchart} 

An algorithm of a construction of a process 
defined by a flowchart has the following form.

\bn 
\i At every edge of the flowchart it 
   is selected a point. 
\i For 
   \bi 
   \i each node $n$ of the flowchart, 
      which has no the type
      \re{pussiii} or \re{sorepussiii}, 
      and 
   \i each pair $F_1,F_2$ 
      of edges of the flowchart such that 
      $F_1$ is incoming in $n$, 
      and $F_2$ is outgoing from $n$
   \ei 
   the following actions are performed: 
   \bn 
   \i it is drawn an arrow $f$ from a point
      on $F_1$ to a point on $F_2$
   \i it is drawn a label $label(f)$
       on the arrow $f$, defined as follows: 
       \bn 
        \i if $op(n)$ has the type \re{priss}, 
           then 
           $$label(f) \eam (x:=e)$$
        \i if $op(n)$ has the type
           \re{pproo}, and an 
            edge outgoing from $n$, 
            has a label ``$+$'', then
            $$label(f) \eam \langle b\rangle$$ 
        \i if $op(n)$ has the type 
           \re{pproo}, and an 
           edge outgoing from $n$, 
           has a label ``$-$'', then
           $$label(f) \eam\;\; 
           \langle \neg b\rangle$$
        \i if $op(n)$ has the type
           \re{possllwe} or \re{polusp},
            then $label(f)=op(n)$. 
        \en 
   \en 
\i For each node $n$ of the type \re{pussiii} 
   and each edge $F$ outgoing from $n$,
   the following actions are performed.
   Let 
   \bi
   \i $p$ be a point on an edge incoming to 
      $n$, 
   \i $p'$ be a point on $F$,
   \i $n'$ be an end of $F$, and
   \i $p''$ be a poing on an edge outgoing from
      $n'$.
   \ei
   Then 
   \bi
   \i an arrow from $p'$ to $p''$ is replaced
      on an arrow from $p$ to $p''$ 
      with the same label, and 
   \i the point $p'$ is removed.
   \ei

\i For each node $n$ of the type \re{sorepussiii} 
   and each edge $F$ incoming from $n$,
   the following actions are performed.
   Let 
   \bi
   \i $p$ be a point on an edge outgoing from 
      $n$, 
   \i $p'$ be a point on $F$,
   \i $n'$ be a start of $F$, and
   \i $p''$ be a poing on an edge incoming to
      $n'$.
   \ei
   Then 
   \bi
   \i an arrow from $p''$ to $p'$ is replaced
      on an arrow from $p''$ to $p$ 
      with the same label, and 
   \i the point $p'$ is removed.
   \ei

\i States of a constructed process 
   are remaining points. 
\i An initial state $s^0_P$ is defined 
   as follows. 
   \bi 
   \i If a point which was selected on an 
      edge outgoing from a start node 
      of the flowchart 
      was not removed, 
      then $s^0_P$ is this point.

   \i If this point was removed, then
      an end of an edge outgoing from 
      a start note of the flowchart is a node
      $n$ of the type \re{sorepussiii}.
      In this case, $s^0_P$ is a point 
      on an edge outgoing from $n$. 
   \ei 

\i Transitions of the process correspond to 
   the pictured arrows:
   for each such arrow $f$ the process 
   contains a transition 
   $$\diagrw{s_1&\pright{label(f)}&s_2}$$
   where $s_1$ and $s_2$ are
   a start and an end of the arrow $f$ 
   respectively. 

\i A set of variables of the process 
    consists of 
    \bi     
    \i all variables occurred in any 
       operator of the flowchart, and
   \i the variable $at_P$. 
   \ei 
\i An initial condition of the process 
   coincides with the initial condition 
   $Init$ of the flowchart. 
\en 

\section{An example of a process 
with a message passing}

In this section we consider a process ``buffer''
as an example of a process with a message 
passing:
\bi 
\i at first, we define this process 
   as a flowchart, and 
\i then we transform this flowchart 
   to a standard graph representation 
   of a process.
\ei 

\subsection{The concept of a buffer}

A {\bf buffer} is a system which has 
the following properties. 
\bi 
\i It is possible to input messages to a buffer.

   A message which is entered to the buffer
   is stored in the buffer.

   Messages which are stored in a buffer 
   can be extracted from the buffer. 

   We assume that a buffer can store
   not more than a given number 
   of messages. If $n$ is a such number, 
   then we shall denote the buffer as
   ${\it Buffer}_n$.

\i At each time 
   a list of messages 
   \be{sofewr}
   c_1,\ldots, c_k\qquad(0\leq k 
   \leq n)\ee 
   stored in ${\it Buffer}_n$
   is called a {\bf content of the buffer}. 

   The number $k$ in \re{sofewr}
   is called a {\bf size}
   of this content. 

   The case $k = 0$ corresponds 
   to the situation when a content 
   of the buffer  is empty.
\i If at current time
   a content of ${\it Buffer}_n$ 
   has the form 
   \re{sofewr}, and $k <n$, 
   then \bi\i the buffer can accept   
   any message, and \i
   after an execution
   of the action of an input of a message $c$
   a content of the buffer becomes 
   $$c_1,\ldots, c_k,c$$\ei
\i If at current time
    a content of ${\it Buffer}_n$ 
    has the form 
   \re{sofewr}, and $k>0$, then 
    \bi\i it is possible
    to extract the message $c_1$ 
    from the buffer, 
   and \i after an execution of this operation
   a content of the 
   buffer becomes $$c_2,\ldots, c_k$$\ei
\ei 

Thus, at each time
a content of a buffer is a queue 
of messages, and 
\bi 
\i each action of an input of a message to a
   buffer adds this message to an end of the 
   queue, and
\i each action of an output of a message 
   from the buffer
   \bi 
   \i extracts a first message of this queue, 
      and 
   \i removes this message from the queue.
   \ei 
\ei 

A queue with the above operations
is called {\bf a queue of the type FIFO} 
(First Input - First Output). 

\subsection{Representation 
of a buffer by a flowchart}

In this section we present a formal description 
of the concept of a buffer by a flowchart.

In this flowchart
\bi 
\i an operation of an input of a message
   to a buffer is represented
   by an action with the name $In$, and 
\i an operation of an output of a message
   from a buffer is represented
   by an action with the name $Out$.
\ei 

The flowchart has the following variables: 
\bi 
\i the variable $n$ of the type
   \verb"int", its value 
   does not change, it is equal to
   the maximal size of a content
   of the buffer 
\i the variable $k$ of the type \verb"int", 
   its value is equal to 
   a size of a content of the buffer
   at current time
\i the variable $f$ of the type \verb'mes',
   this variable will store messages 
   that will come to the buffer
\i the variable $q$ of the type \verb'list', 
   this variable will store a content of the buffer.
\ei 

A flowchart representing a behavior 
of a buffer 
has the following form:\\
(notations used in this flowchart
were defined in section \ref{vyrazhenija})

{\def\arraystretch{1}

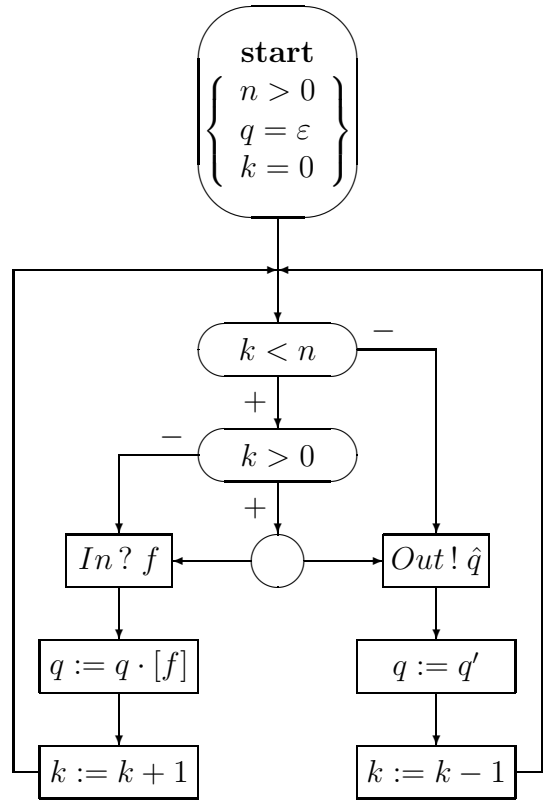
$$
\begin{picture}(0,295)

\put(0,250){\oval(60,80)}
\put(0,250){\makebox(0,0)[c]{$\begin{array}
{c}{\bf start}\\
\c{n>0\\q=\varepsilon\\ k=0}\ey$}}

\put(0,160){\oval(60,20)}
\put(0,160){\makebox(0,0)[c]{$k<n$}}

\put(0,120){\oval(60,20)}
\put(0,120){\makebox(0,0)[c]{$k>0$}}
 
\put(60,80){\pu{20}{10}}
\put(60,80){\makebox(0,0)[c]{$Out\,!\;
   \hat q$}}

\put(60,40){\pu{30}{10}}
\put(60,40){\makebox(0,0)[c]{$q:=q'$}}

\put(60,0){\pu{30}{10}}
\put(60,0){\makebox(0,0)[c]{$k:=k-1$}}

\put(-60,80){\pu{20}{10}}
\put(-60,80){\makebox(0,0)[c]{$In\,?\;f$}}

\put(-60,40){\pu{30}{10}}
\put(-60,40){\makebox(0,0)[c]{$q:=
   q\cdot [f]$}}

\put(-60,0){\pu{30}{10}}
\put(-60,0){\makebox(0,0)[c]{$k:=k+1$}}

\put(0,80){\oval(20,20)}
\put(10,80){\vector(1,0){30}}
\put(-10,80){\vector(-1,0){30}}

\put(0,210){\vector(0,-1){40}}

\put(0,150){\vector(0,-1){20}}
\put(0,110){\vector(0,-1){20}}

\put(60,160){\vector(0,-1){70}}
\put(-60,120){\vector(0,-1){30}}

\put(60,70){\vector(0,-1){20}}
\put(-60,70){\vector(0,-1){20}}

\put(30,160){\line(1,0){30}}
\put(-30,120){\line(-1,0){30}}
\put(40,163){\makebox(0,0)[b]{$-$}}
\put(-40,123){\makebox(0,0)[b]{$-$}}

\put(-4,100){\makebox(0,0)[r]{$+$}}
\put(-4,140){\makebox(0,0)[r]{$+$}}

\put(-100,0){\line(0,1){190}}
\put(-100,190){\vector(1,0){100}}
\put(100,0){\line(0,1){190}}
\put(100,190){\vector(-1,0){100}}

\put(-100,0){\line(1,0){10}}
\put(100,0){\line(-1,0){10}}

\put(60,30){\vector(0,-1){20}}
\put(-60,30){\vector(0,-1){20}}

\end{picture}
$$
}

\subsection{Representation 
of a buffer as a process} 
\label{bufferprocess}

To construct a process ${\it Buffer}_n$, 
which corresponds to the above flowchart, 
we select points at its edges: 

{\def\arraystretch{1}
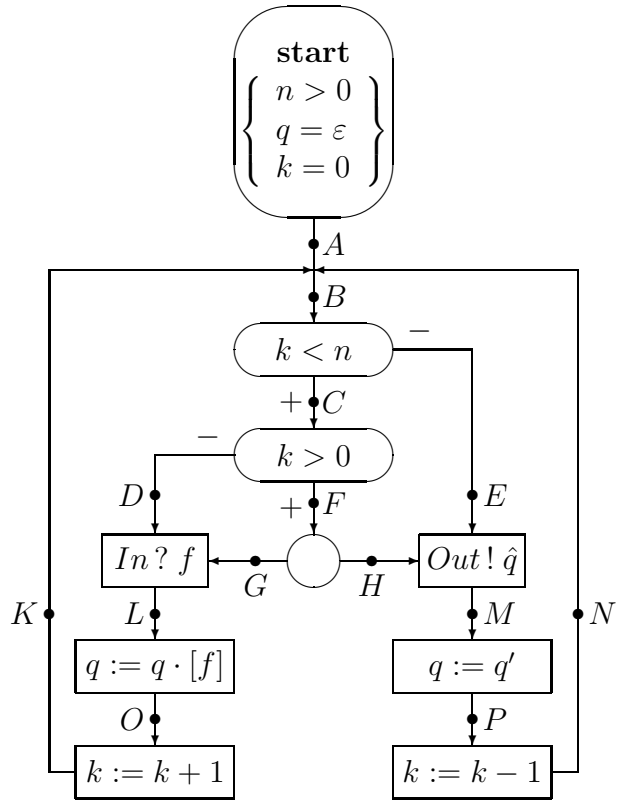
$$
\begin{picture}(0,295)

\put(0,200){\circle*{4}}
\put(3,200){\makebox(0,0)[l]{$A$}}

\put(0,180){\circle*{4}}
\put(3,180){\makebox(0,0)[l]{$B$}}

\put(0,140){\circle*{4}}
\put(3,140){\makebox(0,0)[l]{$C$}}

\put(0,102){\circle*{4}}
\put(3,102){\makebox(0,0)[l]{$F$}}

\put(-22,80){\circle*{4}}
\put(-22,75){\makebox(0,0)[t]{$G$}}

\put(22,80){\circle*{4}}
\put(22,75){\makebox(0,0)[t]{$H$}}

\put(-60,105){\circle*{4}}
\put(-64,105){\makebox(0,0)[r]{$D$}}

\put(60,105){\circle*{4}}
\put(64,105){\makebox(0,0)[l]{$E$}}

\put(-60,60){\circle*{4}}
\put(-64,60){\makebox(0,0)[r]{$L$}}

\put(60,60){\circle*{4}}
\put(64,60){\makebox(0,0)[l]{$M$}}

\put(-100,60){\circle*{4}}
\put(-104,60){\makebox(0,0)[r]{$K$}}

\put(100,60){\circle*{4}}
\put(104,60){\makebox(0,0)[l]{$N$}}

\put(-60,20){\circle*{4}}
\put(-64,20){\makebox(0,0)[r]{$O$}}

\put(60,20){\circle*{4}}
\put(64,20){\makebox(0,0)[l]{$P$}}

\put(0,250){\oval(60,80)}
\put(0,250){\makebox(0,0)[c]{$\begin{array}{c}{\bf start}\\
\c{n>0\\q=\varepsilon\\ k=0}\ey$}}

\put(0,160){\oval(60,20)}
\put(0,160){\makebox(0,0)[c]{$k<n$}}

\put(0,120){\oval(60,20)}
\put(0,120){\makebox(0,0)[c]{$k>0$}}
 
\put(60,80){\pu{20}{10}}
\put(60,80){\makebox(0,0)[c]{$Out
\,!\;\hat q$}}

\put(60,40){\pu{30}{10}}
\put(60,40){\makebox(0,0)[c]{$q:=q'$}}

\put(60,0){\pu{30}{10}}
\put(60,0){\makebox(0,0)[c]{$k:=k-1$}}

\put(-60,80){\pu{20}{10}}
\put(-60,80){\makebox(0,0)[c]{$In
\,?\;f$}}

\put(-60,40){\pu{30}{10}}
\put(-60,40){\makebox(0,0)[c]{$q
:=q\cdot [f]$}}

\put(-60,0){\pu{30}{10}}
\put(-60,0){\makebox(0,0)[c]{$k:=k+1$}}

\put(0,80){\oval(20,20)}
\put(10,80){\vector(1,0){30}}
\put(-10,80){\vector(-1,0){30}}

\put(0,210){\vector(0,-1){40}}

\put(0,150){\vector(0,-1){20}}
\put(0,110){\vector(0,-1){20}}

\put(60,160){\vector(0,-1){70}}
\put(-60,120){\vector(0,-1){30}}

\put(60,70){\vector(0,-1){20}}
\put(-60,70){\vector(0,-1){20}}

\put(30,160){\line(1,0){30}}
\put(-30,120){\line(-1,0){30}}
\put(40,163){\makebox(0,0)[b]{$-$}}
\put(-40,123){\makebox(0,0)[b]{$-$}}

\put(-4,100){\makebox(0,0)[r]{$+$}}
\put(-4,140){\makebox(0,0)[r]{$+$}}

\put(-100,0){\line(0,1){190}}
\put(-100,190){\vector(1,0){100}}
\put(100,0){\line(0,1){190}}
\put(100,190){\vector(-1,0){100}}

\put(-100,0){\line(1,0){10}}
\put(100,0){\line(-1,0){10}}

\put(60,30){\vector(0,-1){20}}
\put(-60,30){\vector(0,-1){20}}

\end{picture}
$$
}

In a construction of a process defined 
by this flowchart, 
the points $A$, $G$, $H$, $K$ and $N$
will be removed.

A standard graph representation of the process
${\it Buffer}_n$ is the following.

{\def\arraystretch{1}

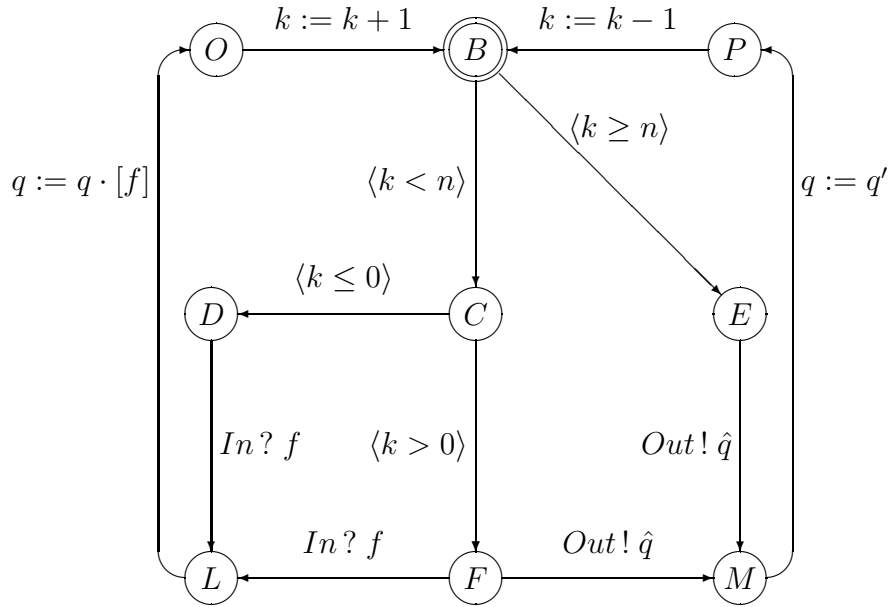
$$\by
\begin{picture}(0,220)

\put(0,200){\oval(20,20)}
\put(0,200){\oval(24,24)}
\put(0,200){\makebox(0,0)[c]{$B$}}

\put(-100,100){\oval(20,20)}
\put(-100,100){\makebox(0,0)[c]{$D$}}

\put(0,100){\oval(20,20)}
\put(0,100){\makebox(0,0)[c]{$C$}}

\put(100,100){\oval(20,20)}
\put(100,100){\makebox(0,0)[c]{$E$}}

\put(-100,0){\oval(20,20)}
\put(-100,0){\makebox(0,0)[c]{$L$}}

\put(0,0){\oval(20,20)}
\put(0,0){\makebox(0,0)[c]{$F$}}

\put(100,0){\oval(20,20)}
\put(100,0){\makebox(0,0)[c]{$M$}}

\put(0,188){\vector(0,-1){78}}
\put(9,191){\vector(1,-1){84}}
\put(0,90){\vector(0,-1){80}}
\put(-100,90){\vector(0,-1){80}}
\put(0,90){\vector(0,-1){80}}
\put(100,90){\vector(0,-1){80}}
\put(-10,100){\vector(-1,0){80}}
\put(-10,0){\vector(-1,0){80}}
\put(10,0){\vector(1,0){80}}

\put(-110,10){\oval(20,20)[lb]}
\put(-120,10){\line(0,1){180}}
\put(-110,190){\oval(20,20)[lt]}
\put(-110,200){\vector(1,0){2}}
\put(-98,200){\oval(20,20)}
\put(-98,200){\makebox(0,0)[c]{$O$}}
\put(98,200){\makebox(0,0)[c]{$P$}}

\put(-88,200){\vector(1,0){76}}

\put(110,10){\oval(20,20)[rb]}
\put(120,10){\line(0,1){180}}
\put(110,190){\oval(20,20)[rt]}
\put(110,200){\vector(-1,0){2}}
\put(98,200){\oval(20,20)}
\put(88,200){\vector(-1,0){76}}

\put(-50,207){\makebox(0,0)[b]{$k:=k+1$}}
\put(50,207){\makebox(0,0)[b]{$k:=k-1$}}

\put(-50,7){\makebox(0,0)[b]{$In\,?\;f$}}
\put(50,7){\makebox(0,0)[b]{$Out\,!\;\hat q$}}

\put(-123,150){\makebox(0,0)[r]{$q:=q\cdot [f]$}}
\put(123,150){\makebox(0,0)[l]{$q:=q'$}}

\put(-97,50){\makebox(0,0)[l]{$In\,?\;f$}}
\put(97,50){\makebox(0,0)[r]{$Out\,!\;\hat q$}}

\put(-3,50){\makebox(0,0)[r]{$\langle k>0\rangle$}}
\put(-3,150){\makebox(0,0)[r]{$\langle k<n\rangle$}}

\put(-50,107){\makebox(0,0)[b]{$\langle k\leq 0\rangle$}}

\put(35,170){\makebox(0,0)[l]{$\langle k\geq n\rangle$}}

\end{picture}
\ey
$$
}

\section{Operations on processes 
with a message passing} 
\label{erqwr2343242}

Operations on processes with a 
message passing 
are similar to operations 
which are considered 
in chapter \ref{operatsii}. 

\subsection{Prefix action}

Let $P$ be a process, and 
$op$ be an operator.

The process $op.\,P$ is 
obtained from $P$ by an adding
\bi 
\i a new state $s$, which is
   an initial state of $op.\,P$, 
\i a new transition 
   $\diagrw{s&\pright{op}&s^0_P}$, and
\i all variables from $op$.
\ei 

\subsection{Alternative composition}

Let $P_1, P_2$ be processes such that
$S_{P_1}\cap S_{P_2} =\emptyset$.

Define a process 
$P_1 + P_2$, which is called an
{\bf alternative composition} 
of $P_1$ and $P_2$, as follows.
\bi 
\i sets of its states, transitions, 
   and an initial state are 
   determined by the same way 
   as corresponding components of
   an alternative composition 
   in chapter \ref{operatsii}
   (section \ref{sgjkdsflgertyyyy})
\i $X_{P_1+P_2}\eam X_{P_1}\cup X_{P_2}$
\i $I_{P_1+P_2}\eam I_{P_1}\wedge I_{P_2}$
\ei

If $S_{P_1}\cap S_{P_2}\neq \emptyset$, 
then for a construction of the process 
$P_1+P_2$ it is necessary 
\bi
\i to replace in $S_{P_2}$ those states 
   that are also in $P_1$ on new states, and 
\i modify accordingly other components 
   of $P_2$. 
\ei

\subsection{Parallel composition}

Let $P_1$ and $P_2$ be processes such that
$X_{P_1}\cap X_{P_2} =\emptyset$.

Define a process 
$P_1 \,|\,P_2$, which is called a 
{\bf parallel composition} of $P_1$ and $P_2$,
as follows:
\bi
\i a set of its states and its initial state
   are defined by the same way 
   as are defined the corresponding components
   of the process $P_1\,|\,P_2$ in chapter
   \ref{operatsii}
\i $X_{P_1+P_2}\eam X_{P_1}\cup X_{P_2}$
\i $I_{P_1+P_2}\eam I_{P_1}\wedge I_{P_2}$
\i the set of transitions of the process 
   $P_1\,|\,P_2$
   is defined as follows:
   \bi
   \i for
      \bi
      \i each transition 
         $\diagrw{s_1&\pright{op}&s'_1}$
         of the process ${P_1}$, and
      \i each state $s$ of the process ${P_2}$
      \ei
      the process ${P_1\,|\,P_2}$ contains 
      the transition
      $$\diagrw{(s_1,s)&
      \pright{op}&(s'_1,s)}$$
   \i for
\bi
\i each transition 
   $\diagrw{s_2&\pright{op}&s'_2}$
   of the process ${P_2}$, and
\i each state $s$ of the process ${P_1}$
\ei
the process 
${P_1\,|\,P_2}$ contains 
      the transition
$$\diagrw{(s,s_2)&\pright{op}&(s,s'_2)}$$
   \i for each pair of transitions of the form
$$\by
\diagrw{s_1&\pright{op_1}&s'_1}\quad\in R_{P_1}\\
\diagrw{s_2&\pright{op_2}&s'_2}\quad\in R_{P_2}\ey$$
where 
\bi \i one of the operators $op_1,op_2$
has the form $\alpha\,?\,x$,\i
and another has the form $\alpha\,!\,e$, 
where $t(x)=t(e)$\\
(names in both the operators are equal)
\ei
the process ${P_1\,|\,P_2}$ 
contains the transition
$$\diagrw{(s_1,s_2)&\pright{x\;:=\;e}&
(s'_1,s'_2)}$$
   \ei
\ei

If $X_{P_1}\cap X_{P_2} \neq\emptyset$,
then before a construction of the process
$P_1\pa P_2$ it is necessary to replace
variables which occur in both processes
on new variables.

\subsection{Restriction and renaming}

Definition of there operations 
is the same as definition of corresponding
operations in chapter \ref{operatsii}.

\section{Equivalence of processes}

\subsection{The concept 
of a concretization of a process} 
\label{dfgjkl3e5tiu3iout3}

Let $P$ be a process. 

We shall denote by $Conc(P)$
a process in the original sense 
of this concept 
(see section \ref{asdfasdf3rt4wertweywey}), 
which 
is called a {\bf concretization} of the
    process $P$, and 
has the following components. 

\bn
\i States of $Conc(P)$ are \bi\i    all evaluations 
from $Eval(X_P)$, and
\i an additional state 
   $s^0$, which is an initial state of   $Conc(P)$\ei
\i For \bi\i each transition
   $\diagrw{s_1&\pright{op}&s_2}$ of the
   process $P$, and
   \i each evaluation $\sigma\in Eval(X_P)$,
   such that $$\sigma(at_P)=s_1$$\ei
   $Conc(P)$ has a transition
   $$\diagrw{\sigma&\pright{a}&\sigma'}$$
   if $\sigma'(at_P)=s_2$, and one of the following
   conditions is satisfied:
   \bi
   \i \bi
\i $op = \alpha\,?\,x,\quad
a=\alpha\,?\,v$, where $v\in D_{t(x)}$
\i $\sigma'(x)=v,\quad
\forall y\in X_P\setminus \{x,at_P\}\qquad
\sigma'(y)=\sigma(y)$
\ei
   \i \bi
\i $op = \alpha\,!\,e,\quad a=\alpha\,!\,\sigma(e)$
\i $\forall x\in X_P\setminus \{at_P\}\qquad
\sigma'(x)=\sigma(x)$
\ei
   \i \bi
\i $op=\;(x:=e),\quad  a=\tau$
\i $\sigma'(x)=\sigma(e),\quad
\forall y\in X_P\setminus \{x,at_P\}\qquad
\sigma'(y)=\sigma(y)$
\ei
   \i \bi
\i $op=\langle b\rangle,\quad \sigma(b)=1,\quad a=\tau$
\i $\forall x\in X_P\setminus \{at_P\}\qquad
\sigma'(x)=\sigma(x)$
\ei
   \ei
\i For
   \bi
   \i each evaluation $\sigma\in Eval(X_P)$,
     such that $$\sigma(I_P)=1$$ 
   \i and each transition of $Conc(P)$ 
      of the form
     $\diagrw{\sigma&\pright{a}&\sigma'}$
   \ei
   $Conc(P)$ has the transition
   $\diagrw{s^0&\pright{a}&\sigma'}$
\en

From the definitions of 
\bi 
\i the concept of an execution
   of a process with a message 
   passing (see section
   \ref{sadfwasf43355}), and 
\i the concept of an execution
   of a process in the original sense 
   (see section   \ref{asdfasdf3rt4wertweywey}) 
\ei 
it follows that there 
is a one-to-one correspondence between 
\bi 
\i the set of all 
   variants of an execution
   of the process $P$, and 
\i the set of all 
   variants of an execution 
   of $Conc(P)$. 
\ei 

A reader is invited to investigate 
the commutativity property 
of the mapping $Conc$ 
with respect to 
the operations on processes 
i.e. to check statements of the form
$$Conc(P_1\pa P_2)=Conc(P_1)\pa Conc(P_2)
$$
etc. 

\subsection{Definition 
of equivalences of processes}

We define that every pair $(P_1,P_2)$
of processes with a message passing 
is in the same equivalence
($\sim$, $\approx$, $\oc$, $\ldots$),
in which is a pair of concretizations of 
these processes, i.e. 
$$
P_1\sim P_2\quad\Leftrightarrow
\quad
Conc(P_1)\sim Conc(P_2),
\quad\mbox{etc.}$$

A reader is invited to
\bi 
\i explore a relationship 
   of the operations on processes 
   with various equivalences
   ($\approx,\oc,\ldots$), 
   i.e. to establish properties, 
    which are similar to the properties
   presented in sections
   \ref {archnatella},
   \ref{dopsvojstva},
   \ref{aldffrrghghggdf},
   \ref{sdfwerr4543534ergteg}

\i formulate and prove necessary 
   and sufficient conditions 
   of equivalence 
   ($\approx,\oc,\ldots$) 
   of processes 
   that do not use the concept 
   of a concretization of a process. 
\ei 

\section{Processes with composite operators}

\subsection{A motivation of the concept 
of a process with composite operators}

A complexity of the problem of an analysis of 
a process essentially depends on a size 
of its description (in particular, on a 
number of its states). 
Therefore, for a construction of efficient 
algorithms of an analysis of processes 
it is required a search of methods 
to decrease a complexity of a description 
of analyzed processes. 
In this section we consider one 
of such methods. 

In this section
we generalize the concept of a process
to the concept of a process with 
composite operators. 
A composite operator is a sequential 
composition of several operators. 
Due to the fact that we combine 
a sequence of operators in a single composite 
operator, 
we are able to exclude from a description 
of a process those states which 
are at intermediate locations
of this sequence of operators.

Also in this section
we define the concept of a reduction 
of processes with composite operators 
in such a way that a reduced process
\bi 
\i has a less complicated description
   than an original process, and 
\i is equivalent (in some sense) to
   an original process. 
\ei 

With use of the above concepts, 
the problem of an analysis of a process 
can be solved as follows. 
\bn 
\i First, we transform an original process $P$ 
   to a process $P'$ with composite operators, 
   which is similar to $P$.
\i Then we reduce $P'$, 
   getting a process $P''$, 
   whose complexity can be significantly less 
   than a complexity of the original process 
   $P$. 
\i After this, we 
   \bi
   \i perform an analysis of $P''$, and 
   \i use results of this analysis for
      drawing a conclusion about 
      properties of the original process $P$. 
   \ei
\en 

\subsection{A concept of a composite operator}

A {\bf composite  operator (CO)} 
is a finite sequence $Op$ of operators 
\be{soasdhfjksla}
Op=(op_1,\ldots, op_n)\qquad(n\geq 1)\ee
which has the following properties.
\bn 
\i $op_1$ is a conditional operator. 
\i The sequence $(op_2,\ldots, op_n)$
   \bi 
   \i does not contain conditional operators, and 
   \i contains no more than one input or       
      output operator. 
   \ei 
\en 

If $Op$ is a CO of the form \re{soasdhfjksla}, 
then we shall denote by $$cond\,(Op)$$ 
a formula $b$ such that 
$op_1 = \langle b \rangle$.

Let $Op$ be a CO.
\bi 
\i $Op$ is said to be an {\bf input CO} 
   (or an {\bf output CO}), 
   if among operators belonging to $Op$, 
   there is an input (or an output) operator.
\i $Op$  is said to be an {\bf internal CO}, 
   if all operators belonging to $Op$   
   are internal. 
\i If $Op$ is an {\bf input CO} 
   (or an {\bf output CO}), then
   the notation $$name \, (Op)$$ 
   denotes a name occurred in $Op$. 
\i If $\sigma$ is an evaluation of variables
   occurred in $cond\,(Op)$, 
   then we say that 
   {\bf $Op$ is open on $\sigma$}, if 
   $$\sigma(cond\,(Op))=1$$
\ei 

\subsection{A concept of a process with COs}

A concept of a {\bf process with COs}
differs from the concept of a process 
in section \ref{asdffqe23rq23rtgtg} 
only in the following: labels of 
transitions of a process with COs are COs.

\subsection{An execution 
of a process with COs}

An execution of a process with COs
\bi 
\i is defined in much the same 
   as an execution of a process is defined 
   in section \ref{sadfwasf43355}, 
   and 
\i is also a bypass of a set of its states, 
   \bi
   \i starting from an initial state, and
   \i with an execution of COs which are
      labels of visited transitions.
   \ei
\ei 

Let $P = (X_P, I_P, S_P, s^0_P, R_P)$ be a 
process with COs. 

At each step $i\geq 0$ of an execution of $P$
\bi 
\i  the process $P$ is located at some state  $s_i$ 
    ($s_0=s^0_P$)
\i  there is defined an evaluation
    $\sigma_i$ of variables from $X_P$\\
   ($\sigma_0(I_P)=1, \quad \sigma_i(at_P)=s_i$)
\i if there is a transition from $R_P$,
   starting at $s_i$,
   then the process
   \bi
   \i selects 
       a transition starting at $s_i$,
       which is labelled by a CO $Op_i$ 
       with the 
       following properties:
      \bi
\i $Op_i$ is open on $\sigma_i$
\i if among operators occurred in $Op_i$
there is an operator of the form
$$\alpha\,?\,x\quad \mbox{or}\quad \alpha\,!
\,e$$
then at current time 
the process $P$ can execute an action of the form 
$$\alpha\,?\,v\quad \mbox{or}\quad \alpha\,!
\,v$$ respectively
\ei
(if there is no such transitions, then 
the process $P$ suspends until 
such transition will appear)

   \i executes sequentially all 
operators occurred in $Op_i$, 
with a corresponding modification 
of current evaluation 
after an execution of 
each operator occurred in $Op_i$, 
and thereafter 
\i turns to the state $s_{i+1}$, 
which is an end of the selected transition 
   \ei 
\i if there is no a transition in $R_P$ 
    starting at $s_i$,    then the process 
completes its work. 
\ei 

\subsection{Operations on processes with COs}

\label{sdffwe3453476347893}

Definitions of operations on processes with
COs almost coincide with corresponding 
definitions in section \ref{erqwr2343242}, 
so we only point out the differences 
in these definitions. 
\bi 
\i In definitions of all operations 
   COs are mentioned instead of operators. 
\i Definitions of the operation ``$\pa$''
   differ only in the item, 
   which is related to a description of 
   ``diagonal'' transitions. 

   For processes with COs this item 
   has the following form: 
   for each pair of transitions of the form 
   $$\by
   \diagrw{s_1&\pright{Op_1}&s'_1}
   \quad\in R_{P_1}\\
   \diagrw{s_2&\pright{Op_2}&s'_2}
   \quad\in R_{P_2}\ey$$
   where one of the COs $Op_1,Op_2$
   has the form
   $$(op_1,\,\ldots, \,op_{i},
   \,\alpha\,?\,x,\,op_   {i+1},\,\ldots, \,op_n)
   $$
   and another of the COs has the form
   $$(op'_1,\,\ldots, \,op'_{j},\,\alpha\,!   
   \,e,\,op'_{j+1},\,\ldots,\, op'_m)
   $$
   where \bi\i $t(x)=t(e)$,
    \i the subsequences
   \begin{center}
   $(op_{i+1},\,\ldots, \,op_n)$
   and $(op'_{j+1},\,\ldots, \,op'_m)$
   \end{center}
   may be empty 
   \ei 
the process ${P_1\,|\,P_2}$ 
has the transition 
$$\diagrw{(s_1,s_2)&\pright{Op}&(s'_1,s'_2)}
$$
where $Op$ has the form
$$\b{ \langle cond\,(Op_1)\wedge 
cond\,(Op_2)\rangle,\\
op_2,\,\ldots, \,op_{i},\\
op'_2,\,\ldots, \,op'_{j},\\
(x:=e),\\
op_{i+1},\,\ldots, \,op_{n},\\
op'_{j+1},\,\ldots, \,op'_{m}}
$$
\ei 

\subsection{Transformation of processes 
with a message passing 
to processes with COs}

Each process with a message passing 
can be transformed to a process with COs 
by a replacement of labels of its transitions: 
for each transition 
$$\by s_1&\pright{op}&s_2\ey$$
its label $op$ is replaced by a CO $Op$, 
defined as follows. 
\bi 
\i If $op$ is a conditional operator, 
   then $$Op\eam(op)$$ 
\i If $op$ is \bi\i an assignment operator, or 
   \i an input or output operator\ei
   then $Op\eam (\langle \top\rangle\,,op)$\\
   (remind that $\top$ is a true formula)
\ei 

For each process with a message passing $P$ 
we denote the corresponding 
process with COs by the same symbol $P$. 

\subsection{Sequential composition of COs}

In this section, we introduce the concept 
of a {\bf sequential composition} of COs: 
for some pairs $(Op_1, Op_2)$ of COs 
we define a CO, which is denoted as
\be{sadfsdfwertrety} Op_1 \cdot Op_2 \ee 
and is called a {\bf sequential composition} 
of the COs $Op_1$ and $Op_2$. 

A necessary condition of a possibility 
to define a sequential composition \re{sadfsdfwertrety}
is the condition that at least one of the COs
$Op_1$, $Op_2$ is internal. 

Below we shall use the following notations. 
\bn 
\i For 
   \bi 
   \i each CO $Op=(op_1,\ldots, op_n)$, and 
   \i each assignment operator $op$ 
   \ei 
   the notation $Op\cdot op$ denotes the 
   CO 
   \be{sdwetrwertwer} 
   (op_1, \ldots, op_n, op) 
   \ee 
\i For \bi \i each internal CO 
   $Op=(op_1,\ldots, op_n)$, and 
   \i each input or output operator $op$ \ei 
   the notation $Op \cdot op$ denotes CO
   \re{sdwetrwertwer}
\i For \bi \i each CO 
   $Op=(op_1,\ldots, op_n)$, 
   and \i each conditional   operator 
   $op = \langle b\rangle$ \ei 
   the notation $Op \cdot op$ denotes 
   an object that 
   \bi 
   \i either is a CO 
   \i or is not defined. 
   \ei 
   This object is defined recursively as follows. 

   If $n = 1$, then $$Op\cdot op \eam 
   (\langle cond\, (Op)\wedge b\rangle)$$

   If $n>1$, then
   \bi 
   \i if $op_n$ is an assignment operator of the
      form $(x: = e)$, then
     $$Op\cdot op\eam
  \underbrace{((op_1,\ldots, op_{n-1})\cdot op_n(op))}_{(*)}
  \cdot op_n$$
where 
\bi 
\i $op_n(op)$ is a conditional operator, which 
is obtained from $op$ by a replacement
of all occurrences of the variable $x$ on the 
expression $e$ 
\i if the object $(*)$ is undefined, 
then $Op \cdot op$ also is undefined 
\ei 
   \i if $op_n$ is an output operator, then 
   $Op \cdot op$ is the CO
\be{gdfgewrtwertwer65}
((op_1,\ldots, op_{n-1})
 \cdot op)\cdot op_n\ee

   \i if $op_n$ is an input operator, 
and has the form  $\alpha\,?\,x$, 
then $Op \cdot op$ 
\bi 
\i is undefined, if $op$ depends on $x$, and
\i is equal to CO \re{gdfgewrtwertwer65}, 
otherwise. 
\ei 
   \ei 
\en 

Now we can formulate a definition of a sequential 
composition of COs. 

Let $Op_1,Op_2$ be COs, and
$Op_2$ has the form 
$$Op_2=(op_1,\ldots, op_n)$$

We shall say that {\bf there is defined 
a sequential composition} of $Op_1$ 
and $Op_2$, if the following conditions are 
met: 
\bi 
\i at least one of the COs $Op_1, Op_2$ 
   is internal
\i there is no undefined objects 
    in the parentheses in the expression 
   \be{werftwqfsdfgsdfghwer}
   (\ldots((Op_1\cdot op_1)\cdot op_2)
   \cdot\ldots)\cdot op_n
   \ee
\ei 
If these conditions are met, 
then a {\bf sequential composition} 
$Op_1$ and $Op_2$ is a value of expression 
\re{werftwqfsdfgsdfghwer}. 
This CO is denoted by
$$Op_1 \cdot Op_2$$ 

\subsection{Reduction of processes with COs}

Let $P$ be a  process with COs. 

A {\bf reduction} of $P$ is a sequence 
\be{ghghretty4}\by P=P_0&\pright{}
&P_1&\pright{}&
\ldots&\pright{}&P_n\ey\ee
of transformations of this process, 
each of which is performed 
according to any of the 
reduction rules described below. 
Each of these transformations 
(except the first) 
is made on the result 
of the previous transformation. 

A {\bf result} of the reduction \re{ghghretty4}
is a result of the last transformation
(i.e. the process $P_n$). 

Reduction rules have the following form. 

\begin{description}
\i[Rule 1 (sequential composition).]$\;$\\
  Let $s$ be a state of a process with COs, 
   which is not an initial state, and
  \bi
  \i a set of all transitions 
     of this process 
     with an end $s$ 
     has the form
     $$\by s_1&\pright{Op_1}
     &s,\quad\ldots\;,
     s_n&\pright{Op_n}&s\ey$$
  \i a set of all transitions 
     of this process 
     with a  start $s$ 
     has the form
     $$\by s&\pright{Op'_1}&
     s'_1,\quad\ldots\;,
     s&\pright{Op'_m}&s'_m\ey$$
  \i $s\not\in\{s_1,\ldots,
     s_n,s'_1,\ldots,s'_m\}$
  \i for each $i=1,\ldots, n$ 
     and each $j=1,\ldots, m$
     there is defined the sequential composition
     $$Op_i\cdot Op_j$$
  \ei
  Then this process can be transformed 
   to a process 
  \bi
  \i states of which are states of the original
     process, with the exception of $s$
  \i transitions of which are 
     \bi
     \i transitions of the original
        process, a start or an end of which 
        is not $s$, and
     \i transitions of the form
        $$\by s_i&
        \pmiddleright{Op_i\cdot Op'_j}
        &s'_j\ey$$
        for each $i=1,\ldots, n$ and 
        each $j=1,\ldots, m$
     \ei
  \i \bi\i an initial state of which,\i
 a set of variables, and \i an initial 
  condition\ei
coincide with the corresponding components 
of the original process. 
  \ei

\i[Rule 2 (gluing).]$\;$\\
  Let $P$ be a process with COs, 
  which has two transitions 
  with a common start and a common end:
  \be{fggty5w34}\by s_1&\pright{Op}
   &s_2,\qquad s_1&\pright{Op'}&s_2
  \ey\ee
  and labels of these transitions 
  differ only in first components, 
  i.e. $Op$ and $Op'$ have the form 
  $$\by Op=(op_1,op_2,\ldots, op_n)\\
  Op'=(op'_1,op_2,\ldots, op_n)\ey$$

  Rule 2 is a replacement of the 
  pair of transitions \re{fggty5w34}
  on a transition 
  $$\by s_1&\pright{Op}&s_2\ey$$
  where $Op=(\langle 
   cond\,(Op)\vee cond\,(Op')\rangle,
   op_2,\ldots, op_n)$

\i[Rule 3 (removal of inessential    
assignments).]$\;$\\

  Let 
  \bi
  \i $P$ be a process with COs, and 
  \i $op(P)$ be a set of all operators, 
      occurred in COs of $P$.
  \ei
  A variable $x\in X_P$ is said to be 
  {\bf inessential}, if
  \bi
  \i $x$ does not occur in 
     \bi
     \i conditional operators, and
     \i output operators 
     \ei 
     in $op(P)$, 
  \i if $x$ has an occurrence in right size
     of any assignment operator from $op(P)$
     of the form $(y:=e)$, then the variable $y$
     is inessential.
  \ei
  Rule 3 is a removal from all COs 
  of all assignment operators 
  of the form $(x:=e)$, 
  where the variable $x$ is inessential.

\end{description}

\subsection{An example of a reduction}

In this section we consider a reduction 
of the process ${\it Buffer}_n$, 
the graph representation of which 
is given in section \ref{bufferprocess}. 

Below we use the following agreements.
\bi 
\i If $Op$ is a CO such that 
   $$cond\,(Op) =\top$$
   then the first operator in this CO
   will be omitted.
\i Operators in COs
   can be placed vertically.
\i Brackets, which embrace a sequence of 
   operators consisting in a CO, can be omitted. 
\ei 

The original process 
${\it Buffer}_n$ has the following form: 

{\def\arraystretch{1}

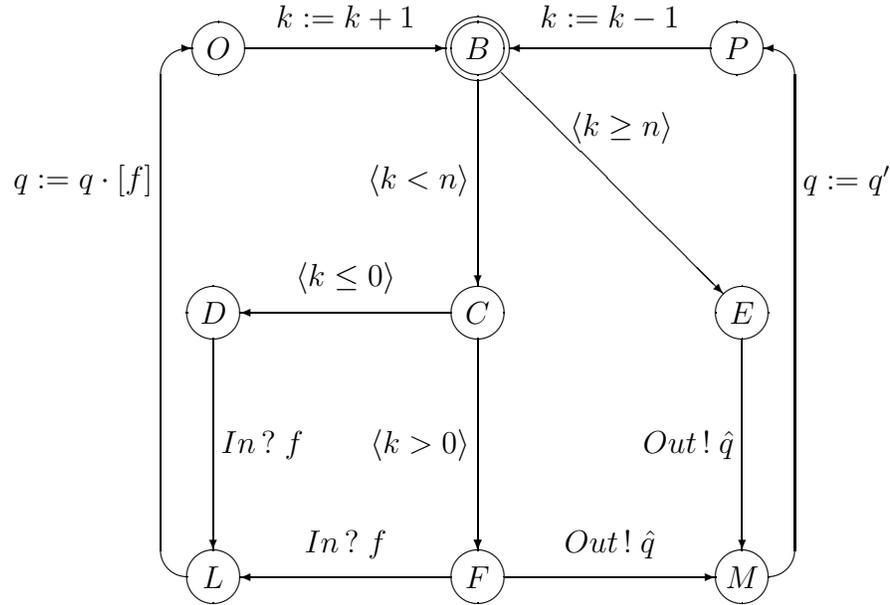
$$\by
\begin{picture}(0,220)

\put(0,200){\oval(20,20)}
\put(0,200){\oval(24,24)}
\put(0,200){\makebox(0,0)[c]{$B$}}

\put(-100,100){\oval(20,20)}
\put(-100,100){\makebox(0,0)[c]{$D$}}

\put(0,100){\oval(20,20)}
\put(0,100){\makebox(0,0)[c]{$C$}}

\put(100,100){\oval(20,20)}
\put(100,100){\makebox(0,0)[c]{$E$}}

\put(-100,0){\oval(20,20)}
\put(-100,0){\makebox(0,0)[c]{$L$}}

\put(0,0){\oval(20,20)}
\put(0,0){\makebox(0,0)[c]{$F$}}

\put(100,0){\oval(20,20)}
\put(100,0){\makebox(0,0)[c]{$M$}}

\put(0,188){\vector(0,-1){78}}
\put(9,191){\vector(1,-1){84}}
\put(0,90){\vector(0,-1){80}}
\put(-100,90){\vector(0,-1){80}}
\put(0,90){\vector(0,-1){80}}
\put(100,90){\vector(0,-1){80}}
\put(-10,100){\vector(-1,0){80}}
\put(-10,0){\vector(-1,0){80}}
\put(10,0){\vector(1,0){80}}

\put(-110,10){\oval(20,20)[lb]}
\put(-120,10){\line(0,1){180}}
\put(-110,190){\oval(20,20)[lt]}
\put(-110,200){\vector(1,0){2}}
\put(-98,200){\oval(20,20)}
\put(-98,200){\makebox(0,0)[c]{$O$}}
\put(98,200){\makebox(0,0)[c]{$P$}}

\put(-88,200){\vector(1,0){76}}

\put(110,10){\oval(20,20)[rb]}
\put(120,10){\line(0,1){180}}
\put(110,190){\oval(20,20)[rt]}
\put(110,200){\vector(-1,0){2}}
\put(98,200){\oval(20,20)}
\put(88,200){\vector(-1,0){76}}

\put(-50,207){\makebox(0,0)[b]{$k:=k+1$}}
\put(50,207){\makebox(0,0)[b]{$k:=k-1$}}

\put(-50,7){\makebox(0,0)[b]{$In\,?\;f$}}
\put(50,7){\makebox(0,0)[b]{$Out\,!\;\hat 
q$}}

\put(-123,150){\makebox(0,0)[r]{$q:=q\cdot 
[f]$}}
\put(123,150){\makebox(0,0)[l]{$q:=q'$}}

\put(-97,50){\makebox(0,0)[l]{$In\,?\;f$}}
\put(97,50){\makebox(0,0)[r]{$Out\,!\;\hat 
q$}}

\put(-3,50){\makebox(0,0)[r]{$\langle
k>0\rangle$}}
\put(-3,150){\makebox(0,0)[r]{$\langle 
k<n\rangle $}}

\put(-50,107){\makebox(0,0)[b]{$\langle k
\leq 0\rangle$}}

\put(35,170){\makebox(0,0)[l]{$\langle 
k\geq n\rangle$}}

\end{picture}
\\
\vspace{0mm}
\ey
$$
}

First reduction step 
is a removing of the state $C$ 
(we apply rule 1 for $s = C$): 

{\def\arraystretch{1}

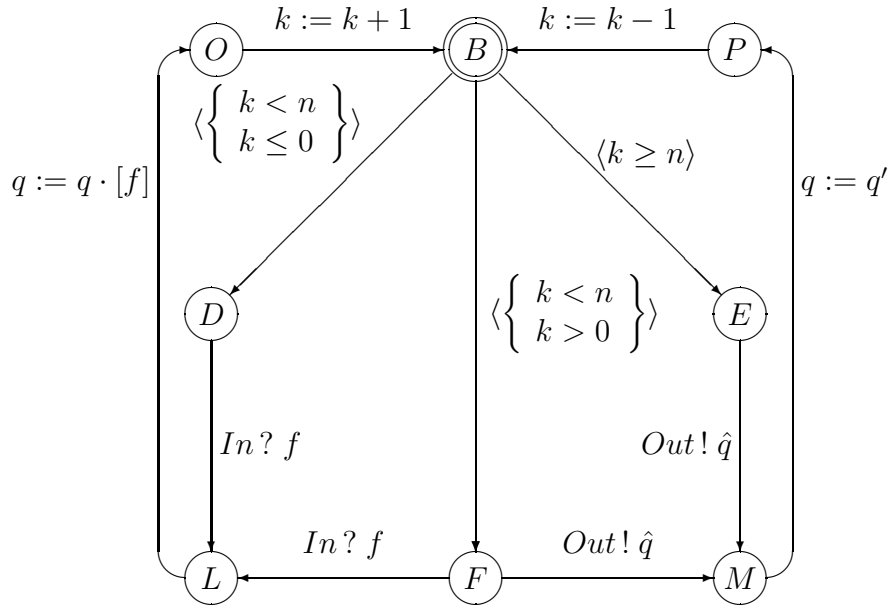
$$\by
\begin{picture}(0,220)

\put(0,200){\oval(20,20)}
\put(0,200){\oval(24,24)}
\put(0,200){\makebox(0,0)[c]{$B$}}

\put(-100,100){\oval(20,20)}
\put(-100,100){\makebox(0,0)[c]{$D$}}

\put(100,100){\oval(20,20)}
\put(100,100){\makebox(0,0)[c]{$E$}}

\put(-100,0){\oval(20,20)}
\put(-100,0){\makebox(0,0)[c]{$L$}}

\put(0,0){\oval(20,20)}
\put(0,0){\makebox(0,0)[c]{$F$}}

\put(100,0){\oval(20,20)}
\put(100,0){\makebox(0,0)[c]{$M$}}

\put(0,188){\vector(0,-1){178}}
\put(9,191){\vector(1,-1){84}}

\put(-9,191){\vector(-1,-1){84}}

\put(0,90){\vector(0,-1){80}}
\put(-100,90){\vector(0,-1){80}}
\put(0,90){\vector(0,-1){80}}
\put(100,90){\vector(0,-1){80}}
\put(-10,0){\vector(-1,0){80}}
\put(10,0){\vector(1,0){80}}

\put(-110,10){\oval(20,20)[lb]}
\put(-120,10){\line(0,1){180}}
\put(-110,190){\oval(20,20)[lt]}
\put(-110,200){\vector(1,0){2}}
\put(-98,200){\oval(20,20)}
\put(-98,200){\makebox(0,0)[c]{$O$}}
\put(98,200){\makebox(0,0)[c]{$P$}}

\put(-88,200){\vector(1,0){76}}

\put(110,10){\oval(20,20)[rb]}
\put(120,10){\line(0,1){180}}
\put(110,190){\oval(20,20)[rt]}
\put(110,200){\vector(-1,0){2}}
\put(98,200){\oval(20,20)}
\put(88,200){\vector(-1,0){76}}

\put(-50,207){\makebox(0,0)[b]{$k:=k+1$}}
\put(50,207){\makebox(0,0)[b]{$k:=k-1$}}

\put(-50,7){\makebox(0,0)[b]{$In\,?\;f$}}
\put(50,7){\makebox(0,0)[b]{$Out\,!\;\hat 
q$}}

\put(-123,150){\makebox(0,0)[r]{$q:=q\cdot 
[f]$}}
\put(123,150){\makebox(0,0)[l]{$q:=q'$}}

\put(-97,50){\makebox(0,0)[l]{$In\,?\;f$}}
\put(97,50){\makebox(0,0)[r]{$Out\,!\;\hat 
q$}}

\put(5,100){\makebox(0,0)[l]{$\langle\c
{k<n\\k>0}\rangle $}}

\put(-75,157){\makebox(0,0)[b]{$\langle \c
{k<n\\k\leq 0}\rangle$}}

\put(45,160){\makebox(0,0)[l]{$\langle 
k\geq n\rangle$}}

\end{picture}
\\
\vspace{0mm}
\ey
$$
}

Since $n>0$, then the formula 
$(k<n)\wedge(k\leq 0)$ 
in the label of the transition 
from $B$ to $D$ 
can be replaced 
by the equivalent formula $k\leq 0$. 

Second and third reduction steps 
are removing of states $O$ and $P$: 

{\def\arraystretch{1}

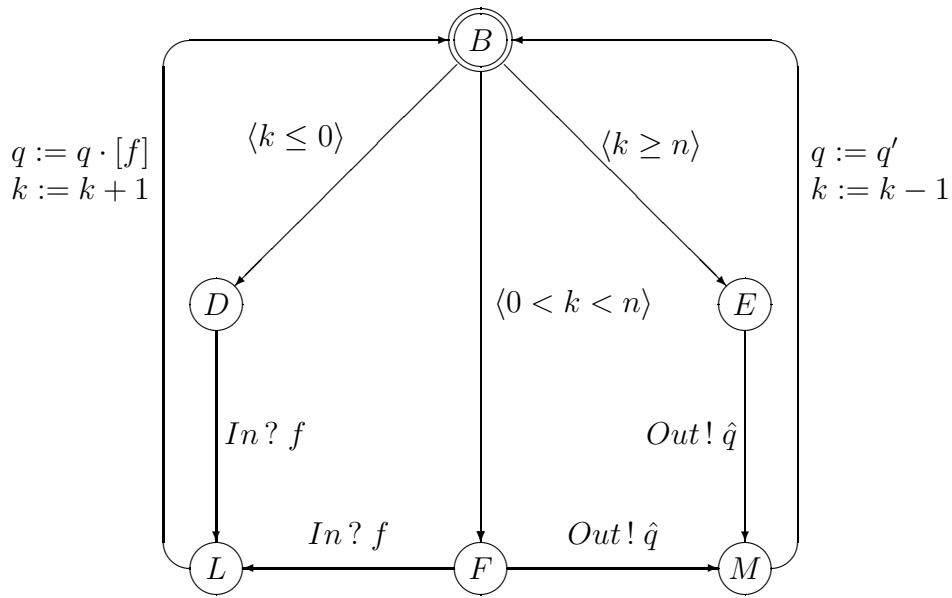
$$\by
\begin{picture}(0,220)

\put(0,200){\oval(20,20)}
\put(0,200){\oval(24,24)}
\put(0,200){\makebox(0,0)[c]{$B$}}

\put(-100,100){\oval(20,20)}
\put(-100,100){\makebox(0,0)[c]{$D$}}

\put(100,100){\oval(20,20)}
\put(100,100){\makebox(0,0)[c]{$E$}}

\put(-100,0){\oval(20,20)}
\put(-100,0){\makebox(0,0)[c]{$L$}}

\put(0,0){\oval(20,20)}
\put(0,0){\makebox(0,0)[c]{$F$}}

\put(100,0){\oval(20,20)}
\put(100,0){\makebox(0,0)[c]{$M$}}

\put(0,188){\vector(0,-1){178}}
\put(9,191){\vector(1,-1){84}}

\put(-9,191){\vector(-1,-1){84}}

\put(0,90){\vector(0,-1){80}}
\put(-100,90){\vector(0,-1){80}}
\put(0,90){\vector(0,-1){80}}
\put(100,90){\vector(0,-1){80}}
\put(-10,0){\vector(-1,0){80}}
\put(10,0){\vector(1,0){80}}

\put(-110,10){\oval(20,20)[lb]}
\put(-120,10){\line(0,1){180}}
\put(-110,190){\oval(20,20)[lt]}
\put(-110,200){\vector(1,0){98}}

\put(110,10){\oval(20,20)[rb]}
\put(120,10){\line(0,1){180}}
\put(110,190){\oval(20,20)[rt]}
\put(110,200){\vector(-1,0){98}}

\put(-50,7){\makebox(0,0)[b]{$In\,?\;f$}}
\put(50,7){\makebox(0,0)[b]{$Out\,!\;
\hat q$}}

\put(-120,150){\makebox(0,0)[r]{$\by 
q:=q\cdot [f]\\
k:=k+1\ey$}}
\put(120,150){\makebox(0,0)[l]{$\by 
q:=q'\\k:=k-1
\ey$}}

\put(-97,50){\makebox(0,0)[l]{$In\,?\;f$}}
\put(97,50){\makebox(0,0)[r]{$Out\,!\;\hat 
q$}}

\put(5,100){\makebox(0,0)[l]{$\langle 
 0<k<n \rangle$}}

\put(-70,157){\makebox(0,0)[b]{$\langle 
k\leq 0\rangle$}}

\put(45,160){\makebox(0,0)[l]{$\langle 
k\geq n\rangle $}}

\end{picture}
\\
\vspace{0mm}
\ey
$$
}

Fourth and fifth reduction steps 
are removing of the states $D$ and $E$: 

{\def\arraystretch{1}

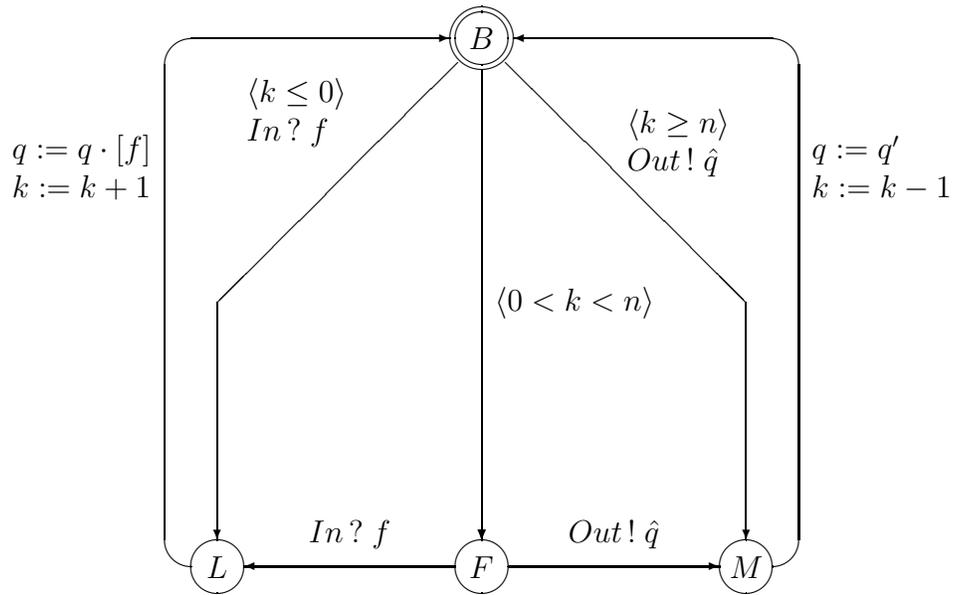
$$\by
\begin{picture}(0,220)

\put(0,200){\oval(20,20)}
\put(0,200){\oval(24,24)}
\put(0,200){\makebox(0,0)[c]{$B$}}

\put(-100,0){\oval(20,20)}
\put(-100,0){\makebox(0,0)[c]{$L$}}

\put(0,0){\oval(20,20)}
\put(0,0){\makebox(0,0)[c]{$F$}}

\put(100,0){\oval(20,20)}
\put(100,0){\makebox(0,0)[c]{$M$}}

\put(0,188){\vector(0,-1){178}}
\put(9,191){\line(1,-1){91}}
\put(-9,191){\line(-1,-1){91}}

\put(0,90){\vector(0,-1){80}}
\put(-100,100){\vector(0,-1){90}}
\put(0,90){\vector(0,-1){80}}
\put(100,100){\vector(0,-1){90}}
\put(-10,0){\vector(-1,0){80}}
\put(10,0){\vector(1,0){80}}

\put(-110,10){\oval(20,20)[lb]}
\put(-120,10){\line(0,1){180}}
\put(-110,190){\oval(20,20)[lt]}
\put(-110,200){\vector(1,0){98}}

\put(110,10){\oval(20,20)[rb]}
\put(120,10){\line(0,1){180}}
\put(110,190){\oval(20,20)[rt]}
\put(110,200){\vector(-1,0){98}}

\put(-50,7){\makebox(0,0)[b]{$In\,?\;f$}}
\put(50,7){\makebox(0,0)[b]{$Out\,!\;\hat 
q$}}

\put(-120,150){\makebox(0,0)[r]{$\by 
q:=q\cdot [f]\\
k:=k+1\ey$}}
\put(120,150){\makebox(0,0)[l]{$\by 
q:=q'\\k:=k-1
\ey$}}

\put(5,100){\makebox(0,0)[l]{$\langle 
0<k<n\rangle$}}

\put(-70,157){\makebox(0,0)[b]{$\by
\langle k\leq 0\rangle \\
In\,?\;f
\ey$}}

\put(50,160){\makebox(0,0)[l]{$\by
\langle k\geq n\rangle \\Out\,!\;\hat q
\ey$}}

\end{picture}
\\
\vspace{0mm}
\ey
$$
}

Sixth reduction step is  
removing of the state $F$: 

{\def\arraystretch{1}

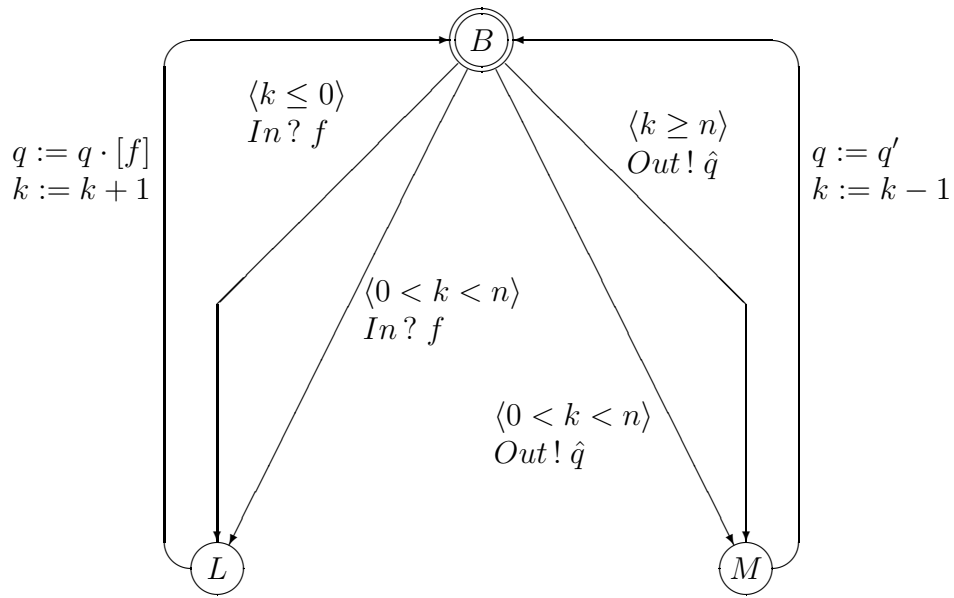
$$\by
\begin{picture}(0,220)

\put(0,200){\oval(20,20)}
\put(0,200){\oval(24,24)}
\put(0,200){\makebox(0,0)[c]{$B$}}

\put(-100,0){\oval(20,20)}
\put(-100,0){\makebox(0,0)[c]{$L$}}

\put(100,0){\oval(20,20)}
\put(100,0){\makebox(0,0)[c]{$M$}}

\put(-5.5,189){\vector(-1,-2){90}}
\put(5.5,189){\vector(1,-2){90}}

\put(9,191){\line(1,-1){91}}
\put(-9,191){\line(-1,-1){91}}

\put(-100,100){\vector(0,-1){90}}
\put(100,100){\vector(0,-1){90}}

\put(-110,10){\oval(20,20)[lb]}
\put(-120,10){\line(0,1){180}}
\put(-110,190){\oval(20,20)[lt]}
\put(-110,200){\vector(1,0){98}}

\put(110,10){\oval(20,20)[rb]}
\put(120,10){\line(0,1){180}}
\put(110,190){\oval(20,20)[rt]}
\put(110,200){\vector(-1,0){98}}

\put(-120,150){\makebox(0,0)[r]{$\by 
q:=q\cdot [f]\\
k:=k+1\ey$}}
\put(120,150){\makebox(0,0)[l]{$\by 
q:=q'\\k:=k-1
\ey$}}

\put(-50,97){\makebox(0,0)[l]{$\by 
\langle 0<k<n\rangle \\
In\,?\;f\ey$}}

\put(70,50){\makebox(0,0)[r]{$\by
\langle 0<k<n\rangle \\
Out\,!\;\hat q\ey$}}

\put(-70,157){\makebox(0,0)[b]{$\by
\langle k\leq 0\rangle \\
In\,?\;f
\ey$}}

\put(50,160){\makebox(0,0)[l]{$\by
\langle k\geq n\rangle \\Out\,!\;\hat q
\ey$}}

\end{picture}
\\
\vspace{0mm}
\ey
$$
}

Seventh and eighth reduction steps 
consist of an application of rule 2 
to the transitions from 
$B$ to $L$ and from $B$ to $M$. 
In the resulting process, we replace 
\bi 
\i the formula  $(0<k<n)\vee (k\leq 0)$ 
   on the equivalent formula $k <n$, and
\i the formula $(0<k<n)\vee (k\geq n)$ 
   on the equivalent formula $k> 0$.
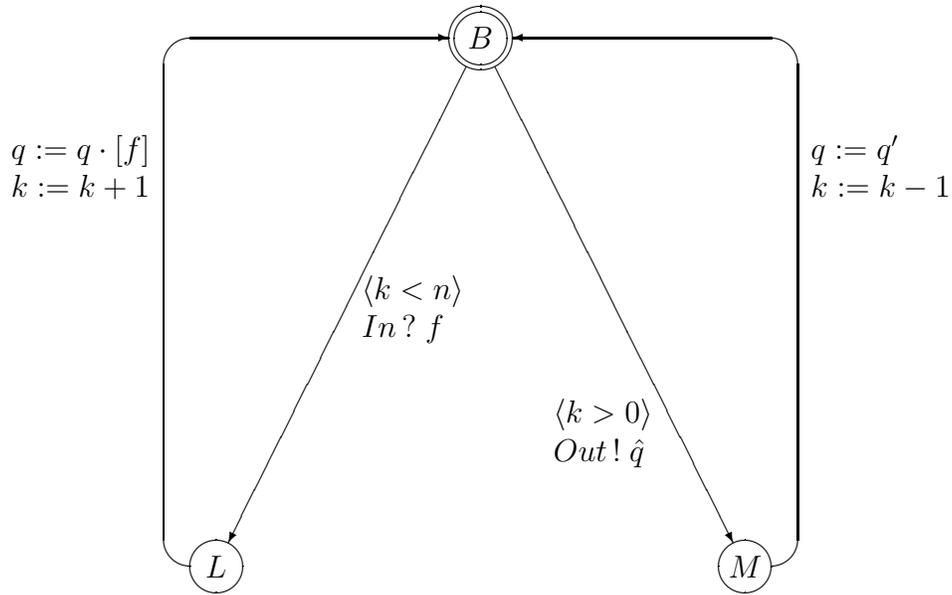
\ei 

{\def\arraystretch{1}

$$\by
\begin{picture}(0,220)

\put(0,200){\oval(20,20)}
\put(0,200){\oval(24,24)}
\put(0,200){\makebox(0,0)[c]{$B$}}

\put(-100,0){\oval(20,20)}
\put(-100,0){\makebox(0,0)[c]{$L$}}

\put(100,0){\oval(20,20)}
\put(100,0){\makebox(0,0)[c]{$M$}}

\put(-5.5,189){\vector(-1,-2){90}}
\put(5.5,189){\vector(1,-2){90}}

\put(-110,10){\oval(20,20)[lb]}
\put(-120,10){\line(0,1){180}}
\put(-110,190){\oval(20,20)[lt]}
\put(-110,200){\vector(1,0){98}}

\put(110,10){\oval(20,20)[rb]}
\put(120,10){\line(0,1){180}}
\put(110,190){\oval(20,20)[rt]}
\put(110,200){\vector(-1,0){98}}

\put(-120,150){\makebox(0,0)[r]{$\by 
q:=q\cdot [f]\\
k:=k+1\ey$}}
\put(120,150){\makebox(0,0)[l]{$\by 
q:=q'\\k:=k-1
\ey$}}

\put(70,50){\makebox(0,0)[r]{$\by 
\langle k>0\rangle \\
Out\,!\;\hat q\ey$}}

\put(-50,97){\makebox(0,0)[l]{$\by 
\langle k<n\rangle \\
In\,?\;f\ey$}}

\end{picture}
\\
\vspace{0mm}
\ey
$$
}

Ninth and tenth reduction steps 
are removing of states 
$L$ and $M$. 

{\def\arraystretch{1}

{\small
\be{firllfgksa44}\by
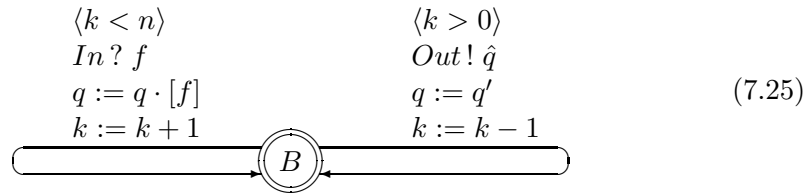
\begin{picture}(0,70)

\put(0,0){\oval(20,20)}
\put(0,0){\oval(24,24)}
\put(0,0){\makebox(0,0)[c]{$B$}}

\put(-100,0){\oval(10,10)[l]}
\put(-11,5){\line(-1,0){89}}
\put(-100,-5){\vector(1,0){89}}

\put(-60,5){\makebox(0,0)[b]{
$\by\langle k<n\rangle \\
In\,?\;f\\q:=q\cdot [f]\\k:=k+1\ey
$}}

\put(100,0){\oval(10,10)[r]}
\put(11,5){\line(1,0){89}}
\put(100,-5){\vector(-1,0){89}}

\put(70,5){\makebox(0,0)[b]{$
\by \langle k>0\rangle \\
Out\,!\;\hat q\\q:=q'\\k:=k-1 \ey
$}}

\end{picture}
\\
\vspace{0mm}
\ey
\ee } 
}

The last process is the result of the 
reduction of ${\it Buffer}_n$.

\subsection{A concretization of processes
with COs}

A concept of a concretization of processes
with COs is similar to the concept of a 
concretization of processes with a message 
passing (see section \ref{dfgjkl3e5tiu3iout3}). 

Let $P$ be a process with COs.
The notation $Conc(P)$ denotes 
a process in the original sense of this concept 
(see section \ref{asdfasdf3rt4wertweywey}), 
which is called a {\bf concretization} 
of the process $P$, 
and has the following components. 

\bn
\i States of $Conc(P)$ are
   \bi
   \i all evaluations from $Eval(X_P)$, and
   \i an additional state $s^0$, which is 
     an initial state of $Conc(P)$
   \ei
\i For
   \bi
   \i each transition
      $\diagrw{s_1&\pright{Op}&s_2}$ 
      of the process $P$, and
   \i each evaluation $\sigma\in Eval(X_P)$,
      such that
      \bi
      \i $\sigma(at_P)=s_1$, and
      \i $Op$ is open on $\sigma$
      \ei
   \ei
   $Conc(P)$ has the transition
   $$\diagrw{\sigma&\pright{a}&\sigma'}$$
   if $\sigma'(at_P)=s_2$, and
   one of the following cases hold:
   \bn
   \i $Op$ is internal, $a=\tau$,
      and the following statement holds:
      $$\diagrw{\sigma&\pright{Op}&\sigma'}$$
      which means the following:
      if $Op$ has the form 
      $$(op_1,\ldots, op_n)$$
      then there is a sequence 
      $\sigma_1,\ldots, \sigma_n$ of evaluations from
      $Eval(X_P)$, such that
\bi
\i $\forall\, x\in X_P\setminus\{at_P\}\quad
\sigma(x)=\sigma_1(x),\quad \sigma'(x)=\sigma_n(x)$, and
\i
$\forall\,i=2,\ldots, n$,
if $op_i$ has the form $(x:=e)$, then
$$\sigma_i(x)=\sigma_{i-1}(e),\quad
\forall y\in X_P\setminus \{x,at_P\}\quad
\sigma_i(y)=\sigma_{i-1}(y)$$
\ei
   \i \bi\i $Op=Op_1\cdot(\alpha\,?\,x)\cdot 
Op_2$, \i $a=\alpha\,?\,v$,
where $v \in D_{t(x)}$, and \i
there are evaluations 
$\sigma_1$ and $\sigma_2$ from $Eval(X_P)$,
such that
$$\by \diagrw{\sigma&\pright{Op_1}
&\sigma_1},\qquad
\diagrw{\sigma_2&\pright{Op_2}&\sigma'}\\
\sigma_2(x)=v,\quad
\forall y\in X_P\setminus \{x,at_P\}\qquad
\sigma_2(y)=\sigma_1(y)\ey$$
\ei
   \i \bi
\i $Op=Op_1\cdot(\alpha\,!\,e)\cdot Op_2$,
\i there is an evaluation $\sigma_1$ from $Eval
(X_P)$,
such that
$$\by \diagrw{\sigma&\pright{Op_1}
&\sigma_1},\qquad
\diagrw{\sigma_1&\pright{Op_2}&\sigma'},\qquad
a=\alpha\,!\,\sigma_1(e)
\ey$$
\ei
   \en
\i For
   \bi
   \i each evaluation $\sigma\in Eval(X_P)$,
such that $$\sigma(I_P)=1$$
   \i and each transition of $Conc(P)$ 
of the form
$\diagrw{\sigma&\pright{a}&\sigma'}$
   \ei
   $Conc(P)$ has the transition
   $\diagrw{s^0&\pright{a}&\sigma'}$.
\en

A reader is invited to investigate 
a relationship between 
\bi 
\i a concretization of an arbitrary process 
   with a message passing $P$, and\i a 
concretization of a process with COs, 
   which is derived by a reduction 
   of the process $P$. 
\ei 

\subsection{Equivalences 
on processes with COs}

Let $P_1$ and $P_2$ be processes with COs. 

We shall say that $P_1$ and $P_2$ are 
{\bf observationally equivalent}
and denote this fact by 
$$P_1 \approx P_2$$
if the concretizations $Conc(P_1)$ 
and $Conc(P_2)$ are 
observationally equivalent 
in the original sense of this concept
(see section \ref{defequisdfgval43}). 

Similarly, the equivalence $\oc$ 
is defined on processes with COs. 

Using the concept of a reduction 
of processes with COs, 
it is possible to define 
another equivalence on the set 
of processes with COs. 
This equivalence 
\bi 
\i is denoted by $\redeq$, and 
\i is a minimal congruence on the set 
   of processes with COs, 
   with the following property: 
   if $P'$ is derived from $P$ 
   by any reduction 
   rule, then $P\redeq P'$ 
\ei 
(i.e. $\redeq$ is the intersection 
of all congruences on the set of processes 
with COs, which have the above property). 

A reader is invited 
\bi 
\i to investigate
   a relation between \bi\i operations 
   on processes with COs, and
   \i the equivalences 
   $\approx$ and $\oc$ \ei
   i.e. to establish properties, 
   which are similar to properties
   represented in sections
   \ref{archnatella}, 
   \ref{dopsvojstva}, 
   \ref{aldffrrghghggdf}, 
   \ref{sdfwerr4543534ergteg} 
\i to formulate and justify 
   necessary and sufficient conditions 
   of observational equivalence 
   of processes with COs, 
   without use of the concept of a 
   concretization 
\i explore a relationship between 
   the equivalences $\approx$, $\oc$ 
   and $\redeq$\i find reduction rules such that
   $$\redeq\;\subseteq \; \oc$$
\ei 

\subsection{A method of a proof 
of observational equivalence 
of processes with COs}

One of possible methods 
of a proof of observational equivalence 
of processes with COs is based on 
theorem \ref{mueqyuivr}
presented below. 

To formulate this theorem, we introduce 
auxiliary concepts and notations. 
\bn 
\i Let $P$ be a process with COs. 

A {\bf composite transition (CT)} in $P$ is
a (possibly empty) sequence $CT$ 
of transitions of the process 
$P$ of the form 
\be{asff2323}CT=\quad
\diagrw{s_0&\pright{Op_1}&s_1&\pright{Op_2}&
\ldots&\pright{Op_n}&s_n}\qquad(n\geq 0)\ee
such that 
\bi 
\i among the COs $Op_1,\ldots, Op_n$ 
   there is no more than one    input or output CO
\i there is defined the sequential composition 
   $$(\ldots(Op_1\cdot Op_2)\cdot\ldots)\cdot Op_n$$
   which will be denoted 
   by the same symbol $CT$. 
\ei 

If sequence \re{asff2323} is empty,
then its sequential composition $CT$
by a definition is the CO $(\langle \top\rangle)$.

The state $s_0$ is said to be a 
{\bf start} of CT \re{asff2323},
and the state $s_n$ is said to be
an  {\bf end} of this CT.

The notation $\diagrw{s_0&\pright{CT}&s_n}$ 
is an abridged record 
of the statement  that  $CT$ 
\bi \i is a CT with the start $s_0$ and the end 
$s_n$, and also 
\i is a CO that corresponds to this CT.
\ei

\i Let $\varphi$ and $\psi$ be formulas.

   The notation $\varphi\leq \psi$
   is an abridged record of the statement
   that the formula   $\varphi\to\psi$  is true.

\i Let $Op=(op_1,\ldots, op_n)$ be an
   internal CO, and $\varphi$ be a formula.

The notation $Op(\varphi)$
denotes a formula defined recursively: 
 $$Op(\varphi)\eam\left\{\by
   cond\,(Op)\to\varphi,&\mbox{ if } n=1\\
   (op_1,\ldots, op_{n-1})\,(op_n
(\varphi)),&\mbox{ if } n>1
   \ey\right.$$
   where $op_n(\varphi)$
denotes the following formula: 
if $op_n\;=\; (x:=e)$,
then $op_n(\varphi)$ is obtained from
$\varphi$ by a replacement of each occurrence 
of the variable $x$ on the expression $e$.
\i Let $\varphi, \psi$ be formulas, and 
$Op_1, Op_2$ be COs.

We shall say that the following diagram is correct
{\def\arraystretch{1}
{\small
\be{1sdff3436636775}\by
\begin{picture}(100,160)

\put(0,130){\makebox(0,0)[c]{$A$}}
\put(10,130){\line(1,0){80}}
\put(100,130){\makebox(0,0)[c]{$B$}}
\put(50,140){\makebox(0,0)[b]{$
\varphi
$}}
\put(0,120){\vector(0,-1){80}}
\put(-5,80){\makebox(0,0)[r]{$
Op_1
$}}
\put(100,120){\vector(0,-1){80}}
\put(105,80){\makebox(0,0)[l]{$
Op_2
$}}
\put(0,30){\makebox(0,0)[c]{$C$}}
\put(100,30){\makebox(0,0)[c]{$D$}}
\put(10,30){\line(1,0){80}}
\put(50,20){\makebox(0,0)[t]{$
\psi
$}}

\end{picture}
\ey
\ee } 
}

if one of the following conditions is met.
   \bn\i $Op_1$ and $Op_2$ are internal COs, 
and the following inequality holds:$$\varphi\leq 
(Op_1\cdot Op_2)(\psi)$$

\i $Op_1$ and $Op_2$ can be represented
as sequential compositions
$$\by Op_1=Op_3\cdot (\alpha\,?\,x)\cdot 
Op_4\\
Op_2=Op_5\cdot (\alpha\,?\,y)\cdot Op_6
\ey$$
where $Op_3$, $Op_4$, $Op_5$, $Op_6$ are 
internal COs,
and the following inequality holds
$$\varphi\leq
(Op_1'\cdot Op_2')(\psi)$$
where
\bi
\i $Op'_1=Op_3\cdot (x:=z)\cdot Op_4$
\i $Op'_2=Op_5\cdot (y:=z)\cdot Op_6$
\i $z$ is a new variable (i.e. $z$ does not
occur in $\varphi$, $\psi$, $Op_1$, $Op_2$)
\ei

\i $Op_1$ and $Op_2$ can be represented
as sequential compositions
$$\by Op_1=Op_3\cdot (\alpha\,!\,e_1)\cdot 
Op_4\\
Op_2=Op_5\cdot (\alpha\,!\,e_2)\cdot Op_6
\ey$$
where 
$Op_3$, $Op_4$, $Op_5$, $Op_6$ are internal 
COs,
and the following inequality holds:
$$\varphi\leq
 \c{
(Op_3\cdot Op_5)(e_1=e_2)\\
(Op_3\cdot Op_4\cdot Op_5\cdot Op_6)(\psi)}
$$
\en
\en

\refstepcounter{theorem}
{\bf Theorem \arabic{theorem}\label{mueqyuivr}}.

Let $P_1$ and $P_2$ be processes with COs
$$P_i=(X_{P_i},I_{P_i},S_{P_i},s^0_{P_i},R_
{P_i})\qquad
(i=1,2)$$
which have no common states 
and common variables. 

Then $P_1 \approx P_2$, 
if there is a function $\mu$ of the form
$$\mu: S_{P_1}\times S_{P_2}\to Fm$$
which has the following properties.
\bn
\i $I_{P_1}\wedge I_{P_2}\leq \mu(s^0_{P_1},s^0_{P_2})$.
\i For
   \bi\i each pair 
    $(A_1,A_2)\in S_{P_1}\times S_{P_2}$,
   and \i each transition
   $\diagrw{A_1&\pright{Op}&A'_1}$
   of the process $P_1$, such that
   \be{gfsdfgsdfswr2332}cond\,(Op)\;\wedge 
\;\mu(A_1,A_2)\neq \bot\ee
   \ei
   there is a set of CTs of the process $P_2$ 
    starting from $A_2$
\be{gfeeeg5433}\{\diagrw{A_2&\pright{CT_i}
&A_2^i}
\mid i\in \Im\}\ee
satisfying the following conditions: 
\bn
\i the following inequality holds:
\be{bigveeee}cond\,(Op)\;\wedge \;\mu
(A_1,A_2)\;\leq\;\bigvee\limits_{i\in\Im}cond\,
(CT_i)\ee
\i for each $i\in\Im$ the following diagram 
   is correct:
   {\def\arraystretch{1}
{\small
\be{sdff3436636775}\by
\begin{picture}(100,160)

\put(0,130){\makebox(0,0)[c]{$A_1$}}
\put(10,130){\line(1,0){80}}
\put(100,130){\makebox(0,0)[c]{$A_2$}}
\put(50,140){\makebox(0,0)[b]{$
\mu(A_1,A_2)
$}}
\put(0,120){\vector(0,-1){80}}
\put(-5,80){\makebox(0,0)[r]{$
Op
$}}
\put(100,120){\vector(0,-1){80}}
\put(105,80){\makebox(0,0)[l]{$
CT_i
$}}
\put(0,30){\makebox(0,0)[c]{$A_1'$}}
\put(100,30){\makebox(0,0)[c]{$A_2^i$}}
\put(10,30){\line(1,0){80}}
\put(50,20){\makebox(0,0)[t]{$
\mu(A'_1,A_2^i)
$}}
\end{picture}
\ey
\ee }
}
   \en
\i The property symmetrical to previous:
   for
   \bi\i each pair $(A_1,A_2)\in S_{P_1}
   \times S_{P_2}$,
   and \i each transition
   $\diagrw{A_2&\pright{Op}&A'_2}$
   of the process $P_2$, such that
   \re{gfsdfgsdfswr2332} holds
   \ei
   there is a set of CTs of the process $P_1$ 
starting from $A_1$
\be{hgyt5333788j}\{\diagrw{A_1&\pright{CT_i}&A_1^i}
\mid i\in \Im\}\ee
satisfying the following conditions: 

\bn
\i inequality \re{bigveeee} holds
\i for each $i\in\Im$ the following diagram 
    is correct:
   {\def\arraystretch{1}
{\small
\be{sdff34dfsg36636775}\by
\begin{picture}(100,160)

\put(0,130){\makebox(0,0)[c]{$A_1$}}
\put(10,130){\line(1,0){80}}
\put(100,130){\makebox(0,0)[c]{$A_2$}}
\put(50,140){\makebox(0,0)[b]{$
\mu(A_1,A_2)
$}}
\put(0,120){\vector(0,-1){80}}
\put(-5,80){\makebox(0,0)[r]{$
CT_i
$}}
\put(100,120){\vector(0,-1){80}}
\put(105,80){\makebox(0,0)[l]{$
Op
$}}
\put(0,30){\makebox(0,0)[c]{$A_1^i$}}
\put(100,30){\makebox(0,0)[c]{$A'_2$}}
\put(10,30){\line(1,0){80}}
\put(50,20){\makebox(0,0)[t]{$
\mu(A^i_1,A'_2)
$}}
\end{picture}
\ey
\ee }
}
\en
\en

\subsection{An example 
of a proof of observational equivalence 
of processes with COs}

As an example of a use of theorem 
\ref{mueqyuivr} prove that 
$${\it Buffer}_1\approx {\it Buf}$$
where
\bi
\i ${\it Buffer}_1$
   is a considered above 
   process ${\it Buffer}_n$
   (see \re{firllfgksa44})
   for $n = 1$, 
   i.e. a process of the form 
{\def\arraystretch{1}
{\small
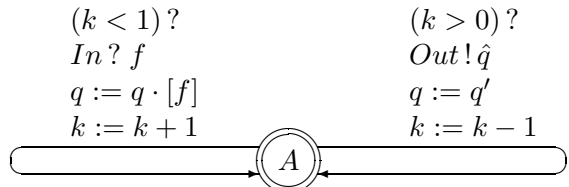
$$\by
\begin{picture}(0,70)

\put(0,0){\oval(20,20)}
\put(0,0){\oval(24,24)}
\put(0,0){\makebox(0,0)[c]{$A$}}

\put(-100,0){\oval(10,10)[l]}
\put(-11,5){\line(-1,0){89}}
\put(-100,-5){\vector(1,0){89}}

\put(-60,5){\makebox(0,0)[b]{
$\by(k<1)\,?\\
In\,?\;f\\q:=q\cdot [f]\\k:=k+1\ey$}}
\put(100,0){\oval(10,10)[r]}
\put(11,5){\line(1,0){89}}
\put(100,-5){\vector(-1,0){89}}

\put(70,5){\makebox(0,0)[b]{$
\by (k>0)\,?\\
Out\,!\,\hat q\\q:=q'\\k:=k-1 \ey
$}}

\end{picture}
\ey
$$ } 
}
its initial condition is 
$(k=0)\wedge (q=\varepsilon)$,
and
\i ${\it Buf}$ is a process of the form
   {\small
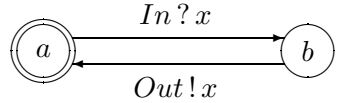
$$\by
\begin{picture}(100,20)

\put(100,0){\oval(20,20)}
\put(0,0){\oval(24,24)}
\put(0,0){\oval(20,20)}

\put(100,0){\makebox(0,0)[c]{$b$}}
\put(0,0){\makebox(0,0)[c]{$a$}}

\put(91,-5){\vector(-1,0){80}}
\put(11,5){\vector(1,0){80}}
\put(50,-10){\makebox(0,0)[t]{$Out\,!\,x$}}
\put(50,10){\makebox(0,0)[b]{$In\,?\,x$}}

\end{picture}
\ey
$$ } 

The initial condition of this process is $\top$.
\ei

Define a function 
$\mu:\{A\}\times \{a,b\}\to Fm$
as follows:
$$\by
\mu(A,a)\eam\;{(k=0)\wedge(q=\varepsilon)}
\\
\mu(A,b)\eam\;{(k=1)\wedge(q=[x])}\ey$$

Check properties 1, 2, and 3 
for the function  $\mu$.
\bn
\i Property 1 in this case is the inequality 
   $$((k=0)\wedge(q=\varepsilon))\wedge \top 
\leq ((k=0)\wedge(q=\varepsilon))$$
   which is obviously true. 
\i Check property 2.

\bi\i
  For the pair $(A,a)$  we have to consider 
   left transition in the process
  ${\it Buffer}_1$
   (because \re{gfsdfgsdfswr2332} does not satisfied
   for right transition).

   As \re{gfeeeg5433} we take the
   set consisting of a single transition 
   from $a$ to $b$.

   Diagram \re{sdff3436636775} in this case
   has the form
{\def\arraystretch{1}
{\small
\be{sdfasdfg45434}\by
\begin{picture}(100,160)

\put(0,130){\makebox(0,0)[c]{$A$}}
\put(10,130){\line(1,0){80}}
\put(100,130){\makebox(0,0)[c]{$a$}}
\put(50,145){\makebox(0,0)[b]{$
(k=0)\wedge(q=\varepsilon)
$}}
\put(0,120){\vector(0,-1){80}}
\put(-5,80){\makebox(0,0)[r]{$
\by k<1\\In\,?\,f\\q:=q\cdot [f]\\k:=k+1\ey
$}}
\put(100,120){\vector(0,-1){80}}
\put(105,80){\makebox(0,0)[l]{$
In\,?\,x
$}}
\put(0,30){\makebox(0,0)[c]{$A$}}
\put(100,30){\makebox(0,0)[c]{$b$}}
\put(10,30){\line(1,0){80}}
\put(50,20){\makebox(0,0)[t]{$
(k=1)\wedge(q=[x])
$}}

\end{picture}
\ey
\ee } 
}

Using the fact that 
\be{twerte336}\forall\; \varphi,\psi,\theta \in 
Fm\qquad
(\varphi\leq 
\psi\to\theta\quad\Leftrightarrow\quad
\varphi\wedge \psi\leq\theta)\ee
write an inequality corresponding to this 
diagram in the form 
\be{gdfg456467888}\c{k=0
\\q=\varepsilon\\k<1}\leq
\c{k+1=1\\q\cdot[z]=[z]}\ee
Clearly, this inequality is true. 

\i For the pair $(A,b)$ we have to consider 
   only right transition in the process
   ${\it Buffer}_1$
   (because condition \re{gfsdfgsdfswr2332}
   does not    satisfied for   left transition).

   As set \re{gfeeeg5433} in this case we take a
   set consisting of a single 
   transition from $b$ to $a$.

   Diagram \re{sdff3436636775} 
in this case has the form
{\def\arraystretch{1}
{\small
\be{gsdfghsghgfhj555}\by
\begin{picture}(100,160)

\put(0,130){\makebox(0,0)[c]{$A$}}
\put(10,130){\line(1,0){80}}
\put(100,130){\makebox(0,0)[c]{$b$}}
\put(50,145){\makebox(0,0)[b]{$
(k=1)\wedge(q=[x])
$}}
\put(0,120){\vector(0,-1){80}}
\put(-5,80){\makebox(0,0)[r]{$
\by k>0\\Out\,!\,\hat q\\q:=q'\\k:=k-1\ey
$}}
\put(100,120){\vector(0,-1){80}}
\put(105,80){\makebox(0,0)[l]{$
Out\,!\,x
$}}
\put(0,30){\makebox(0,0)[c]{$A$}}
\put(100,30){\makebox(0,0)[c]{$a$}}
\put(10,30){\line(1,0){80}}
\put(50,20){\makebox(0,0)[t]{$
(k=0)\wedge(q=\varepsilon)
$}}

\end{picture}
\ey
\ee } 
}

Using 
\re{twerte336},
write the inequality corresponding to this 
diagram in the form 
\be{4r4twergfghtr}\c{k=1\\q=[x]\\k>0}\leq
\c{\hat q=x\\
k-1=0\\q'=\varepsilon}\ee
Obviously, this inequality is true. 
\ei

\i Check property 3. 

\bi\i  For the pair $(A,a)$ and for a single 
   transition from $a$ to $b$
   as 
   \re{hgyt5333788j}
   we take a set, 
   consisting of left transition 
   from $A$ to $A$.

   Diagram \re{sdff34dfsg36636775} 
in this case has the form 
\re{sdfasdfg45434}. 
As already established, this diagram is correct. 

\i For the pair $(A,b)$ and for a single transition 
from $b$ to $a$
   as \re{hgyt5333788j}
   we take a set, 
   consisting of right transition from 
   $A$ to $A$.

   Daigram \re{sdff34dfsg36636775} 
   in this case has the form 
   \re{gsdfghsghgfhj555}. As already    
justified, this diagram is correct. 
\ei

\en

\subsection{Additional remarks} 
\label{vjbkldgfrw}

To improve a usability of theorem
\ref{mueqyuivr}
you can use the following notions and 
statements. 

\subsubsection{Invariants of processes}

 Let $P$ be a process with COs.

   A formula $Inv$ with variables from $X_P$
   is said to be an {\bf invariant}
   of the process $P$, 
   if it has the following properties.
\bi
\i $I_P\leq Inv$
\i for each transition 
   $\diagrw{s&\pright{Op}&s'}$
   of the process $P$
   \bi
   \i if $Op$ is internal, then 
   $Inv\leq Op (Inv)$
   \i if $Op$ is an input CO of the form
   $Op_1\cdot (\alpha\,?\,x)\cdot Op_2$,
then
$$Inv\leq (Op_1\cdot (x:=z)\cdot Op_2)(Inv)$$
where $z$ is a variable which does not
belong to $X_P$
   \i if $Op$ is an output CO of the form
$Op_1\cdot (\alpha\,!\,e)\cdot Op_2$,
then
$$Inv\leq (Op_1\cdot Op_2)(Inv)$$
   \ei
\ei

Using the concept of an invariant, 
theorem \ref{mueqyuivr} can be modified
as follows.\\

\refstepcounter{theorem}
{\bf Theorem \arabic{theorem}
\label{fdgmueqyuivrdf}}.

Let
\bi\i $P_1$ and $P_2$ be two processes 
    with COs: 
$$P_i=(X_{P_i},I_{P_i},S_{P_i},s^0_{P_i},
R_{P_i})\qquad
(i=1,2)$$
which have no common states 
and common variables, and 
\i formulas $Inv_1$ and $Inv_2$ are 
invariants of the   processes 
$P_1$ and $P_2$ respectively. 
\ei

Then $P_1\approx P_2$, if 
there is a function $\mu$ of the form
$$\mu: S_{P_1}\times S_{P_2}\to Fm$$
with the following properties. 
\bn
\i $I_{P_1}\wedge I_{P_2}\leq 
\mu(s^0_{P_1},s^0_{P_2})$.
\i For
   \bi\i each pair 
   $(A_1,A_2)\in S_{P_1}\times S_{P_2}$,
   and \i  each transition
   $\diagrw{A_1&\pright{Op}&A'_1}$
   of the process $P_1$, such that 
   \be{gsdgfsdfgsdfswr2332}
   \c{cond\,(Op)\\ \mu(A_1,A_2)
   \\ Inv_1\\ Inv_2}
   \neq\bot\ee
   \ei
   there is a set of CTs of the process $P_2$ 
    with the start $A_2$
\be{gfeehfdeg5433}\{\diagrw{A_2&
\pright{CT_i}&A_2^i} \mid i\in \Im\}\ee
satisfying the following conditions: 
\bn
\i the following inequality holds:
\be{bigvhgeeeee}\c{cond\,(Op)\\ \mu(A_1,A_2)
\\ Inv_1\\Inv_2}
\;\leq\;\bigvee\limits_{i\in\Im}cond\,(CT_i)\ee
\i for each $i\in\Im$ the following diagram
is correct

   {\def\arraystretch{1}
{\small
\be{sdff3hreh436636775}\by
\begin{picture}(100,180)

\put(0,130){\makebox(0,0)[c]{$A_1$}}
\put(10,130){\line(1,0){80}}
\put(100,130){\makebox(0,0)[c]{$A_2$}}
\put(50,140){\makebox(0,0)[b]{$
\c{\mu(A_1,A_2) \\ Inv_1 \\Inv_2}
$}}
\put(0,120){\vector(0,-1){80}}
\put(-5,80){\makebox(0,0)[r]{$
Op
$}}
\put(100,120){\vector(0,-1){80}}
\put(105,80){\makebox(0,0)[l]{$
CT_i
$}}
\put(0,30){\makebox(0,0)[c]{$A_1'$}}
\put(100,30){\makebox(0,0)[c]{$A_2^i$}}
\put(10,30){\line(1,0){80}}
\put(50,20){\makebox(0,0)[t]{$
\mu(A'_1,A_2^i)
$}}
\end{picture}
\ey
\ee } 
}

\en

\i The property, which is symmetrical 
   to the previous one:    
   for
   \bi\i  each pair $(A_1,A_2)\in S_{P_1}    \times S_{P_2}$,
   and \i each transition
   $\diagrw{A_2&\pright{Op}&A'_2}$
   of the process 
   $P_2$, such that
   \re{gsdgfsdfgsdfswr2332} holds, 
   \ei
there is a set of CTs of the process $P_1$ 
with the start $A_1$
\be{hgythjg5333788j}\{\diagrw{A_1&\pright{CT_i}&A_1^i}
\mid i\in \Im\}\ee
satisfying the following conditions: 
\bn
\i the inequality \re{bigvhgeeeee} holds
\i for each $i\in\Im$ the following diagram 
is correct

   {\def\arraystretch{1}
{\small
\be{sdff34fjdfsg36636775}\by
\begin{picture}(100,180)

\put(0,130){\makebox(0,0)[c]{$A_1$}}
\put(10,130){\line(1,0){80}}
\put(100,130){\makebox(0,0)[c]{$A_2$}}
\put(50,140){\makebox(0,0)[b]{$
\c{\mu(A_1,A_2)\\ Inv_1\\ Inv_2}
$}}
\put(0,120){\vector(0,-1){80}}
\put(-5,80){\makebox(0,0)[r]{$
CT_i
$}}
\put(100,120){\vector(0,-1){80}}
\put(105,80){\makebox(0,0)[l]{$
Op
$}}
\put(0,30){\makebox(0,0)[c]{$A_1^i$}}
\put(100,30){\makebox(0,0)[c]{$A'_2$}}
\put(10,30){\line(1,0){80}}
\put(50,20){\makebox(0,0)[t]{$
\mu(A^i_1,A'_2)
$}}
\end{picture}
\ey
\ee } 
}
\en
\en

\subsubsection{Composition of diagrams}

\refstepcounter{theorem}
{\bf Theorem \arabic{theorem}
\label{fdgmueqyuivrdfdsfgf2df}}.

Let
\bi
\i $\varphi$, $\psi$, $\theta$ be formulas
\i $Op_1$, $Op_2$ be internal COs, such
that the following diagram is correct

{\def\arraystretch{1}
{\small
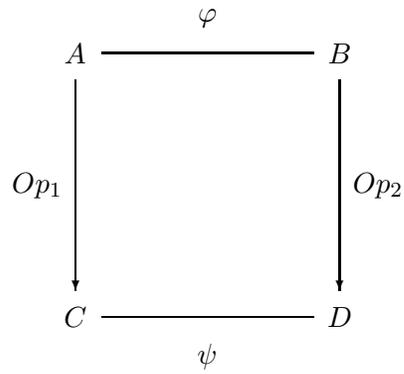
$$\by
\begin{picture}(100,160)

\put(0,130){\makebox(0,0)[c]{$A$}}
\put(10,130){\line(1,0){80}}
\put(100,130){\makebox(0,0)[c]{$B$}}
\put(50,140){\makebox(0,0)[b]{$
\varphi
$}}
\put(0,120){\vector(0,-1){80}}
\put(-5,80){\makebox(0,0)[r]{$
Op_1
$}}
\put(100,120){\vector(0,-1){80}}
\put(105,80){\makebox(0,0)[l]{$
Op_2
$}}
\put(0,30){\makebox(0,0)[c]{$C$}}
\put(100,30){\makebox(0,0)[c]{$D$}}
\put(10,30){\line(1,0){80}}
\put(50,20){\makebox(0,0)[t]{$
\psi
$}}

\end{picture}
\ey
$$ } 
}

\i $Op'_1$, $Op'_2$ be COs such that the 
following diagram is correct

{\def\arraystretch{1}
{\small
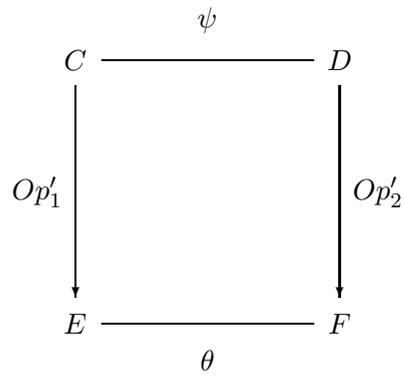
$$
\by
\begin{picture}(100,160)

\put(0,130){\makebox(0,0)[c]{$C$}}
\put(10,130){\line(1,0){80}}
\put(100,130){\makebox(0,0)[c]{$D$}}
\put(50,140){\makebox(0,0)[b]{$
\psi
$}}
\put(0,120){\vector(0,-1){80}}
\put(-5,80){\makebox(0,0)[r]{$
Op'_1
$}}
\put(100,120){\vector(0,-1){80}}
\put(105,80){\makebox(0,0)[l]{$
Op'_2
$}}
\put(0,30){\makebox(0,0)[c]{$E$}}
\put(100,30){\makebox(0,0)[c]{$F$}}
\put(10,30){\line(1,0){80}}
\put(50,20){\makebox(0,0)[t]{$
\theta
$}}

\end{picture}
\ey
$$ } 
}

\i $\{Op_1, Op'_1\}$ and $\{Op_2, Op'_2\}$ 
have no common variables. 
\ei

Then the following diagram is correct

{\def\arraystretch{1}
{\small
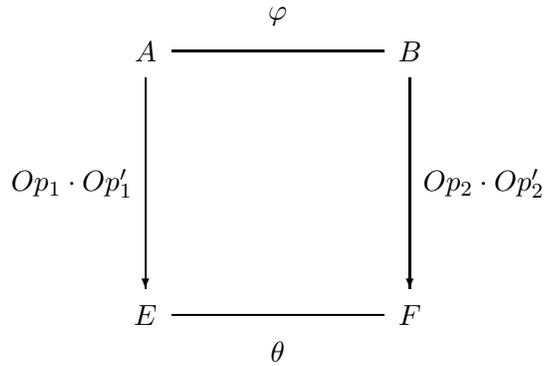
$$
\by
\begin{picture}(100,160)

\put(0,130){\makebox(0,0)[c]{$A$}}
\put(10,130){\line(1,0){80}}
\put(100,130){\makebox(0,0)[c]{$B$}}
\put(50,140){\makebox(0,0)[b]{$
\varphi
$}}
\put(0,120){\vector(0,-1){80}}
\put(-5,80){\makebox(0,0)[r]{$
Op_1\cdot Op'_1
$}}
\put(100,120){\vector(0,-1){80}}
\put(105,80){\makebox(0,0)[l]{$
Op_2\cdot Op'_2
$}}
\put(0,30){\makebox(0,0)[c]{$E$}}
\put(100,30){\makebox(0,0)[c]{$F$}}
\put(10,30){\line(1,0){80}}
\put(50,20){\makebox(0,0)[t]{$
\theta
$}}

\end{picture}
\ey
$$ } 
}

\subsection{Another example 
of a proof of observational equivalence 
of processes with COs}

As an example of a use of theorems from section 
\ref{vjbkldgfrw}
prove an observational equivalence of
\bi
\i the process 
   \be{sdfgwertyet34}(
{\it Buffer}_{n_1}[Pass/Out]\pa 
{\it Buffer}_{n_2}[Pass/In])\setminus\{Pass\}
\ee where
   $Pass\not\in\{In,\,Out\}$, and
\i the process ${\it Buffer}_{n_1+n_2}$.
\ei

Process \re{sdfgwertyet34}
is a sequential composition of two buffers, 
size of which is $n_1$ and $n_2$ respectively.

A flow graph of this process has the form

{\small
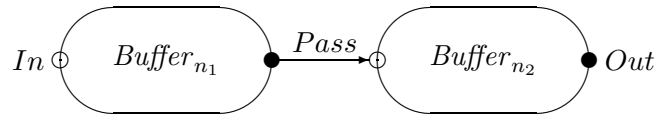
$$\by
\begin{picture}(0,50)

\put(60,20){\oval(80,40)}
\put(-60,20){\oval(80,40)}

\put(-100,20){\circle{6}}
\put(100,20){\circle*{6}}

\put(-20,20){\circle*{6}}
\put(20,20){\circle{6}}

\put(-17,20){\vector(1,0){34}}

\put(-60,20){\makebox(0,0)[c]{${\it Buffer}_{n_1}$}}
\put(60,20){\makebox(0,0)[c]{${\it Buffer}_{n_2}$}}
\put(-106,20){\makebox(0,0)[r]{$In$}}
\put(106,20){\makebox(0,0)[l]{$Out$}}

\put(0,23){\makebox(0,0)[b]{$Pass$}}

\end{picture}
\\
\vspace{0mm}
\ey
$$ } 

According to the definition of operations on 
processes with COs
(see section \ref{sdffwe3453476347893}),
a graph representation of the process
\re{sdfgwertyet34} has the form

{\def\arraystretch{1}
{\small
\be{gsdf3w456u}\by
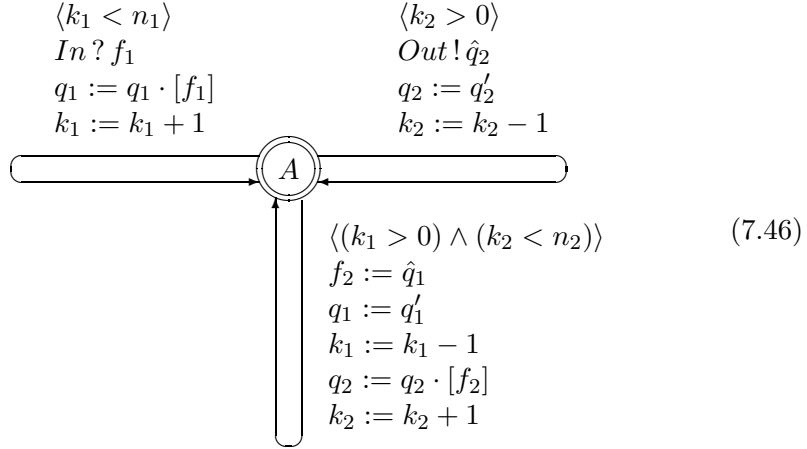
\begin{picture}(0,170)

\put(0,100){\oval(20,20)}
\put(0,100){\oval(24,24)}
\put(0,100){\makebox(0,0)[c]{$A$}}

\put(-100,100){\oval(10,10)[l]}
\put(-11,105){\line(-1,0){89}}
\put(-100,95){\vector(1,0){89}}

\put(-60,110){\makebox(0,0)[b]{
$\by
\langle k_1<n_1\rangle \\
In\,?\,f_1\\q_1:=q_1\cdot [f_1]\\k_1:=k_1+1
\ey
$}}
\put(100,100){\oval(10,10)[r]}
\put(11,105){\line(1,0){89}}
\put(100,95){\vector(-1,0){89}}

\put(70,110){\makebox(0,0)[b]{$
\by \langle k_2>0 \rangle \\
Out\,!\,\hat q_2\\q_2:=q_2'\\k_2:=k_2-1 \ey
$}}
\put(0,0){\oval(10,10)[b]}
\put(5,-1){\line(0,1){89}}
\put(-5,0){\vector(0,1){89}}

\put(10,40){\makebox(0,0)[l]{$\by
\langle (k_1>0)\wedge (k_2<n_2)
\rangle \\
f_2:=\hat q_1\\q_1:=q_1'\\k_1:=k_1-1
\\q_2:=q_2\cdot [f_2]\\k_2:=k_2+1
\ey $}}

\end{picture}
\\
\vspace{0mm}
\ey
\ee } 
}

An initial condition of the process
\re{gsdf3w456u}
is the formula
$$\c{
(n_1>0)\;\wedge\;(k_1=0)\;\wedge\;
(q_1=\varepsilon)\\
(n_2>0)\;\wedge\;(k_2=0)\;\wedge\;
(q_2=\varepsilon)
}$$

A graph representation of the process 
${\it Buffer}_{n_1+n_2}$
has the form

{\def\arraystretch{1}
{\small
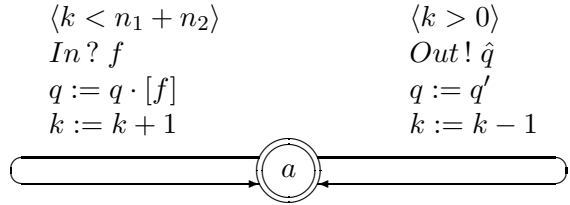
$$\by
\begin{picture}(0,70)

\put(0,0){\oval(20,20)}
\put(0,0){\oval(24,24)}
\put(0,0){\makebox(0,0)[c]{$a$}}

\put(-100,0){\oval(10,10)[l]}
\put(-11,5){\line(-1,0){89}}
\put(-100,-5){\vector(1,0){89}}

\put(-60,10){\makebox(0,0)[b]{
$\by\langle k<n_1+n_2\rangle \\
In\,?\;f\\q:=q\cdot [f]\\k:=k+1\ey
$}}
\put(100,0){\oval(10,10)[r]}
\put(11,5){\line(1,0){89}}
\put(100,-5){\vector(-1,0){89}}

\put(70,10){\makebox(0,0)[b]{$
\by \langle k>0\rangle \\
Out\,!\;\hat q\\q:=q'\\k:=k-1 \ey
$}}

\end{picture}
\\
\vspace{0mm}
\ey
$$ } 
}

An initial condition of the process
${\it Buffer}_{n_1+n_2}$
is the formula
$$
(n_1+n_2>0)\;\wedge\;(k=0)\;\wedge\;
(q=\varepsilon)
$$

It is easy to verify that the formula 
$$Inv\eam \;\;\c{0\leq k_1\leq n_1\\
|q_1|=k_1\\
0\leq k_2\leq n_2\\
|q_2|=k_2\\
n_1>0\\
n_2>0}$$
is an invariant of the process
\re{gsdf3w456u}.
This fact follows, in particular, from 
the statement
$$
\left\{
\by  |u|>0\quad\Rightarrow\quad|u'|=|u|-1\\
|u\cdot [a]| = |[a]\cdot u| = |u|+1
\ey
\right.
$$
which hold for each list $u$ 
and each message $a$.

As an invariant of the second process we take 
the formula $\top$. 

Define a function $\mu:\{A\}\times \{a\}\to 
Fm$
as follows:
$$\mu(A,a)\eam \c{q=q_2\cdot q_1
\\k=k_2+k_1
}$$

Check properties 1, 2, and 3 
for the function $\mu$. 

\bn
\i Property 1 in this case is the inequality 
   $$\c{
(n_1>0)\;\wedge\;(k_1=0)\;\wedge\;
(q_1=\varepsilon)\\
(n_2>0)\;\wedge\;(k_2=0)\;\wedge\;
(q_2=\varepsilon)\\
(n_1+n_2>0)\;\wedge\;(k=0)\;\wedge\;
(q=\varepsilon)
}
  \leq
\c{q=q_2\cdot q_1\\k=k_2+k_1
}   $$
   which is obviously true. 

\i Check property 2. 

   \bi
   \i For left transition of the process
   \re{gsdf3w456u}
   inequality \re{gsdgfsdfgsdfswr2332} holds.
As 
\re{gfeehfdeg5433}
we take the set, the only element of which 
is left transition of the process
${\it Buffer}_{n_1+n_2}$.

Inequality \re{bigvhgeeeee}  
in this case has the form 
$$\c{k_1<n_1\\q=q_2\cdot q_1\\k=k_2+k_1\\ 
Inv}\leq (k<n_1+n_2)$$
that is obviously true. 

Using 
\re{twerte336},
write an inequality corresponding to diagram 
\re{sdff3hreh436636775}
for this case as 
\be{hgythjg5333788j111}
\c{q=q_2\cdot q_1\\k=k_2+k_1\\Inv\\
k_1<n_1\\k<n_1+n_2
}\leq
\c{q\cdot[z]=q_2\cdot q_1\cdot[z]
\\k+1=k_2+k_1+1}
\ee
It is easy to check 
that the last inequality is true. 

\i For the middle (internal) transition 
   of the process \re{gsdf3w456u} 
   inequality \re{gsdgfsdfgsdfswr2332} holds.
As \re{gfeehfdeg5433}
we take the set, the only element of which 
is an empty CT of the process
${\it Buffer}_{n_1+n_2}$.

Inequality \re{bigvhgeeeee} in this case 
holds for the trivial reason: 
its right side is $\top$. 

Using statement \re{twerte336},
write an inequality corresponding 
to diagram \re{sdff3hreh436636775}
for this case, in the form 
\be{hgythjg5333788j112}
\c{q=q_2\cdot q_1\\k=k_2+k_1\\
Inv\\
k_1>0\\ k_2<n_2
}\leq
\c{q=(q_2\cdot [\hat q_1])\cdot q'_1
\\k=k_2+1+k_1-1
}\ee
This inequality follows from 
\bi
\i the associativity property of 
   of a concatenation, and 
\i the statement 
   $$|u|>0\quad
   \Rightarrow\quad u=[\hat u]\cdot 
   u'$$
   which holds for each list $u$.
\ei
   \i For right transition of the process
    \re{gsdf3w456u} 
     inequality \re{gsdgfsdfgsdfswr2332} 
     holds.
A \re{gfeehfdeg5433} we take the set, 
the only element of which is right 
transition of the process
${\it Buffer}_{n_1+n_2}$.

Inequality \re{bigvhgeeeee}  
in this case has the form 
$$\c{k_2>0 \\q=q_2\cdot q_1\\k=k_2+k_1
\\Inv
}\leq (k>0)$$
that is obviously true. 

Using the  statement \re{twerte336}, 
we write the inequality which 
corresponds to diagram
\re{sdff3hreh436636775}
for this case, in the form 
\be{hgythjg5333788j113}\c{q=q_2\cdot q_1
\\k=k_2+k_1\\
Inv\\
k_2>0\\k>0}
\leq \c{\hat q_2=\hat q\\q'=q'_2\cdot q_1\\
k-1=k_2-1+k_1}\ee

This inequality follows from the statement
$$|u|>0\quad\Rightarrow\quad \c{
(u\cdot v)\hat{\;}=\hat u\\
(u\cdot v)'=u'\cdot v}$$
which holds for each pair of lists $u,v$.
\ei

\i Check property 3. 

   \bi
   \i For left transition of the process
    ${\it Buffer}_{n_1+n_2}$ inequality 
    \re{gsdgfsdfgsdfswr2332} holds.
As \re{hgythjg5333788j}
we take the set, consisting of two CTs: 
\bi
\i left transition of the process
   \re{gsdf3w456u}, and
\i  the sequence, which consists of 
    a pair of transitions 
\bi\i  the first element of which 
     is the middle (internal) transition 
   of the process \re{gsdf3w456u},
\i and the second is the  left transition 
of the process 
\re{gsdf3w456u}\ei
\ei

Inequality \re{bigvhgeeeee} 
in this case has the form 
$$\c{k<n_1+n_2\\q=q_2\cdot q_1
\\k=k_2+k_1\\
Inv
}
\leq (k_1<n_1)\;\vee\; \c{k_1>0\\k_2<n_2
\\k_1-1<n_1}$$

This inequality is true, and in 
the proof of this inequality 
the conjunctive term $n_1> 0$
(contained in $Inv$) is used.

The inequalities which correspond to 
diagrams
\re{sdff34fjdfsg36636775}
for both elements of the set
\re{hgythjg5333788j},
follow from \re{hgythjg5333788j111},
\re{hgythjg5333788j112} and theorem 
\ref{fdgmueqyuivrdfdsfgf2df}.

\i For right transition of the process
${\it Buffer}_{n_1+n_2}$
inequality \re{gsdgfsdfgsdfswr2332} holds.
As \re{hgythjg5333788j}
we take the set, consisting of two CTs: 
\bi
\i right transition of the process
   \re{gsdf3w456u}, and
\i   the sequence which consists of 
    a pair of transitions,
\bi\i the first element of which is the middle   
(internal) transition of the process
   \re{gsdf3w456u}, and
\i the second is right transition of the 
process \re{gsdf3w456u}\ei
\ei

Inequality \re{bigvhgeeeee} 
in this case has the form 
$$\c{k>0\\q=q_2\cdot q_1\\k=k_2+k_1\\
Inv
}
\leq (k_2>0)\;\vee\; \c{k_1>0\\k_2<n_2
\\k_2+1>0}$$

This inequality is true, and in 
the proof of this inequality 
the conjunctive term $n_2> 0$
(contained in $Inv$) is used.

The inequalities corresponding to 
diagrams
\re{sdff34fjdfsg36636775}
for both elements of the set 
\re{hgythjg5333788j},
follow from \re{hgythjg5333788j112},
\re{hgythjg5333788j113} and theorem 
\ref{fdgmueqyuivrdfdsfgf2df}.

\ei

\en

\section{Recursive definition of processes with a message passing}

A concept of a {\bf recursive definition}
of processes with a message passing 
is similar to a concept of a RD
presented in chapter \ref{recur}. 

A concept of a RD is based on a
concept of a {\bf process expression (PE)} 
which is analogous to the 
corresponding concept in section 
\ref{ppprocvyr}, 
so we only point out differences 
in definitions of these concepts. 
\bi 
\i In all PEs operators are used
   (instead of actions). 
\i Each process name $A$ has a {\bf type}
   $t(A)$ of the form
   $$t(A)=(t_1,\ldots, t_n)\qquad(n\geq 0)$$
   where $\forall\, i=1,\ldots, n\quad 
    t_i\in Types$
\i Each process name $A$ occurs in each PE
   only together with a list of expressions 
   of corresponding types, i.e. 
   each occurrence of $A$ 
   in each PE $P$ is contained 
   in a subexpression of $P$ of the form 
   $$A(e_1,\ldots, e_n)$$
   where
   \bi\i $\forall\,i=1,\ldots, n\quad
   e_i\in {\cal E}$\i
   $(t(e_1),\ldots, t(e_n))=t(A)$\ei
\ei 

For each PE $P$ the notation $fv(P)$ 
denotes a set of {\bf free variables} 
of $P$, which consists of all variables from 
$X_P$ having free occurrences in $P$. 

A concepts of a free occurrence
and a bound occurrence 
of a variable in a PE is 
similar to an analogous 
concept in predicate logic.
Each free occurrence 
of a variable $x$ in a PE $P$ 
becomes bound in 
the PEs $(\alpha? x). P$ and $(x: = e). P$. 

A {\bf recursive definition (RD) of processes}
is a list of formal equations 
of the form 
\be{ssfdgdpu}\left\{\by
A_1(x_{11},\ldots, x_{1k_1}) = P_1 \\ \ldots \\
A_n(x_{n1},\ldots, x_{nk_n}) = P_n
\ey \right.\ee
where 
\bi
\i $A_1,\ldots, A_n$ are process names,
\i for each $i=1,\ldots, n$ the list
   $(x_{i1},\ldots, x_{ik_i})$ in left side 
   of $i$--th equality consists of different 
    variables 
\i $P_1,\ldots, P_n$ are PEs, which satisfy 
   \bi 
   \i the conditions set out in the definition 
      of a RD in section \ref{gjklefgmsdfgsdgrrr}, and 
   \i the following condition: 
   $$\forall \,i=1,\ldots, n\quad
   fv (P_i) = \{x_{i1},\ldots, x_{ik_i}\}$$
\ei 
\ei 

We shall assume that for each process
name $A$ there is a unique RD such that
$A$ has an occurrence in this RD.

RD \re{ssfdgdpu} 
can be interpreted as a functional program, 
consisting of functional definitions.
For each $ i = 1, \ldots, n $ the variables 
$x_{i1}$, $\ldots$, $x_{ik_i}$ 
can be regarded as 
formal parameters of the function
$A_i(x_{i1},\ldots, x_{ik_i})$.

A reader is requested to define
a correspondence, which associates
with each PE of the form
$A(x_1,\ldots,x_n)$, where 
\bi 
\i $A$ is a process name, and 
\i $x_1,\ldots, x_n$ is a list of 
   different variables of appropriate types 
\ei 
the process \be{fdsvsdetttttv}
[\![A(x_1,\ldots, x_n)]\!]\ee

Also a reader is invited to investigate
the following problems.
\bn
\i Construction of minimal processes 
   which are equivalent ($\approx, \oc, \ldots$) to
   processes of the form 
   \re{fdsvsdetttttv}.
\i Recognition of equivalence 
   of processes of the form \re{fdsvsdetttttv}.
\i Finding necessary and sufficient conditions 
   of uniqueness of the list 
   of processes defined by a RD. 
\en
\chapter{Examples of processes with a message passing}

\section{Separation of sets}

\subsection{The problem of separation of sets}

Let $U,V$ be a pair of finite disjoint sets, 
and 
each element $x \in U \cup V$ 
is associated with an integer $weight(x)$, 
called a {\bf weight} of this element. 

It is required to convert this pair 
to a pair of sets $U', V'$, so that
\bi
\i $|U|=|U'|,\quad
   |V|=|V'|$\\  (for each finite set $M$ 
the notation $|M|$ denotes a number 
of elements in $M$) 
\i for each $u \in U'$ and each $v \in V'$ 
   the following inequality holds:
   $$weight(u) \leq weight(v)$$ 
\ei 

Below we shall call the sets $U$ and $V$ as
{\bf left set} and {\bf right set}, respectively.

\subsection{Distributed algorithm 
of separation of sets}

The problem of separation of sets 
can be solved by an execution of 
several sessions of exchange elements 
between these sets. 
Each session consists of the following actions: 
\bi 
\i find an element $mx$ with 
   a maximum weight in left set 
\i find an element $mn$ with minimum weight 
   in right set 
\i transfer 
   \bi 
   \i $mx$ from left set to right set, and 
   \i $mn$ from right set to left set. 
   \ei 
\ei 

To implement this idea a distributed algorithm is proposed.
This algorithm is
defined as a process of the form 
\be{rrwarefgtee333}(Small\mid Large)
    \setminus  \{\alpha,\beta\}\ee
where 
\bi 
\i a process $Small$ executes 
   operations associated with 
   left set, and 
\i a process $Large$ executes
   operations associated with 
   right set. 
\ei 

A flow graph
corresponding to this process
has the form 
{\small
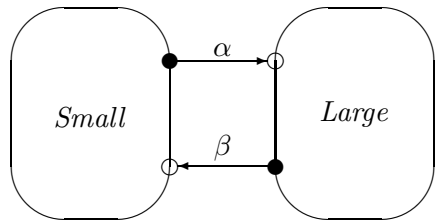
$$\by
\begin{picture}(0,100)

\put(50,50){\oval(60,80)}
\put(-50,50){\oval(60,80)}
\put(20,70){\circle{6}}
\put(20,30){\circle*{6}}
\put(-20,70){\circle*{6}}
\put(-20,30){\circle{6}}

\put(-17,70){\vector(1,0){34}}
\put(17,30){\vector(-1,0){34}}

\put(0,72){\makebox(0,0)[b]{$\alpha$}}
\put(0,32){\makebox(0,0)[b]{$\beta$}}

\put(-50,45){\makebox(0,0)[b]{${\it Small}$}}
\put(50,45){\makebox(0,0)[b]{${\it Large}$}}

\end{picture}
\ey
$$ } 

Below we shall use the following notations: 
\bi 
\i for each subset $W\subseteq U\cup V$
   the notations\begin{center}
   $\max(W)$ and $\min(W)$
   \end{center}
   denote an element of $W$ 
   with maximum and minimum weight, respectively, 
\i for \bi \i any subsets 
$W_1,W_2 \subseteq U\cup V$, and \i any 
   $u\in U\cup V$ \ei 
   the notations
$$W_1\leq u,\quad u\leq W_1,\quad W_1\leq W_2$$
   are shorthand expressions
   $$\by \forall x\in W_1\quad weight(x)\leq weight(u)\\
   \forall x\in W_1\quad weight(u)\leq weight(x)\\
   \forall x\in W_1,\; \forall y\in W_2\quad weight(x)
   \leq weight(y)
   \ey$$
   respectively.
\ei 
A similar meaning have the expressions 
$$\max(W),\quad \min(W),
\quad W\leq u,\quad u\leq W,\quad
W_1\leq W_2$$
in which the symbols 
$W$, $W_i$ and $u$ denote 
variables whose values are 
\bi \i subsets of the set $U \cup V$, 
and \i elements  of the set $U \cup V$ 
\ei respectively. 

\subsection{Processes {\it Small} 
and {\it Large}}

Processes {\it Small} and {\it Large} can be 
\bi
\i defined in terms of flowcharts, 
\i which then are transformed to processes 
   with COs, and reduced. 
\ei
We will not describe these flowcharts and their 
transformations and reductions, 
we present only reduced COs. 

A reduced process {\it Small} 
has the following form.

$Init= (S=U)$.

{\def\arraystretch{1}

{\small
\be{script343}\by
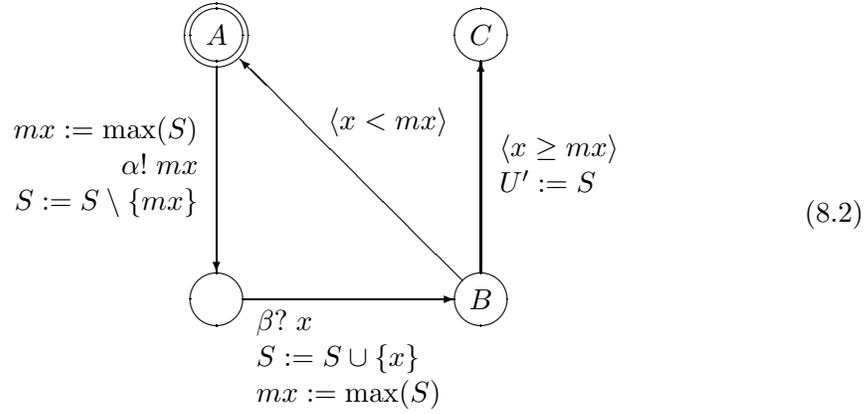
\begin{picture}(100,105)

\put(0,100){\oval(20,20)}
\put(0,100){\oval(24,24)}
\put(0,0){\oval(20,20)}
\put(100,100){\oval(20,20)}
\put(100,0){\oval(20,20)}

\put(0,100){\makebox(0,0){$A$}}
\put(100,0){\makebox(0,0){$B$}}
\put(100,100){\makebox(0,0){$C$}}

\put(0,88){\vector(0,-1){78}}
\put(10,0){\vector(1,0){80}}
\put(100,10){\vector(0,1){80}}
\put(93,7){\vector(-1,1){84}}

\put(-2,50){\makebox(0,0)[r]{$\begin{array}
{r}
mx:=\max(S)\\
\alpha!\;mx\\
S:=S \setminus \{mx\}
\ey$}}

\put(50,-2){\makebox(0,0)[t]{$\begin{array}{l}
\beta?\;x\\
S:=S\cup \{x\}\\
mx:=\max(S)
\ey$}}

\put(102,50){\makebox(0,0)[l]{$
\by \l{x \geq mx}\\ U':=S\ey
$}}

\put(65,67){\makebox(0,0){$\l{x < mx}$}}

\end{picture}
\\
\vspace{8mm}
\ey
\ee } 
}

The reduced process {\it Large}
has the following form.

$Init = (L=V)$.

{\def\arraystretch{1}

{\small

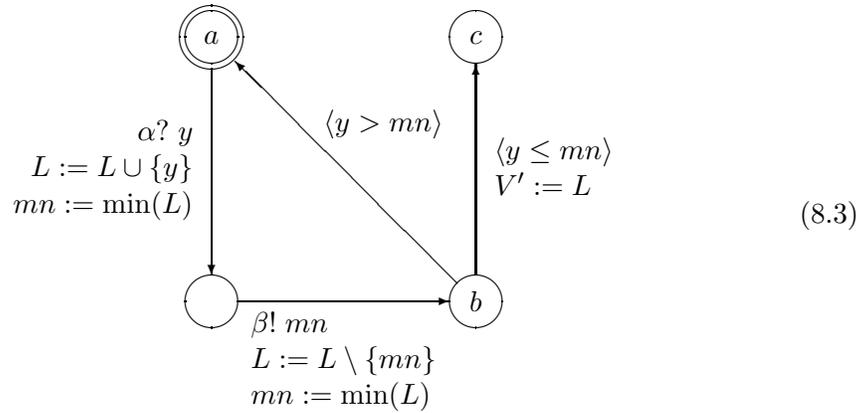
\be{large343}\by

\begin{picture}(100,105)

\put(0,100){\oval(20,20)}
\put(0,100){\oval(24,24)}
\put(0,0){\oval(20,20)}
\put(100,100){\oval(20,20)}
\put(100,0){\oval(20,20)}

\put(0,100){\makebox(0,0){$a$}}
\put(100,0){\makebox(0,0){$b$}}
\put(100,100){\makebox(0,0){$c$}}

\put(0,88){\vector(0,-1){78}}
\put(10,0){\vector(1,0){80}}
\put(100,10){\vector(0,1){80}}
\put(93,7){\vector(-1,1){84}}

\put(-2,50){\makebox(0,0)[r]{$\begin{array}{r}
\alpha?\;y\\
L:=L\cup \{y\}\\
mn:=\min(L)
\ey$}}

\put(50,-2){\makebox(0,0)[t]{$\begin{array}{l}
\beta!\;mn\\
L:=L\setminus \{mn\}\\
mn:=\min(L)
\ey$}}

\put(102,50){\makebox(0,0)[l]{$
\by \l{y\leq mn}\\V':=L\ey
$}}

\put(65,67){\makebox(0,0){$\l{y> mn}$}}

\end{picture}
\\
\vspace{8mm}
\ey
\ee
 } 
}

\subsection{An analysis of the algorithm 
of separation of sets}

A process described by expression 
\re{rrwarefgtee333}, is obtained by 
\bi 
\i a performing of operations of parallel 
   composition  and restrictions
   on processes  \re{script343} 
   and \re{large343}, 
   in accordance with definition 
   \re{rrwarefgtee333}, and 
\i a reduction of a resulting process. 
\ei 
The reduced process has the following form:

{\def\arraystretch{1}
{\small
\be{gff54444777}\by
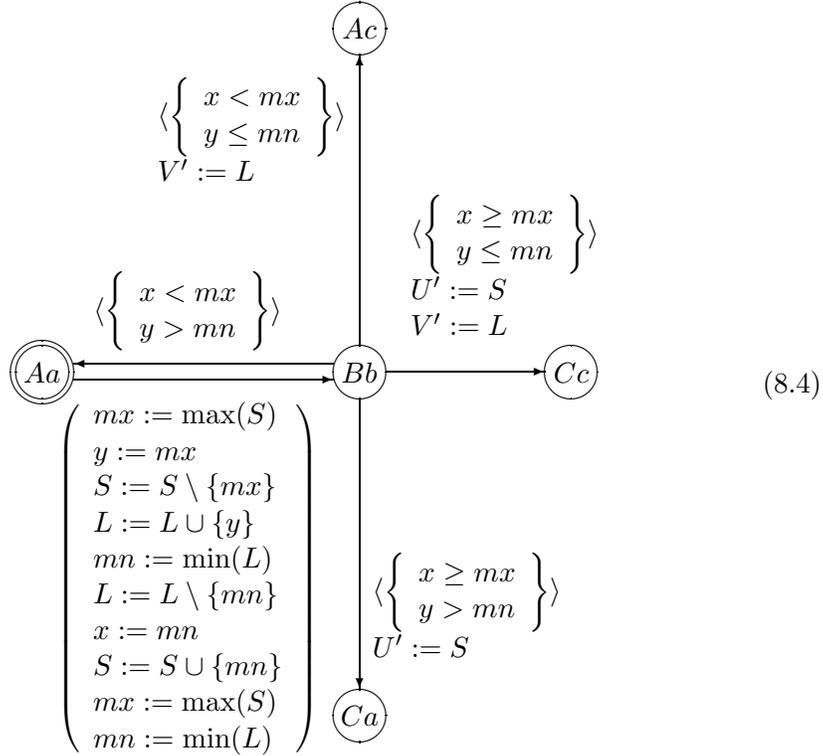
\begin{picture}(200,250)

\put(0,100){\oval(20,20)}
\put(0,100){\oval(24,24)}
\put(120,230){\oval(20,20)}
\put(120,100){\oval(20,20)}
\put(120,-30){\oval(20,20)}
\put(200,100){\oval(20,20)}

\put(0,100){\makebox(0,0){$Aa$}}
\put(120,-30){\makebox(0,0){$Ca$}}
\put(120,100){\makebox(0,0){$Bb$}}
\put(120,230){\makebox(0,0){$Ac$}}
\put(200,100){\makebox(0,0){$Cc$}}

\put(12,97){\vector(1,0){98}}
\put(110,103){\vector(-1,0){98}}

\put(120,110){\vector(0,1){110}}
\put(120,90){\vector(0,-1){110}}
\put(130,100){\vector(1,0){60}}

\put(55,90){\makebox(0,0)[t]{$\b{
mx:=\max(S)\\
y:=mx\\
S:=S\setminus \{mx\}\\
L:=L\cup\{y\}\\
mn:=\min(L)\\
L:=L\setminus \{mn\}\\
x:=mn\\
S:=S\cup\{mn\}\\
mx:=\max(S)\\
mn:=\min(L)
}$}}

\put(55,108){\makebox(0,0)[b]{$\l{\c{
x< mx\\y> mn
}}$}}

\put(175,110){\makebox(0,0)[b]{$\by \l{\c{
x\geq mx\\y\leq mn
}}\\ U':=S \\ V':=L \ey$}}

\put(120,10){\makebox(0,0)[l]{$\by \l{\c{
x\geq mx\\y> mn
}}\\U':=S\ey$}}

\put(120,190){\makebox(0,0)[r]{$\by\l{\c{
x<mx\\y\leq mn
}}\\V':=L\ey$}}

\end{picture}
\\
\vspace{15mm}
\ey
\ee
 } 
}

This diagram shows that 
there are states of
process \re{gff54444777}
(namely, $Ac$ and $Ca$)
with the following properties:
\bi 
\i there is no transitions starting 
   at these states \\
   (such states are said to be {\bf terminal})
\i but falling into these states 
   is not a normal 
   completion of the process. 
\ei 
The situation when a process falls  
in one of such states 
is called a {\bf deadlock}.

Process \re{rrwarefgtee333} can indeed 
fall in one of such states, for example, 
in the case when
$$U=\{3\}\quad\mbox{and} 
\quad V=\{1,2\}$$
where a weight of each integer is equal to
its value. 

Nevertheless, process 
\re{rrwarefgtee333}
has the following properties: 
\bi 
\i this process always terminates 
   (i.e., falls into one of 
   terminal states - 
   $Ac$, $Cc$ or $Ca$) 
\i after a termination of the process, 
   the following statements hold:
   \be{asdfasfadsfxzcvzcx}\left.
   \by S\cup L = U\cup V\\
   |S| = |U|,\;\;|L| = |V|\\
   S\leq L\ey\right\}
   \ee
\ei 

To justify these properties, we shall 
use the function
$$f(S,L)\;\eam\; \mid\{(s,l)\in S\times L\mid
weight(s)>weight(l)\}\mid$$

Furthermore, for an analyzing 
of a sequence of assignment operators 
performed during the transition from $Aa$ to 
$Bb$, it is convenient 
to represent this sequence 
schematically as 
a sequence of the following actions: 
\bn\i
$\by
S&\pmiddleright{y:=\max(S)}& L\ey$\\
(transfer of an element 
$y:=\max(S)$ from $S$ to$L$)
\i $\by L&\pmiddleright{x:=\min(L)}& S\ey$
\i $mx:=\max(S)$
\i $mn:=\min(L)$
\en

It is not so difficult to prove the
following statements. 
\bn 
\i If at current time $i$ 
\bi 
\i the process 
   is located at the state $Aa$, and 
\i values $S_i, L_i$ of the
   variables $S$ and $L$ 
   at this time
   satisfy the equation 
   $$f(S_i,L_i)=0$$
   i.e. the inequality
   $S_i\leq L_i$ holds
\ei 
then $S_{i+1}=S_i$ and $L_{i+1}=L_i$.

Furthermore, after an execution of 
the transition from $Aa$ to $Bb$ 
values of the variables $x$, $y$, $mx$ 
and $mn$ will satisfy the following statement:
$$y=x=mx\leq mn$$
and, thus, a next transition 
will be the transition from 
$Bb$ to state $Cc$, 
i.e. the process normally completes its work. 

Herewith 
\bi 
\i values of the variables $U'$ and $V'$ 
   will be equal to 
   $S_i$ and $L_i$, respectively, 
\i and, consequently, values of the
variables $U'$ and $V'$ 
will meet the required conditions
$$|U|=|U'|,\quad |V|=|V'|,\quad 
U'\leq V'
$$ \ei 

\i If at current time $i$ 
\bi 
\i the process 
   is located at the state $Aa$, and 
\i values $S_i, L_i$  of the 
   variables $S$ and $L$ 
   satisfy the inequality 
   $$f(S_i, L_i)> 0$$ 
\ei 
then after an execution of the transition 
from $Aa$ to $Bb$ (i.e., at the time $i+1$) 
new values $S_{i+1},L_{i+1}$ 
of the variables $S$ and $L$ 
will satisfy the inequality 
\be{adfasdfwerwewt5}f(S_{i+1},L_{i+1})<
f(S_i,L_i)\ee

In addition, the variables 
$x, y, mx, mn$ at the time $i +1$ 
will satisfy 
$$\by y=\max(S_i),\quad x=\min(L_i)\\
   mx=\max(S_{i+1}),\quad mn=\min(L_{i+1})\\
x<y,\quad x\leq mx,\quad mn\leq y\ey$$

It follows that if at the time $i +1$ 
the process will move from $Bb$ 
to one of the terminal states 
($Ac$, $Cc$ or $Ca$), then 
it is possible 
\bi
\i[(a)] either if $x=mx$
\i[(b)] or if $y=mn$
\ei

In the case (a) the following statement holds:
$$S_{i+1}\leq mx=x\leq L_i$$
whence, using 
$$x<y\quad\mbox{and}\quad L_{i+1}\subseteq L_i\cup \{y\}$$
we obtain: 
\be{sdfdasfweee}S_{i+1}\leq L_{i+1}\ee

In the case (b) the following statement holds:
$$S_i\leq y=mn\leq L_{i+1}$$
whence, using 
$$x<y\quad\mbox{and}\quad S_{i+1}
\subseteq S_i\cup \{x\}$$
we obtain \re{sdfdasfweee}. 

Thus, if the process is in a terminal state,
then $S\leq L$.

Other statements listed in 
\re{asdfasfadsfxzcvzcx} 
are proved directly. 
\en 

First and second statements imply that
this process can not be endless, because 
an infinite loop is possible only in the case 
when 
\bi 
\i the process infinitely many times 
   falls into the state $Aa$, and 
\i every time when the process is located at the state
   $Aa$, a value of the function $f$ 
   on current values 
   of the variables $S, T$ is positive. 
\ei 

An impossibility of this situation 
follows from \bi 
\i inequality \re{adfasdfwerwewt5}, and 
\i the founding property of  the set 
   of integers \\
   (there is no an infinite descending chain
   of integers).
\ei 

A reader is requested 
\bi 
\i to find necessary and sufficient conditions 
   to be met by the shared sets $U$ and $V$,
   that there is no a deadlock situation 
   in an execution of process 
   \re{gff54444777}
   (i.e. the process terminates 
   in the state $Cc$)
   with these $U$ and $V$, and 
\i develop an algorithm for separation of sets 
   that would work without a deadlock 
   on any shared sets $U$ and $V$. 
\ei 

\section{Calculation of a square} 
\label{umnozhenianadva}

Suppose we have a system ``multiplier'', 
which has 
\bi 
\i two input ports with names 
   $In_1$ and $In_2$, and 
\i one output port with name $Out$. 
\ei 

An execution of the multiplier is that it 
\bi 
\i receives on its input ports two values, and 
\i gives their product on the output port. 
\ei 
A behavior of the multiplier 
is described by the process $Mul$: 

{\def\arraystretch{1}
{\small
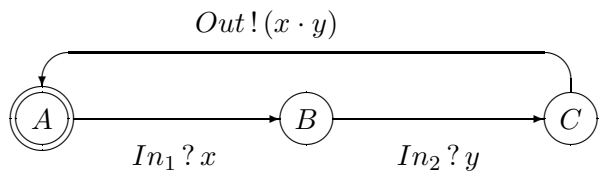
$$\by
\begin{picture}(210,40)

\put(0,0){\oval(20,20)}
\put(0,0){\oval(24,24)}
\put(0,0){\makebox(0,0)[c]{$A$}}

\put(100,0){\oval(20,20)}
\put(100,0){\makebox(0,0)[c]{$B$}}

\put(200,0){\oval(20,20)}
\put(200,0){\makebox(0,0)[c]{$C$}}

\put(12,0){\vector(1,0){78}}
\put(110,0){\vector(1,0){80}}

\put(50,-10){\makebox(0,0)[t]{$In_1\,?\,x$}}
\put(150,-10){\makebox(0,0)[t]{$In_2\,?\,y$}}

\put(0,15){\vector(0,-1){3}}
\put(200,15){\line(0,-1){5}}
\put(10,15){\oval(20,20)[tl]}
\put(190,15){\oval(20,20)[tr]}
\put(190,25){\line(-1,0){180}}
\put(85,30){\makebox(0,0)[b]{
$Out\,!\,(x\cdot y)$
}}

\end{picture}
\\
\vspace{0mm}
\ey
$$
 } 
}

Using this multiplier, we want to 
build a system ``a calculator of a square'', 
whose behavior is described by 
the process $Square\_Spec$: 

{\small
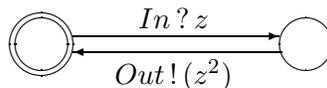
$$\by
\begin{picture}(100,20)

\put(0,0){\oval(20,20)}
\put(0,0){\oval(24,24)}
\put(100,0){\oval(20,20)}

\put(12,3){\vector(1,0){78}}
\put(91,-3){\vector(-1,0){79}}

\put(50,6){\makebox(0,0)[b]{$In\,?\,z$}}
\put(50,-6){\makebox(0,0)[t]{$Out\,!\,(z^2)$}}

\end{picture}
\\
\vspace{0mm}
\ey
$$ } 

We shall build a desired system as a composition of
\bn
\i an auxiliary system ``duplicator'', 
   which has
   \bi 
   \i an input port $In$, and 
   \i output ports $Out_1$ and $Out_2$ 
   \ei 
   and behavior of which 
   is described by the process $Dup$: 

{\def\arraystretch{1}
{\small
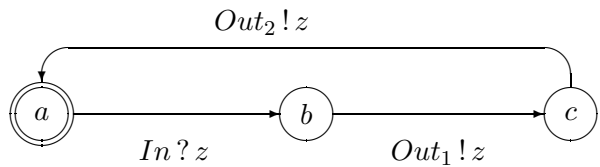
$$\by
\begin{picture}(210,40)

\put(0,0){\oval(20,20)}
\put(0,0){\oval(24,24)}
\put(0,0){\makebox(0,0)[c]{$a$}}

\put(100,0){\oval(20,20)}
\put(100,0){\makebox(0,0)[c]{$b$}}

\put(200,0){\oval(20,20)}
\put(200,0){\makebox(0,0)[c]{$c$}}

\put(12,0){\vector(1,0){78}}
\put(110,0){\vector(1,0){80}}

\put(50,-10){\makebox(0,0)[t]{$In\,?\,z$}}
\put(150,-10){\makebox(0,0)[t]{$Out_1\,!\,z$}}

\put(0,15){\vector(0,-1){3}}
\put(200,15){\line(0,-1){5}}
\put(10,15){\oval(20,20)[tl]}
\put(190,15){\oval(20,20)[tr]}
\put(190,25){\line(-1,0){180}}
\put(85,30){\makebox(0,0)[b]{$Out_2\,!\,z$
}}

\end{picture}
\\
\vspace{0mm}
\ey
$$
 } 
}
   i.e. the duplicator 
   copies its input to 
   two outputs, and 
\i the multiplier, which receives 
   on its input ports 
   those values that duplicator gives. 
\en

A process $Square$, 
corresponding to such a composition 
is determined as follows: 
$$\by Square\eam \\ \eam \b{
Dup[pass_1/Out_1, pass_2/Out_2]\;\mid\\\mid
Mul[pass_1/In_1, pass_2/In_2]}
\setminus\{pass_1,pass_2\}\ey$$

A flow graph of the process $Square$ 
has the form 

{\small
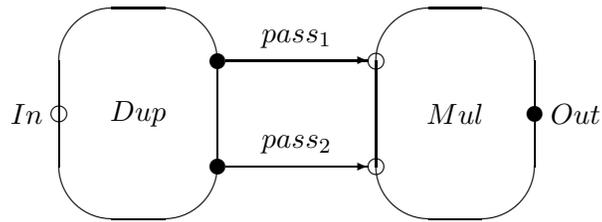
$$\by
\begin{picture}(0,100)

\put(60,50){\oval(60,80)}
\put(-60,50){\oval(60,80)}

\put(-90,50){\circle{6}}
\put(90,50){\circle*{6}}

\put(-30,70){\circle*{6}}
\put(-30,30){\circle*{6}}
\put(30,70){\circle{6}}
\put(30,30){\circle{6}}

\put(-27,70){\vector(1,0){54}}
\put(-27,30){\vector(1,0){54}}

\put(-60,50){\makebox(0,0)[c]{$Dup$}}
\put(60,50){\makebox(0,0)[c]{$Mul$}}
\put(-96,50){\makebox(0,0)[r]{$In$}}
\put(96,50){\makebox(0,0)[l]{$Out$}}

\put(0,75){\makebox(0,0)[b]{$pass_1$}}
\put(0,35){\makebox(0,0)[b]{$pass_2$}}

\end{picture}
\ey
$$ } 

However, the process $Square$ 
does not meet the specification 
$Square\_Spec$.
This fact is easy to detect
by a construction of a graph 
representation of $Square$, 
which has the following form:

{\def\arraystretch{1}
{\small
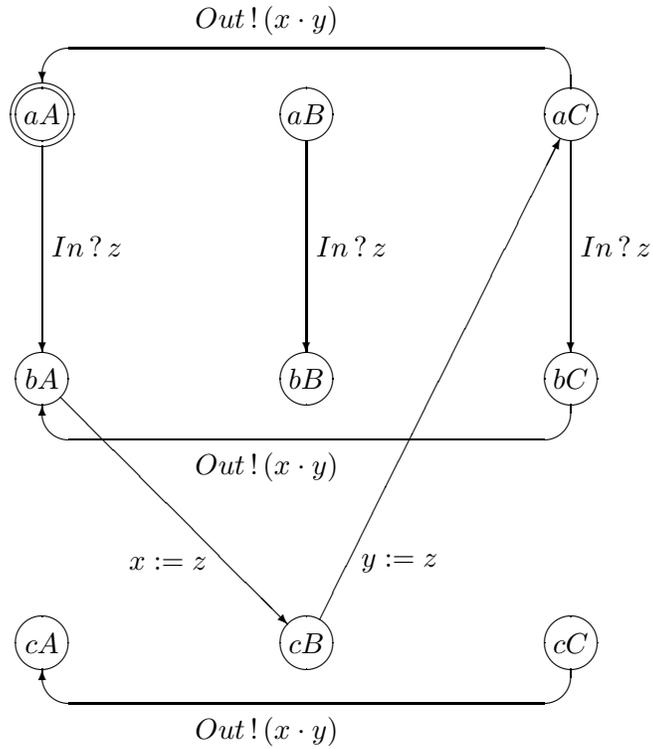
$$\by
\begin{picture}(210,240)

\put(0,200){\oval(20,20)}
\put(0,200){\oval(24,24)}
\put(0,200){\makebox(0,0)[c]{$aA$}}

\put(0,100){\oval(20,20)}
\put(0,100){\makebox(0,0)[c]{$bA$}}

\put(0,0){\oval(20,20)}
\put(0,0){\makebox(0,0)[c]{$cA$}}

\put(100,200){\oval(20,20)}
\put(100,200){\makebox(0,0)[c]{$aB$}}

\put(100,100){\oval(20,20)}
\put(100,100){\makebox(0,0)[c]{$bB$}}

\put(100,0){\oval(20,20)}
\put(100,0){\makebox(0,0)[c]{$cB$}}

\put(200,200){\oval(20,20)}
\put(200,200){\makebox(0,0)[c]{$aC$}}

\put(200,100){\oval(20,20)}
\put(200,100){\makebox(0,0)[c]{$bC$}}

\put(200,0){\oval(20,20)}
\put(200,0){\makebox(0,0)[c]{$cC$}}

\put(0,-13){\vector(0,1){3}}
\put(200,-13){\line(0,1){3}}
\put(10,-13){\oval(20,20)[bl]}
\put(190,-13){\oval(20,20)[br]}
\put(190,-23){\line(-1,0){180}}
\put(85,-28){\makebox(0,0)[t]{
$Out\,!\,(x\cdot y)$
}}

\put(0,87){\vector(0,1){3}}
\put(200,87){\line(0,1){3}}
\put(10,87){\oval(20,20)[bl]}
\put(190,87){\oval(20,20)[br]}
\put(190,77){\line(-1,0){180}}
\put(85,72){\makebox(0,0)[t]{
$Out\,!\,(x\cdot y)$
}}

\put(0,215){\vector(0,-1){3}}
\put(200,215){\line(0,-1){5}}
\put(10,215){\oval(20,20)[tl]}
\put(190,215){\oval(20,20)[tr]}
\put(190,225){\line(-1,0){180}}
\put(85,230){\makebox(0,0)[b]{
$Out\,!\,(x\cdot y)$
}}

\put(0,150){\makebox(0,0)[l]{
$In\,?\,z$
}}

\put(100,150){\makebox(0,0)[l]{
$In\,?\,z$
}}

\put(200,150){\makebox(0,0)[l]{
$In\,?\,z$
}}

\put(0,188){\vector(0,-1){78}}
\put(100,190){\vector(0,-1){80}}
\put(200,190){\vector(0,-1){80}}

\put(7,93){\vector(1,-1){86}}

\put(105,9){\vector(1,2){91}}

\put(66,30){\makebox(0,0)[r]{
$x:=z$
}}

\put(117,30){\makebox(0,0)[l]{
$y:=z$
}}

\end{picture}
\\
\vspace{15mm}
\ey
$$
 } 
}

After a reduction of this process we obtain 
the diagram

{\small
\be{sdfkjsadlfjasd1}\by
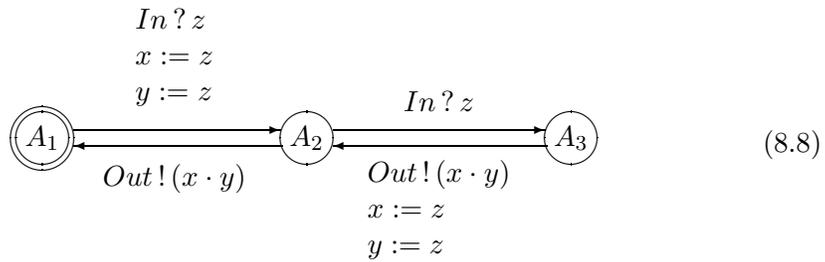
\begin{picture}(200,55)

\put(0,0){\oval(20,20)}
\put(0,0){\oval(24,24)}
\put(100,0){\oval(20,20)}
\put(200,0){\oval(20,20)}

\put(0,0){\makebox(0,0)[c]{$A_1$}}
\put(100,0){\makebox(0,0)[c]{$A_2$}}
\put(200,0){\makebox(0,0)[c]{$A_3$}}

\put(12,3){\vector(1,0){78}}
\put(91,-3){\vector(-1,0){79}}

\put(50,10){\makebox(0,0)[b]{$\by In\,?\,z\\ 
x:=z\\y:=z\ey$}}
\put(50,-10){\makebox(0,0)[t]{$Out\,!\,(x\cdot 
y)$}}

\put(110,3){\vector(1,0){80}}
\put(191,-3){\vector(-1,0){81}}

\put(150,10){\makebox(0,0)[b]{$In\,?\,z$}}
\put(150,-7){\makebox(0,0)[t]{$\by Out\,!\,
(x\cdot y)\\x:=z\\y:=z\ey$}}

\end{picture}
\\
\vspace{15mm}
\ey
\ee } 

which shows that 
\bi 
\i the process $Square$ can execute 
   two input actions together (i.e.
   without an execution of an output action
   between them), and\i the process 
   $Square\_Spec$ can not do so.
\ei

The process $Square$ meets another 
specification: 
$$Square\_Spec' \eam
\b{{\it Buf} [pass/Out]\;\mid\\\mid 
Square\_Spec[pass/In]}\setminus \{pass\}$$
where ${\it Buf}$ is a 
buffer which can store one message,
whose behavior is represented 
by the diagram 
{\small
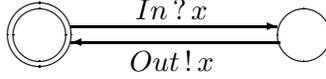
$$\by
\begin{picture}(100,30)

\put(0,0){\oval(20,20)}
\put(0,0){\oval(24,24)}
\put(100,0){\oval(20,20)}

\put(12,3){\vector(1,0){78}}
\put(91,-3){\vector(-1,0){79}}

\put(50,6){\makebox(0,0)[b]{$In\,?\,x$}}
\put(50,-6){\makebox(0,0)[t]{$Out\,!\,x$}}

\end{picture}
\\
\vspace{10mm}
\ey
$$ } 

A flow graph of $Square\_Spec'$ 
has the form
{\small
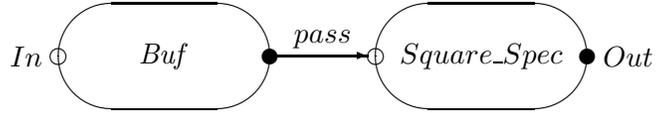
$$\by
\begin{picture}(0,50)

\put(60,20){\oval(80,40)}
\put(-60,20){\oval(80,40)}

\put(-100,20){\circle{6}}
\put(100,20){\circle*{6}}

\put(-20,20){\circle*{6}}
\put(20,20){\circle{6}}

\put(-17,20){\vector(1,0){34}}

\put(-60,20){\makebox(0,0)[c]{${\it Buf}$}}
\put(60,20){\makebox(0,0)[c]
{$Square\_Spec$}}
\put(-106,20){\makebox(0,0)[r]{$In$}}
\put(106,20){\makebox(0,0)[l]{$Out$}}

\put(0,23){\makebox(0,0)[b]{$pass$}}

\end{picture}
\ey
$$ } 

A reduced process $Square\_Spec'$
has the form

{\small
\be{sdfkjsadlfjasd2}\by
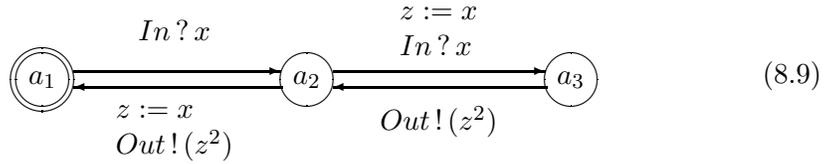
\begin{picture}(200,35)

\put(0,0){\oval(20,20)}
\put(0,0){\oval(24,24)}
\put(100,0){\oval(20,20)}
\put(200,0){\oval(20,20)}

\put(0,0){\makebox(0,0)[c]{$a_1$}}
\put(100,0){\makebox(0,0)[c]{$a_2$}}
\put(200,0){\makebox(0,0)[c]{$a_3$}}

\put(12,3){\vector(1,0){78}}
\put(91,-3){\vector(-1,0){79}}

\put(50,15){\makebox(0,0)[b]{$ In\,?\,x $}}
\put(50,-5){\makebox(0,0)[t]{$\by 
z:=x\\Out\,!\,(z^2)\ey$}}

\put(110,3){\vector(1,0){80}}
\put(191,-3){\vector(-1,0){81}}

\put(150,5){\makebox(0,0)[b]{$\by 
z:=x\\In\,?\,x\ey$}}
\put(150,-10){\makebox(0,0)[t]{$Out\,!\,(z^2)$}}

\end{picture}
\\
\vspace{5mm}
\ey
\ee } 

The statement that $Square$ meets the 
specification $Square\_Spec'$ 
can be formalized as 
\be{zfdgdsfgsdfg}
\re{sdfkjsadlfjasd1}\;\approx 
\re{sdfkjsadlfjasd2}\ee

We justify \re{zfdgdsfgsdfg} 
with use of theorem \ref{mueqyuivr}. 
At first, we rename variables 
of the process \re{sdfkjsadlfjasd2}, 
i.e. instead of \re{sdfkjsadlfjasd2}
we shall consider the process
{\small
\be{sdfkjsadlfjasd3}\by
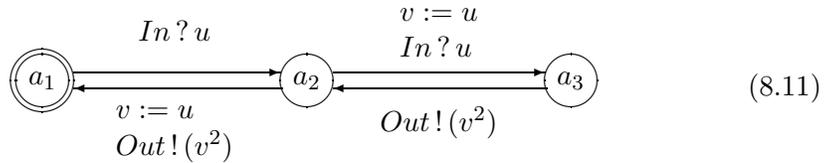
\begin{picture}(200,40)

\put(0,0){\oval(20,20)}
\put(0,0){\oval(24,24)}
\put(100,0){\oval(20,20)}
\put(200,0){\oval(20,20)}

\put(0,0){\makebox(0,0)[c]{$a_1$}}
\put(100,0){\makebox(0,0)[c]{$a_2$}}
\put(200,0){\makebox(0,0)[c]{$a_3$}}

\put(12,3){\vector(1,0){78}}
\put(91,-3){\vector(-1,0){79}}

\put(50,15){\makebox(0,0)[b]{$ In\,?\,u $}}
\put(50,-5){\makebox(0,0)[t]{$\by 
v:=u\\Out\,!\,(v^2)\ey$}}

\put(110,3){\vector(1,0){80}}
\put(191,-3){\vector(-1,0){81}}

\put(150,5){\makebox(0,0)[b]{$\by 
v:=u\\In\,?\,u\ey$}}
\put(150,-10){\makebox(0,0)[t]{$Out\,!\,
(v^2)$}}

\end{picture}
\\
\vspace{10mm}
\ey
\ee } 

To prove 
$\re{sdfkjsadlfjasd1}\;\approx \re{sdfkjsadlfjasd3}$
with use of  theorem \ref{mueqyuivr}
we define the function 
$$\mu:\{A_1,A_2,A_3\}
\times\{a_1,a_2,a_3\} \to Fm$$
as 
follows: 
\bi
\i $\mu(A_i,a_j)\eam \bot$, if $i\neq j$
\i $\mu(A_1,a_1)\eam \top$
\i $\mu(A_2,a_2)\eam (x=y=z=u)$
\i $\mu(A_3,a_3)\eam \c{x=y=v\\z=u}$
\ei

Detailed verification of correctness of 
corresponding diagrams 
left to a reader as a simple exercise. 

\section{Petri nets}

One of mathematical models to describe 
a behavior of distributed systems is a 
{\bf Petri net}. 

A {\bf Petri net} is a directed graph, 
whose set of nodes is divisible 
in two classes: 
places ($V$) and transitions ($T$). 
Each edge connects a place with
a transition. 

Each transition $t \in T$ 
is associated with two sets of places: 
\bi
\i $in(t)\eam \{v\in V\mid \mbox{
   there is an edge  from $v$ to $t$}\}$
\i $out(t)\eam \{v\in V\mid \mbox{there 
   is an edge  from  $t$ to $v$}\}$
\ei

A {\bf marking} of a Petri net 
is a mapping $\sigma$ of the form 
$$\sigma: V\to \{0,1,2,\ldots\}$$

An {\bf execution} of a Petri net 
is  a transformation of its marking
which occurs as a result 
of an execution of transitions. 

A marking $\sigma_0$ at time 0 
is assumed to be given. 

If a net has a marking $\sigma_i$
at a time $i$, then any of transition 
$t\in T$, which satisfies the condition 
$$\forall \;v\in in(t)\quad\sigma_i(v)>0$$
can be executed at time $i$.

If a transition $t$ was executed at time $i$ , 
then a marking $\sigma_{i+1}$
at time $i+1$ is defined as follows: 
$$\by \forall\; v\in in(t)\quad\sigma_{i+1}(v):=
\sigma(v)-1\\
\forall\; v\in out(t)\quad\sigma_{i+1}(v):=\sigma(v)+1
\\
\forall\; v\in V\setminus (in(t)\cup out(t))
\quad\sigma_{i+1}(v):=\sigma(v)
\ey$$

Each Petri net ${\cal N}$ can be 
associates with a process $P_{\cal N}$, 
which simulates a behavior of this net. 
Components of the process 
$P_{\cal N}$ are as follows. 
\bi
\i\bi
   \i $X_{P_{\cal N}}\eam 
       \{x_v\mid v\in V\}$,
   \i $I_{P_{\cal N}}\eam 
       \bigwedge\limits_{v\in V}
       (x_v=\sigma_0(v))$,
   \i $S_{P_{\cal N}}\eam  
       \{s^0\}$
   \ei
\i Let $t$ be a transition of the net
${\cal N}$, and the sets
$in(t)$ and $out(t)$ have the form
$\{u_1,\ldots, u_n\}$ and
$\{v_1,\ldots, v_m\}$
respectively. 

Then the process $P_{\cal N}$
has a transition from $s^0$ to $s^0$
with the label
$$\b{\l{(x_{u_1}>0)\wedge\ldots\wedge(x_{u_n}>0)}\\
x_{u_1}:=x_{u_1}-1,\ldots,x_{u_n}:=
x_{u_n}-1\\
x_{v_1}:=x_{v_1}+1,\ldots,x_{v_m}:=
x_{v_m}+1}$$
\ei 

\chapter{Communication protocols}

In this chapter we consider an application 
of the theory of processes to the problem
of modeling and verification 
of communication protocols
(which are called below {\bf protocols}).

\section{The concept of a protocol}

A {\bf protocol} is a distributed system
which consists of several 
interacting components, including
\bi 
\i components that perform a formation, sending, receiving
   and processing of messages \\ 
   (such components are called {\bf agents}, 
   and messages sent from one agent to another, 
   a called {\bf frames})
\i components of an environment, 
   through which frames are forwarded\\
   (usually such components are called 
   {\bf communication channels}). 
\ei 

There are several layers of protocols. 
In this chapter we consider 
data link layer protocols.

\section{Frames}

\subsection{The concept of a frame}

Each frame is a string of bits. 

When a frame is passed through an environment, 
it may be distorted or lost
(a distortion of a frame is an inverting
of some bits of this frame). 
Therefore, each frame must contain 
\bi 
\i not only an information which one agent 
   wishes to transfer to another agent, 
   but 
\i means allowing to a recipient 
   of the frame to find out 
   whether this frame 
   is distorted during a transmission. 
\ei 

Below we consider some methods 
of  detection of distortions in frames. 
These methods are divided into two classes: 
\bn 
\i methods which allow
   \bi 
   \i not only detect distortions of frames, 
   \i but also determine distorted bits 
      of a frame and fix them 
   \ei 
   (discussed in section \ref{jvklsf}), and 
\i methods to determine only a fact of    
   a    
   distortion of a frame, without correction of this distortion
   (discussed in section \ref{jvklsfq}). 
\en 

\subsection{Methods for correcting of 
distortions in  frames}
\label{jvklsf}

Methods of detection of distortion in frames, 
which allow 
\bi 
\i not only detect the fact of a distortion, but 
\i determine indexes of distorted bits 
\ei 
are used in such situations, 
when a probability that each 
transmitted frame will be distorted 
in a transmission of this frame, is high. 
For example, such a situation occurs 
in wireless communications. 

If you know a maximum number 
of bits of a frame which can be inverted, 
then for a recognition of inverted bits
and their correction 
methods of {\bf error correction coding}
can be used. 
These methods constitute one of 
directions of the {\bf coding theory}. 

In this section we consider 
an encoding method with correction 
of errors in a simplest case, when 
in a frame no more than one bit
can be inverted. 
This method is called 
a {\bf Hamming code}
to correct one error 
(there are Hamming codes 
to fix an arbitrary number of errors). 

The idea of this method is that bits 
of a frame are divided into two classes: 
\bi 
\i information bits (which contain an 
   information 
   which a sender of the frame 
   wants to convey 
   to the recipient), and 
\i control bits (values of which are computed 
   on values of information bits). 
\ei 

Let 
\bi
\i $f$ be a frame of the form 
   $(b_1, \ldots, b_n)$
\i $k$ is a number of information bits in $f$ 
\i $r$ is a number of control bits in $f$ \\
   (i.e. $n=k+r$)
\ei

Since a sender can place his information 
in $k$ information bits, 
then we can assume that an information 
that a sender sends to a recipient 
in a frame $f$, is a string $M$, 
which consists of $k$ bits.

A frame which is derived from the string $M$ 
by addition of control bits, 
we denote by $\varphi (M)$. 

For each frame $f$ denote by $\okr{f}$ 
the set of all frames obtained from $f$ 
by inversion of no more than one bit. 
Obviously, a number of elements of $\okr{f}$ 
is equal to $n+1$. 

The assumption that during a transmission
of the frame $\varphi(M)$ 
no more than one bit of this frame
can be inverted, 
can be reformulated as follows: 
a recipient can receive instead 
of $\varphi(M)$ any frame from 
the set $\okr{\varphi (M)}$.

It is easy to see that 
the following conditions are
equivalent:
\bn
\i for each $M\in \{0,1\}^k$
   a recipient 
   can uniquely reconstruct $M$
   having an arbitrary frame from
   $\okr{\varphi (M)}$
\i the family 
   \be{sagfvsadgvfsd}
   \{\okr{\varphi (M)}\mid M\in \{0,1\}^k\}\ee
   of subsets of $\{0,1\}^n$
   consists of disjoint subsets. 
\en

Since 
\bi 
\i family \re{sagfvsadgvfsd} consists of 
   $2^k$ subsets, and
\i each of these subsets 
   consists of $n +1$ elements 
\ei 
then a necessary condition of disjointness
of subsets from \re{sagfvsadgvfsd} 
is the inequality
$$(n+1)\cdot 2^k\leq 2^n$$
which can be rewritten as 
\be{f4sdafs}(k+r+1)\leq 2^r\ee

It is easy to prove that for every fixed 
$k>0$ the inequality \re{f4sdafs} 
(where $r$ is assumed to be positive) 
is equivalent to the inequality
$$r_0\leq r$$
where $r_0$ depends on $k$, 
and is a lower bound 
on the number of control bits. 

It is easy to calculate $r_0$, 
when $k$ has the form 
\be{dfgdfgertyht}k=2^m-m-1,\qquad
\mbox{where $m\geq 1$}\ee 
in this case \re{f4sdafs} 
can be rewritten as the inequality
\be{f4srdaf3s}2^m-m\leq 2^r-r\ee
which is equivalent to the inequality
$m\leq r$ (because the function 
$2^x-x$ is monotone for $x\geq 1$). 

Thus, in this case 
a lower bound 
of a number of control bits is $m$. 

Below we present a coding method with 
correction of one error,
in which a number $r$ of control bits 
is equal to the minimum possible value $m$. 

If $k$ has the form \re{dfgdfgertyht}, 
and $r=r_0=m$,
then $n=2^m-1$, i.e. 
indices of bits of the frame 
$f=(b_1,\ldots, b_n)$
can be identified with
$m$--tuples from $\{0,1\}^m$:
each index  $i\in\{1,\ldots, n\}$
is identified with
a binary record of $i$ 
(which is complemented 
by zeros to the left, if it is necessary).

By definition, indices of control bits 
are $m$--tuples of the form
\be{sfvsfvfsdgfgg} (0 \,\ldots\, 0\, 1\,
0\, \ldots\, 0) \quad
(\mbox{1 is at $j$--th position})\ee
where $j=1,\ldots, m$. 

For each $j=1,\ldots, m$
a value of a
control bit which has an index
\re{sfvsfvfsdgfgg}
is equal to the sum modulo 2 
values of information bits, 
indices of which 
contain 1 at $j$-th position. 

When a receiver gets a frame 
$(b_1,\ldots, b_n)$
he checks $m$ equalities
\be{44467326tgye}\sum\limits_{i_j=1}b_{i_1
\ldots i_m}=0\qquad(j=1,\ldots, m)\ee
(the sum is modulo 2). 

The following cases are possible. 
\bi 
\i The frame is not distorted.\\
   In this case, all the equalities
   \re{44467326tgye} are correct. 
\i A control bit which has the index
   \re{sfvsfvfsdgfgg} is distorted.

   In this case only $j$--th equality in 
   \re{44467326tgye} is incorrect.

\i An information bit 
   \re{sfvsfvfsdgfgg} is distorted.

   Let an index of this bit 
   contains 1 at the positions 
   $j_1$, $\ldots$, $j_l$.

   In this case among equalities 
   \re{44467326tgye}
   only equalities with numbers 
   $j_1$, $\ldots$, $j_l$
   are incorrect.
\ei 

Thus, in all cases, we can 
\bi 
\i detect it whether a frame is 
   distorted, and
\i calculate an index of a distorted bit, 
   if a frame is distorted.
\ei 

\subsection{Methods for detection 
of distortions in frames} 
\label{jvklsfq}

Another class of methods 
for detection of distortions in frames 
is related to a detection of 
only a fact of a distortion. 

The problem of a calculation of 
indices of distorted bits has 
high complexity. 
Therefore, if a probability of a distortion
in transmitted frames is low 
(that occurs when a copper 
or fibre communication channel is used), 
then more effective is a re-sending 
of distorted frames: 
if a receiver detects that a received
frame is distorted, then he requests 
a sender to send the frame again. 

For a comparison of a complexity 
of the problems of 
\bi\i correcting of distortions, 
and \i detection of distortions 
  (without  correcting)\ei
consider the following example. 
Suppose that no more than one bit of 
a frame can be distorted. 
If a size of this frame is 1000, then 
\bi 
\i for a {\it correction} of such distortion 
   it is needed 10 control bits, but
\i for a {\it detection} of such distortion 
   it is enough 1 control bit, whose value 
   is assumed equal 
   to a parity of a number 
   of units in remaining bits 
   of the frame.
\ei 

One method of coding to detection 
of distortion is the following: 
\bi 
\i a frame is divided into $k$ parts, and 
\i in each part it is assigned 
   one control bit, 
   whose value is assumed equal 
   to a parity of a number 
   of units in remaining bits 
   of this part. 
\ei 

If bits of the frame 
are distorted equiprobably
and independently, 
then for each such part 
of the frame the probability that 
\bi 
\i this part is distorted, and 
\i nevertheless, its parity is correct 
   (i.e., we consider it as undistorted) 
\ei 
is less than 1/2, therefore a probability 
of undetected distortion is less 
than $2^{-k}$.

Another method of coding 
to detection of distortions 
is a {\bf polynomial code}
(which is called
{\bf Cyclic Redundancy Check, CRC}). 

This method is based on a consideration 
of bit strings as polynomials over the field 
${\bf Z}_2=\{0,1\}$:
a bit string of the form 
$$(b_k,b_{k-1},\ldots, b_1,b_0)$$
is regarded as the polynomial 
$$b_k\cdot x^k+b_{k-1}\cdot x^{k-1}
+\ldots+ b_1\cdot x+b_0$$

Suppose you need to transfer frames 
of size $m+1$. 
Each such frame is considered 
as a polynomial $M(x)$ of a degree $\leq m$. 

To encode these frames there are selected 
\bi 
\i a number $r<m$, and 
\i a polynomial $G(x)$ of degree $r$, 
   which has the form  $$x^r+\ldots +1$$ 
\ei 
The polynomial $G(x)$ is called 
a {\bf generator polynomial}. 

For each frame $M(x)$ its code $T(x)$ 
is calculated as follows. 
The polynomial $x^r\cdot M(x)$
is divided on $G(x)$ with a remainder: 
$$x^r\cdot M(x)=G(x)\cdot Q(x)+R(x)$$
where $R(x)$ is a remainder
(a degree of $R(x)$ is less than $r$). 

A code of the frame $M(x)$ 
is the polynomial 
$$T(x)\eam G(x)\cdot Q(x)$$

It is easy to see that a size of $T(x)$ is larger 
than a size of $M(x)$ on $r$. 

Detection of a distortion 
in a transmission of the frame $T(x)$ is 
produced by a dividing a received frame 
$T'(x)$ on $G(x)$: 
we consider that the frame $T(x)$ was 
transmitted without a distortion 
(i.e. a received frame $T'(x)$ 
coincides with $T(x)$), 
if $T'(x)$ is divisible on $G(x)$ 
(i.e. $T'(x)$
has the form $G(x)\cdot Q'(x)$, 
where $Q'(x)$ is a polynomial).

If the frame $T(x)$ was transmitted 
without a distortion, then
the original frame $M(x)$ 
can be recovered 
by a representation of $T(x)$ as a sum 
$$T(x)=x^r\cdot M(x)+R(x)$$
where $R(x)$ consists of all monomials in $T(x)$ 
of a degree $<r$. 

A relation between \bi\i an original frame
$T(x)$, and \i a received frame $T'(x)$ \ei
can be represented as
$$T'(x)=T(x)+E(x)$$
where $E(x)$ is a polynomial which
\bi\i is called a {\bf polynomial of distortions},
and \i corresponds to a string of bits
each component of which is equal to
\bi
\i 1 if the corresponding bit
   of the frame $T(x)$ has been distorted, and 
\i 0, otherwise. 
\ei 
\ei

Thus 
\bi 
\i if $T(x)$ has been distorted in a single bit, 
   then $E(x)=x^i$ 
\i if $T(x)$ has been distorted in two bits, 
   then $E(x)=x^i+x^j$, \i etc. 
\ei 

From the definitions of $T'(x)$ and $E(x)$ 
it follows that 
$T'(x)$ is divisible on $G(x)$ 
if and only if 
$E(x)$ is divisible on $G(x)$. 

Therefore, a distortion 
corresponding to the polynomial $ E(x)$, 
can be detected if and only if 
$E(x)$ is not divisible on $G(x)$. 

Let us consider the question 
of what kinds of distortions 
can be detected using this method. 

\bn 
\i A single-bit distortion can be detected
   always, 
   because the polynomial 
   $E(x) = x^i$ is not divisible    on $G(x)$. 
\i A double-byte distortion can not 
   be detected in the case when 
   the corresponding polynomial 
   $$E(x)=x^i+x^j=x^j\cdot(x^{i-j}+1)
   \quad   (i> j)$$
   is divisible on $G(x)$: 
   \be{sdfg35678}\exists\,Q(x):\;\;
   x^j\cdot(x^{i-j}+1) = G(x) \cdot Q(x)
   \ee

On the reason of a uniqueness 
of factorization of polynomials 
over a field, statement 
\re{sdfg35678} 
implies the statement
\be{sdf5tg35678}\exists\,Q_1(x):\;\;
x^{i-j}+1 = 
G(x)\cdot Q_1(x)\ee

The following fact holds: if 
\be{5tert3wtwer}G(x)= 
x^{15}+x^{14}+1\ee
then
for each $k=1,\ldots, 32768$
the polynomial $x^k+1$
is not divisible on $G(x)$.

Therefore the generator polynomial 
\re{5tert3wtwer} can detect
a double-byte distortion 
in frames of a size $\leq 32768$. 

\i Consider the polynomial of distortions 
   $E(x)$ as a product of the form
   \be{gffgegt44467}
   E(x)=x^j\cdot (x^{k-1}   +\ldots +1)\ee

The number $k$ in \re{gffgegt44467}
is called a {\bf size of a packet of errors}. 
$k$ is equal to the size of a substring 
of a string of distortions 
(which corresponds to $E(x)$), 
which is bounded from left and right 
by the bits ``1''. 

Let $E_1(x)$ be the second factor in 
\re{gffgegt44467}. 

On the reason of a uniqueness 
of factorization of polynomials 
over a field we get that 
\bi
\i a distortion corresponding 
   to the polynomial
   \re{gffgegt44467} is not detected 
   if and only if  
\i $E_1(x)$ is divisible on $G(x)$. 
\ei

Consider separately the following cases.

\bn
\i $k\leq r$, i.e. $k-1<r$.

  In this case $E_1(x)$ is not divisible 
  on $G(x)$, because 
  a degree of $E_1(x)$ 
  is less than a degree of $G(x)$.

Thus, in this case we can detect 
any distortion.

\i $k=r +1$.

   In this case the polynomial $E_1(x)$ 
   is divisible on $G(x)$ if and only if 
   $E_1(x)=G(x)$.

   The probability of such coincidence 
   is equal to $2^{-(r-1)}$.

   Thus, a probability that such 
   distortion will not be detected
   is equal to $2^{-(r-1)}$.
\i $k>r +1$.

   It can be proved that in this case
   a probability that such 
   distortion will not be detected
   is less that $<2^{-r}$.
\en

\i If 
   \bi 
   \i an odd number of bits is distorted, i.e. 
   $E(x)$ has an odd number of monomials, 
   and 
  \i $G(x)=(x+1)\cdot G_1(x)$ \ei 
  then such a distortion can be detected, 
  because if for some polynomial $Q(x)$
  $$
  E(x)=G(x)\cdot Q(x)$$
  then, in particular 
  \be{sdfff367u}E(1)=G(1)\cdot Q(1)\ee
  that is wrong, 
  since \bi\i left side of \re{sdfff367u} 
  is equal to 1,  
  and \i right side of \re{sdfff367u} 
  is equal to  0. \ei
\en 

In standard IEEE 802 the following 
generator polynomial $G(x)$ is used:
$$\by
G(x)=&x^{32}+x^{26}+x^{23}+x^{22}
+x^{16}+x^{12}+
x^{11}+\\&+x^{10}+x^{8}+x^{7}+x^{5}
+x^{4}+x^{2}+x+1
\ey$$
This polynomial can detect a distortion, 
in which 
\bi
\i a size of a packet of errors is no more 
   than   $32$, or 
\i it is distorted an odd number of bits. 
\ei

\section{Protocols of one-way transmission}

\subsection{A simplest protocol 
of one-way transmission}
\label{sdfjksldjgjkghsdjklfghs}

A protocol which is considered in this section
consists 
of the following agents: 
\bi
\i a sender, 
\i a timer (which is used by a sender), 
\i a receiver, and
\i a channel.
\ei
The purpose of the protocol 
is a delivery of frames from 
a sender to a receiver 
via a channel. 
A channel is assumed to be unreliable, 
it can distort and lose transmitted frames. 

A protocol works as follows. 
\bn
\i A sender receives a message 
   (which is called a {\bf packet})
   from an agent which is not included 
   in the protocol. This agent 
   is called a {\bf sender's 
   network agent (SNA)}. 

   A purpose of a sender is 
   a cyclic execution of the
   following sequence of actions: 
\bi 
\i get a packet from a SNA
\i build a frame, which is 
      obtained by an applying 
      of a encoding function $\varphi$
      to the packet, 
\i send this frame to the channel 
   and switch-on the timer 
\i if the signal {\it timeout}  came
   from the timer, 
   which means that 
   \bi\i the waiting time    
    of a confirmation of the sent 
   frame has ended, and \i apparently 
   this frame is not received
   by the receiver\ei
   then send the frame again 
\i if  a confirmation signal came
   from the receiver, then\bi\i this means that 
   the current frame is 
   successfully accepted by 
   the receiver, 
   and \i the sender can 

   \bi \i 
   get the next packet from the
   SNA, 
\i build a frame from this packet, 
   \i etc. \ei \ei
\ei 

A flowchart representing this
behavior has the following form: 

{\small
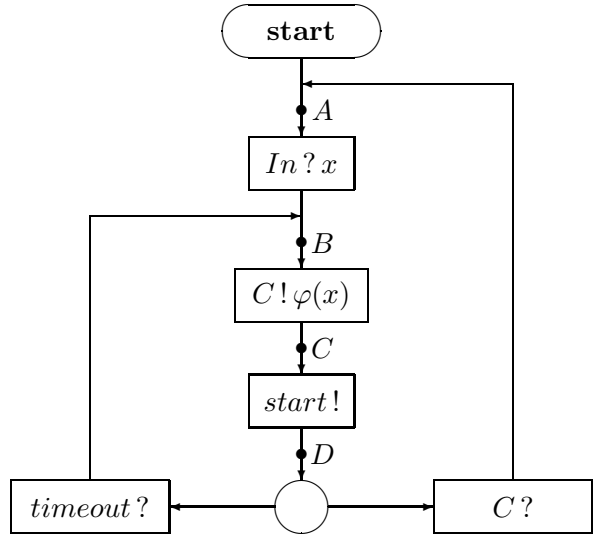
$$\begin{picture}(0,210)

\put(0,200){\oval(60,20)}
\put(0,200){\makebox(0,0){${\bf start}$}}

\put(0,190){\vector(0,-1){30}}

\put(0,170){\circle*{4}}
\put(8,170){\makebox(0,0)[c]{$A$}}
\put(0,120){\circle*{4}}
\put(8,120){\makebox(0,0)[c]{$B$}}
\put(0,80){\circle*{4}}
\put(8,80){\makebox(0,0)[c]{$C$}}
\put(0,40){\circle*{4}}
\put(8,40){\makebox(0,0)[c]{$D$}}

\put(0,150){\pu{20}{10}}
\put(0,150){\makebox(0,0)[c]{$In\,?\,x$}}

\put(0,140){\vector(0,-1){30}}

\put(0,100){\pu{25}{10}}
\put(0,100){\makebox(0,0)[c]{$C\,!\,\varphi (x)$}}

\put(0,90){\vector(0,-1){20}}

\put(0,60){\pu{20}{10}}
\put(0,60){\makebox(0,0)[c]{$start\,!$}}

\put(0,50){\vector(0,-1){20}}

\put(-80,20){\pu{30}{10}}
\put(-80,20){\makebox(0,0)[c]{$timeout\,?$}}

\put(80,20){\pu{30}{10}}
\put(80,20){\makebox(0,0)[c]{$C\,?$}}

\put(0,20){\oval(20,20)}
\put(-10,20){\vector(-1,0){40}}
\put(10,20){\vector(1,0){40}}

\put(-80,130){\vector(1,0){80}}
\put(-80,130){\line(0,-1){100}}

\put(80,30){\line(0,1){150}}
\put(80,180){\vector(-1,0){80}}

\end{picture}
$$
 } 

Operators belonging to this flowchart 
have the following meanings. 
\bi 
\i $In\,?\,x$ is a receiving 
   a packet from the SNA, 
   and record this packet 
   to the variable $x$ 
\i $C\,!\,\varphi (x)$ is a sending 
   the frame $\varphi (x)$ 
   to the channel 
\i $start\,!$ is a switching-on of the timer 
\i $timeout\,?$ is a receiving of a signal 
   ``timeout'' from the timer 
\i $C\,?$ is a receiving a confirmation signal 
   from the channel.
\ei 

The process represented by this flowchart, 
is denoted by $Sender$ and has the 
following form: 

{\def\arraystretch{1}

{\small
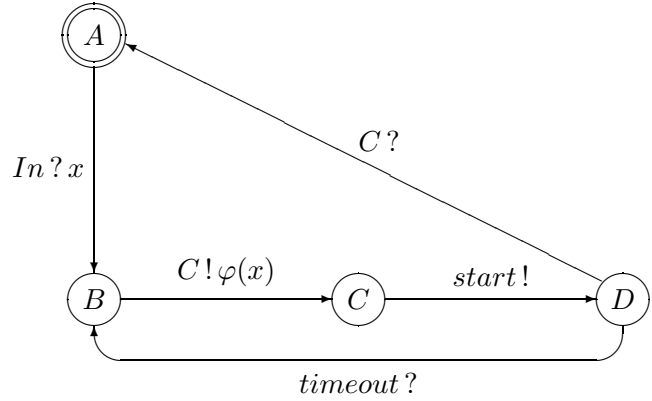
$$\by
\begin{picture}(0,110)

\put(-100,0){\oval(20,20)}
\put(-100,100){\oval(24,24)}
\put(0,0){\oval(20,20)}
\put(100,0){\oval(20,20)}
\put(-100,100){\oval(20,20)}

\put(-100,100){\makebox(0,0)[c]{$A$}}
\put(-100,0){\makebox(0,0)[c]{$B$}}
\put(0,0){\makebox(0,0)[c]{$C$}}
\put(100,0){\makebox(0,0)[c]{$D$}}

\put(-100,88){\vector(0,-1){78}}
\put(-90,0){\vector(1,0){80}}
\put(10,0){\vector(1,0){80}}

\put(-100,50){\makebox(0,0)[r]{
$In\,?\,x$
}}

\put(50,5){\makebox(0,0)[b]{
$start\,!$
}}

\put(-100,-13){\vector(0,1){3}}
\put(100,-13){\line(0,1){3}}
\put(-90,-13){\oval(20,20)[bl]}
\put(90,-13){\oval(20,20)[br]}
\put(90,-23){\line(-1,0){180}}

\put(0,-28){\makebox(0,0)[t]{
$timeout\,?$
}}

\put(-50,5){\makebox(0,0)[b]{
$C\,!\,\varphi (x)$
}}

\put(92,7){\vector(-2,1){180}}

\put(0,60){\makebox(0,0)[l]{$C\,?$}}

\end{picture}
\\
\vspace{10mm}
\ey
$$ } 
}

The behavior of the timer 
is represented by the process $Timer$
having the form

{\def\arraystretch{1}

{\small
\be{fghsdjklfeeyyerr}\by
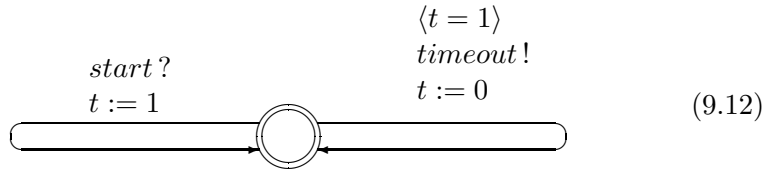
\begin{picture}(0,40)

\put(0,0){\oval(20,20)}
\put(0,0){\oval(24,24)}

\put(-100,0){\oval(10,10)[l]}
\put(-11,5){\line(-1,0){89}}
\put(-100,-5){\vector(1,0){89}}

\put(-60,5){\makebox(0,0)[b]{$\by start\,?

\\t:=1\ey$}}

\put(100,0){\oval(10,10)[r]}
\put(11,5){\line(1,0){89}}
\put(100,-5){\vector(-1,0){89}}

\put(70,10){\makebox(0,0)[b]{$\by
\l{t=1}\\timeout\,!\\t:=0\ey$}}

\end{picture}
\\
\vspace{0mm}
\ey
\ee } 
}

An initial condition of $Timer$ is $t = 0$. 

In this model we do not detail 
a magnitude of an interval between 
\bi 
\i a switching-on of the timer 
   (the action $start\,?$), and
\i a switching-off of the timer 
   (the action $ timeout \,!$).
\ei 

\i A {\bf channel} at each time can 
   contain no more than one 
   frame or signal. 

   It can execute the following actions: 
\bi 
\i receiving a frame from the sender, and 
   \bi 
   \i sending this frame to the receiver, or 
   \i sending a distorted frame to the receiver, or 
   \i loss of the frame 
   \ei 
\i receivng a confirmation signal 
   from the receiver, and 
   \bi 
   \i sending this signal to the sender, or 
   \i loss of the signal. 
   \ei 
\ei 

The behavior of the channel 
is described by the following process: 

{\def\arraystretch{1}

{\small
\be{dvfsdvgsdsdg}\by
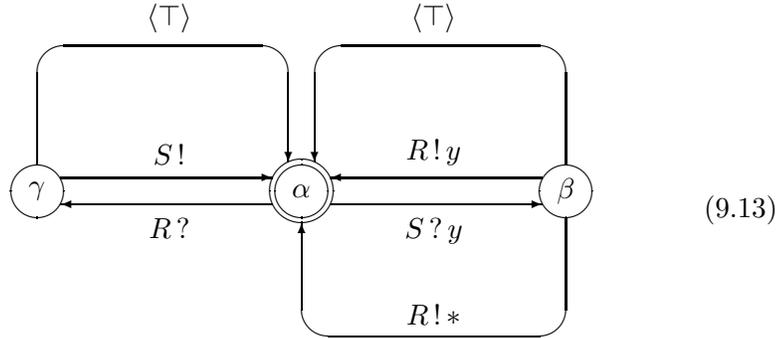
\begin{picture}(200,60)

\put(100,0){\oval(20,20)}
\put(100,0){\oval(24,24)}
\put(200,0){\oval(20,20)}
\put(0,0){\oval(20,20)}

\put(100,0){\makebox(0,0)[c]{$\alpha$}}
\put(200,0){\makebox(0,0)[c]{$\beta$}}
\put(0,0){\makebox(0,0)[c]{$\gamma$}}

\put(111,-5){\vector(1,0){80}}
\put(191,5){\vector(-1,0){80}}
\put(150,-10){\makebox(0,0)[t]{$S\,?\,y$}}
\put(150,10){\makebox(0,0)[b]{$R\,!\,y$}}

\put(89,-5){\vector(-1,0){80}}
\put(9,5){\vector(1,0){80}}
\put(50,-10){\makebox(0,0)[t]{$R\,?$}}
\put(50,10){\makebox(0,0)[b]{$S\,!$}}

\put(105,45){\vector(0,-1){34}}
\put(115,45){\oval(20,20)[lt]}
\put(190,45){\oval(20,20)[tr]}
\put(200,45){\line(0,-1){35}}
\put(190,55){\line(-1,0){75}}
\put(150,60){\makebox(0,0)[b]{$\l{\top}$}}

\put(100,-45){\vector(0,1){33}}
\put(110,-45){\oval(20,20)[lb]}
\put(190,-45){\oval(20,20)[br]}
\put(200,-45){\line(0,1){35}}
\put(190,-55){\line(-1,0){80}}
\put(150,-43){\makebox(0,0)[t]{$R\,!\,*$}}

\put(95,45){\vector(0,-1){34}}
\put(85,45){\oval(20,20)[tr]}
\put(10,45){\oval(20,20)[lt]}
\put(0,45){\line(0,-1){35}}
\put(10,55){\line(1,0){75}}
\put(50,60){\makebox(0,0)[b]{$\l{\top}$}}


\end{picture}
\\
\vspace{20mm}
\ey
\ee } 
}

In this process, we use the following 
abstraction: the symbol `$*$' means 
a ``distorted frame''. 
We do not specify exactly, how frames
can be distorted in the channel. 

Each frame which has been received 
by the channel 
\bi 
\i either is transferred from the channel 
   to the receiver 
\i or is transformed to the abstract value `$*$', 
   and this value is transferred 
   from the channel to receiver 
\i or disappears, which is expressed 
   by the transition of the process
   \re{dvfsdvgsdsdg} with the label
   $\l{\top}$ 
\ei 

\i The {\bf receiver} 
   executes the following actions: 
\bi 
\i receiving a frame from the channel 
\i checking of a distortion of the frame 
\i if the frame is not distorted, then \bi \i 
   extracting a packet from the frame 
   \i sending this packet to a process 
   called a {\bf receiver's network 
   agent (RNA)}\\
   (this process is not included in the protocol) 
   \i sending a confirmation signal 
      to the sender through the channel 
   \ei 
\i if the frame is distorted, then 
   the receiver ignores it 
   (assuming that the sender will be tired 
    to wait a confirmation signal, 
   and will send the frame again) 
\ei 

A flowchart representing the above
behavior has the following form:

{\small
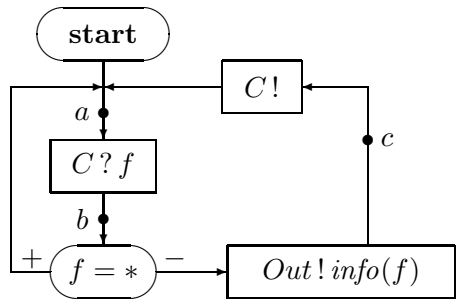
$$\begin{picture}(100,130)

\put(0,110){\oval(50,20)}
\put(0,110){\makebox(0,0){${\bf start}$}}

\put(0,100){\vector(0,-1){30}}
\put(0,50){\vector(0,-1){20}}

\put(0,60){\pu{20}{10}}
\put(0,60){\makebox(0,0)[c]{$C\,?\,f$}}

\put(0,20){\oval(40,20)}
\put(0,20){\makebox(0,0)[c]{$f=*$}}

\put(90,20){\pu{43}{10}}
\put(90,20){\makebox(0,0)[c]{$Out\,!\, {\it 
info} (f)$}}

\put(60,90){\pu{15}{10}}
\put(60,90){\makebox(0,0)[c]{$C\,!$}}

\put(45,90){\vector(-1,0){45}}
\put(100,30){\line(0,1){60}}
\put(100,90){\vector(-1,0){25}}
\put(-35,20){\line(0,1){70}}
\put(-35,90){\vector(1,0){35}}
\put(-35,20){\line(1,0){15}}
\put(20,20){\vector(1,0){27}}

\put(27,25){\makebox(0,0)[c]{$-$}}
\put(-27,25){\makebox(0,0)[c]{$+$}}

\put(0,80){\circle*{4}}
\put(-8,80){\makebox(0,0)[c]{$a$}}
\put(0,40){\circle*{4}}
\put(-8,40){\makebox(0,0)[c]{$b$}}
\put(100,70){\circle*{4}}
\put(105,70){\makebox(0,0)[l]{$c$}}

\end{picture}
$$
 } 

Operators belonging to this flowchart have the 
following meanings. 
\bi 
\i $C\,?\,f$ is a receiving of a frame
   from the channel, and a record it
   to the variable $f$ 
\i $(f=*)$ is a checking of a distortion 
   of the frame $f$ 
\i $Out\,!\,{\it info}(f)$ is a sending
  of the packet ${\it info}(f)$, 
   extracted from the frame $f$, 
   to the RNA 
\i $C\,!$ is a sending of the 
   confirmation signal 
\ei 

The process represented by this flowchart, 
is denoted as $Receiver$
and has the following form: 

{\def\arraystretch{1}

{\small
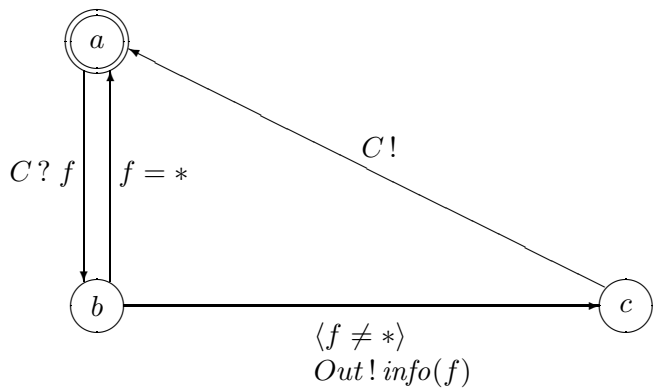
$$\by
\begin{picture}(200,120)

\put(0,100){\oval(20,20)}
\put(0,100){\oval(24,24)}
\put(0,0){\oval(20,20)}
\put(200,0){\oval(20,20)}

\put(0,100){\makebox(0,0)[c]{$a$}}
\put(0,0){\makebox(0,0)[c]{$b$}}
\put(200,0){\makebox(0,0)[c]{$c$}}

\put(-5,89){\vector(0,-1){80}}
\put(5,9){\vector(0,1){80}}
\put(-8,50){\makebox(0,0)[r]{$C\,?\;f$}}
\put(8,50){\makebox(0,0)[l]{$f=*$}}

\put(10,0){\vector(1,0){180}}

\put(192,7){\vector(-2,1){180}}

\put(100,60){\makebox(0,0)[l]{$C\,!$}}

\put(110,-5){\makebox(0,0)[t]{
$\by \l{f\neq *}\\
Out\,!\,{\it info}(f)\ey$}}

\end{picture}
\\
\vspace{0mm}
\ey
$$ } 
}
\en 

The process {\it Protocol}, 
corresponding to the whole system, 
is defined as a parallel composition 
(with restriction and renaming) 
of the above processes: 

\be{sdfffew56766}{\it Protocol}\eam\;\;\b{
Sender\,[S/C]\pa \\
Timer\pa\\
Channel\pa\\
Receiver\,[R/C]
}\setminus\{S,R,start,timeout\}\ee

A flow graph of the process $Protocol$ 
has the form

{\small
\be{podfkgnsdw}\by 
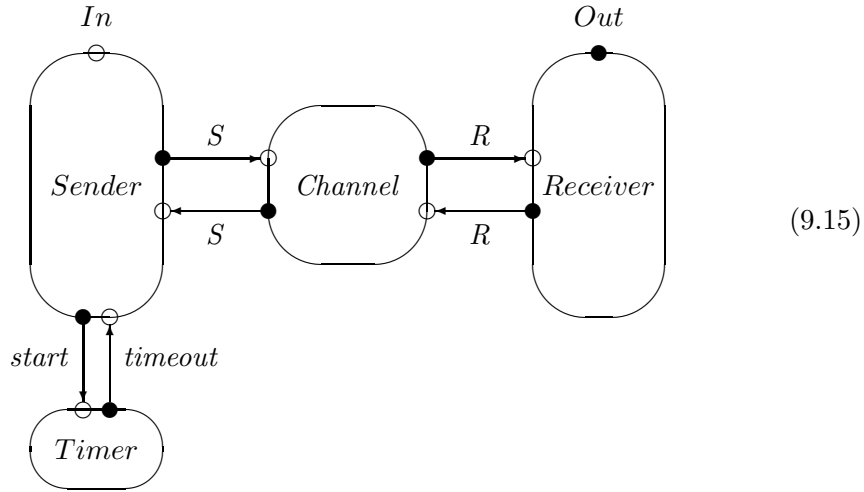
\begin{picture}(0,130)


\put(-30,60){\circle{6}}
\put(30,60){\circle*{6}}
\put(-30,40){\circle*{6}}
\put(30,40){\circle{6}}
\put(-70,60){\circle*{6}}
\put(70,60){\circle{6}}
\put(-70,40){\circle{6}}
\put(70,40){\circle*{6}}

\put(-95,100){\circle{6}}
\put(95,100){\circle*{6}}

\put(-100,-35){\circle{6}}
\put(-90,-35){\circle*{6}}

\put(-100,0){\circle*{6}}
\put(-90,0){\circle{6}}

\put(-100,-3){\vector(0,-1){29}}
\put(-90,-32){\vector(0,1){29}}

\put(-105,-15){\makebox(0,0)[r]{${\it start}
$}}
\put(-85,-15){\makebox(0,0)[l]{${\it timeout}
$}}

\put(-50,65){\makebox(0,0)[b]{${\it S}$}}
\put(-50,35){\makebox(0,0)[t]{${\it S}$}}

\put(50,65){\makebox(0,0)[b]{${\it R}$}}
\put(50,35){\makebox(0,0)[t]{${\it R}$}}

\put(0,50){\oval(60,60)}
\put(0,50){\makebox(0,0){${\it Channel}$}}

\put(-95,50){\oval(50,100)}
\put(-95,50){\makebox(0,0){$Sender$}}

\put(-95,-50){\oval(50,30)}
\put(-95,-50){\makebox(0,0){$Timer$}}

\put(95,50){\oval(50,100)}
\put(95,50){\makebox(0,0){$Receiver$}}

\put(-67,60){\vector(1,0){34}}
\put(-33,40){\vector(-1,0){34}}
\put(33,60){\vector(1,0){34}}
\put(67,40){\vector(-1,0){34}}
\put(-95,110){\makebox(0,0)[b]{$In$}}
\put(95,110){\makebox(0,0)[b]{$Out$}}

\end{picture}
\\
\vspace{14mm}\ey
\ee
 } 
$$\;$$

In order to be able to analyze the correctness 
of this protocol is necessary to determine 
a specification which he must meet. 

If we want to specify only properties 
of external actions executed
by the protocol (i.e., actions of the form 
$In\,?\,v$ and $Out\,!\,v$), 
then the specification can be
as follows: 
the behavior of this protocol coincides
with the behavior of the buffer of the size 1, 
i.e. the process {\it Protocol}
is observationally equivalent to the 
process ${\it Buf}$, which has the form 

{\small
\be{gjkdflsgjeeetyy}\by
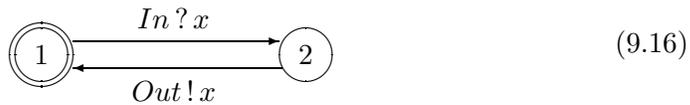
\begin{picture}(100,25)

\put(0,0){\oval(20,20)}
\put(0,0){\oval(24,24)}
\put(100,0){\oval(20,20)}

\put(0,0){\makebox(0,0)[c]{$1$}}
\put(100,0){\makebox(0,0)[c]{$2$}}

\put(12,5){\vector(1,0){78}}
\put(91,-5){\vector(-1,0){79}}

\put(50,10){\makebox(0,0)[b]{$In\,?\,x$}}
\put(50,-10){\makebox(0,0)[t]{$Out\,!\,x$}}

\end{picture}
\\
\vspace{0mm}
\ey
\ee } 

After a reduction of 
the graph representation
of the process $Protocol$ 
we get the diagram 

{\small
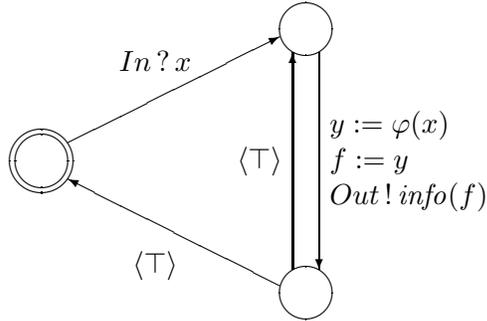
$$\by
\begin{picture}(100,120)

\put(0,50){\oval(20,20)}
\put(0,50){\oval(24,24)}

\put(100,100){\oval(20,20)}

\put(100,0){\oval(20,20)}

\put(95,9){\vector(0,1){82}}
\put(105,91){\vector(0,-1){82}}
\put(10,57){\vector(2,1){80}}
\put(90,3){\vector(-2,1){80}}

\put(43,84){\makebox(0,0)[b]{$In\,?\,x$}}
\put(43,16){\makebox(0,0)[t]{$\l{\top}$}}

\put(91,50){\makebox(0,0)[r]{$\l{\top}$}}
\put(104,50){\makebox(0,0)[l]{$
\by
y:=\varphi (x)\\
f:=y\\
Out\,!\,{\it info}(f)
\ey
$}}
\end{picture}
\\
\vspace{0mm}
\ey
$$} 

which is observationally equivalent 
to the diagram 
{\small
\be{zfgzxgfsdgsdfgds}
\by
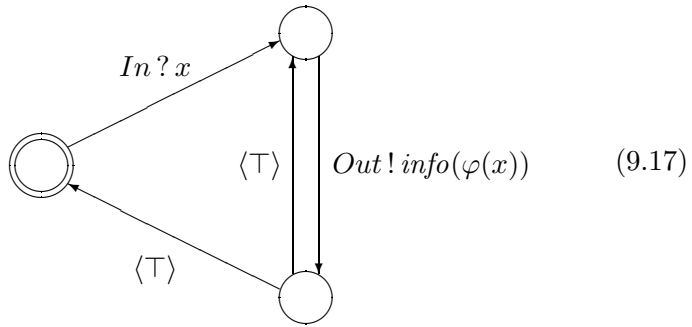
\begin{picture}(100,120)

\put(0,50){\oval(20,20)}
\put(0,50){\oval(24,24)}

\put(100,100){\oval(20,20)}

\put(100,0){\oval(20,20)}

\put(95,9){\vector(0,1){82}}
\put(105,91){\vector(0,-1){82}}
\put(10,57){\vector(2,1){80}}
\put(90,3){\vector(-2,1){80}}

\put(43,84){\makebox(0,0)[b]{$In\,?\,x$}}
\put(43,16){\makebox(0,0)[t]{$\l{\top}$}}

\put(91,50){\makebox(0,0)[r]{$\l{\top}$}}
\put(110,50){\makebox(0,0)[l]{$
Out\,!\,{\it info}(\varphi (x))
$}}
\end{picture}
\\
\vspace{0mm}
\ey
\ee } 

We assume that the function 
{\it info} of extracting of packets 
from frames is inverse to $\varphi$, 
i.e. for each packet $x$ 
$$
{\it info}(\varphi (x)) = x
$$
therefore the diagram 
\re{zfgzxgfsdgsdfgds} 
can be redrawn as follows: 

{\small
\be{gjkdflsgjeeetyy11}
\by
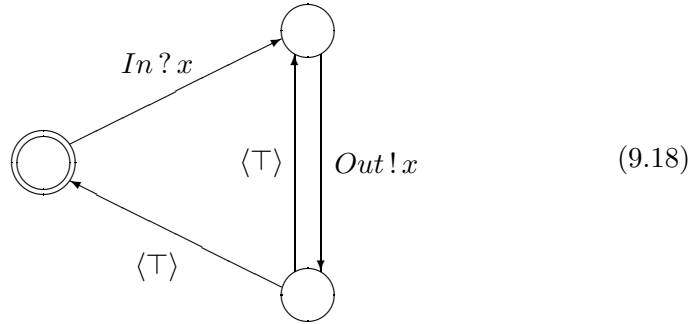
\begin{picture}(100,120)

\put(0,50){\oval(20,20)}
\put(0,50){\oval(24,24)}

\put(100,100){\oval(20,20)}

\put(100,0){\oval(20,20)}

\put(95,9){\vector(0,1){82}}
\put(105,91){\vector(0,-1){82}}
\put(10,57){\vector(2,1){80}}
\put(90,3){\vector(-2,1){80}}

\put(43,84){\makebox(0,0)[b]{$In\,?\,x$}}
\put(43,16){\makebox(0,0)[t]{$\l{\top}$}}

\put(91,50){\makebox(0,0)[r]{$\l{\top}$}}
\put(110,50){\makebox(0,0)[l]{$
Out\,!\,x
$}}
\end{picture}
\\
\vspace{0mm}
\ey
\ee } 

The process \re{gjkdflsgjeeetyy11}
can be reduced, resulting in the process 

{\small
\be{zfgtr3zxgfsdgsdfgds}
\by
\begin{picture}(150,25)

\put(0,0){\oval(20,20)}
\put(0,0){\oval(24,24)}
\put(100,0){\oval(20,20)}

\put(12,5){\vector(1,0){79}}
\put(91,-5){\vector(-1,0){79}}

\put(109,5){\line(1,0){41}}
\put(150,-5){\vector(-1,0){41}}
\put(150,0){\oval(10,10)[r]}

\put(50,10){\makebox(0,0)[b]{$In\,?\,x$}}
\put(50,-10){\makebox(0,0)[t]{$Out\,!\,x$}}

\put(130,-10){\makebox(0,0)[t]{$Out\,!\,x$}}

\end{picture}
\\
\vspace{5mm}
\ey
\ee } 

After a comparing of the processes 
\re{zfgtr3zxgfsdgsdfgds} and 
\re{gjkdflsgjeeetyy} we conclude 
that these processes can not be 
equivalent in any acceptable way. 
For example, 
\bi 
\i the process \re{gjkdflsgjeeetyy}
   after receiving the packet $x$ can only 
   \bi 
   \i send this packet to the RNA, and 
   \i move to the state of waiting 
       of another packet 
   \ei 
\i while the process \re{zfgtr3zxgfsdgsdfgds}
   after receiving the packet $x$ can 
   send this packet to the RNA several times. 
\ei 

Such retransmission 
can occur, for example, 
in the following version of an execution
of the protocol. 
\bi 
\i First frame which is sent by the sender, 
   reaches the receiver successfully. 
\i The receiver 
   \bi 
   \i sends the packet, extracted from 
      this frame, to the RNA, and 
   \i sends a confirmation to the sender 
      through the channel. 
  \ei 
\i This confirmation is lost in the channel. 
\i The sender does not received a 
   confirmation, and
   sends this frame again, 
   and this frame again goes well.

\i The receiver perceives this frame 
   as a new one. 
   He \bi \i sends the packet, 
   extracted from this frame, 
   to the RNA, and \i 
   sends the confirmation signal 
   to the sender through the channel.\ei 
\i This confirmation again is lost in the channel. 
\i etc. 
\ei 

This situation may arise because 
in this protocol there is no a mechanism 
through which the receiver can distinguish:
\bi
\i is a received frame a new one, or 
\i this frame was transmitted before. 
\ei

In section \ref{sfvgsadgewgtwe} we consider
a protocol which has such mechanism. 
For this protocol it is possible 
to prove formally 
its compliance with the specification
\re{gjkdflsgjeeetyy}. 

\subsection{One-way alternating bit protocol}
\label{sfvgsadgewgtwe}

The protocol described in this section 
is called the {\bf one-way alternating bit protocol}, 
or, in an abbreviated notation, {\bf ABP}.

The protocol ABP
is designed to solve the same problem as the 
protocol in section \ref{sdfjksldjgjkghsdjklfghs}: 
delivery of frames from the sender 
to the receiver via an unreliable 
channel (which can distort and lose 
transmitted frames). 

The protocol ABP
\bi
\i consists of the same agents 
   as the protocol in section 
   \ref{sdfjksldjgjkghsdjklfghs}
   (namely: the sender, 
   the timer, 
   the receiver, and
   the channel), and
\i has the same flow graph.
\ei

A mechanism by which the receiver 
can distinguish new frames from 
retransmitted ones, 
is implemented in this protocol as follows: 
among the variables of the sender and 
the receiver there are boolean variables 
$s$ and $r$, respectively, 
values which have the following meanings: 
\bi 
\i a value of $s$ 
   is equal to a parity of an index
   of a current frame, which is trying 
   to be sent by the sender, and 
\i a value of $r$ is equal to a parity 
   of an index of a 
   frame, which is expected by the receiver. 
\ei 

At the initial time values of $s$ and $r$ are
equal to 0 (the first frame has an index 0). 

As in the protocol in section \ref{sdfjksldjgjkghsdjklfghs}, 
the abstract value 
``$*$'' is used in this protocol, 
this value denotes a distorted frame. 

The protocol works as follows. 

\bn 
\i The {\bf sender} gets a packet 
   from the SNA, and
  \bi 
  \i records this packet
     to the variable $x$, 
   \i builds the frame, which is 
      obtained by an applying 
      of a coding function $\varphi$
      to the pair $(x, s)$, 
   \i sends the frame 
     to the channel, \i 
      starts the timer, and then
   \i expects 
 a confirmation of the frame
which has been sent. \ei 

   If \bi \i the sender gets from the times
the signal {\it timeout}, 
and 
   \i he does not received yet
an acknowledgment from the receiver
\ei
then the sender retransmits this frame.

If the sender receives from the channel
an undistorted frame, 
which contains a boolean value, 
then the sender 
analyzes this value: if it 
coincides with the current value of $s$, 
then   the sender 
   \bi 
   \i inverts the value of the variable $s$ 
(using the function 
$Inv(x)$ = $1-x$), and 
   \i starts a new cycle of his work. 
   \ei 
   Otherwise, he sends the frame again.

The flowchart representing this behavior 
has the following form: 

{\small
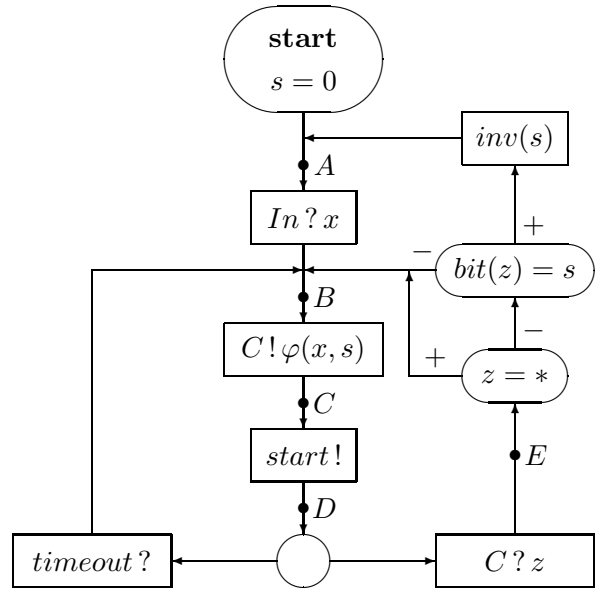
$$\begin{picture}(0,240)

\put(0,210){\oval(60,40)}
\put(0,218){\makebox(0,0){${\bf start}$}}
\put(0,202){\makebox(0,0)[c]{$s=0$}}

\put(0,190){\vector(0,-1){30}}

\put(0,170){\circle*{4}}
\put(8,170){\makebox(0,0)[c]{$A$}}
\put(0,120){\circle*{4}}
\put(8,120){\makebox(0,0)[c]{$B$}}
\put(0,80){\circle*{4}}
\put(8,80){\makebox(0,0)[c]{$C$}}
\put(0,40){\circle*{4}}
\put(8,40){\makebox(0,0)[c]{$D$}}
\put(80,60){\circle*{4}}
\put(88,60){\makebox(0,0)[c]{$E$}}

\put(0,150){\pu{20}{10}}
\put(0,150){\makebox(0,0)[c]{$In\,?\,x$}}

\put(0,140){\vector(0,-1){30}}

\put(0,100){\pu{30}{10}}
\put(0,100){\makebox(0,0)[c]{$C\,!\,\varphi (x,s)$}}

\put(0,90){\vector(0,-1){20}}

\put(0,60){\pu{20}{10}}
\put(0,60){\makebox(0,0)[c]{$start\,!$}}

\put(0,50){\vector(0,-1){20}}

\put(-80,20){\pu{30}{10}}
\put(-80,20){\makebox(0,0)[c]{$timeout\,?$}}

\put(80,20){\pu{30}{10}}
\put(80,20){\makebox(0,0)[c]{$C\,?\,z$}}

\put(0,20){\oval(20,20)}
\put(-10,20){\vector(-1,0){40}}
\put(10,20){\vector(1,0){40}}

\put(80,180){\pu{20}{10}}
\put(80,180){\makebox(0,0)[c]{$inv(s)$}}
\put(60,180){\vector(-1,0){60}}

\put(80,130){\oval(60,20)}
\put(80,130){\makebox(0,0)[c]{$bit(z)=s$}}
\put(50,130){\vector(-1,0){50}}

\put(80,90){\oval(40,20)}
\put(80,90){\makebox(0,0)[c]{$z=*$}}
\put(60,90){\line(-1,0){20}}
\put(40,90){\vector(0,1){40}}

\put(-80,130){\vector(1,0){80}}
\put(-80,130){\line(0,-1){100}}

\put(80,30){\vector(0,1){50}}
\put(80,100){\vector(0,1){20}}
\put(80,140){\vector(0,1){30}}

\put(87,107){\makebox(0,0)[c]{$-$}}
\put(87,147){\makebox(0,0)[c]{$+$}}

\put(50,97){\makebox(0,0)[c]{$+$}}
\put(45,137){\makebox(0,0)[c]{$-$}}

\end{picture}
$$
 } 

The process, which corresponds to this 
flowchart, is denoted by $Sender$, 
and has the following form: 

$Init=(s=0)$.

{\def\arraystretch{1}

{\small
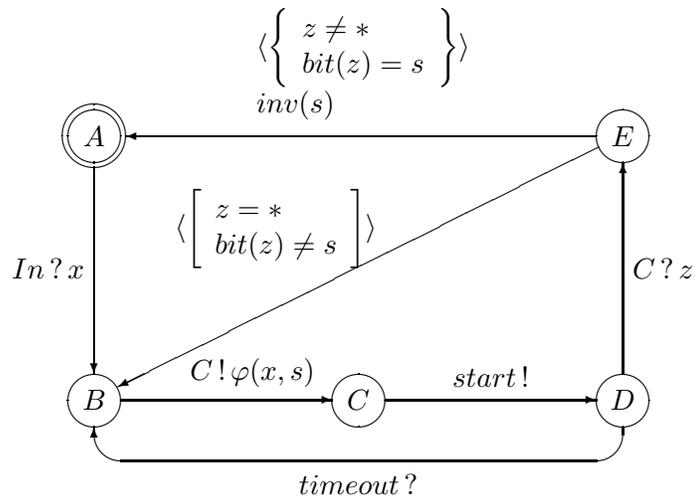
$$\by
\begin{picture}(0,140)

\put(-100,0){\oval(20,20)}
\put(-100,100){\oval(24,24)}
\put(0,0){\oval(20,20)}
\put(100,0){\oval(20,20)}
\put(-100,100){\oval(20,20)}
\put(100,100){\oval(20,20)}

\put(-100,100){\makebox(0,0)[c]{$A$}}
\put(-100,0){\makebox(0,0)[c]{$B$}}
\put(0,0){\makebox(0,0)[c]{$C$}}
\put(100,0){\makebox(0,0)[c]{$D$}}
\put(100,100){\makebox(0,0)[c]{$E$}}

\put(-100,88){\vector(0,-1){78}}
\put(-90,0){\vector(1,0){80}}
\put(10,0){\vector(1,0){80}}
\put(100,10){\vector(0,1){80}}
\put(90,100){\vector(-1,0){178}}
\put(91,96){\vector(-2,-1){182}}

\put(0,105){\makebox(0,0)[b]{
$\by \l{\c{z\neq *\\ bit(z)=s}}
\\inv(s)\ey$}}

\put(7,65){\makebox(0,0)[r]{
$\l{\d{z= *\\bit(z)\neq s}}$}}

\put(-100,50){\makebox(0,0)[r]{
$In\,?\,x$
}}

\put(100,50){\makebox(0,0)[l]{
$C\,?\,z$
}}

\put(-40,5){\makebox(0,0)[b]{
$C\,!\,\varphi (x,s)$
}}

\put(50,5){\makebox(0,0)[b]{
$start\,!$
}}

\put(-100,-13){\vector(0,1){3}}
\put(100,-13){\line(0,1){3}}
\put(-90,-13){\oval(20,20)[bl]}
\put(90,-13){\oval(20,20)[br]}
\put(90,-23){\line(-1,0){180}}

\put(0,-28){\makebox(0,0)[t]{
$timeout\,?$
}}

\end{picture}
\\
\vspace{10mm}
\ey
$$ } 
}

\i The {\bf channel} can 
   contain no more than one frame. 

   It can execute the following actions: 
\bi 
\i receive a frame from the sender, and 
   \bi 
   \i either send this frame to the receiver, 
   \i or send a distorted frame to the receiver, 
   \i or lose the frame 
   \ei 
\i receive a confirmation frame 
   from the receiver, and 
   \bi 
   \i either send this frame to the sender, 
   \i or send the distorted frame to 
      the sender, 
   \i or lose the frame. 
   \ei 
\ei 

The behavior of the channel 
is represented by the following process: 

{\def\arraystretch{1}

{\small
\be{1dvfsdvgsdsdg}\by
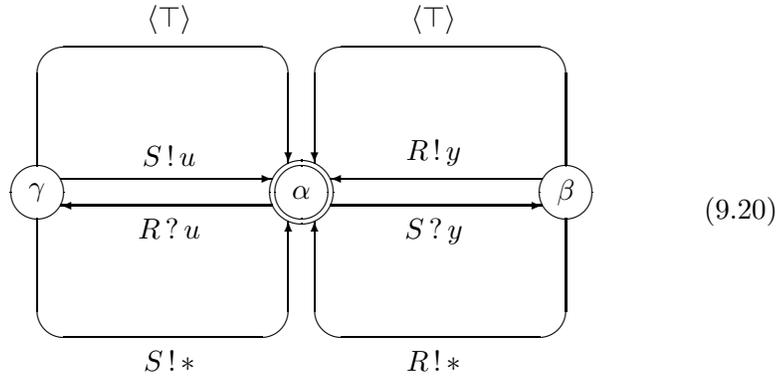
\begin{picture}(200,60)

\put(100,0){\oval(20,20)}
\put(100,0){\oval(24,24)}
\put(200,0){\oval(20,20)}
\put(0,0){\oval(20,20)}

\put(100,0){\makebox(0,0)[c]{$\alpha$}}
\put(200,0){\makebox(0,0)[c]{$\beta$}}
\put(0,0){\makebox(0,0)[c]{$\gamma$}}

\put(111,-5){\vector(1,0){80}}
\put(191,5){\vector(-1,0){80}}
\put(150,-10){\makebox(0,0)[t]{$S\,?\,y$}}
\put(150,10){\makebox(0,0)[b]{$R\,!\,y$}}

\put(89,-5){\vector(-1,0){80}}
\put(9,5){\vector(1,0){80}}
\put(50,-10){\makebox(0,0)[t]{$R\,?\,u$}}
\put(50,10){\makebox(0,0)[b]{$S\,!\,u$}}

\put(105,45){\vector(0,-1){34}}
\put(115,45){\oval(20,20)[lt]}
\put(190,45){\oval(20,20)[tr]}
\put(200,45){\line(0,-1){35}}
\put(190,55){\line(-1,0){75}}
\put(150,60){\makebox(0,0)[b]{$\l{\top}$}}

\put(105,-45){\vector(0,1){34}}
\put(115,-45){\oval(20,20)[lb]}
\put(190,-45){\oval(20,20)[br]}
\put(200,-45){\line(0,1){35}}
\put(190,-55){\line(-1,0){75}}
\put(150,-60){\makebox(0,0)[t]{$R\,!\,*$}}

\put(95,45){\vector(0,-1){34}}
\put(85,45){\oval(20,20)[tr]}
\put(10,45){\oval(20,20)[lt]}
\put(0,45){\line(0,-1){35}}
\put(10,55){\line(1,0){75}}
\put(50,60){\makebox(0,0)[b]{$\l{\top}$}}

\put(95,-45){\vector(0,1){34}}
\put(85,-45){\oval(20,20)[br]}
\put(10,-45){\oval(20,20)[bl]}
\put(0,-45){\line(0,1){35}}
\put(10,-55){\line(1,0){75}}
\put(50,-60){\makebox(0,0)[t]{$S\,!\,*$}}

\end{picture}
\\
\vspace{20mm}
\ey
\ee } 
}

\i The {\bf receiver} 
  upon receiving of a frame      from the channel 
   \bi 
   \i checks whether the frame is      distorted, 
  \i and if the frame is not distorted, then 
     the receiver extracts from the frame
    a packet and a boolean value 
   using functions {\it info} and {\it bit}, 
   with the following properties: 
   $$
   {\it info}(\varphi (x,b))=x,\quad {\it bit}   
(\varphi (x,b))=b
   $$\ei

   The receiver checks whether 
   the boolean value extracted from the frame
   coincides with the expected value, 
   which is contained in the variable $r$, and
\bn\i
   if the checking gave a positive result, 
   then the receiver 
   \bi 
   \i transmits the packet extracted from
      this frame to the RNA
   \i inverts the value of $r$, and 
   \i sends the confirmation frame
     to the sender through 
     the channel. 
   \ei 
   \i if the checking gave a negative result, 
   then the receiver sends a confirmation frame 
   with an incorrect boolean value
   (which will cause the sender 
   to send its current frame again). 
\en

   If the frame is distorted, 
   then the receiver ignores this frame
   (assuming that the sender
   will send this frame again
   on the reason of receiving of the signal
   {\it timeout} from the timer).

The flowchart representing the above
behavior has the following form:

{\small
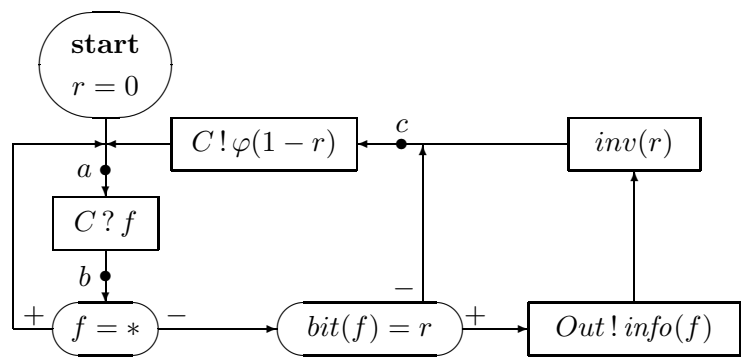
$$\begin{picture}(180,150)

\put(0,120){\oval(50,40)}
\put(0,128){\makebox(0,0){${\bf start}$}}
\put(0,112){\makebox(0,0)[c]{$r=0$}}

\put(0,100){\vector(0,-1){30}}
\put(0,50){\vector(0,-1){20}}

\put(0,60){\pu{20}{10}}
\put(0,60){\makebox(0,0)[c]{$C\,?\,f$}}

\put(0,20){\oval(40,20)}
\put(0,20){\makebox(0,0)[c]{$f=*$}}

\put(100,20){\oval(70,20)}
\put(100,20){\makebox(0,0)[c]{$bit(f)=r$}}

\put(200,20){\pu{40}{10}}
\put(200,20){\makebox(0,0)[c]{$Out\,!\, {\it 
info} (f)$}}

\put(200,90){\pu{25}{10}}
\put(200,90){\makebox(0,0)[c]{$inv(r)$}}

\put(60,90){\pu{35}{10}}
\put(60,90){\makebox(0,0)[c]{$C\,!\,\varphi 
(1-r)$}}

\put(25,90){\vector(-1,0){25}}
\put(175,90){\vector(-1,0){80}}
\put(200,30){\vector(0,1){50}}
\put(120,30){\vector(0,1){60}}
\put(-35,20){\line(0,1){70}}
\put(-35,90){\vector(1,0){35}}
\put(-35,20){\line(1,0){15}}
\put(20,20){\vector(1,0){45}}
\put(135,20){\vector(1,0){25}}

\put(27,25){\makebox(0,0)[c]{$-$}}
\put(-27,25){\makebox(0,0)[c]{$+$}}
\put(140,25){\makebox(0,0)[c]{$+$}}
\put(113,35){\makebox(0,0)[c]{$-$}}

\put(0,80){\circle*{4}}
\put(-8,80){\makebox(0,0)[c]{$a$}}
\put(0,40){\circle*{4}}
\put(-8,40){\makebox(0,0)[c]{$b$}}
\put(112,90){\circle*{4}}
\put(112,97){\makebox(0,0)[c]{$c$}}

\end{picture}
$$
 } 

The process represented by this flowchart, 
is denoted by $Receiver$
and has the following form: 

$Init = (r=0)$

{\def\arraystretch{1}

{\small
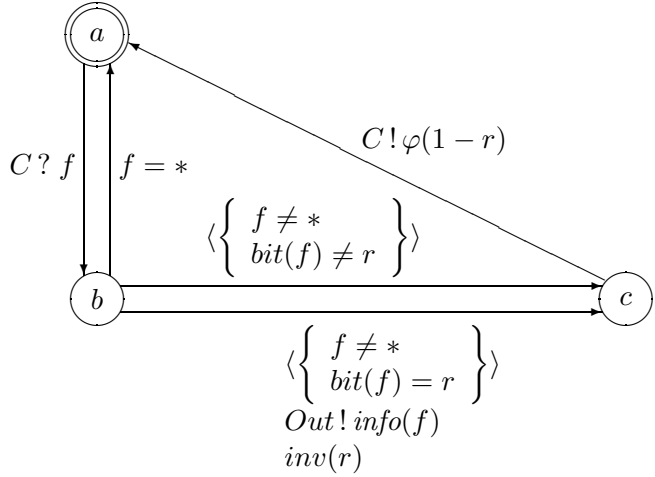
$$\by
\begin{picture}(200,120)

\put(0,100){\oval(20,20)}
\put(0,100){\oval(24,24)}
\put(0,0){\oval(20,20)}
\put(200,0){\oval(20,20)}

\put(0,100){\makebox(0,0)[c]{$a$}}
\put(0,0){\makebox(0,0)[c]{$b$}}
\put(200,0){\makebox(0,0)[c]{$c$}}

\put(-5,89){\vector(0,-1){80}}
\put(5,9){\vector(0,1){80}}
\put(-8,50){\makebox(0,0)[r]{$C\,?\;f$}}
\put(8,50){\makebox(0,0)[l]{$f=*$}}

\put(9,5){\vector(1,0){182}}
\put(9,-5){\vector(1,0){182}}

\put(192,7){\vector(-2,1){180}}

\put(100,60){\makebox(0,0)[l]{$C\,!\,\varphi 
(1-r)$}}

\put(80,8){\makebox(0,0)[b]{
$\l{\c{f\neq *\\ bit(f)\neq r}}$}}

\put(110,-10){\makebox(0,0)[t]{
$\by \l{\c{f\neq *\\bit(f)= r}}\\
Out\,!\,\mbox{{\it info}}(f)\\inv(r)\ey$}}

\end{picture}
\\
\vspace{15mm}
\ey
$$ } 
}

\en 

The process {\it Protocol}, 
which
corresponds to the whole protocol ABP, 
is defined in the same manner as 
in section \ref{sdfjksldjgjkghsdjklfghs}, 
by the expression \re{sdfffew56766}. 
The flow graph of this process has the 
form \re{podfkgnsdw}. 

The specification of the protocol ABP
also has the same form as 
in section \ref{sdfjksldjgjkghsdjklfghs}, 
i.e. is defined as the process 
\re{gjkdflsgjeeetyy}. 

The reduced process {\it Protocol}
has the form

{\small
\be{zfrgtfgr3zxgfsdgsdfgrhgds}
\by
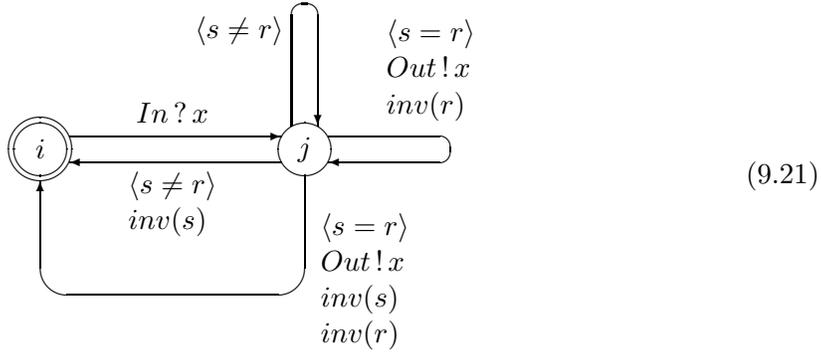
\begin{picture}(200,60)

\put(0,0){\oval(20,20)}
\put(0,0){\oval(24,24)}
\put(100,0){\oval(20,20)}

\put(0,0){\makebox(0,0)[c]{$i$}}
\put(100,0){\makebox(0,0)[c]{$j$}}

\put(11,5){\vector(1,0){80}}
\put(91,-5){\vector(-1,0){80}}
\put(50,-7){\makebox(0,0)[t]{$\by\l{s\neq r}
\\inv(s)\ey$}}
\put(50,10){\makebox(0,0)[b]{$In\,?\,x$}}

\put(0,-45){\vector(0,1){33}}
\put(10,-45){\oval(20,20)[lb]}
\put(90,-45){\oval(20,20)[br]}
\put(100,-45){\line(0,1){35}}
\put(90,-55){\line(-1,0){80}}

\put(109,5){\line(1,0){41}}
\put(150,-5){\vector(-1,0){41}}
\put(150,0){\oval(10,10)[r]}

\put(95,9){\line(0,1){41}}
\put(105,50){\vector(0,-1){41}}
\put(100,50){\oval(10,10)[t]}

\put(92,45){\makebox(0,0)[r]{$
\l{s\neq r}
$}}

\put(101,-50){\makebox(0,0)[l]{$
\by \l{s = r}\\
Out\,!\,x\\
inv(s)\\
inv(r)
\ey$}}

\put(126,30){\makebox(0,0)[l]{$
\by\l{s=r}\\
Out\,!\,x\\
inv(r)
\ey$}}

\end{picture}
\\
\vspace{22mm}
\ey
\ee } 

The statement 
$$\re{gjkdflsgjeeetyy} \approx 
\re{zfrgtfgr3zxgfsdgsdfgrhgds}$$
can be proven, for example, with use
of  theorem \ref{mueqyuivr}, 
defining the function 
$\mu$ of the form
$$\mu:\{1,2\}\times\{i,j\}\to Fm$$
as follows: 
$$\left\{\by
\mu(1,i)\eam (s=r)\\
\mu(2,i)\eam \bot \\
\mu(1,j)\eam (s\neq r)\\
\mu(2,j)\eam (s=r)
\ey\right.$$

\section{Two-way alternating bit protocol}
\label{fgdfhk56768899}

The above protocols 
implement a data transmission 
(i.e. a transmission of frames 
with packets from a NA)
only in one direction.

In most situations, a data transmission 
must be implemented in both directions, 
i.e. each agent, which communicates 
with a channel, must act as a sender and 
as a receiver simultaneously.

Protocols which implement
a data transmission in both directions, 
are called {\bf duplex} protocols, or
protocols of two-way transmission.

In protocols of two-way transmission
a sending of confirmations can be combined 
with a sending of data frames
(i.e. frames which contain packets from a NA):
if an agent $B$ has successfully received 
a data frame $f$ from an agent $A$, 
then he may send a confirmation 
of receipt of the frame $f$ not separately, 
but as part of his data frame.

In this section we consider 
the simplest correct protocol 
of two-way transmission. 

This protocol \bi\i is a generalization 
of ABP (which is considered in section
\ref{sfvgsadgewgtwe}), and \i is denoted
as ABP-2.\ei

ABP-2 also involves two agents, 
but behavior of each agent is described 
by the same process, 
which combines the processes 
$Sender$ and $Receiver$
from ABP. 

Each frame $f$, which is sent by any 
of these agents, contains 
\bi
\i a packet $x$, and 
\i two boolean values: 
   $s$ and $r$, where 
\bi 
\i $s$ has the same meaning 
   as in ABP: 
   this is a boolean value 
    associated with the packet $x$, and
\i $r$ is a boolean value 
    associated with a packet in 
    the last received undistorted frame. 
\ei \ei
To build a frame,  the encoding 
function $\varphi$ is used. 

To extract a packet and boolean values
$s$ and $r$ from a frame the functions 
{\it info}, {\it seq} and {\it ack}
are used. These functions have the 
following properties:
$$\begin{array}{rllll}
{\it info}(\varphi (x,s,r))&=&x\\
{\it seq}(\varphi (x,s,r))&=&s\\
{\it ack}(\varphi (x,s,r))&=&r\ey$$

Also, agents use the inverting function 
{\it inv} to invert values 
of the boolean variables. 

Each sending/receiving agent
is associated with a timer.
A behavior of the timer 
is described by the process $Timer$, 
which is represented by the diagram
\re{fghsdjklfeeyyerr}. 

A flow graph of the protocol is as follows: 

{\small
\be{podffkgnsdw}\by 
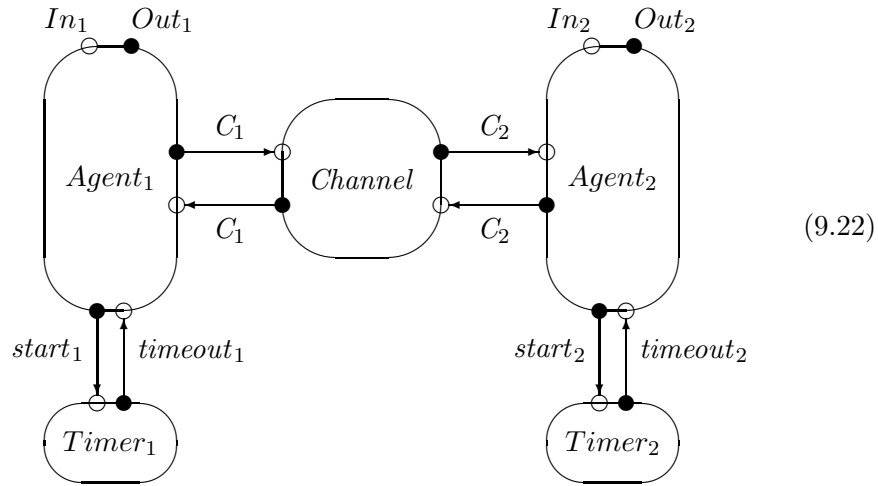
\begin{picture}(0,120)


\put(-30,60){\circle{6}}
\put(30,60){\circle*{6}}
\put(-30,40){\circle*{6}}
\put(30,40){\circle{6}}
\put(-70,60){\circle*{6}}
\put(70,60){\circle{6}}
\put(-70,40){\circle{6}}
\put(70,40){\circle*{6}}

\put(-103,100){\circle{6}}
\put(-87,100){\circle*{6}}

\put(87,100){\circle{6}}
\put(103,100){\circle*{6}}

\put(-100,-35){\circle{6}}
\put(-90,-35){\circle*{6}}

\put(-100,0){\circle*{6}}
\put(-90,0){\circle{6}}

\put(-100,-3){\vector(0,-1){29}}
\put(-90,-32){\vector(0,1){29}}

\put(-105,-15){\makebox(0,0)[r]{${\it start}_1$}}
\put(-85,-15){\makebox(0,0)[l]{${\it timeout}_1$}}

\put(90,-35){\circle{6}}
\put(100,-35){\circle*{6}}

\put(90,0){\circle*{6}}
\put(100,0){\circle{6}}

\put(90,-3){\vector(0,-1){29}}
\put(100,-32){\vector(0,1){29}}

\put(85,-15){\makebox(0,0)[r]{${\it start}_2$}}
\put(105,-15){\makebox(0,0)[l]{${\it timeout}_2$}}

\put(-50,65){\makebox(0,0)[b]{${\it C}_1$}}
\put(-50,35){\makebox(0,0)[t]{${\it C}_1$}}

\put(50,65){\makebox(0,0)[b]{${\it C}_2$}}
\put(50,35){\makebox(0,0)[t]{${\it C}_2$}}

\put(0,50){\oval(60,60)}
\put(0,50){\makebox(0,0){${\it Channel}$}}

\put(-95,50){\oval(50,100)}
\put(-95,50){\makebox(0,0){$Agent_1$}}

\put(-95,-50){\oval(50,30)}
\put(-95,-50){\makebox(0,0){$Timer_1$}}

\put(95,-50){\oval(50,30)}
\put(95,-50){\makebox(0,0){$Timer_2$}}

\put(95,50){\oval(50,100)}
\put(95,50){\makebox(0,0){$Agent_2$}}

\put(-67,60){\vector(1,0){34}}
\put(-33,40){\vector(-1,0){34}}
\put(33,60){\vector(1,0){34}}
\put(67,40){\vector(-1,0){34}}
\put(-103,110){\makebox(0,0)[r]{$In_1$}}
\put(-87,110){\makebox(0,0)[l]{$Out_1$}}

\put(87,110){\makebox(0,0)[r]{$In_2$}}
\put(103,110){\makebox(0,0)[l]{$Out_2$}}

\end{picture}
\\
\vspace{14mm}\ey
\ee
 } 
$$\;$$

The process describing the behavior 
of sending/receiving agents, 
is represented by the following flowchart: 


{\small
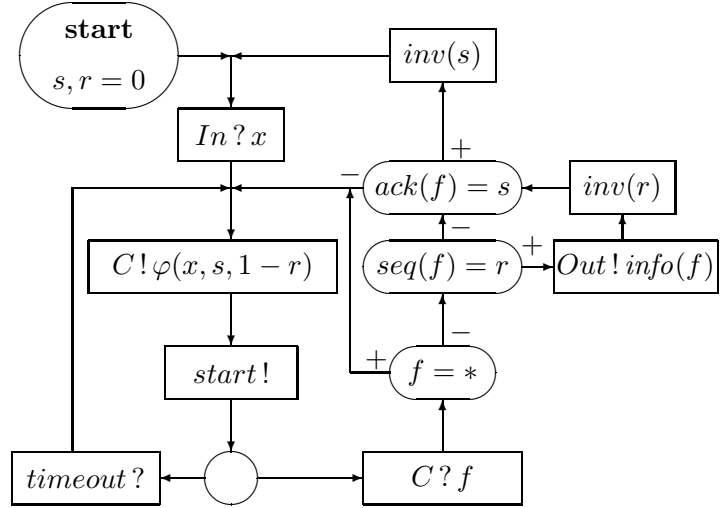
$$\begin{picture}(104,200)

\put(-50,180){\oval(60,40)}
\put(-50,190){\makebox(0,0){${\bf start}$}}
\put(-50,170){\makebox(0,0)[c]{$s,r=0$}}

\put(-20,180){\vector(1,0){20}}

\put(0,180){\vector(0,-1){20}}

\put(0,150){\pu{20}{10}}
\put(0,150){\makebox(0,0)[c]{$In\,?\,x$}}

\put(0,140){\vector(0,-1){30}}

\put(-7,100){\pu{47}{10}}
\put(-7,100){\makebox(0,0)[c]{$C\,!\,\varphi (x,s,1-r)$}}

\put(0,90){\vector(0,-1){20}}

\put(0,60){\pu{25}{10}}
\put(0,60){\makebox(0,0)[c]{$start\,!
$}}

\put(0,50){\vector(0,-1){20}}

\put(-55,20){\pu{28}{10}}
\put(-55,20){\makebox(0,0)[c]{$timeout\,?$}}

\put(80,20){\pu{30}{10}}
\put(80,20){\makebox(0,0)[c]{$C\,?\,f$}}

\put(0,20){\oval(20,20)}
\put(-10,20){\vector(-1,0){17}}
\put(10,20){\vector(1,0){40}}

\put(80,180){\pu{20}{10}}
\put(80,180){\makebox(0,0)[c]{$inv(s)$}}
\put(60,180){\vector(-1,0){60}}

\put(80,100){\oval(60,20)}
\put(80,100){\makebox(0,0)[c]{$ seq (f)=r$}}
\put(80,110){\vector(0,1){10}}
\put(110,100){\vector(1,0){12}}

\put(80,130){\oval(60,20)}
\put(80,130){\makebox(0,0)[c]{$ ack (f)=s$}}

\put(50,130){\vector(-1,0){50}}

\put(80,60){\oval(40,20)}
\put(80,60){\makebox(0,0)[c]{$f=*$}}
\put(60,60){\line(-1,0){15}}
\put(45,60){\vector(0,1){70}}

\put(-60,130){\vector(1,0){60}}
\put(-60,130){\line(0,-1){100}}

\put(80,30){\vector(0,1){20}}
\put(80,70){\vector(0,1){20}}
\put(80,140){\vector(0,1){30}}

\put(148,110){\vector(0,1){10}}

\put(128,130){\vector(-1,0){18}}


\put(153,100){\pu{31}{10}}
\put(153,100){\makebox(0,0)[c]{$Out\,!\, {\it info} (f)$}}

\put(148,130){\pu{20}{10}}
\put(148,130){\makebox(0,0)[c]{$inv(r)$}}

\put(55,65){\makebox(0,0)[c]{$+$}}
\put(45,135){\makebox(0,0)[c]{$-$}}
\put(114,108){\makebox(0,0)[c]{$+$}}

\put(87,145){\makebox(0,0)[c]{$+$}}
\put(87,115){\makebox(0,0)[c]{$-$}}
\put(87,75){\makebox(0,0)[c]{$-$}}

\end{picture}
$$
 } 

This flowchart shows that 
the agent sends a frame with its next packet
only after receiving a confirmation 
of receiving of its current packet. 

The flowchart describing the behavior 
of a specific agent (i.e. $Agent_1$ 
or $Agent_2$), 
is obtained from this flowchart 
by assigning the corresponding 
index ($1$ or $2$) to the variables 
and names, included in this flowchart. 

The behavior of the channel is described 
by the process 
$$\re{1dvfsdvgsdsdg}\,[\,C_1/S,\, C_2/R\,]$$

The reader is requested 
\bi 
\i to define the process $Spec$, 
   which is a specification of this protocol, 
    and 
\i to prove that this protocol meets the 
   specification $Spec$.
\ei

\section{Two-way sliding window protocols} 

ABP-2 is practically 
acceptable only when a duration 
of a frame transmission 
through the channel is negligible. 

If a duration of a frame transmission 
through the channel is large, 
then it is better to use 
a {\bf conveyor transmission}, 
in which the sender may send 
several frames in a row, 
without waiting their confirmation. 

Below we consider two protocols 
of two-way conveyor transmission, 
called {\bf sliding window protocols (SWPs)}.

These protocols are extensions
of ABP-2. They
\bi 
\i also involve two sending/receiving
   agents, 
   and behavior of each of these agent
   is described by the same process, 
   combining functions 
   of a sender and a receiver

\i an analog of a boolean value 
   associated with each frame 
   is an element of the set 
   $${\bf Z}_n=\{0,\ldots,n-1\}$$
   where $n$ is a fixed integer 
   of the form $2^k$. 
\ei 

An element of the set ${\bf Z}_n$, 
associated with a frame, 
is called a {\bf number} of this frame.

\subsection{The sliding window protocol 
using go back $n$}
\label{fkivy4499}

The first SWP
is called {\bf SWP using go back $n$}.

The process which describes
a behavior of a sending/receiving agent
of this protocol, 
has the array $x[n]$ among its variables.
Components of this array 
may contain packets which are sent, 
but not yet confirmed. 

A set of components of the array $x$, 
which contain such packets 
at the current time, is called a {\bf window}. 

Three variables of the process
are related to the window: 
\bi 
\i $b$ (a lower bound of the  window) 
\i $s$ (an upper bound of the window), and 
\i $w$ (a number of packets in the window). 
\ei 
Values of the variables 
$b$, $s$ and $w$ belong 
to the set ${\bf Z}_n$.

At the initial time 
\bi
\i the window is empty, and 
\i values of the variables $b$, $s$ and $w$ 
   are equal to 0. 
\ei

Adding a new packet to the window 
is performed by execution of 
the following actions: 
\bi 
\i this packet is written 
   in the component 
   $x[s]$,
   and it is assumed that 
   the number $s$ is associated 
   with this packet 
\i upper bound of the window $s$ 
   increases by 1 modulo $n$, 
   i.e. new value of $s$ is assumed to be 
   \bi\i $s+1$, if $s<n-1$, and\i
   $0$, if $s=n-1$,\ei
   and
\i $w$ (the number of packets 
   in the window)  is increased by 1. 
\ei 
Removing a packet from the window 
is performed by execution of 
the following operations: 
\bi 
\i $b$  (the lower bound of the window)
   is increased by 1 modulo $n$, and 
\i $w$ (the number of packets 
   in the window) is decreased by 1
\ei 
i.e. it is removed a packet 
whose number is equal to 
the lower bound of the window. 

To simplify an understanding 
of the operations with a window 
you can use the following 
figurative analogy: 
\bi 
\i the set of components of the array 
   $x$ can be regarded as a ring \\
   (i.e. after the component $x[n-1]$ 
   is the component $x[0]$) 
\i at each time the window is a
   connected subset of this ring, 
\i during the execution of the 
   process this window is moved 
   on this ring in the same direction. 
\ei 

If the window size reaches its maximum 
value $(n-1)$, 
then the agent does not accept 
new packets from his NA until 
the window size is not reduced. 

An ability to receive a new packet 
is defined by the boolean variable $enable$: 
\bi
\i if the value is 1, then the agent 
   can receive new packets from his   NA, and 
\i if 0, then he can not do receive 
   new packets. 
\ei

If the agent receives 
an acknowledgment 
of a packet whose number is equal
to the lower bound of the window, 
then this packet is removed from the window. 

Each component $x[i]$ of the array $x$ is 
associated with a timer, which determines
a duration of waiting of confirmation 
from another agent
of a receiving of the packet 
contained in the component $x[i]$. 
The combination of these timers 
is considered as one process $Timers$, 
which has an array of $t\,[n]$ of boolean 
variables. 
This process is defined as follows: 

$Init=(t=(0,\ldots,0))$

{\small
{\def\arraystretch{1}
\be{fgdfhjtyjktjuiop}\by
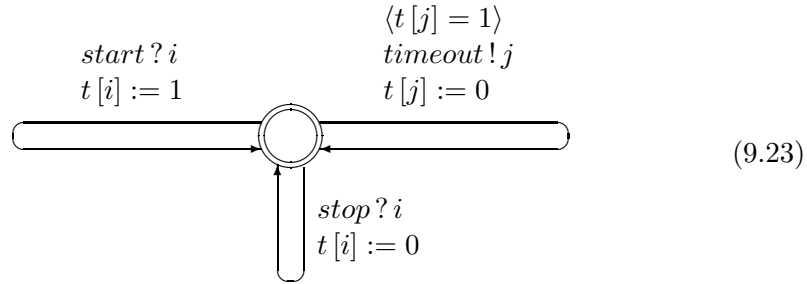
\begin{picture}(0,90)

\put(0,50){\oval(20,20)}
\put(0,50){\oval(24,24)}

\put(0,0){\oval(10,10)[b]}
\put(5,-1){\line(0,1){39}}
\put(-5,0){\vector(0,1){39}}

\put(5,15){\makebox(0,0)[l]{$\by
stop\,?\,i\\
t\,[i]:=0
\ey $}}

\put(-100,50){\oval(10,10)[l]}
\put(-11,55){\line(-1,0){89}}
\put(-100,45){\vector(1,0){89}}

\put(-60,60){\makebox(0,0)[b]{$\by start\,?
\,i\\
t\,[i]:=1
\ey $}}

\put(100,50){\oval(10,10)[r]}
\put(11,55){\line(1,0){89}}
\put(100,45){\vector(-1,0){89}}

\put(60,60){\makebox(0,0)[b]{$\by
\l{t\,[j]=1}\\
timeout\,!\,j\\
t\,[j]:=0
\ey$}}

\end{picture}
\ey
\ee } 
}

The right arrow in this diagram 
is the abbreviation for a set 
of $n$ transitions with labels
$$\by
\l{t\,[0]=1}\\
timeout\,!\,0\\
t\,[0]:=0
\ey\qquad\ldots\qquad
\by
\l{t\,[n-1]=1}\\
timeout\,!\,(n-1)\\
t\,[n-1]:=0
\ey
$$

Note that in this process there is 
the operator  $stop\,?\,i$, 
an execution of which 
prematurely terminates 
a corresponding timer. 

The protocol has the following features
\bi 
\i If a sending/receiving agent has received 
   a signal {\it timeout} from any timer, 
   then the agent sends again all 
   packets from his window. 
\i If an agent has received 
   a confirmation of a packet, 
   then 
   all previous packets in the window
   are considered also as confirmed 
   (even if their confirmations 
   were not received). 
\ei 

Each frame $f$, which is 
sent by any of the sending/receiving 
agents of this protocol, 
contains 
\bi\i a packet $x$, 
\i a number $s$, which is associated with 
   the packet $x$\\ 
   (by definition, $s$ is also associated with
    the frame $f$)
\i a number $r$, which is a number 
associated with 
a last received undistorted frame. 
\ei 

To build a frame,  the encoding 
function $\varphi$ is used. 

To extract the components 
from the frames, the functions 
{\it info}, {\it seq} and {\it ack},
are used.
These functions have 
the following properties:
$$\begin{array}{rllll}
{\it info}(\varphi (x,s,r))&=&x\\
{\it seq}(\varphi (x,s,r))&=&s\\
{\it ack}(\varphi (x,s,r))&=&r\ey$$

The description of the process, 
representing the behavior of an agent 
of the protocol, 
we give in a flowchart form, 
which easily can be transformed 
to a flowchart.

In this description we use 
the following notations. 
\bi 
\i The symbols $\modnop{+}$ 
   and $\modnop{-}$
   denote addition and subtraction modulo $n$. 
\i The symbol $r$ denotes 
   a variable with has values at
   ${\bf Z}_n$.

   A value of $r$ is equal to a number of 
   an expected frame. 

   The agent sends to his NA a packet, 
   extracted  from such a frame $f$, 
   whose number ${\it seq}(f)$ 
   coincides with a value of the
   variable $r$. 

   If a frame $f$ is such that 
   ${\it seq}(f)\neq r$, 
   then
   \bi
   \i the packet ${\it info}(f)$
      in this frame is ignored, and
   \i it is taken into account only 
      the   component $ack(f)$.
   \ei

\i The notation $send$ is the abbreviation
   of the following group of operators: 
   $$send=\c{C\,! \,\varphi (x[s], s, 
   r\modnop{-}1)\\start\;!\;s\\
   s:=s\modnop{+}1}$$

\i The notation $$between(a,b,c)$$
   is the abbreviation of the formula 
   \be{betrreeet}
   \b{a\leq b<c}
   \vee \b{c<a\leq b} \vee 
   \b{b<c<a}\ee
\i The expression $(w<n-1)$ in  the 
   operator $$enable:=(w<n-1)$$ 
    has a value \bi\i $1$, if the inequality
   $w<n-1$ holds, and \i 0, otherwise.\ei
\ei

The process representing 
the behavior of a sending/receiveng agent
of this protocos is the following:

{\small
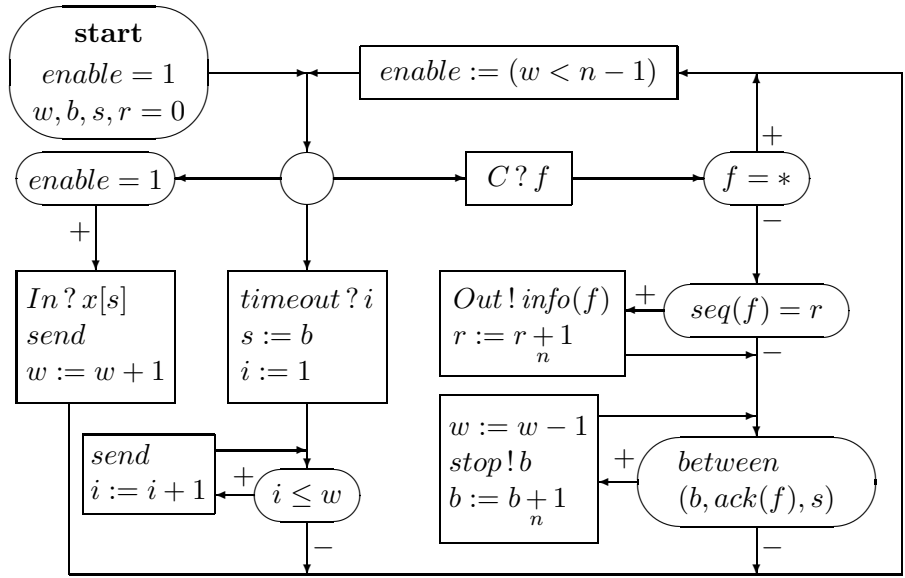
$$\by
\begin{picture}(110,190)

\put(-75,160){\oval(75,50)}
\put(-75,175){\makebox(0,0){${\bf start}$}}
\put(-75,160){\makebox(0,0)[c]{$enable=1$}}
\put(-75,145){\makebox(0,0)[c]{$w,b,s,r=0$}}

\put(0,60){\pu{30}{25}}
\put(0,60){\makebox(0,0){$\by timeout\,?\,i\\
s:=b\\i:=1\ey $}}

\put(-80,120){\oval(60,20)}
\put(-80,120){\makebox(0,0){$enable=1$}}

\put(-80,60){\pu{30}{25}}
\put(-80,60){\makebox(0,0){$\by In\,?\,x[s]
\\send\\
w:=w+1\ey $}}

\put(-60,8){\pu{25}{15}}
\put(-60,8){\makebox(0,0){$\by send\\ i:=i+1
\ey $}}

\put(0,0){\oval(40,20)}
\put(0,0){\makebox(0,0){$i\leq w$}}

\put(0,120){\oval(20,20)}

\put(85,65){\pu{35}{20}}
\put(85,65){\makebox(0,0){$\by Out\,!\, {\it info} (f)\\
r:=r\modnop{+}1
\ey $}}

\put(80,10){\pu{30}{28}}
\put(80,10){\makebox(0,0){$\by w:=w-1

\\stop\,!\,b\\b:=b\modnop{+}1
\ey $}}

\put(170,5){\oval(90,34)}
\put(170,5){\makebox(0,0){$\by between \\(b, ack (f),s)
\ey $}}

\put(170,70){\oval(70,20)}
\put(170,70){\makebox(0,0){$
 seq (f)=r
$}}

\put(170,120){\oval(40,20)}
\put(170,120){\makebox(0,0){$f=*
$}}

\put(80,120){\pu{20}{10}}
\put(80,120){\makebox(0,0){$C\,?\,f
$}}

\put(80,160){\pu{60}{10}}
\put(80,160){\makebox(0,0){$enable:=(w<n-1)
$}}

\put(-37.5,160){\vector(1,0){37.5}}

\put(0,160){\vector(0,-1){30}}
\put(0,110){\vector(0,-1){25}}
\put(0,35){\vector(0,-1){25}}
\put(0,-10){\vector(0,-1){20}}

\put(-90,-30){\line(1,0){315}}
\put(-90,-30){\line(0,1){65}}
\put(225,-30){\line(0,1){190}}
\put(225,160){\vector(-1,0){85}}
\put(20,160){\vector(-1,0){20}}
\put(-10,120){\vector(-1,0){40}}
\put(10,120){\vector(1,0){50}}
\put(100,120){\vector(1,0){50}}

\put(170,130){\vector(0,1){30}}
\put(170,110){\vector(0,-1){30}}
\put(-80,110){\vector(0,-1){25}}
\put(170,60){\vector(0,-1){38}}
\put(170,-12){\vector(0,-1){18}}

\put(125,5){\vector(-1,0){15}}

\put(110,30){\vector(1,0){60}}

\put(135,70){\vector(-1,0){15}}
\put(120,53){\vector(1,0){50}}

\put(-35,17){\vector(1,0){35}}

\put(-20,0){\vector(-1,0){15}}

\put(176,136){\makebox(0,0){$+$}}
\put(176,104){\makebox(0,0){$-$}}
\put(176,54){\makebox(0,0){$-$}}
\put(176,-20){\makebox(0,0){$-$}}
\put(6,-20){\makebox(0,0){$-$}}
\put(-86,100){\makebox(0,0){$+$}}
\put(-24,7){\makebox(0,0){$+$}}
\put(120,14){\makebox(0,0){$+$}}
\put(128,76){\makebox(0,0){$+$}}

\end{picture}\\
\vspace{10mm}
\ey
$$ } 

The reader is requested 
\bi 
\i to define a process ``channel''
   for this protocol \\
   (channel contains an ordered
   sequence of frames, 
   which may distort and disappear) 
\i to define a specification $Spec$ of 
   this protocol, and
\i to prove that the protocol 
   meets the specification $Spec$.
\ei 

In conclusion, we note that this protocol
is ineffective if a number of distortions
in the frame transmission is large. 

\subsection{The sliding window protocol 
using selective repeat} 

The second SWP differs from the 
previous one in the following: 
an agent of this protocol has two windows. 

\bn 
\i First window has the same function, 
   as a window of the first SWP
   (this window is called a 
   {\bf sending window}). 

   The maximum size of the sending window 
   is $m\eam n/2$, 
   where $n$ has the same status as 
   described in section \ref{fkivy4499} 
   (in particular, frame numbers 
   are elements of ${\bf Z}_n$).

\i Second window (called a 
   {\bf receiving window}) 
   is designed to accommodate packets 
   received from another agent, 
   which can not yet be transferred to a NA, 
   because some packets with smaller 
   numbers have not received yet. 

   A size of the receiving window is
   $m=n/2$.
\en 

Each frame $f$, 
which is sent by a sending/receiving 
agent of this protocol, has 4 components: 

\bn 
\i $k$ is a type of the frame, \\ 
   this component can have one 
   of the following three values: 
   \bi 
   \i $data $ (data frame)
   \i $ack$  (frame containing 
       only a confirmation)
   \i $nak$ (frame containing a request 
      for retransmission) \\
      (``nak'' is an abbreviation of 
     ``negative acknowledgment'')
   \ei 
\i $x$ is a packet 
\i $s$ is  a number associated with the frame 
\i $r$ is  a number associated with the 
   last received undistorted packet. 
\en 

If a type of a frame is 
$ack$ or $nak$, then 
second and third components 
of this frame are fictitious. 

To build a frame,  the encoding 
function $\varphi$ is used. 

To extract the components from 
the frames, 
the functions {\it kind}, {\it info}, 
{\it seq} and 
{\it ack} are used.
These functions have the 
following properties:
$$\begin{array}{rllll}
{\it kind}(\varphi (k,x,s,r))&=&k\\
{\it info}(\varphi (k,x,s,r))&=&x\\
{\it seq}(\varphi (k,x,s,r))&=&s\\
{\it ack}(\varphi (k,x,s,r))&=&r\ey$$

The process describing the behavior 
of a sending/receiveng
agent has the following variables. 

\bn 
\i Arrays $x[m]$ and $y[m]$,
   designed to accommodate 
   the sending window 
   and the receiving window, 
   respectively. 
\i Variables  $enable,b,s,w$,
   having
   \bi 
   \i the same sets of values, and 
   \i the same meaning 
   \ei 
   as they have in the previous protocol. 
\i Variables $r, u$, values of which 
   \bi
   \i belong to  ${\bf Z}_n$, and
   \i are equal to 
   lower and upper bounds respectively
   of the receiving window.
   \ei

   If these is a packet in 
   the receiving window, 
   a number of which 
   is equal to the lower boundary 
   receiving window (i.e. $r$), 
   then the agent 
   \bi 
   \i transmits this packet to his NA, and 
   \i increases by 1 (modulo $n$) 
       values of $r$ and $u$. 
   \ei 

\i Boolean array $$arrived[m]$$
   whose components 
   have the following    meaning: 
   $arrived[i]=1$ 
   if and only if 
   an $i$--th component 
   of the receiving window    contains a packet which is 
   not yet transmitted to the NA. 
\i Boolean variable $no\_nak$,  which is   
   used with the following purpose. 

   If the agent receives 
   \bi 
   \i a distorted frame, or 
   \i a frame, which has a number
      different from 
        the lower boundary of the receiving 
       window (i.e. $r$)
   \ei 
   then 
   he sends to his colleague 
   a request for retransmission 
   of a frame whose number is $r$. 

   This request is called a 
   {\bf Negative Acknowledgement (NAK)}.

   The boolean variable $no\_nak$ is used 
   to avoid multiple requests 
   for a retransmission of the same frame: 
   This variable is set to 1, 
   if NAK for a frame 
   with the number $r$ 
   has not yet been sent. 
\en

When a sending/receiveng agent 
gets an undistorted frame $f$ of the type 
$data$, it performs the following actions. 
\bi 
\i If the number ${\it seq}(f)$ falls into    
    the receiving window, 
   i.e. the following statement holds:
   $$between(r,{\it seq}(f),u)$$
   where the predicate symbol $between$ 
   has the same meaning as
   in the previous protocol (see \re{betrreeet}), 
   then the agent 
   \bi 
   \i extracts a packet from this frame, and 
   \i puts the packet in its receiving window. 
   \ei 
\i If the condition from the previous item 
   does not satisfied 
   (i.e. the number ${\it seq}(f)$ of the
    frame $f$ does not 
    fall into the receiving window) 
     then \bi \i a packet in this frame 
    is ignored, and     \i only the component $ack(f)$
   of this frame is taken into account. \ei 
\ei

The following timers are used by 
the sending/receiving agent. 
\bn 
\i An array of $m$ timers,  
   whose behavior is described by
   the process $Timers$ 
   (see \re{fgdfhjtyjktjuiop}, 
   with the replacement of $n$ on $m$).

   Each timer from this array 
   is intended to alert the 
   sending/receiving agent 
   that 
   \bi 
   \i a waiting of a confirmation of a 
   packet from the sending window 
   with the corresponding number
   is over, and 
   \i it is necessary to
   send a frame with
   this packet again \ei 
\i Additional timer, whose behavior 
   is described by the following process: 

   $Init=(t=0)$

{\small
{\def\arraystretch{1}
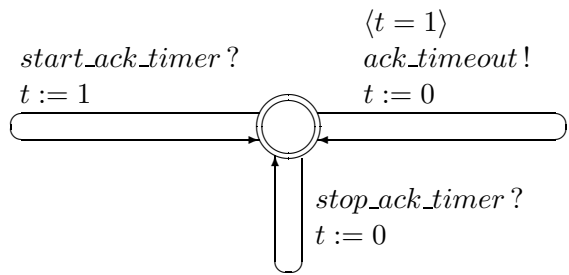
$$\by
\begin{picture}(0,90)

\put(0,50){\oval(20,20)}
\put(0,50){\oval(24,24)}

\put(0,0){\oval(10,10)[b]}
\put(5,-1){\line(0,1){39}}
\put(-5,0){\vector(0,1){39}}

\put(5,15){\makebox(0,0)[l]{$\by
stop\_ack\_timer\,?\\
t:=0
\ey $}}

\put(-100,50){\oval(10,10)[l]}
\put(-11,55){\line(-1,0){89}}
\put(-100,45){\vector(1,0){89}}

\put(-60,55){\makebox(0,0)[b]{$\by 
start\_ack\_timer\,?
\\
t:=1
\ey $}}

\put(100,50){\oval(10,10)[r]}
\put(11,55){\line(1,0){89}}
\put(100,45){\vector(-1,0){89}}

\put(60,55){\makebox(0,0)[b]{$\by
\l{t=1}\\
ack\_timeout\,!\\
t:=0
\ey$}}

\end{picture}
\ey
$$ } 
}

This timer is used with the following purpose. 

A sending by an agent of confirmations of 
frames received from another agent 
can be done as follows: 
the confirmation is sent 
\bn 
\i as a part of a data frame, or 
\i as a special frame of the type $ack$. 
\en 

When the agent should send 
a confirmation {\bf conf}, he 

\bi 
\i starts the auxiliary timer 
   (i.e. executes the action 
   $start\_ack\_timer\,!$), 
\i if the agent has received a new packet
   from his NA
   before a receiving of the signal {\it timeout}
   from the auxiliary timer, then the agent
   \bi 
   \i builds a frame of the type $data$, 
      with consists of 
      \bi
      \i this packet, and
      \i the confirmation {\bf conf} 
         as the component  $ack$
      \ei
   \i sends this frame to the colleague 
   \ei 
\i if after an expiration 
   of the auxiliary timer 
   (i.e., after receiving 
   the signal $ack\_timeout$)
   the agent has not yet received 
   a new packet from his NA, then
   he sends the confirmation {\bf conf}  by 
   a separate frame of the type $ack$. 
\ei

\en

The description of the process, 
representing the behavior of an agent 
of the protocol, 
we give in a flowchart form, 
which easily can be transformed 
to a flowchart.

In this description we use 
the following notations and 
agreements. 
\bn 
\i If $i$ is an integer, 
   then the notation $i\%m$ denotes a
   remainder of the division of $i$ on $m$. 
\i If 
   \bi 
   \i $mass$ is a name of an array 
      of $m$ components 
   (i.e. $x$, $y$, $arrived$, etc.) 
   and 
 \i $i$ is an integer \ei then the 
    notation $mass[i]$ denotes 
    the element $mass[i\%m]$. 
\i A notation of the form
   $send(kind,i)$ is the abbreviation of
   the following group of operators: 
   $$send(kind,i)=\c{C\,!\,\varphi
	(kind, x[i], i, r\modnop{-}1)\\
   {\bf if}\;\;(kind= nak)\;\:{\bf then}
\;\;no\_nak:=0\\
   {\bf if}\;\;(kind=data)\;\;{\bf then}
\;\;start\,!\,(i\%m)\\
   stop\_ack\_timer\,!}$$

\i The notation 
   $between (a, b, c)$
   has the same meaning 
   as in the previous protocol. 
\i If any oval contains several formulas, then
   we assume that these formulas 
   are connected by the conjunction ($\wedge$). 
\i In order to save a space, 
    some expressions of the form 
   $$f(e_1,\ldots, e_n)$$
   are written in two lines
   ($f$ in the first line, and the list
   $(e_1,\ldots, e_n)$ in the second line)
\en 

The process which represents 
a behavior of an agent of this protocol, 
has the following form: 

{\small
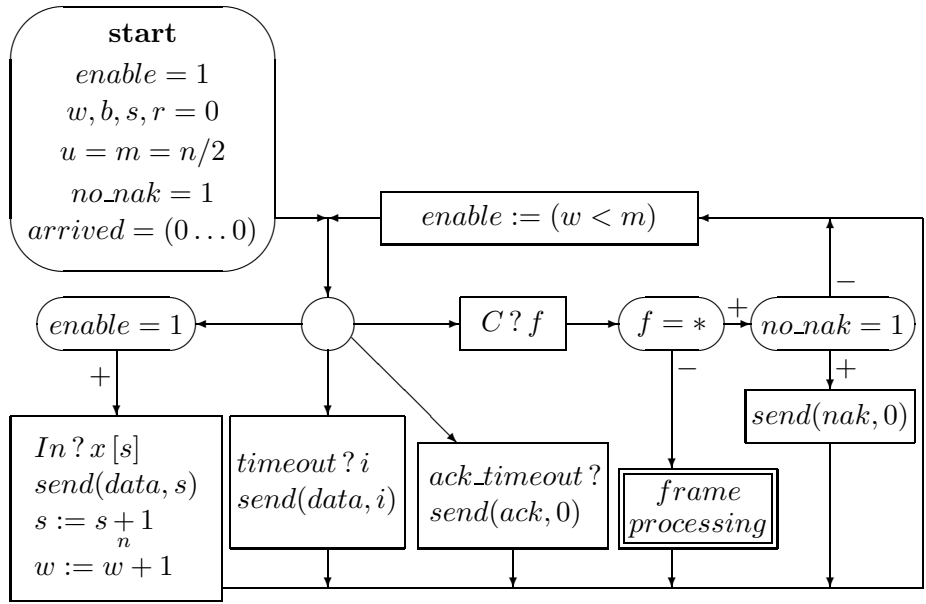
$$
\begin{picture}(108,220)

\put(-70,170){\oval(100,100)}
\put(-70,210){\makebox(0,0){${\bf start}$}}
\put(-70,195){\makebox(0,0)[c]{$enable=1$}}
\put(-70,180){\makebox(0,0)[c]{$w,b,s,r=0$}}
\put(-70,165){\makebox(0,0)[c]{$u=m=n/2$}}
\put(-70,150){\makebox(0,0)[c]{$no\_nak=1$}}
\put(-70,135){\makebox(0,0)[c]{$arrived=(0\ldots 0)$}}

\put(-4,40){\pu{33}{25}}
\put(-4,40){\makebox(0,0){$\by timeout\,?\,i\\
send(data,i)
\ey$}}

\put(0,15){\vector(0,-1){15}}

\put(-80,100){\oval(60,20)}
\put(-80,100){\makebox(0,0){$enable=1$}}

\put(-80,30){\pu{40}{35}}
\put(-80,30){\makebox(0,0){$\by In\,?\,x\,[s]\\
send(data, s)\\
s:=s\modnop{+}1\\
w:=w+1
\ey $}}

\put(0,100){\oval(20,20)}

\put(70,35){\pu{36}{20}}
\put(71,35){\makebox(0,0){$\by ack\_timeout\,?\\
send(ack,0)
\ey $}}

\put(9,95){\vector(1,-1){40}}
\put(70,15){\vector(0,-1){15}}

\put(130,100){\oval(40,20)}
\put(130,100){\makebox(0,0){$f=*
$}}

\put(130,90){\vector(0,-1){45}}

\put(140,30){\pu{30}{15}}
\put(140,30){\pu{28}{13}}
\put(140,36){\makebox(0,0){$frame$}}
\put(140,24){\makebox(0,0){$processing$}}

\put(130,15){\vector(0,-1){15}}

\put(191,100){\oval(62,20)}
\put(191,100){\makebox(0,0){$no\_nak=1$}}

\put(190,65){\pu{32}{10}}
\put(190,65){\makebox(0,0){$send(nak,0)
$}}

\put(150,100){\vector(1,0){10}}
\put(190,110){\vector(0,1){30}}

\put(190,55){\vector(0,-1){55}}

\put(70,100){\pu{20}{10}}
\put(70,100){\makebox(0,0){$C\,?\,f
$}}

\put(80,140){\pu{60}{10}}
\put(80,140){\makebox(0,0){$enable:=(w<m)
$}}

\put(0,140){\vector(0,-1){30}}
\put(0,90){\vector(0,-1){25}}

\put(-40,0){\line(1,0){265}}
\put(225,0){\line(0,1){140}}
\put(225,140){\vector(-1,0){85}}
\put(20,140){\vector(-1,0){20}}
\put(-10,100){\vector(-1,0){40}}

\put(-20,140){\vector(1,0){20}}

\put(10,100){\vector(1,0){40}}
\put(90,100){\vector(1,0){20}}

\put(190,90){\vector(0,-1){15}}
\put(-80,90){\vector(0,-1){25}}

\put(-86,80){\makebox(0,0){$+$}}

\put(155,106){\makebox(0,0){$+$}}
\put(196,116){\makebox(0,0){$-$}}
\put(136,83){\makebox(0,0){$-$}}
\put(196,83){\makebox(0,0){$+$}}

\end{picture}
$$ }

The fragment {\it frame processing}
in this diagram has the following form. 

{\small
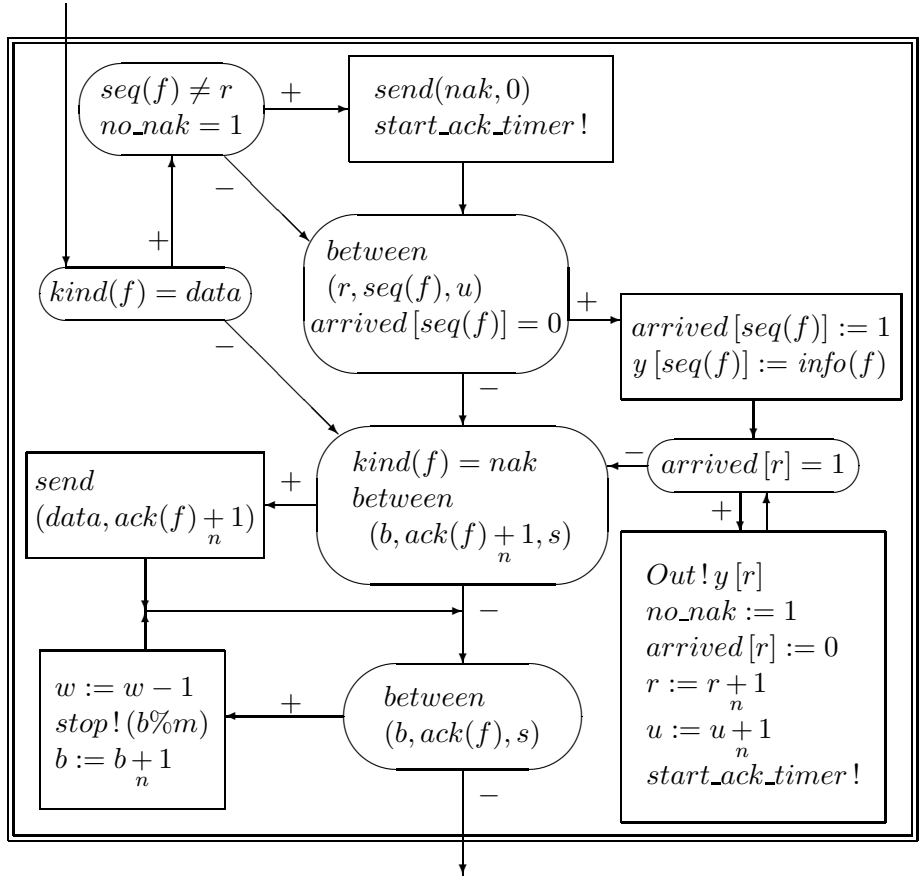
$$\by
\begin{picture}(0,280)

\put(0,125){\pu{170}{150}}
\put(0,125){\pu{172}{152}}

\put(0,70){\vector(0,-1){30}}
\put(0,0){\vector(0,-1){40}}
\put(-55,100){\vector(-1,0){20}}
\put(-45,20){\vector(-1,0){45}}
\put(-120,60){\vector(1,0){120}}
\put(-120,80){\vector(0,-1){20}}
\put(-120,45){\vector(0,1){15}}
\put(110,140){\vector(0,-1){15}}
\put(105,105){\vector(0,-1){15}}
\put(115,90){\vector(0,1){15}}
\put(70,115){\vector(-1,0){15}}
\put(-110,190){\vector(0,1){42}}
\put(-150,290){\vector(0,-1){100}}
\put(-90,170){\vector(1,-1){43}}
\put(-90,232){\vector(1,-1){32}}
\put(-75,250){\vector(1,0){32}}
\put(0,230){\vector(0,-1){20}}
\put(0,150){\vector(0,-1){20}}
\put(40,170){\vector(1,0){20}}

\put(-115,200){\makebox(0,0){$+$}}
\put(-90,160){\makebox(0,0){$-$}}
\put(-65,256){\makebox(0,0){$+$}}
\put(-90,220){\makebox(0,0){$-$}}
\put(47,177){\makebox(0,0){$+$}}
\put(10,144){\makebox(0,0){$-$}}
\put(65,120){\makebox(0,0){$-$}}
\put(98,98){\makebox(0,0){$+$}}
\put(-65,109){\makebox(0,0){$+$}}
\put(10,60){\makebox(0,0){$-$}}
\put(-65,25){\makebox(0,0){$+$}}
\put(10,-10){\makebox(0,0){$-$}}

\put(-120,180){\oval(80,20)}
\put(-120,180){\makebox(0,0){$ kind (f)=data
$}}

\put(0,100){\oval(110,60)}
\put(0,100){\makebox(0,0){$\by kind (f)=nak\\
 between \\
\mbox{$\;$}\;(b, ack (f)\modnop{+}1, s)\ey$}}

\put(0,20){\oval(90,40)}
\put(0,20){\makebox(0,0){$\by between \\(b, ack (f), s)\ey$}}

\put(-120,100){\pu{45}{20}}
\put(-120,100){\makebox(0,0){$\by send\\(data,	ack (f)\modnop{+}1)\ey$}}

\put(-125,15){\pu{35}{30}}
\put(-125,15){\makebox(0,0){$\by w:=w-1\\stop\,!\,(b\%m)\\b:=b\modnop{+}1\ey$}}

\put(-110,250){\oval(70,35)}
\put(-110,250){\makebox(0,0){$\by seq (f)\neq r\\
no\_nak=1\ey$}}

\put(7,250){\pu{50}{20}}
\put(7,250){\makebox(0,0){$\by
send(nak,0)\\start\_ack\_timer\,!
\ey$}}

\put(-10,180){\oval(100,60)}
\put(-10,182){\makebox(0,0){$\by
\mbox{$\;$}\; between \\\mbox{$\;$}\;(r, seq (f), u)\\
arrived\,[ seq (f)]=0
\ey
$}}

\put(113,160){\pu{53}{20}}
\put(113,160){\makebox(0,0){$\by
arrived\,[ seq (f)]:=1\\{\it y}\,
[ seq (f)]:= {\it info} (f)\ey
$}}

\put(110,115){\oval(80,20)}
\put(110,115){\makebox(0,0){$arrived\,[r] = 1
$}}

\put(110,35){\pu{50}{55}}
\put(110,35){\makebox(0,0){$\by
Out\,!\,y\,[r]\\
no\_nak:=1\\
arrived\,[r]:=0\\
r:=r\modnop{+}1\\
u:=u\modnop{+}1\\
start\_ack\_timer\,!
\ey$}}

\end{picture}
\\\vspace{10mm}
\ey
$$ } 

The reader is requested 
\bi 
\i to define a process ``channel''
   for this protocol \\
   (channel contains an ordered
   sequence of frames, 
   which may distort and disappear) 
\i to define a specification $Spec$ of 
   this protocol, and
\i to prove that the protocol 
   meets the specification $Spec$.
\ei 
\chapter{History and overview 
of the current state of the art}

Theory of processes 
combines several
research areas, each of which 
reflects a certain approach 
to modeling and analysis of processes. 
Below we consider the largest of these 
directions. 

\section{Robin Milner}

The largest contribution to the theory of 
processes was made by outstanding 
English mathematician and 
computer scientist {\bf Robin Milner}
(see \cite{mil1} - \cite{mil6}).
He was  born 13 January 1934 
near Plymouth, 
in the family of military officer, 
and died 20 March 2010 in Cambridge.

Since 1995 Robin Milner 
worked as a professor 
of computer science 
at University of Cambridge 
(\verb"http://www.cam.ac.uk"). 
From January 1996 to October 1999 
Milner served as a head of 
Computer Lab at University of Cambridge. 

In 1971-1973, Milner worked
in the Laboratory of Artificial 
Intelligence at Stanford University. 
From 1973 to 1995 he worked at
Computer Science Department
of University of Edinburgh 
(Scotland), where in 1986 
he founded the Laboratory 
for Foundation of Computer Science. 

From 1971 until 1980, when he worked
at Stanford and then in Edinburgh, 
he made a research
in the area of automated reasoning. 
Together with colleagues 
he developed a Logic for 
Computable Functions (LCF), 
which 
\bi
\i is a generalization of D. Scott's approach 
   to the concept of computability, and 
\i is designed for an automation 
   of formal reasoning. 
\ei
This work formed the basis 
for applied systems 
developed under the leadership of Milner. 

In 1975-1990 
Milner led the team which developed the 
Standard ML (ML is an abbreviation of 
``Meta-language''). 
ML is a widely used in industry 
and education Programming Language.
A semantics of this language 
has been fully formalized.
In the language Standard ML 
it was first implemented 
an algorithm for inference 
of polymorphic types. 
The main advantages of Standard ML are
\bi
\i an opportunity of operating 
   with logic proofs, and
\i means of an automation of 
   a construction of  logical proofs. 
\ei

Around 1980 Milner developed 
his main scientific contribution - 
a Calculus of Communicating Systems (CCS, 
see section \ref{fsvgdfsgseger}).
CCS is one of the first algebraic calculi 
for an analysis of parallel processes. 

In late 1980, together with two colleagues 
he developed a $\pi$-calculus, 
which is the main model of the behavior 
of mobile interactive systems. 

In 1988, Milner was elected a 
Fellow of the Royal Society. 
In 1991 he was awarded 
by A. M. Turing Award -- 
the highest award in the area 
of Computer Science.

The main objective of his scientific activity 
Milner himself defined as a building of 
a theory unifying the concept of a computation 
with the concept of an interaction. 

\section{A Calculus of Communicating
Systems (CCS)}
\label{fsvgdfsgseger}

A Calculus of Communicating
Systems (CCS) was first published in 1980 
in Milner's book \cite{b72}. 
The standard textbook on CCS is 
\cite{b75}. 

In \cite{b72} presented
the results of Milner's research
during the period from 1973 to 1980.

The main Milner's works on models 
of parallel processes made
at this period: 
\bi 
\i papers \cite{b67}, \cite{b68}, where
   Milner explores the denotational semantics 
   of parallel processes 
\i papers \cite{b66}, \cite{b71}, where
   in particular, 
   it is introduced the concept 
   of a flow graph with synchronized ports 
\i \cite{b69}, \cite{b70}, in these papers 
   the modern CCS was appeared. 
\ei 

The model of interaction of parallel 
processes, which is used in CCS, 
\bi
\i is based on the concept 
   of a message passing, 
   and
\i was taken from the work of Hoare \cite{b54}. 
\ei

In the paper \cite{b49} 
\bi\i a strong and observational equivalences
are studied, 
and \i it is introduced 
the logic of Hennessy-Milner. \ei

The concepts introduced in CCS
were developed in other approaches, 
the most important of them are
\bi 
\i the $\pi$-calculus (\cite{b36}, 
   \cite{b79}, \cite{b77}), and 
\i structural operational semantics (SOS), 
   this approach
   was established by G. Plotkin, 
   and published in the paper \cite{b86}. 
\ei 

More detail historical information 
about  CCS can be found in 
\cite{b87}. 

\section{Theory of communicating
sequential processes (CSP)} 

Theory of 
Communicating Sequential Processes (CSP) 
was developed by English mathematician 
and computer scientist Tony Hoare
(C.A.R. Hoare)
(b. 1934). 
This theory arose in 1976 
and was published in \cite{b54}. 
A more complete summary of CSP is contained in 
the book \cite{b56}. 

In the CSP it is investigated a 
model of communication of parallel 
processes, based on the concept 
of a message passing. 
It is considered a synchronous 
interaction between processes.

One of the key concepts of CSP is the 
concept of a guarded command, 
which is borrowed from 
Dijkstra's work \cite{b35}. 

In \cite{b55} it is considered 
a model of CSP, based on the theory of traces. 
The main disadvantage of this model 
is the lack of methods for studying 
of the deadlock property. 
This disadvantage is eliminated in the other 
model CSP (failure model), 
introduced in \cite{b29}. 

\section{Algebra of communicating processes 
(ACP)}

Jan Bergstra and 
Jan Willem Klop in 1982 introduced 
in \cite{b20} the term 
``process algebra'' 
for the first order theory with equality, 
in which the object variables take values 
in the set of processes. 
Then they have developed approaches 
led to the creation of a new direction 
in the theory of processes - the Algebra 
of Communicating Processes (ACP), 
which is contained in the papers \cite{b22}, 
\cite{b23}, \cite{b15}. 

The main object of study in the ACP 
 logical theories, function symbols of which 
correspond to operations on processes 
($a.$, $+$, etc).

In \cite{b1} a comparative analysis 
of different points of view 
on the concept of a process algebra
can be found. 

\section{Process Algebras} 

The term {\bf process algebra (PA)},
introduced by Bergstra and Klop,
is  used now in two meanings.
\bi 
\i In the first meaning, 
   the term refers to an arbitrary 
   theory of first order with equality, 
   the domain of interpretation of
   which is a set of processes.
\i In the second meaning, the term 
   denotes a large class of directions, 
   each of which is an algebraic theory, 
   which describes properties 
   of processes.

   In this meaning, 
   the term is used, for example, 
   in the title of the book 
   ``Handbook of Process Algebra'' 
   \cite{b25}.
\ei 

Below we list the most
important directions 
related to PA in both meanings 
of this term. 

\bn 
\i Handbook of PA \cite{b25}. 
\i Summary of the main results in the PA: 
   \cite{b1}. 
\i Historical overviews: \cite{b9}, \cite{hist}, 
\cite{hi3}. 
\i Different approaches related to the concept 
   of an equivalence of processes: 
   \cite{b83}, \cite{b42}, 
   \cite{b40}, \cite{b41}, \cite{b39}. 
\i PA with the semantics of partial orders: 
\cite{b27}. 
\i PA with recursion: \cite{b74}, \cite{b30}. 
\i SOS-model for the PA: \cite{b3}, \cite{b21}. 
\i Algebraic methods of verification: \cite{b46}. 
\i PA with data (actions and processes are    
parameterized by elements of the data set) 
   \bi 
   \i PA with data $\mu$-CRL 
   \i \cite{b45} (there is a software 
     tool for verification on the base 
   of presented approach). 
   \i PSF \cite{b62} (there is a software tool). 
   \i Language of formal specifications 
     LOTOS 
    \cite{b28}. 
   \ei 
\i PA with time (actions and processes 
   are parameterized    by times)
   \bi 
   \i PA with time based on CCS: \cite{b96}, 
\cite{b81}. 
   \i PA with time based on CSP: \cite{b89}. 
Textbook: \cite{b91}. 
   \i PA with time on the base of ACP: 
    \cite{b10}. 
   \i Integration of discrete and dense time 
relative and absolute time: \cite{b13}.
 
   \i Theory ATP: \cite{b82}. 
   \i Account of time in a bisimulation: 
    \cite{b14}. 
   \i Software tool UPPAAL \cite{b57}
   \i Software tool KRONOS \cite{b97} 
    (timed automata). 
   \i $\mu$-CRL with time: \cite{b93} 
(equational reasonings). 
   \ei 
\i Probabilistic PA (actions and processes are 
parameterized by probabilities). 

   These PAs are intended for 
   combined systems research, 
   which simultaneously produced 
   verification, and performance analysis. 

   \bi 
   \i Pioneering work: \cite{b47}. 
   \i Probabilistic PA, based on 
CSP: \cite{b59} 
   \i Probabilistic PA, based on 
CCS: \cite{b52}
   \i Probabilistic PA, based on 
ACP: \cite{b12}. 
   \i PA TIPP (and the associated software tool): 
\cite{b43}. 
   \i PA EMPA: \cite{b26}. 
   \i In the works \cite{b3} and \cite{b5}
 it is considered
simultaneous use of conventional and 
probabilistic 
alternative composition of processes. 
   \i In the paper \cite{b34} the concept of an 
approximation of 
probabilistic processes is considered. 
   \ei 
\i Software related to PAs 
   \bi 
   \i Concurrency Workbench \cite{b80} 
(PAs similar to CCS). 
   \i CWB-NC \cite{b98}. 
   \i CADP \cite{b37}. 
   \i CSP: FDR \verb"http://www.fsel.com/" 
   \ei 
\en 

\section{Mobile Processes} \label{mobsdfasdf} 

Mobile processes describe a behavior 
of distributed systems, which may change 
\bi 
\i a configuration of connections
   between their components, and 
\i structure of these components
\ei 
during their functioning.

Main sources: 
\bn 
\i the $\pi$-calculus (Milner and others): 
   \bi 
   \i the old handbook: \cite{b36}, 
   \i standard reference: \cite{b79}, 
   \i textbooks: \cite{b77}, \cite{pitut}, 
   \cite{san}, \cite{parrow} 
   \i page on Wikipedia: \cite{hi2} 
   \i implementation 
  of the $\pi $-calculus on a distributed 
computer system: \cite{wish}. 
   \i application of the $\pi$-calculus to 
modeling and verification 
of security protocols: 
\cite{spiag}.
   \ei 
\i The ambient calculus: \cite{b31}. 
\i Action calculus (Milner): \cite{b76}
\i Bigraphs: \cite{b78}, \cite{bigra}. 
\i Review of the literature on mobile processes: 
\cite{bibmob}. 
\i Software tool: Mobility Workbench \cite{b94}. 
\i Site \verb"www.cs.auc.dk/mobility" 
\en 

Other sources: 

\bi 
\i R. Milner's lecture 
   ``Computing in Space'' \cite{gates}, 
   which he 
   gave at the opening of the building named by B.Gates 
   built for the Computer Lab of 
   Cambridge University, May 1, 2002.

   In the lecture the concepts of 
an ``ambient'' and a ``bigraph''
are introduced. 

\i R. Milner's lecture 
   ``Turing, Computing and Communication'' 
   \cite{turing}. 
\ei 

\section{Hybrid Systems} 

A hybrid system is a system, in which 
\bi 
\i values of some variables change discretely, 
and 
\i values of other variables are changed 
continuously. 
\ei 

Modeling of a behavior of such systems 
is produced by using 
of differential and algebraic equations. 

The main 
approaches: 
\bi 
\i Hybrid Process Algebras: \cite{b24}, 
\cite{b32}, \cite{b95}. 
\i Hybrid automata: \cite{b4} \cite{b60}. 
\ei 

For simulation and verification 
of hybrid systems it is 
developed a software tool HyTech \cite{b51}. 

\section{Other mathematical theories and 
software tools, 
associated with a modeling 
and an analysis of processes} 

\bn
\i Page in Wikipedia on the theory of processes 
   \cite{hi1}. 
\i Theory of Petri nets \cite{b85}. 
\i Theory of partial orders \cite{b63}. 
\i Temporal logic and model checking \cite{b88},
 \cite{modelchecking}. 
\i Theory of traces \cite{b90}. 
\i Calculus of invariants \cite{b6}. 
\i Metric approach (which studies the concept 
   of a distance between processes):
   \cite{b16}, \cite{b17}. 
\i SCCS \cite{b73}. 
\i CIRCAL \cite{b65}. 
\i MEIJE \cite{b7}. 
\i Process algebra of Hennessy \cite{b48}. 
\i Models of processes with infinite 
   sets of states: 
   \cite{e}, \cite{abd}, \cite{mc}, 
   \cite{iss}. 
\i Synchronous interacting machines: 
   \cite{nep1}, 
   \cite{nep2}, \cite{nep3}. 
\i Asynchronous  interacting extended 
   machines: 
   \cite{nep4} - \cite{nep8}. 
\i Formal languages SDL \cite{nep9}, 
   Estelle \cite{nep10}, 
   LOTOS \cite{nep11}. 
\i The formalism of Statecharts, introduced by 
   D. Harel \cite{nep12}, \cite{nep13} and 
  used in the design of the language 
   UML. 
\i A model of communicating extended 
   timed automata CETA 
   \cite{nep14} - \cite{nep18}. 
\i A Calculus of Broadcasting Systems \cite{hi5}, 
\cite{hi6}. 
\en 

\section{Business Processes}

\bn 
\i BPEL 
   (Business process execution 
   language) 
   \cite{bpm11}. 

\i BPML (Business Process Modeling Language) 
   \cite{hi4}, \cite{bmpl}.

\i The article ``Does Better Math 
Lead to Better 
Business Processes?'' 
   \cite{bpm5}. 

\i The web-page
   ``$\pi$-calculus and Business Process 
Management'' 
   \cite{bpm6}. 

\i The paper ``Workflow is just a 
$\pi$-process'', 
   Howard Smith and Peter Fingar, October 2003 
   \cite{bpm7}. 

\i ``Third wave'' 
   in the modeling of business processes: 
\cite{bpm8}, 
   \cite{bpm8a}. 

\i The paper ``Composition of executable 
business process 
   models by combining business rules and 
process flows'' 
   \cite{bpm9}. 

\i Web services 
   choreography description language 
   \cite{bpm10}. 
\en

\end{document}